\documentclass[twoside,a4paper,12pt,openright]{book}
\usepackage{setspace}

\usepackage[numbers,sort&compress]{natbib}
        \usepackage{graphicx}
        \usepackage{amsmath}
	\usepackage{float}
\usepackage{array}
\usepackage{booktabs}
\usepackage[a4paper,inner=3cm,outer=2cm,vmargin=2cm,includeheadfoot]{geometry}
        \usepackage{fancyhdr}
        \usepackage{textcomp}
        \usepackage{ifpdf}
\usepackage{multicol}
\usepackage{bbm}
\usepackage{psfrag}
\usepackage{dcolumn}
\usepackage{color}
\usepackage{mathtools}
\usepackage{tabu}
    
\definecolor{dullpurple}{rgb}{0.431,0.188,0.534}
\definecolor{darkgreen}{rgb}{0.133,0.545,0.133}
\definecolor{dullred}{rgb}{0.706,0.208,0.192}
\RequirePackage[colorlinks=true
,urlcolor=dullpurple
,anchorcolor=blue
,citecolor=dullred
,filecolor=blue
,linkcolor=darkgreen
,menucolor=blue
,linktocpage=true
,pdfproducer=medialab
]{hyperref}        
\usepackage[noabbrev]{cleveref}

\newcommand\note[2][]{%
\if!#1!%
\stepcounter{footnote}\footnotetext{#2}%
\else%
{\renewcommand\thefootnote{#1}%
\footnotetext{#2}}%
\fi}

\usepackage{picinpar}
\usepackage{bm}
\usepackage{latexsym}
\usepackage{colortbl}
\usepackage{multirow}
\usepackage{array}
\usepackage{booktabs}
\usepackage{emptypage}
        \definecolor{dullpurple}{rgb}{0.431,0.188,0.534}
\hypersetup{urlcolor=dullpurple}
\hypersetup{citecolor=dullpurple}

\usepackage{bm}
\usepackage{amsopn}
\usepackage{latexsym}
\usepackage{colortbl}
\usepackage{multirow}
\usepackage{array}
\usepackage{booktabs}
\usepackage{fancyhdr}
\usepackage{graphicx}
\usepackage[utopia]{mathdesign}
\usepackage{microtype}
\usepackage{enumerate}       

\graphicspath{ {./figures/} }

\newcommand{\HRule}{\rule{\linewidth}{0.5mm}}

\newcommand{\sh}{{\mathcal{H}}}
\newcommand{\gc}{G_c}
\newcommand{\pmvec}{{(\pm1)}}
\newcommand{\Mp}{M_\mathrm{Pl}}
\newcommand{\Ae}{\AE}

\newcommand{\lp}{\left(}
\newcommand{\rp}{\right)}

\newcommand{\al}{\alpha}
\newcommand{\sdg}{\sqrt{-g}}
\newcommand{\sdf}{\sqrt{-f}}
\newcommand{\sdh}{\sqrt{-h}}
\newcommand{\mn}{{\mu\nu}}
\newcommand{\ab}{{\alpha\beta}}

\newcommand{\ag}{\alpha_g}
\newcommand{\af}{\alpha_f}

\newcommand{\sdgb}{\sqrt{-g_\eff}}
\newcommand{\calL}{\mathcal{L}_m}
\newcommand{\eff}{\mathrm{eff}}
\newcommand{\mc}{\mathcal{C}}

\newcommand{\creflastconjunction}{, and\nobreakspace}
\makeatletter
\DeclareRobustCommand{\rcite}[1]{%
  \rcite@aux#1,\@nil{#1}%
}
\def\rcite@aux#1,#2\@nil#3{%
  \if\relax#2\relax
    Ref.~\cite{#3}%
  \else
    Refs.~\cite{#3}%
  \fi
}
\makeatother
\allowdisplaybreaks[1]

\usepackage[avantgarde]{quotchap}

		\makeindex
		
\begin{document}
		\begin{titlepage}
			\hspace*{-0,8cm} 	   \includegraphics[width=2.5cm]{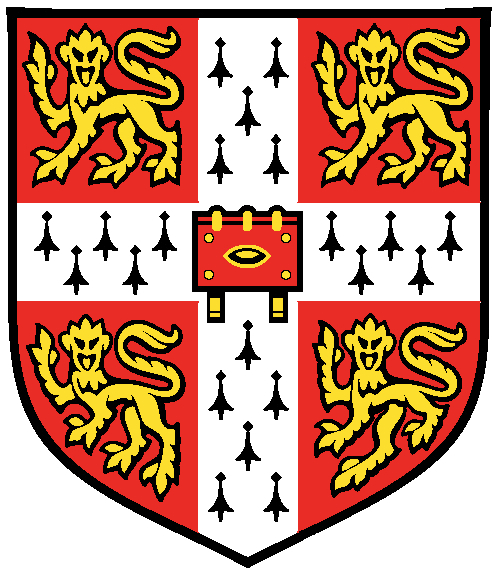}
				   \vspace*{0.1cm}

\noindent\begin{Large}University of Cambridge
\end{Large}\\
\begin{large}
Department of Applied Mathematics 

\vspace*{-0.1cm}

\noindent and Theoretical Physics
\end{large}

\vspace*{2cm}

\noindent\HRule   \HRule
		\vspace*{0,5cm}
		{\Huge \bfseries COSMOLOGY BEYOND EINSTEIN
		\\}
		\vspace*{0,3cm}

\begin{Large}
\noindent\textit{Adam Ross Solomon}
\end{Large}

\noindent\HRule \HRule
\vspace*{0.5cm}

		\begin{flushright} 
\begin{small}
\noindent This thesis is submitted to the University of Cambridge\\
		for the degree of Doctor of Philosophy
\end{small}
\end{flushright}

				\vspace*{1.3cm}

\begin{flushright} 
		
						\hspace*{-0,7cm}   \includegraphics[width=2cm]{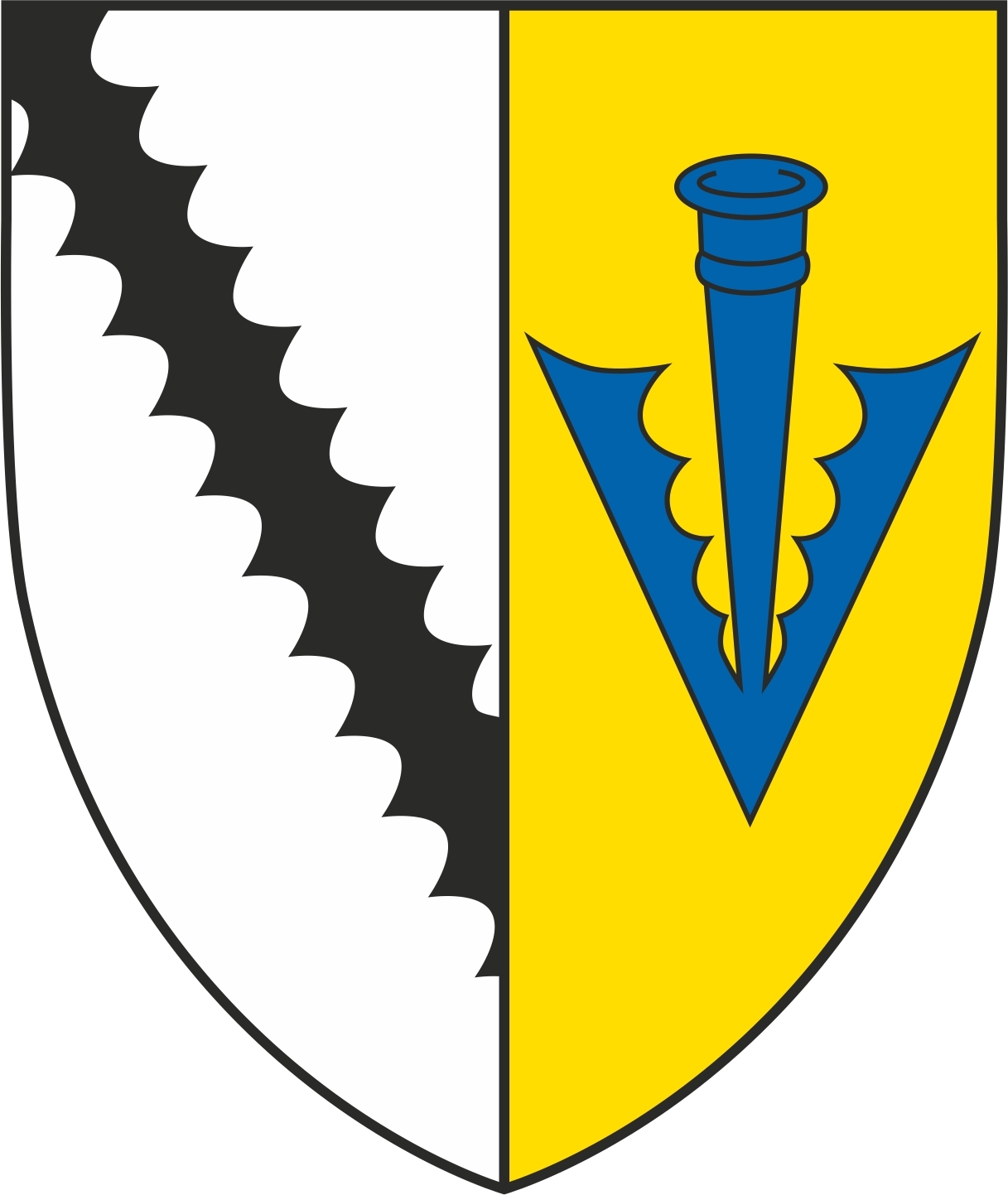}
						   
\noindent\begin{large}Sidney Sussex College
\end{large}\\
	 		\noindent\begin{normalsize} Cambridge, United Kingdom \\ April 2015
	\end{normalsize}
	\end{flushright}
	
		\end{titlepage}

	\newpage
		\pagestyle{fancy}

	\thispagestyle{empty}
	\begin{center}
	\vspace*{\fill}
	PhD Dissertation, April 2015 \\
	 \textbf{\textit{Adam Ross Solomon}}\\ Department of Applied Mathematics and Theoretical Physics\\ \& \\ Sidney Sussex College\\
	University of Cambridge, United Kingdom
	\end{center}
	
	\newpage
	\frontmatter
	\thispagestyle{empty}

 	\vspace*{5cm}
 	
\begin{centering}
 \begin{large}
  \textit{This thesis is dedicated to my parents, \\Scott and Edna Solomon, \\
 	without whose love and support\\ I could never have made it this far.}
 \end{large}
\vfill

\end{centering}

\small\normalsize

	\newpage
\thispagestyle{empty}

\newpage

\chapter*{Abstract}
\addcontentsline{toc}{chapter}{Abstract} \markboth{Abstract}{}

The accelerating expansion of the Universe poses a major challenge to our understanding of fundamental physics. One promising avenue is to modify general relativity and obtain a new description of the gravitational force. Because gravitation dominates the other forces mostly on large scales, cosmological probes provide an ideal testing ground for theories of gravity. In this thesis, we describe two complementary approaches to the problem of testing gravity using cosmology.

In the first part, we discuss the cosmological solutions of massive gravity and its generalisation to a bimetric theory. These theories describe a graviton with a small mass, and can potentially explain the late-time acceleration in a technically-natural way. We describe these self-accelerating solutions and investigate the cosmological perturbations in depth, beginning with an investigation of their linear stability, followed by the construction of a method for solving these perturbations in the quasistatic limit. This allows the predictions of stable bimetric models to be compared to observations of structure formation. Next, we discuss prospects for theories in which matter ``doubly couples'' to both metrics, and examine the cosmological expansion history in both massive gravity and bigravity with a specific double coupling which is ghost-free at low energies.

In the second and final part, we study the consequences of Lorentz violation during inflation. We consider Einstein-aether theory, in which a vector field spontaneously breaks Lorentz symmetry and couples nonminimally to the metric, and allow the vector to couple in a general way to a scalar field. Specialising to inflation, we discuss the slow-roll solutions in background and at the perturbative level. The system exhibits a severe instability which places constraints on such a vector-scalar coupling to be at least five orders of magnitude stronger than suggested by other bounds. As a result, the contribution of Lorentz violation to the inflationary dynamics can only affect the cosmic microwave background by an unobservably small amount.

\chapter*{Declaration}
\addcontentsline{toc}{chapter}{Declaration}\markboth{Declaration}{}

This dissertation is the result of my own work and includes nothing which is the outcome of work done in collaboration except where specifically indicated in the text. Nevertheless, following the tendency of modern research in theoretical physics, most of the material discussed in this dissertation is the result of research in a collaboration network. In particular, \cref{chap:bigravity-stability,chap:bigravity-subhorizon,chap:dc-finsler,chap:dc-background,chap:dc-drgt} were based on work done in collaboration with Yashar Akrami, Luca Amendola, Jonas Enander, Tomi Koivisto, Frank K\"onnig, Edvard M\"ortsell, and Mariele Motta, published in \rcite{Akrami:2014lja,Solomon:2014dua,Konnig:2014xva,Enander:2014xga,Solomon:2014iwa}, while \cref{chap:aether} is the result of work done in collaboration with John Barrow, published as \rcite{Solomon:2013iza}. I have made major contributions to the above, both in terms of results and writing.

\vspace*{0.5cm}

 \noindent I hereby declare that my thesis entitled\\
\centerline{\textit{Cosmology Beyond Einstein}}\\
is not substantially the same as any that I have submitted or is being concurrently submitted for a degree, diploma, or other qualification at the University of Cambridge or any other University or similar institution. I further state that no substantial part of my dissertation has already been submitted or is being concurrently submitted for any such degree, diploma, or other qualification at the University of Cambridge or any other University of similar institution.



	\newpage
\thispagestyle{empty}

\newpage

\tableofcontents
\listoffigures
\listoftables

	\newpage
\thispagestyle{empty}

\newpage

	\renewcommand\chapterheadstartvskip{\vspace*{-5\baselineskip}}
\mainmatter  

\clearpage

\newpage
\thispagestyle{empty}
\newpage
\clearpage

\newpage

\thispagestyle{empty}

 \vspace*{\fill}

\begin{quote}
\textit{The effort to understand the universe is one of the very few things which lifts human life a little above the level of farce and gives it some of the grace of tragedy.}
\qauthor{Steven Weinberg, \textit{The First Three Minutes}}
\end{quote}

 \vspace*{\fill}

\newpage
\thispagestyle{empty}

\newpage

\begin{savequote}[30pc]
I am always surprised when a young man tells me he wants to work at cosmology;\\I think of cosmology as something that happens to one, not something one can choose.
\qauthor{Sir William McCrea, \textit{Presidential Address, Royal Astronomical Society}}
\end{savequote}

\chapter{Introduction}
\label{chap:intro}
\hrule
\vspace*{1cm}

One of the driving aims of modern cosmology is to turn the Universe into a laboratory. By studying cosmic history at both early and late times, we have access to a range of energy scales far exceeding that which we can probe on Earth. It falls to us only to construct the experimental tools for gathering data and the theoretical tools for connecting them to fundamental physics.

The most obvious application of this principle is to the study of gravitation. Gravity is by far the weakest of the fundamental forces, yet on sufficiently large distance scales it is essentially the only relevant player; we can understand the motion of the planets or the expansion of the Universe to impressive precision without knowing the details of the electromagnetic, strong, or weak nuclear forces.\footnote{Modulo the fact that we need, as input, to know which matter gravitates, and that the quantum field theories describing these forces are essential to understanding precisely which matter we have.} As a result, we expect the history and fate of our Universe to be intimately intertwined with the correct description of gravity. For nearly a century, the consensus best theory has been Einstein's remarkably simple and elegant theory of general relativity. This consensus is not without reason: practically all experiments and observations have lent increasing support to this theory, from classical weak-field observations such as the precession of Mercury's perihelion and the bending of starlight around the Sun, to the loss of orbital energy to gravitational waves in binary pulsar systems, observations remarkable both for their precision and for their origin in the strongest gravitational fields we have ever tested.

Nevertheless, there are reasons to anticipate new gravitational physics beyond general relativity. In the ultraviolet (UV), i.e., at short distances and high energies, it is well known that general relativity is nonrenormalisable and hence cannot be extended to a quantum theory. It must be replaced at such scales by a \emph{UV-complete} theory which possesses better quantum behaviour. The focus of this thesis is on the infrared (IR), i.e., long distances and low energies. While general relativity is a theoretically-consistent IR theory, the discovery in 1998 that the expansion of the Universe is accelerating presents a problem for gravitation at the longest distances \cite{Riess:1998cb,Perlmutter:1998np}. The simplest explanation mathematically for this acceleration is a cosmological constant, which is simply a number that we can introduce into general relativity without destroying any of its attractive classical features. However, from a quantum-mechanical point of view, the cosmological constant is highly unsatisfactory. The vacuum energy of matter is expected to gravitate, and it would mimic a cosmological constant; however, the value it would generate is as much as $10^{120}$ times larger than the value we infer from observations \cite{Weinberg:1988cp,Martin:2012bt,Burgess:2013ara}. Therefore, the ``bare'' cosmological constant which appears as a free parameter in general relativity would need to somehow know about this vacuum energy, and cancel it out almost \emph{but not quite} exactly. Such a miraculous cancellation has no known explanation. Alternatively, one could imagine that the vacuum energy is somehow either rendered smaller than we expect, or does not gravitate---and theories which achieve this behaviour are known---but we would then most likely need a separate mechanism to explain what drives the current small but nonzero acceleration.

For these reasons, it behooves us to consider the possibility that general relativity may not be the final description of gravity on large scales. To put the problem in historical context, we may consider the story of two planets: Uranus and Mercury. In the first half of the nineteenth century, astronomers had mapped out the orbit of Uranus, then the farthest-known planet, to heroic precision. They found anomalies in the observed orbit when compared to the predictions made by Newtonian gravity, then the best understanding of gravitation available. Newton's theory had not yet been tested at distances larger than the orbit of Uranus: it was, for all intents and purposes, the boundary of the known universe. A natural explanation was therefore that Newtonian gravity simply broke down at such unimaginably large distances, to be replaced by a different theory. In 1846, French astronomer Urbain Le Verrier put forth an alternative proposal: that there was a new planet beyond Uranus' orbit, whose gravitational influence led to the observed discrepancies. Le Verrier predicted the location of this hitherto-unseen planet, and within weeks the planet Neptune was unveiled.

Buoyed by his success, Le Verrier turned his sights to another planet whose orbit did not quite agree with Newtonian calculations: Mercury, the closest to the Sun. As is now famous, the perihelion of Mercury's orbit precessed at a slightly faster rate than was predicted. Le Verrier postulated another new planet, Vulcan, within Mercury's orbit. However, the hypothesised planet was never found, and in the early parts of the twentieth century, Einstein demonstrated that general relativity accounted precisely for the perihelion precession. In the case of Mercury, it was a modification to the laws of gravity, rather than a new planet, which provided the solution.

We find ourselves in a similar position today. Our best theory of gravity, general relativity, combined with the matter we believe is dominant, mostly cold dark matter, predict a decelerating expansion, yet we observe something different. One possibility is that there is new matter we have not accounted for, such as a light, slowly-rolling scalar field. However, we must also consider that the theory of gravity we are using is itself in need of a tune-up.

The project of modifying gravity leads immediately to two defining questions: what does a good theory of modified gravity look like, and how can we test such theories against general relativity? This thesis aims to address both questions, although any answers we find necessarily comprise only a small slice of a deep field of research.

Einstein's theory is a paragon of elegance. It is practically inevitable that this is lost when generalising to a larger theory. Indeed, it is not easy to even define elegance once we leave the cosy confines of Einstein gravity. Consider, as an example, two equivalent definitions of general relativity, each of which can be used to justify the claim that GR is the simplest possible theory of gravity. First we can say that general relativity is the theory whose Lagrangian,
\begin{equation}
 \mathcal{L} = \sdg R, \label{eq:EH-L}
\end{equation}
is the simplest diffeomorphism-invariant Lagrangian that can be constructed out of the metric tensor and its derivatives. Alternatively, we could look at general relativity as being the unique Lorentz-invariant theory of a massless spin-2 field, or \emph{graviton} \cite{Gupta:1954zz,Weinberg:1965rz,Deser:1969wk,Boulware:1974sr,Feynman:1996kb}.

These serve equally well to tell us why general relativity is so lovely, but they diverge once we move to more general theories. Consider, for example, modifying the Lagrangian (\ref{eq:EH-L}) by promoting the Ricci scalar $R$ to a general function $f(R)$,
\begin{equation}
 \mathcal{L} = \sdg f(R). \label{eq:EH-L-fR}
\end{equation}
This is the defining feature of $f(R)$ gravity, a popular theory of modified gravity \cite{Carroll:2003wy,Sotiriou:2008rp,DeFelice:2010aj}. One can certainly make the argument that this is mathematically one of the simplest possible generalisations of general relativity. However, when considered in terms of its fundamental degrees of freedom, we find a theory in which a spin-0 or scalar field interacts in a highly nonminimal way with the graviton.

Alternatively, one can consider massive gravity, in which the massless graviton of general relativity is given a nonzero mass. While this has a simple interpretation in the particle picture, its mathematical construction is so nontrivial that over seven decades were required to finally find the right answer. The resulting action, given in \cref{eq:sdrgtalpha}, is certainly not something one would have thought to construct had it not been for the guiding particle picture.

There are additional, more practical concerns when building a new theory of gravity. General relativity agrees beautifully with tests of gravity terrestrially and in the solar system, and it is not difficult for modified gravity to break that agreement. While this may be surprising if we are modifying general relativity with terms that should only be important at the largest distance scales, it is not difficult to see that this problem is fairly generic. Any extension of general relativity involves adding new degrees of freedom (even massive gravity has three extra degrees of freedom), and in the absence of a symmetry forbidding such couplings, these will generally couple to matter, leading to gravitational-strength fifth forces. Such extra forces are highly constrained by solar-system experiments. Almost all viable theories of modified gravity therefore possess \emph{screening mechanisms}, in which the fifth force is large cosmologically but is made unobservably small in dense environments. The details of these screening mechanisms are beyond the scope of this thesis, and we refer the reader to the reviews \cite{Jain:2010ka,Joyce:2014kja}.

In parallel with these concerns, we must ask how to experimentally distinguish modified gravity from general relativity. One approach is to use precision tests in the laboratory \cite{Mota:2006ed,Mota:2006fz,Brax:2007vm,Steffen:2010ze,Brax:2010gp,Brax:2011sv,Burrage:2014oza,Jain:2013wgs}. Another is to study the effect of modified gravity on astrophysical objects such as stars and galaxies \cite{Davis:2011qf,Jain:2012tn,Sakstein:2013pda,Vikram:2014uza}. In this thesis we will be concerned with cosmological probes of modified gravity. Because screening mechanisms force these modifications to hide locally (with some exceptions), it is natural to look to cosmology, where the new physics is most relevant. Cosmological tests broadly fall into three categories: background, linear, and nonlinear. Background tests are typically geometrical in nature, and try to distinguish the expansion history of a new theory of gravity from the general relativistic prediction. Considering small perturbations around the background, we obtain predictions for structure formation at linear scales. Finally, on small scales where structure is sufficiently dense, nonlinear theory is required to make predictions, typically using N-body computer simulations.

This thesis is concerned with the construction of theoretically-sensible modified gravity theories and their cosmological tests at the level of the background expansion and linear perturbations. In the first part, we focus on massive gravity and its extension to a \emph{bimetric} theory, or massive bigravity, containing two dynamical metrics interacting with each other. In particular, we derive the cosmological perturbation equations for the case where matter couples to one of the metrics, and study the stability of linear perturbations by deriving a system of two coupled second-order evolution equations describing all perturbation growth and examining their eigenfrequencies. Doing this, we obtain conditions for the linear cosmological stability of massive bigravity, and identify a particular bimetric model which is stable at all times. We next move on to the question of observability, constructing a general framework for calculating structure formation in the quasistatic, subhorizon r\'egime, and then applying this to the stable model.

After this, we tackle the question of matter couplings in massive gravity and bigravity, investigating a pair of theories in which matter is coupled to both metrics. In the first, matter couples minimally to both metrics. We show that there is not a single effective metric describing the geometry that matter sees, and so there is a problem in defining observables. In the second theory, matter does couple to an effective metric. We first study it in the context of bigravity, deriving its cosmological background evolution equations, comparing some of the simplest models to data, and examining in depth some particularly interesting parameter choices. We next examine the cosmological implications of massive gravity with such a matter coupling. Massive gravity normally possesses a no-go theorem forbidding flat cosmological solutions, but coupling matter to both metrics has been shown to overcome this. We examine this theory in detail, finding several stumbling blocks to observationally testing the new massive cosmologies.

The remainder of this thesis examines the question of Lorentz violation in the gravitational sector. We focus on Einstein-aether theory, a vector-tensor model which spontaneously breaks Lorentz invariance. We study the coupling between the vector field, or ``aether,'' and a scalar field driving a period of slow-roll inflation. We find that such a coupling can lead to instabilities which destroy homogeneity and isotropy during inflation. Demanding the absence of these instabilities places a constraint on the size of such a coupling so that it must be at least 5 orders of magnitude smaller than the previous best constraints.

The thesis is organised as follows. In the rest of this chapter, we present background material, discussing the essential ingredients of general relativity and modern cosmology which will be important to understanding what follows. In \cref{chap:mg} we give a detailed description of the modified gravity theories discussed in this thesis, specifically massive gravity, massive bigravity, and Einstein-aether theory, focusing on their defining features and their cosmological solutions. In \cref{chap:bigravity-stability,chap:bigravity-subhorizon} we examine the cosmological perturbation theory of massive bigravity with matter coupled to one of the metrics. In \cref{chap:bigravity-stability} we study the stability of perturbations, identifying a particular bimetric model which is stable at all times, while in \cref{chap:bigravity-subhorizon} we turn to linear structure formation in the quasistatic limit and look for observational signatures of bigravity. In \cref{chap:dc-finsler,chap:dc-drgt,chap:dc-background} we examine generalisations of massive gravity and bigravity in which matter couples to both metrics. \Cref{chap:dc-finsler} focuses on the thorny problem of finding observables in one such theory. In \cref{chap:dc-background} we examine the background cosmologies of a doubly-coupled bimetric theory, and do the same for massive gravity in \cref{chap:dc-drgt}. Finally, in \cref{chap:aether} we study the consequences of coupling a slowly-rolling inflaton to a gravitational vector field, or aether, deriving the strongest bounds to date on such a coupling. We conclude in \cref{chap:conclusion} with a summary of the problems we have addressed and the work discussed, as well as an outlook on the coming years for modified gravity.

\section{Conventions}

Throughout this thesis we will use a mostly-positive (-+++) metric signature. We will denote the flat-space or Minkowski metric by $\eta_\mn$. Greek indices $\mu,\nu,\ldots=(0,1,2,3)$ represent spacetime indices, while Latin indices $i,j,\ldots=(1,2,3)$ are used for spatial indices. Latin indices starting from $a,b,c,\ldots$ are also used for field-space and local Lorentz indices. Partial derivatives are denoted by $\partial$ and covariant derivatives by $\nabla$. Commas and semicolons in indices will occasionally be used to represent partial and covariant derivatives, respectively, i.e., $\phi_{,\mu} \equiv \partial_\mu\phi$ and $\phi_{;\mu} \equiv \nabla_\mu\phi$. Symmetrisation and antisymmetrisation are denoted by
\begin{equation}
S_{(\mn)} \equiv \frac 12 \left(S_\mn + S_{\nu\mu}\right), \qquad A_{[\mn]} \equiv \frac 12 \left(A_\mn - A_{\nu\mu}\right),
\end{equation}
and similarly for higher-rank tensors.
In lieu of the gravitational constant $G$ we will frequently use the Planck mass, $\Mp^2=1/8\pi G$. Cosmic time is denoted by $t$ and its Hubble rate is $H$, while we use $\tau$ for conformal time with the Hubble rate $\sh$. For brevity we will sometimes use abbreviations for common terms, listed in \cref{tab:abrv}.

\begin{table}
\heavyrulewidth=.08em
	\lightrulewidth=.05em
	\cmidrulewidth=.03em
	\belowrulesep=.65ex
	\belowbottomsep=0pt
	\aboverulesep=.4ex
	\abovetopsep=0pt
	\cmidrulesep=\doublerulesep
	\cmidrulekern=.5em
	\defaultaddspace=.5em
	\renewcommand{\arraystretch}{1.6}
\centering
\begin{tabu}{l|[2pt] l}
Abbreviation & Expression   \\
\tabucline[2pt]{-}
BAO & baryon-acoustic oscillations  \\
CDM & cold dark matter \\
CMB & cosmic microwave background \\
FLRW & Friedmann-Lema\^{i}tre-Robertson-Walker \\
GR & general relativity \\
SNe & supernovae \\
VEV & vacuum expectation value
\end{tabu}
\caption[Abbreviations.]{Abbreviations used throughout this thesis.}
\label{tab:abrv}
\end{table}

\section{General Relativity}

This thesis deals with modified gravity. Consequently it behoves us to briefly overview the theory of gravity we will be modifying: Einstein's general relativity. The theory is defined by the Einstein-Hilbert action,
\begin{equation}
S_\mathrm{EH} = \frac{\Mp^2}{2} \int d^4x \sdg R, \label{eq:S-EH}
\end{equation}
where $R=g^\mn R_\mn$ is the Ricci scalar, with $g_\mn$ and $R_\mn$ the metric tensor and Ricci tensor, respectively. Allowing for general matter, represented symbolically by fields $\Phi_i$ with Lagrangians $\mathcal{L}_m$ determined by particle physics, the total action of general relativity is
\begin{equation}
S = S_\mathrm{EH} + \int d^4x\sdg\mathcal{L}_m\left(g, \Phi_i\right). \label{eq:S-total}
\end{equation}
Varying the action $S$ with respect to $g^\mn$ we obtain the gravitational field equation, the Einstein equation,
\begin{equation}
R_\mn - \frac{1}{2}Rg_\mn = 8\pi GT_\mn, \label{eq:einstein-intro}
\end{equation}
where the stress-energy tensor of matter is defined by
\begin{equation}
T_\mn \equiv -\frac{2}{\sdg}\frac{\delta\left(\sdg\mathcal{L}_m\right)}{g^\mn}. \label{eq:stressenergy}
\end{equation}
It is often convenient to define the Einstein tensor,
\begin{equation}
G_\mn \equiv R_\mn - \frac 1 2 Rg_\mn,
\end{equation}
which is conserved as a consequence of the Bianchi identity,
\begin{equation}
\nabla_\mu G^\mu{}_\nu = 0.
\end{equation}
Note that we are raising and lowering indices with the metric tensor, $g_\mn$. The Bianchi identity is a geometric identity, i.e., it holds independently of the gravitational field equations. The stress-energy tensor is also conserved,
\begin{equation}
\nabla_\mu T^\mu{}_\nu = 0.
\end{equation}
This is both required by particle physics and follows from the Einstein equation and the Bianchi identity, which is a good consistency check. A consequence of stress-energy conservation is that particles move on geodesics of the metric, $g_\mn$,
\begin{equation}
\ddot x^\mu + \Gamma^\mu_\ab\dot x^\alpha \dot x^\beta = 0,
\end{equation}
where $x^\mu(\lambda)$ is the position 4-vector of a test particle parametrised with respect to a parameter $\lambda$, an overdot denotes the derivative with respect to $\lambda$, and $\Gamma^\mu_\ab = \frac{1}{2}g^\mn(g_{\alpha\nu,\beta} + g_{\beta\nu,\alpha} - g_{\ab,\nu})$ are the Christoffel symbols.

Einstein's equation relates the curvature of spacetime to the distribution of matter. Freely-falling particles then follow geodesics of the metric. The combination of the Einstein and geodesic equations leads to what we call the gravitational force. John Wheeler's description of gravity's nature is perhaps the most eloquent: ``Spacetime tells matter how to move; matter tells spacetime how to curve'' \cite{Wheeler:1998vs}.

As discussed above, it seems clear to the eye that \cref{eq:S-EH} is the simplest action one can construct for the gravitational sector, if one restricts oneself to scalar curvature invariants. Indeed, the simplicity of general relativity can be phrased in two equivalent ways. Lovelock's theorem states that Einstein's equation is the only gravitational field equation which is constructed solely from the metric, is no more than second order in derivatives, is local, and is derived from an action \cite{Lovelock:1971yv}. Alternatively, as alluded to previously, the same field equations are the unique nonlinear equations of motion for a massless spin-2 particle \cite{Gupta:1954zz,Weinberg:1965rz,Deser:1969wk,Boulware:1974sr,Feynman:1996kb}. There is but one extension to the gravitational action presented which neither violates Lovelock's theorem nor introduces any extra degrees of freedom: a cosmological constant, $\Lambda$, which enters in the action as
\begin{equation}
S = \frac{\Mp^2}{2} \int d^4x \sdg \left(R-2\Lambda\right) + \int d^4x\sdg\mathcal{L}_m\left(g, \Phi_i\right).
\end{equation}

$\Lambda$ represents an infrared, or low-energy, modification to general relativity, as it only becomes important for small curvatures, $R\lesssim 2\Lambda$. Because the smallest spacetime curvatures are found at large distances and late times, the most important effect of a cosmological constant is, as the name suggests, on cosmology. As we will see in the next section, a positive cosmological constant has the predominant effect of causing the expansion of the Universe to accelerate at late times. The latest data suggest that, if a cosmological constant is responsible for the present cosmic speed-up, then it has an incredibly tiny value in Planck units, $\Lambda/\Mp^2 \sim \mathcal{O}(10^{-120})$ \cite{Ade:2013zuv}. Note that if there is a different explanation for the cosmic acceleration, such as dynamical dark energy or modified gravity, then $\Lambda$ is typically assumed to be negligible. Indeed, since it is our aim in much of this thesis to explore modified gravity as an alternative to the cosmological constant, we will set $\Lambda=0$ throughout, except when noted.\footnote{The question of \emph{why} $\Lambda$ should be zero, given that it generically receives large, $\mathcal{O}(\Mp^2)$ quantum corrections, is one of the greatest mysteries in modern physics, but is far beyond the scope of this thesis. Consequently we do not speculate about what mechanisms to remove the cosmological constant may be in play.}

\section{The Cosmological Standard Model}

Observations of the cosmic expansion history are, if we include a small cosmological constant, extremely well described \cite{Ade:2013zuv} by general relativity with a Friedmann-Lema\^{i}tre-Robertson-Walker (FLRW) ansatz for the metric,
\begin{equation}
ds^2 = -N(t)^2dt^2 + a(t)^2\left(\frac{dr^2}{1-\kappa r^2} + r^2d\Omega_2^2\right),
\end{equation}
where we have used spherical polar coordinates, $d\Omega_2^2=d\theta^2 + \sin^2\theta d\phi^2$ is the metric on a 2-sphere, and $\kappa$ parametrises the curvature of spatial sections. The FLRW metric is defined by two functions of time: the lapse, $N(t)$, and the scale factor, $a(t)$. It is a natural metric to use for cosmology as it is the most general metric consistent with spatial homogeneity and isotropy, i.e., the principle that there should be neither a preferred location nor a preferred direction in space. These principles are consistent with observations on scales larger than about 100 megaparsecs.

In general relativity, we have the freedom to choose our coordinate system, and so the lapse can be freely ``set'' to a desired functional form $f(t)$ by rescaling the time coordinate as $dt \to f(t)dt/N(t)$. Two common choices for the time coordinate are cosmic time, $N(t)=1$, and conformal time, $N(t)=a(t)$. Cosmic time is more physical, as it corresponds to the time measured by observers comoving with the cosmic expansion (such as, for example, us). Conformal time is often computationally useful, especially since in those coordinates the metric is conformally related to Minkowski space if $\kappa=0$, $g_\mn = a(t)^2\eta_\mn$; consequently, photons move on flat-space geodesics, and their motion in terms of conformal time can be computed without additionally calculating the cosmic expansion. It is worth keeping the lapse in mind because this thesis deals in large part with theories in which the time-time part of the metric (or metrics) cannot so freely be fixed to a desired value. As long as a single metric couples to matter, however (which is the case everywhere except in \cref{chap:dc-finsler}), the time coordinate defined by $d\tau = N(t)dt$, where $N(t)$ is the lapse of the metric to which matter couples, will function as cosmic time when computing observables.

The dynamical variable determining the expansion of the Universe is the scale factor, $a(t)$. Its evolution is determined by the Einstein equation. At this point we specialise to cosmic time; the conformal-time equivalents of the equations we present can be easily derived by switching the time coordinate from $t$ to $\tau$, where $d\tau = dt/a(t)$. We use as the matter source a perfect fluid with the stress-energy tensor
\begin{equation}
T^\mn = \left(\rho + p\right)u^\mu u^\nu + pg^\mn, \label{eq:perfectfluid}
\end{equation}
where $u^\mu$ is the fluid 4-velocity, $\rho$ is the energy density, and $p$ is the pressure. Then, taking the time-time component of the Einstein equation we obtain the Friedmann equation,
\begin{equation}
H^2 = \frac{8\pi G}{3}\rho - \frac{\kappa}{a^2} + \frac{\Lambda}{3},
\end{equation}
where the Hubble rate is defined by
\begin{equation}
H = \frac{\dot a}{a}.
\end{equation}
Conservation of the stress-energy tensor leads to the fluid continuity equation,
\begin{equation}
\dot\rho + 3H\left(\rho + p\right) = 0.
\end{equation}
Note that, in a Universe with multiple matter species which do not interact, this conservation equation holds both for the total density and pressure, as well as for the density and pressure of each individual component. The spatial part of the Einstein equation---or, equivalently, the trace---will lead to the acceleration equation,
\begin{equation}
\frac{\ddot a}{a} = -4\pi G\left(\rho + 3p\right) + \frac{\Lambda}{3}.
\end{equation}
This can also be derived from the Friedmann and continuity equations, hence its utility is limited for the purposes of this thesis.

In order to close the system comprising the Friedmann and continuity equations, it is typical to specify an equation of state relating the pressure to the density, $p=p(\rho)$, for each individual matter species. The simplest and most commonly-used equation of state is
\begin{equation}
p = w\rho,
\end{equation}
where $w$ is a constant. Examples of perfect fluids obeying such an equation of state include pressureless matter, or ``dust,'' with $w=0$, radiation, with $w=1/3$, and vacuum energy or a cosmological constant, with $w=-1$. Indeed, we only need these three fluids to model the Universe back to about a second after the big bang, so we will focus on them.

Let us briefly discuss some simple properties of the cosmological solutions to this system of equations. Because observations are consistent with a flat Universe, i.e., $\kappa=0$ \cite{Ade:2013zuv}, we will neglect the spatial curvature from here out. With $w=\mathrm{const.}$, the continuity equation is solved by
\begin{equation}
\rho = \rho_0 a^{-3(1+w)},
\end{equation}
where $\rho_0$ is a constant corresponding to the density when $a=1$ (usually taken to be the present day). This leads to the following behaviours for the relevant cosmic fluids:
\begin{align}
&& \rho&\sim a^{-4} && \text{radiation} && \\
&& \rho&\sim a^{-3} && \text{dust} && \\
&& \rho&= \mathrm{const.} && \text{vacuum energy, cosmological constant}. &&
\end{align}
Plugging these into the Friedmann equation, we obtain the following expansion rates during the various cosmic eras:
\begin{align}
&& a(t)&\sim t^{1/2} && \text{radiation-dominated era} && \\
&& a(t)&\sim t^{2/3} && \text{matter-dominated era} && \\
&& a(t)&\sim e^{Ht} && \text{$\Lambda$-dominated era.} &&
\end{align}
In general, a universe dominated by a $w=\mathrm{const.}\neq-1$ perfect fluid will evolve as $a(t)\sim t^{2/3(1+w)}$.

Notice that, as time goes on, the densities of radiation and matter (and any fluid with $w>-1$) will decay, while that of a cosmological constant stays the same (which is sensible, since it has a constant contribution to the Friedmann equation). Therefore, if $\Lambda>0$, there is necessarily a time after which the cosmological constant dominates the Friedmann equation, with $H\sim\mathrm{const.}$ and an exponential expansion. This makes quantitative the claim from above that a cosmological constant leads to late-time cosmic acceleration. Observations show that such a late-time acceleration is happening in our own Universe, and if it is caused by a perfect fluid then its equation of state is consistent with $w=-1$ \cite{Ade:2013zuv}. Following this, our criterion for self-acceleration in a theory of modified gravity will generally be that $H$ tends to a constant at late times.

Finally, we note that it is common to define a density parameter, $\Omega_i$, for each matter species,
\begin{equation}
\Omega_i \equiv \frac{8\pi G\rho_i}{3H^2},
\end{equation}
where the subscript $i$ indexes each matter species. In particular, we will define $\Omega_{\mathrm m}$, $\Omega_{\mathrm b}$, $\Omega_{\mathrm c}$, $\Omega_{\gamma}$, and $\Omega_{\Lambda}$ for all matter (specifically dust), baryons, cold dark matter, radiation, and a putative dark energy, respectively. In terms of the density parameter, the Friedmann equation can be written in the simple and general form
\begin{equation}
\displaystyle\sum_i \Omega_i = 1,
\end{equation}
as long as we define appropriate density parameters for the curvature and cosmological-constant terms. It is also common to parametrise the present-day density of each species in terms of the density parameter evaluated at the present, denoted by $\Omega_{i,0}$. Broadly speaking, observations suggest $\Omega_{\mathrm m,0}\sim0.3$ and $\Omega_{\Lambda,0}\sim0.7$, while all other contributions are negligibly small or vanishing \cite{Ade:2013zuv}. The fact that $\Omega_{\Lambda,0}$ is nonzero tells us that in order to match observations using general relativity, we need to introduce a ``dark energy'' component, of which the simplest example is a cosmological constant. The precise best-fit cosmological parameters from the Planck satellite are presented in \cref{tab:planckcosmo}.

\begin{table}
\centering
\begin{tabular}{|c|c|}
\hline
$\Omega_{\mathrm b,0}h^2$ & $0.022$   \\
\hline
$\Omega_{\mathrm c,0}h^2$ & $0.12$  \\
\hline
$\Omega_{\gamma,0}$ & $\mathcal{O}(10^{-5})$  \\
\hline
$\Omega_{\Lambda,0}$ & $0.68$  \\
\hline
$H_0$ & $68.14$ km/s/Mpc\\
\hline
\end{tabular}
\caption[The Planck best-fit cosmological parameters.]{The Planck best-fit cosmological parameters, taken from \rcite{Ade:2013zuv}. Here $H_0=100h$~km/s/Mpc.}
\label{tab:planckcosmo}
\end{table}

We have progressively constructed the cosmological standard model, or $\Lambda$-cold dark matter ($\Lambda$CDM). Its main ingredients are radiation (which is important mostly at early times), baryons and cold dark matter comprising pressureless dust, and a small cosmological constant. The gravitational theory is general relativity. It is the aim of this thesis to explore alternatives in which the cosmological constant is removed at the expense of introducing a different gravitational theory.

\section{Linear Perturbations around FLRW}

The FLRW metric was constructed to be consistent with spatial homogeneity and isotropy. The Universe is, of course, not really homogeneous and isotropic: in various places it contains stars, planets, galaxies, people, and Cambridge. The FLRW approximation holds on scales of hundreds of megaparsecs and higher, and breaks down at smaller distances. At slightly smaller distance scales, spacetime is well described by linear perturbations to FLRW. That is, taking $\bar g_\mn$ to be an FLRW background metric, we consider
\begin{equation}
g_\mn = \bar g_\mn + \delta g_\mn,
\end{equation}
where $\delta g_\mn \ll 1$ is a small perturbation, add a similar small piece to the matter sector, and calculate the Einstein equations for $\delta g_\mn$. This proves to be a powerful tool for testing gravity: using probes of structure to test gravity at the linear level complements and can even be more constraining than studies of the expansion history which operate at the background level.

Let us be more explicit. We will work in conformal time ($N=a$) and write the perturbed metric as
\begin{equation}
g_\mn dx^\mu dx^\nu = a^2\left\{-\left(1+E\right)d\tau^2 + 2\partial_iFdtdx^i + \left[\left(1+A\right)\delta_{ij} + \partial_i\partial_jB\right]dx^idx^j\right\}.
\end{equation}
We define the perturbed stress-energy tensor by
\begin{align}
&& T^0{}_0 &= -\bar\rho(1+\delta), & T^i{}_0 &= -\left(\bar\rho + \bar P\right)v^i, && \nonumber \\
&& T^0{}_i &= \left(\bar\rho + \bar P\right)\left(v_i + \partial_iF\right), & T^i{}_j &= \left(\bar P + \delta P\right)\delta^i{}_j + \Sigma^i{}_j,&& \Sigma^i{}_i=0,
\end{align}
where $v^i \equiv dx^i/dt$ and barred quantities refer to background values. Let us specialise to pressureless dust ($P=\delta P = \Sigma^i{}_j = 0$).

Because of the coordinate independence of general relativity, not all of the perturbation variables represent genuine degrees of freedom: as we have things currently set up, it is possible for some of the perturbations to be nonzero while the spacetime is still purely FLRW, only written in funny coordinates. This could lead to unphysical modes propagating through the equations of motion. To remove this problem, we choose a coordinate system, or fix a gauge. We will work in conformal Newtonian gauge, in which $F=B=0$. We will also decompose each variable into Fourier modes and suppress the mode index; the only effect of this for our purposes is that we can write $\delta^{ij}\partial_i\partial_j\Phi=-k^2\Phi$ where $\Phi$ represents any of the perturbations.

We are left with four perturbation variables, $A$, $E$, $\delta$, and $\theta\equiv\partial_iv^i$. There is only one dynamical degree of freedom among these; any three of the variables can be written in terms of the fourth, which in turn obeys a second-order evolution equation. We will take this independent degree of freedom to be $\delta$. By taking the off-diagonal part of the space-space Einstein equation, we can find that $A = -E$. The potential $E$ is related to the density perturbation, $\delta$, by the time-time Einstein equation,
\begin{equation}
3\sh\left(\dot E + \sh E\right) + k^2E = -\frac{a^2\bar\rho}{\Mp^2}\delta. \label{eq:GRtimetimepert}
\end{equation}
Most of the modes which we can access observationally are within the horizon, $k>\sh$. To simplify the analysis we can focus on these modes by taking the subhorizon limits, $k^2\Phi \gg \sh^2\Phi$, and the quasistatic limit, $\sh^2\Phi \sim \sh\dot\Phi \sim \ddot\Phi$, where again $\Phi$ represents any of the perturbation variables. In this limit, \cref{eq:GRtimetimepert} takes the simple form
\begin{equation}
k^2E = -\frac{a^2\bar\rho}{\Mp^2}\delta.
\end{equation}
We can recognise this as the Poisson equation; the matter density contrast, $\delta$, sources the gravitational potential, $E$, in a familiar way.

We additionally have the $\nu=0$ and $\nu=i$ components of the stress-energy conservation equation,
\begin{align}
&& \dot\delta + \theta -\frac{3}{2}\dot E &= 0, & \nu&=0, && \\
&& \dot\theta + \sh\theta - \frac{1}{2}k^2E &= 0, & \nu&=i, &&
\end{align}
where $\theta\equiv\partial_iv^i$. In the subhorizon and quasistatic limit, these can be combined, along with the Poisson equation, to obtain a closed evolution equation for the density contrast,\footnote{Outside the quasistatic limit, this evolution equation will be sourced by $E$, which in turn obeys its own closed equation. Notice that there is still only one independent degree of freedom.}
\begin{equation}
\ddot\delta + \sh\dot\delta - \frac{a^2\bar\rho}{2\Mp^2}\delta = 0. \label{eq:GRdeltaeq}
\end{equation}

The evolution equation for $\delta$ can be integrated to obtain the growth rate of structure. While in general this requires numerical integration, as an illustration we can obtain exact solutions in general relativity during the various cosmic eras. During matter domination, we have $\rho \approx 3\Mp^2\sh^2/a^2$, $a\sim\tau^2$, and $\sh = 2/\tau$, so \cref{eq:GRdeltaeq} becomes
\begin{equation}
\ddot\delta + \frac 2 \tau \dot\delta - \frac 6 {\tau^2} \delta = 0,
\end{equation}
which has, in addition to a decaying mode which we ignore, the growing solution
\begin{equation}
\delta \sim \tau^2 \sim a.
\end{equation}
During dark energy domination, $\bar\rho$ (which is the density of matter) becomes negligibly small, and the only solutions to \cref{eq:GRdeltaeq} are a decaying mode and $\delta=\mathrm{const.}$ We see that during the dark energy era, matter stops clustering. This makes intuitive sense: as the expansion of the Universe accelerates, it becomes more and more difficult for matter to gravitationally cluster ``against'' the expansion.

A useful parametrisation for comparison to data is based on the growth rate,
\begin{equation}
f(a,k) \equiv \frac{d\log\delta}{d\log a}.
\end{equation}
In the recent past, the growth rate of solutions to \cref{eq:GRdeltaeq} is well approximated in terms of the matter density parameter defined above,
\begin{equation}
f(a,k) \approx \Omega_\mathrm{m}^\gamma,
\end{equation}
where the growth index $\gamma$ has the value $\gamma \approx 0.545$. The growth index typically deviates from this in theories of modified gravity.

\section{Inflation}

To this point we have discussed some of the essential ingredients of modern cosmology, particularly general relativity and its application to background and linearised FLRW spacetimes. We then used this to discuss the expansion history and the growth of structure in the ``late Universe,'' i.e., during the matter- and dark-energy-dominated eras. The standard cosmological model includes, at earlier times, two other important eras: radiation-domination and, before it, inflation. We will skip the radiation era, as it is not directly relevant to this thesis. This leaves inflation.

The original motivations of inflationary cosmology were that it solves some of the glaring problems with a big-bang cosmology, which can be summarised as problems of initial conditions: a Universe that was always decelerating (until the recent dark-energy era) requires highly tuned initial conditions to be as flat and uniform as we see it. After inflation was initially developed, another significant motivation for inflation arose: quantum fluctuations during inflation are blown up to sizes larger than the cosmic horizon before they can average out, leaving a spectrum of perturbations which would seed the formation of cosmic structure, in excellent agreement with observations. For a comprehensive review of these motivations and inflationary physics, we point the reader to \rcite{Baumann:2009ds}.

The simplest physical model for inflation, and the one with which we are concerned here, is \emph{single-field slow-roll inflation}. In this model, inflation is driven by a canonical scalar field or \emph{inflaton}, $\phi$, with a potential $V(\phi)$. The scalar has units of mass. We incorporate the scalar field by taking the gravitational action (\ref{eq:S-total}) with the matter Lagrangian given by
\begin{equation}
 \mathcal{L}_m = \mathcal{L}_\phi = -\frac{1}{2}g^\mn \partial_\mu\phi \partial_\nu\phi - V(\phi). \label{eq:phi-action}
\end{equation}
We will usually leave the dependence of the potential on $\phi$ implicit and simply write $V$. The Einstein equation (\ref{eq:einstein-intro}) is sourced by the stress-energy tensor, defined in \cref{eq:stressenergy}. For the scalar field action (\ref{eq:phi-action}) this yields
\begin{equation}
T_{\mu\nu} = \nabla_{\mu}\phi\nabla_{\nu}\phi-\left(\frac{1}{2}\nabla_{\alpha}\phi\nabla^{\alpha}\phi+V\right) g_{\mu\nu}.
\end{equation}
Finally, by varying the action with respect to $\phi$ (or, equivalently, by calculating the Euler-Lagrange equation for $\mathcal{L}_\phi$) we obtain the equation of motion for the scalar field, or the \emph{Klein-Gordon equation},
\begin{equation}
 \Box\phi = V_\phi,
\end{equation}
where $\Box\equiv g^\mn \nabla_\mu \nabla_\nu$ is the D'Alembertian operator, and $V_\phi \equiv dV/d\phi$. The Einstein and Klein-Gordon equations completely determine the behaviour of the two dynamical fields, $g_\mn$ and $\phi$.

Now let us specialise to homogeneous and isotropic cosmology. As argued above, the metric must take the FLRW form. The scalar can only depend on time, $\phi(x^\mu)=\phi(t)$; if it were to depend on space, then it would break homogeneity or isotropy (or both) and communicate that violation to the metric, through the Einstein equation. With this metric, the stress-energy tensor can be written in the perfect-fluid form (\ref{eq:perfectfluid}). Doing this we can identify the density and pressure of the scalar field,
\begin{equation}
\rho = \frac{\dot\phi^2}{2N^2}+V, \qquad p = \frac{\dot\phi^2}{2N^2}-V.
\end{equation}

The most essential condition for solving the big-bang initial condition problems and generating the observed cosmic structure is that the spacetime geometry during inflation be very close to \emph{de Sitter space}, the vacuum solution of Einstein's equations with a positive cosmological constant. This corresponds to an FLRW spacetime with a constant Hubble rate in cosmic time, $H=\mathrm{const}$. Therefore in order to be a good driver of inflation, the inflaton, $\phi$, must source the Friedmann equation in a way close to a cosmological constant.\footnote{It cannot be exactly a cosmological constant as there would be no physical clock to distinguish one time from the next, and inflation could never end. This is why we need a scalar field with dynamics, rather than just a cosmological constant term.} Recall from the previous section that a cosmological constant enters the Friedmann equation as a constant, while matter species contribute their density to this equation; therefore the condition for inflation is that the density of the scalar be nearly constant. As we have seen, this requires
\begin{equation}
 w = \frac p \rho = \frac{\frac{\dot\phi^2}{2N^2}-V}{\frac{\dot\phi^2}{2N^2}+V} \approx -1.
\end{equation}
It is clear that this condition is satisfied when $\dot\phi^2/N^2 \ll V$. This is the \emph{slow-roll condition}: the scalar field must be rolling very slowly, compared to the size of its potential. It is not difficult to see that this does what we want: if the scalar does not move quickly, then $V(\phi)$ is very close to a constant, and the scalar field density will look very similar to a cosmological constant term.

We can understand this condition in terms of the scalar field microphysics, i.e., as a condition on the potential, by utilising the Friedmann and Klein-Gordon equations. In the FLRW background these take the forms
\begin{align}
 3H^2 = \frac{1}{\Mp^2}\left(\frac{\dot\phi^2}{2N^2} + V\right), \\
 \ddot\phi + \left(3\frac{\dot a}{a} - \frac{\dot N}{N}\right)\dot\phi + N^2V_\phi = 0,
\end{align}
where $H = \dot a/(aN)$ is the cosmic-time Hubble parameter. At this point we will specialise, for simplicity, to cosmic time ($N=1$), although when relevant we will present results in terms of conformal time ($N=a$) as well. The expressions presented up to this point, keeping $N$ general, will be necessary in \cref{chap:dc-drgt} when we consider a scalar field in a theory where we cannot freely rescale $N$.

By taking a derivative of the Friedmann equation and removing terms using the Klein-Gordon equation, we obtain
\begin{equation}
 \dot H = -\frac{\dot\phi^2}{2\Mp^2}. \label{eq:Hdot}
\end{equation}
This formalises the result we had derived less rigorously earlier: if $\dot\phi$ is small, then $H$ is nearly constant. But $\dot\phi^2$ is dimensionful, so what do we mean by ``small?'' We have already argued that $\dot\phi^2$ should be small compared to the potential, $V$. Moreover, in this slow-roll limit, the Friedmann equation becomes
\begin{equation}
 3H^2 \approx \frac{V}{\Mp^2}. \label{eq:SRfried}
\end{equation}
Using this equation and the expression for $\dot H$ we can write the slow-roll condition as
\begin{equation}
 \varepsilon \equiv -\frac{\dot H}{H^2} \ll 1,
\end{equation}
where $\varepsilon$ is the \emph{slow-roll parameter}.

It turns out not to be enough to demand that $\varepsilon$ be small: it must also be small for a sufficiently long period of time. If it were not, inflation would not last very long, and would be unable to solve the initial-conditions problems or produce cosmic structure over the range of scales that we observe. We formalise this condition by defining a second slow-roll parameter. Much the way that $\varepsilon$ is defined to demand that $H$ be small,\footnote{We define $\varepsilon$ with a minus sign because, in order to satisfy the null-energy condition, we need $\dot H<0$.} we define a new parameter, $\eta$, to measure the smallness of $\varepsilon$,
\begin{equation}
 \eta \equiv \frac{\dot\varepsilon}{H\varepsilon}.
\end{equation}
Note that in conformal time, $\tau$, defining $'\equiv d/d\tau$ and using $\sh$ again for the conformal-time Hubble parameter $\sh \equiv a'/a$, the slow-roll parameters are
\begin{equation}
 \varepsilon = 1 - \frac{\sh'}{\sh^2}, \qquad \eta = \frac{\varepsilon'}{\sh\varepsilon}.
\end{equation}
The full slow-roll limit of inflation can therefore be defined by demanding $\varepsilon,\eta\ll1$. Observations require that inflation last at least 50--60 $e$-folds, or $\ln(a_f/a_i)\gtrsim$50--60, where $a_i$ and $a_f$ are the scale factors at the start and end of inflation, respectively. In the slow-roll limit, both $\varepsilon$ and $\eta$ are constant at first order and we can integrate to find, to first order in $\varepsilon$,
\begin{align}
 a &= e^{\bar Ht}\left(1 - \frac{\bar H^2 t^2}{2}\varepsilon + \mathcal{O}(\varepsilon^2)\right) \\
 H &= \bar H \left(1 - \bar H t \varepsilon + \mathcal{O}(\varepsilon^2)\right),
\end{align}
or, in conformal time,
\begin{align}
a &  =-\frac{1}{\bar H\tau}(1+\varepsilon + \mathcal{O}(\varepsilon^2)),\\
{\mathcal{H}} &  =-\frac{1}{\tau}(1+\varepsilon + \mathcal{O}(\varepsilon^2)),
\end{align}
where $\bar H$ is the Hubble rate of the de Sitter background in the limit $\varepsilon\to0$. Note that in conformal time, $\tau$ runs from $-\infty$ at the big bang to $0$ in the far future. In principle, ``the far future'' actually corresponds to the end of inflation, taken to be when $\varepsilon,\eta\sim1$ and the slow-roll expansion breaks down. This is assumed to be followed by a period of \emph{reheating}, in which the scalar field decays into standard-model particles, and the radiation era thus commences.

We now, finally, have the tools to understand the microphysics of a scalar field satisfying the slow-roll conditions. Using the expression (\ref{eq:Hdot}) for $\dot H$ and the definition of the slow-roll parameters, we can find
\begin{equation}
 \eta = 2\left(\varepsilon + \frac{\ddot \phi}{H\dot \phi}\right).
\end{equation}
Therefore, in addition to the aforementioned condition, $\dot\phi^2\ll V$, we see that we must also demand $\ddot\phi \ll H\dot\phi$. These correspond to $\varepsilon\ll1$ and $\eta\ll1$, respectively. The latter condition implies that the $\ddot\phi$ term must be subdominant in the Klein-Gordon equation, so we can write its slow-roll version as
\begin{equation}
 3H\dot\phi \approx -V_\phi.
\end{equation}
Using this expression and the slow-roll Friedmann equation, \cref{eq:SRfried}, we can write the slow-roll parameter as
\begin{equation}
 \varepsilon = -\frac{\dot H}{H^2} = \frac{\dot\phi^2}{2\Mp^2H^2} \approx \frac{\Mp^2}{2}\left(\frac{V_\phi}{V}\right)^2 \equiv \varepsilon_{\mathrm v}.
\end{equation}
Next, by taking the derivative of the slow-roll Klein-Gordon equation, and using the other slow-roll relations as before, we obtain
\begin{equation}
 \varepsilon - \frac{\ddot\phi}{H\dot\phi} \approx \Mp^2\frac{V_{\phi\phi}}{V} \equiv \eta_{\mathrm v}.
\end{equation}
We have already shown that $\ddot\phi/H\dot\phi\ll1$, so $\eta_{\mathrm v}\ll1$ is implied by slow roll.

\begin{figure}
\begin{centering}
\includegraphics{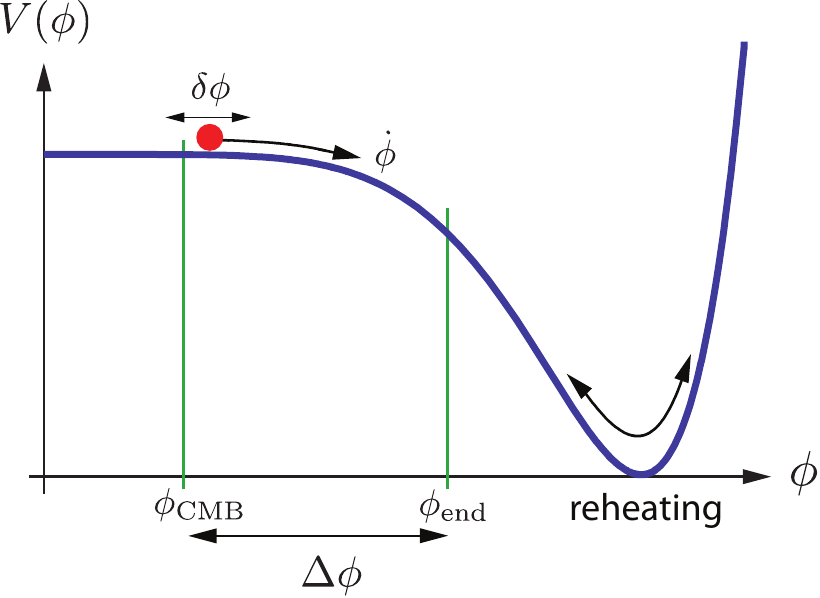}
\par\end{centering}
\caption[A prototypical potential for slow-roll inflation.]{\label{fig:sr-pot} A prototypical potential for slow-roll inflation, taken from \rcite{Baumann:2009ds}.}
\end{figure}

Thus we can equivalently impose slow roll by demanding that the \emph{potential slow-roll parameters}, $\varepsilon_{\mathrm v}$ and $\eta_{\mathrm v}$, be small.\footnote{During slow roll, the potential slow-roll parameters are related to $\varepsilon$ and $\eta$ by $\varepsilon_{\mathrm v}\approx\varepsilon$ and $\eta_{\mathrm v}\approx2\varepsilon-\frac 1 2 \eta$.} This tells us how to construct a good slow-roll potential: it should be very, very flat. Specifically, its first two derivatives with respect to $\phi$ should be small in such a way that $V_\phi \ll V/\Mp$ and $V_{\phi\phi}\ll V/\Mp^2$. A prototypical inflationary potential is shown in \cref{fig:sr-pot}. Popular forms for the potential include $V \sim \phi^2$, $V\sim e^{-\lambda\phi}$, and $V \sim \cos{\phi/f}$.

We conclude by mentioning that this formalism can be applied to the late-time accelerating phase as well. This idea underlies \emph{quintessence} models of dark energy. In quintessence, the cosmological constant is given dynamics by being promoted to a scalar field. This allows for $w$ to vary from $-1$, and can also change the way that structure forms, because unlike a cosmological constant, quintessence can cluster. The mathematical formalism is almost exactly the same as that presented in this section. In the simplest models, the scalar field has the same Lagrangian, hence the same density and pressure, as we have used, and the condition for acceleration is again that the potential dominate over the kinetic energy, now at late rather than early times. The main differences are that we need to include matter (specifically baryons and dark matter) in the matter Lagrangian as well, and that the potential is allowed to satisfy the slow-roll conditions forever, since while we know inflation ended, the same constraint does not apply to dark energy.

\begin{savequote}[30pc]
Is this quintessence o[r] dust?
\qauthor{Hamlet, \textit{Hamlet}, 2.2}
\end{savequote}

\chapter{Gravity Beyond General Relativity}
\label{chap:mg}
\hrule
\vspace*{1cm}

At the core of this thesis is the question of modifying general relativity. In the previous chapter, we introduced general relativity and its cosmological solutions, culminating in a discussion of two key aspects of the cosmological standard model: $\Lambda$CDM at late times and inflation at early times. In this chapter, we extend that discussion to theories of gravity beyond general relativity, and in particular the theories which will receive our attention in this thesis: massive gravity, massive bigravity, and Einstein-aether theory.

General relativity is the unique Lorentz-invariant theory of a massless spin-2 field \cite{Gupta:1954zz,Weinberg:1965rz,Deser:1969wk,Boulware:1974sr,Feynman:1996kb}. To move beyond this theory, we must therefore modify its degrees of freedom. In massive gravity, this is done by endowing the graviton with a small but nonzero mass. Bigravity extends this by giving dynamics to a second tensor field which necessarily appears in the action for massive gravity; its dynamical degrees of freedom are two spin-2 fields, one massive and one massless. Finally, in Einstein-aether theory the massless graviton is supplemented by a vector field. This vector is constrained to always have a timelike vacuum expectation value (vev), and so spontaneously breaks Lorentz invariance by picking out a preferred time direction. It is thus a useful model for low-energy gravitational Lorentz violation.

\section{Massive Gravity and Bigravity}
\label{sec:massgrav}

The history of massive gravity is an old one, dating back to 1939 when the linear theory of Fierz and Pauli was published \cite{Fierz:1939ix}. Studies of interacting spin-2 field theories also have a long history~\cite{Isham:1971gm}. However, there had long been an obstacle to the construction of a fully nonlinear theory of massive gravity in the form of the notorious Boulware-Deser ghost \cite{Boulware:1973my}, a pathological mode that propagates in massive theories at nonlinear order. This ghost mode was thought to be fatal to massive and interacting bimetric gravity until only a few years ago, when a way to avoid the ghost was discovered by utilising a very specific set of symmetric potential terms \cite{deRham:2010ik,deRham:2010kj,deRham:2011rn,deRham:2011qq,Hassan:2011vm,Hassan:2011hr,Hassan:2011tf,Hassan:2011ea}. In this section we review that history, before moving onto the modern formulations of ghost-free massive gravity and bigravity, and elucidating their cosmological solutions. We refer the reader to \rcite{deRham:2014zqa,Hinterbichler:2011tt} for in-depth reviews on massive gravity and its history.

\subsection{Building the Massive Graviton}

\subsubsection{The Fierz-Pauli Mass Term}

Let us begin by considering linearised gravity described by a spin-2 field, $h_\mn$. We will routinely refer to this field as the graviton. The theory of a massless graviton is given by linearising the Einstein-Hilbert action around Minkowski space, i.e., by splitting the metric up as 
\begin{equation}
g_\mn = \eta_\mn + \frac{1}{\Mp} h_\mn,
\end{equation}
where $h_\mn/\Mp \ll 1$, and keeping in the action only terms quadratic in $h_\mn$. Doing this, we obtain the Lagrangian of linearised general relativity,
\begin{equation}
\mathcal{L}_\mathrm{GR,linear} = -\frac 1 4 h^\mn \hat{\mathcal{E}}^\ab_\mn h_\ab, \label{eq:linearizedGR}
\end{equation}
where indices are raised and lowered with the Minkowski metric, $\eta_\mn$, and we have defined the \emph{Lichnerowicz operator}, $\hat{\mathcal E}$, by
\begin{equation}
\hat{\mathcal{E}}^\ab_\mn h_\ab \equiv -\frac 12 \left(\Box h_\mn-2\partial_{(\mu}\partial_\alpha h^\alpha_{\nu)}+\partial_\mu\partial_\nu h-\eta_\mn (\Box h-\partial_\alpha\partial_\beta h^{\alpha\beta})\right)\,,
\end{equation}
where $h\equiv \eta^\mn h_\mn$ is the trace of $h_\mn$. No other kinetic terms are consistent with locality, Lorentz invariance, and gauge invariance under linearised diffeomorphisms,
\begin{equation}
h_\mn \to h_\mn + 2\partial_{(\mu} \xi_{\nu)}. \label{eq:lindiffs}
\end{equation}
Indeed, this uniqueness is a necessary (though not sufficient) part of the aforementioned uniqueness of general relativity as the nonlinear theory of a massless spin-2 field.

The role of gauge invariance is to ensure that there are no \emph{ghosts}, i.e., no degrees of freedom with higher derivatives or wrong-sign kinetic terms. Ostrogradsky's theorem tells us that, up to a technical condition,\footnote{Namely that the Lagrangian be nondegenerate, i.e., that $\partial L / \partial \dot q$ depend on $\dot q$.} a Lagrangian with higher than second derivatives will lead to a Hamiltonian which is unbounded from below, and thus states with arbitrarily negative energy are allowed (for a thorough, modern derivation, see \rcite{Woodard:2006nt}). If we had included in \cref{eq:linearizedGR} other terms that can be constructed out of $h_\mn$ and its first and second derivatives, then the action would no longer be invariant under \cref{eq:lindiffs}. In that case, we could split $h_\mn$ into a transverse piece, $h_\mn^T$, and a vector field, $\chi_\mu$, as
\begin{equation}
h_\mn = h^T_\mn + 2\partial_{(\mu} \chi_{\nu)},
\end{equation}
and any terms not included in the action (\ref{eq:linearizedGR}) would contain pieces with higher derivatives of $\chi_\mu$. Therefore we can see the linearised Einstein-Hilbert term as the kinetic term uniquely set by three requirements: locality, Lorentz invariance, and the absence of a ghost.

We would like to give $h_\mn$ a mass---i.e., add a nonderivative interaction term---while maintaining those three requirements. Unfortunately, it is impossible to construct a local interaction term which is consistent with diffeomorphism invariance (\ref{eq:lindiffs}). Since this was useful in exorcising ghost modes, we will need to take care to ensure that no ghost is introduced by the mass term. At second order, this is not especially difficult as there are only two possible terms we can consider: $h^\mn h_\mn$ and $h^2$. We can then consider a general quadratic mass term,
\begin{equation}
\mathcal{L}_\mathrm{mass} = -\frac 1 8 m^2\left(h_\mn h^\mn - (1-a)h^2\right).
\end{equation}
This leads to a ghostlike, scalar degree of freedom with mass $m_g^2 = \frac{3-4a}{2a}m^2$. The only way to remove the ghost from this theory, besides setting $m=0$, is to set $a=0$. The ghost then has infinite mass and is rendered nondynamical. We see the unique ghost-free action for a massive graviton at quadratic order is the \emph{Fierz-Pauli action},
\begin{equation}
\mathcal{L}_\mathrm{FP} = -\frac 1 4h^\mn \hat{\mathcal{E}}^\ab_\mn h_\ab - \frac{1}{8} m^2 \left(h_\mn h^\mn - h^2\right). \label{eq:fp}
\end{equation}

\subsubsection{\texorpdfstring{The St\"uckelberg ``Trick''}{The St\"uckelberg "Trick"}}

Before moving on to higher orders in $h_\mn$, let us take a moment to count and classify the degrees of freedom it contains at linear order. Recall that a massless graviton contains two polarisations. Because we lose diffeomorphism invariance when we give the graviton a mass, it will contain more degrees of freedom. In fact, there are five in total. In principle, a sixth mode can arise, but it is always ghost-like and must therefore be removed from any healthy theory of massive gravity. To separate the degrees of freedom contained in the massive graviton, we use the St\"uckelberg ``trick.''

St\"uckelberg's idea is based on the observation that a gauge freedom such as diffeomorphism is not a physical property of a theory so much as a redundancy in description, and that redundancy can always be introduced by bringing in redundant variables. Let us consider splitting up $h_\mn$ as
\begin{equation}
h_\mn \to h_\mn + \frac 2m\partial_{(\mu}A_{\nu)} + \frac{2}{m^2}\partial_\mu \partial_\nu \phi.
\end{equation}
Defining the field strength tensor for $A_\mu$ analogously to electromagnetism, $F_\mn \equiv \frac 1 2 \partial_{[\mu}A_{\nu]}$, as well as $\Pi_\mn \equiv \partial_\mu\partial_\nu\phi$ and the trace notation $[A] \equiv \eta^\mn A_\mn$, the Fierz-Pauli action (\ref{eq:fp}) becomes
\begin{align}
\mathcal{L}_\mathrm{FP} &= -\frac 14 h^\mn \hat{\mathcal E}^\ab_\mn h_\ab - \frac 12 h^\mn\left(\Pi_\mn - [\Pi]\eta_\mn\right) - \frac 1 8 F_\mn F^\mn \nonumber \\
&\hphantom{{}=}-\frac 18m^2\left(h_\mn h^\mn - h^2\right) - \frac 12m\left(h^\mn - h\eta^\mn\right)\partial_{(\mu}A_{\nu)}.
\end{align}
This action is invariant under the simultaneous gauge transformations
\begin{equation}
h_\mn \to h_\mn + 2\partial_{(\mu}\xi_{\nu)}, \qquad A_\mu \to A_\mu - \frac m 2 \xi_\mu
\end{equation}
for $h_\mn$ and
\begin{equation}
A_\mu \to A_\mu + \partial_\mu\lambda, \qquad \phi \to \phi - m\lambda
\end{equation}
for $A_\mu$. With these gauge invariances restored, one can find that $h_\mn$ contains the usual two independent components of a spin-2 degree of freedom, $A_\mu$ similarly contains the standard two independent components, and $\phi$ contains one, leading to a total of $2+2+1=5$ degrees of freedom for a Fierz-Pauli massive graviton.

Before moving on, let us briefly consider the limit $m\to0$. Intuitively we would expect this to reduce to general relativity. In this limit, the vector completely decouples from the other two fields, while the scalar remains mixed with the tensor. They can be unmixed by transforming $h_\mn \to h_\mn + \phi \eta_\mn$. However, this transformation introduces a coupling between $\phi$ and the stress-energy tensor for matter (which, for simplicity, we have neglected so far) which does not vanish in the massless limit. This is the origin of the van Dam-Veltman-Zakharov (vDVZ) discontinuity \cite{vanDam:1970vg,Zakharov:1970cc}. While linearised Fierz-Pauli theory with matter indeed does not reduce to general relativity in the limit $m\to0$, nonlinear effects cure this discontinuity: this is the celebrated Vainshtein mechanism which restores general relativity in environments where $\phi$ is nonlinear and allows theories like massive gravity to agree with solar system tests of gravity \cite{Vainshtein:1972sx}. For a modern introduction to the Vainshtein mechanism, see \rcite{Babichev:2013usa}.

\subsubsection{The Boulware-Deser Ghost}
\label{sec:bd}

Upon moving beyond linear order, disaster strikes. While the Fierz-Pauli tuning (expressed above as $a=0$) removes a sixth, ghostlike degree of freedom from the linear theory, Boulware and Deser found that this mode generically reappears at higher orders \cite{Boulware:1973my}. This is the notorious \emph{Boulware-Deser ghost}. It is infinitely heavy on flat backgrounds, as evidenced by its infinite mass at the purely linear level, but can become light around nontrivial solutions \cite{deRham:2010ik}, including cosmological backgrounds \cite{Gabadadze:2003jq} and weak-field solutions around static matter \cite{ArkaniHamed:2002sp,Creminelli:2005qk,Deffayet:2005ys}.

A full history of this ghost mode is beyond the scope of this thesis; we will simply, following the review \cite{deRham:2014zqa}, introduce the simplest nonlinear extension of the Fierz-Pauli mass term and demonstrate the existence of a ghost mode, as an illustration of the fact that nontrivial interaction terms are required beyond the linear order in order to obtain a ghost-free theory of massive gravity. Let us define the matrix
\begin{equation}
 \mathbb M^\mu{}_\nu \equiv g^{\mu\alpha}\eta_{\nu\alpha}.
\end{equation}
Linearising this matrix around $\eta_\mn$ as above, we find
\begin{equation}
 \frac1\Mp h^\mu{}_\nu \approx \delta^\mu{}_\nu - \mathbb M^\mu{}_\nu.
\end{equation}
The Fierz-Pauli term (\ref{eq:fp}) can be obtained by linearising ($g_\mn = \eta_\mn + h_\mn/\Mp$) the nonlinear action \cite{ArkaniHamed:2002sp}
\begin{equation}
 \mathcal{L}_\mathrm{FP,nonlinear} = -m^2\Mp^2\sdg \left\{ [\left(\mathbb I - \mathbb M\right)^2] - \left[\mathbb I - \mathbb M \right]^2\right\},
\end{equation}
where $\mathbb I$ is the identity matrix in four dimensions.

With a candidate nonlinear completion of massive gravity in hand, let us examine the behaviour of the helicity-0 mode, $\phi$, ignoring the helicity-1 and helicity-2 modes for simplicity. The matrix $\mathbb M$ becomes
\begin{equation}
 \mathbb M^\mu{}_\nu = \delta^\mu{}_\nu - \frac{2}{\Mp m^2}\Pi^\mu{}_\nu + \frac{1}{\Mp^2m^4}\Pi^\mu{}_\alpha\Pi^\alpha{}_\nu.
\end{equation}
Plugging this into our nonlinear action, we find
\begin{equation}
 \mathcal{L}_\mathrm{FP,nonlinear} = -\frac{4}{m^2}\left([\Pi^2]-[\Pi]^2\right) + \frac{4}{\Mp m^2}\left([\Pi^3]-[\Pi][\Pi^2]\right) + \frac{1}{\Mp^2m^6}\left([\Pi^4]-[\Pi^2]^2\right).
\end{equation}
This clearly contains higher time derivatives for $\phi$, and will therefore lead to an Ostrogradsky instability and hence to ghosts. In fact, even the first, quadratic term---the most obvious ancestor of the Fierz-Pauli term---has higher derivatives. However, they turn out to be harmless for the reason that this quadratic term is, after integration by parts, a total derivative and therefore does not contribute to the dynamics. This miracle does not extend to any of the higher terms, which lead to a genuine sixth degree of freedom.\footnote{While we have seemingly split the metric up into five degrees of freedom---two tensor, two vector, and one scalar---and focused on the scalar, when $\phi$ has higher time derivatives it in fact contains two degrees of freedom, one of which is generically a ghost.}

The Boulware-Deser ghost is not specific to this particular, simple nonlinear completion of the Fierz-Pauli term. It is a very generic problem, to the point that as of a decade ago it was thought to plague \emph{all} nonlinear massive gravity theories \cite{Creminelli:2005qk}. The ghost can, however, be removed: one can consider general higher-order extensions and, at each order, tune the coefficients to eliminate higher time derivatives by packaging them into total derivative terms \cite{deRham:2010ik}. This led to a ghost-free, fully nonlinear theory of massive gravity in \rcite{deRham:2010kj}, nearly four decades after the discovery of the Boulware-Deser ghost.

\subsubsection{To the Nonlinear Theory}

Taken on its own, linearised gravity in the form (\ref{eq:linearizedGR}) is a perfectly acceptable theory without being seen as a truncation of a nonlinear theory such as general relativity. Indeed, it is even a perfectly fine \emph{gauge theory}, as its linearised diffeomorphism invariance is exact (as long as we include the St\"uckelberg fields, if we take the graviton to be massive). The wrench in the works comes when we add matter in. Unfortunately, the coupling to matter is necessary, as we prefer our theories to communicate with the rest of the Universe and hence be subject to experiment.

The coupling to matter at the linear level is of the form
\begin{equation}
 \mathcal L_\mathrm{matter,linear} = \frac{1}{2\Mp}h_\mn T^\mn_0,
\end{equation}
where $T^\mn_0$ is the stress-energy tensor of our matter source. Diffeomorphism invariance is preserved in the matter sector if stress-energy is conserved, i.e., if $\partial_\mu T^\mn_0=0$. However, the coupling to $h_\mn$ necessarily induces a violation of this conservation. For a simple example of this using a scalar field, see \rcite{deRham:2014zqa}. This problem is fixed by adding nonlinear corrections both to the matter coupling,
\begin{equation}
 \mathcal L_\mathrm{matter,nonlinear} = \frac{1}{2\Mp}h_\mn T^\mn_0 + \frac{1}{2\Mp^2}h_\mn h_\ab T^{\mn\ab}_1,
\end{equation}
for some tensor $T^{\mn\ab}_1$, and to the gauge symmetry, symbolically written as
\begin{equation}
 h \to h + \partial\xi + \frac 1 \Mp \partial(h\xi).
\end{equation}
While this ensures the conservation of the stress-energy tensor at the linear level, it is broken at the next order, and so we must continue this procedure order by order, \emph{ad infinitum}.

For a massless spin-2 field, the end result of this procedure is well-known: it is general relativity. The fully nonlinear gauge symmetry is diffeomorphism invariance, and as long as the matter action is invariant under this symmetry, the stress-energy tensor is covariantly conserved, $\nabla_\mu T^\mn = 0$. The linear action (\ref{eq:linearizedGR}) must be promoted to something which is also consistent with this symmetry, and there is one answer: the Einstein-Hilbert action (\ref{eq:S-EH}).

If we wish to extend this procedure to a massive graviton, i.e., nonlinearly complete the Fierz-Pauli mass term, then, as discussed in \cref{sec:bd}, the Boulware-Deser ghost looms as a pitfall. Demanding that this ghost be absent will severely restrict the allowed potentials to a very specific and special set of functions of $\mathbb M$. However, the gauge invariance can still be restored quite easily by a nonlinear version of the St\"uckelberg trick. Because this elucidates several of the properties of massive gravity, we will review it here.

When constructing a nonlinear candidate theory of massive gravity in \cref{sec:bd}, we employed the matrix $\mathbb M = g^{-1}\eta$, where $g^{-1}$ is the inverse of the dynamical metric and $\eta$ is the Minkowski metric. The appearance of a second, fixed metric in addition to the dynamical one is new to massive gravity. Indeed, it is necessary to have such a second metric, or \emph{reference metric}, in order to give the graviton a mass. A mass term is a nonderivative interaction,\footnote{We could not have kinetic interactions anyway; in four dimensions, the Einstein-Hilbert term is unique \cite{Lovelock:1971yv}.} and the only nonderivative scalars or scalar densities we can construct out of $g_\mn$ alone are $\operatorname{tr} g=4$ and $\det g$. The first possibility is trivial, while it was shown in \rcite{Boulware:1973my} that functions of the metric determinant can only consistently lead to a cosmological constant as well. Therefore we need a reference metric in order to construct a massive graviton.\footnote{We have assumed locality in this discussion. A nonlocal theory can give the graviton a mass without the need for a reference metric \cite{Jaccard:2013gla,Maggiore:2013mea,Nesseris:2014mea}; see \rcite{Maggiore:2014sia,Modesto:2013jea} for studies of two interesting realisations of this idea.}

Note that, while we have so far taken the reference metric to be Minkowski, in principle we could extend the theory to a more general reference metric, $f_\mn$. Consequently, even once we have specified the interaction potential there are many different massive gravity theories, one for each reference metric. Alternatively, $f_\mn$ can be viewed as a ``constant tensor'' which must be specified by hand. Physically, the reference metric corresponds to the background around which linear fluctuations acquire the Fierz-Pauli form \cite{Hassan:2011tf,Guarato:2013gba}. This is why we have naturally discussed theories with a Minkowski reference metric: we began by considering fluctuations around that metric, and so it remains when extending to the nonlinear theory. Note that $f_\mn=\eta_\mn$ is a natural choice, as the theory then possesses a Poincar\'e-invariant preferred metric, allowing us to define mass and spin regardless of the solutions of the theory.\footnote{We thank Claudia de Rham for emphasising this point.}

The nonlinear St\"uckelberg trick is simply to introduce into the reference metric four St\"uckelberg fields, $\phi^a$, as
\begin{equation}
 f_\mn \to \tilde f_\mn \equiv f_{ab}\partial_\mu \phi^a \partial_\nu \phi^b.
\end{equation}
Note that here Latin indices are in field space, not spacetime; in particular, each of the $\phi^a$ fields transforms as a spacetime scalar. Consequently, $\tilde f_\mn$ transforms as a tensor under general coordinate transformations, as well as $\mathbb M^\mu{}_\nu$ and all of the nonlinear completions of the Fierz-Pauli term constructed out of it, as long as we replace $f_\mn$ with $\tilde f_\mn$. If we choose the coordinate axes to align with the St\"uckelberg fields, $x^a = \phi^a$, then we have $\tilde f_\mn = f_\mn$ and recover the previous description. This is called \emph{unitary gauge}. For simplicity we will usually work in unitary gauge when dealing with massive gravity.

\subsection{Ghost-Free Massive Gravity}

\subsubsection{de Rham-Gabadadze-Tolley Massive Gravity}
\label{sec:drgt}

Recent years have seen a breakthrough in massive gravity, stemming from the development of a fully ghost-free and nonlinear theory by de Rham, Gabadadze, and Tolley (dRGT) \cite{deRham:2010ik,deRham:2010kj}, following the important preliminary steps taken in \rcite{Gabadadze:2009ja,deRham:2009rm} using auxiliary extra dimensions. The full proof that the dRGT theory is ghost-free, including for a general reference metric, was completed in \rcite{Hassan:2011hr,Hassan:2011tf,Hassan:2011ea}. There are indications that this theory is the unique ghost-free massive gravity; in particular, new kinetic interactions do not appear to be consistent \cite{deRham:2013tfa,Gao:2014jja}. We will use the formulation of the dRGT interaction potential developed in \rcite{Hassan:2011vm}.

The action for dRGT massive gravity around a general reference metric, $f_\mn$,\footnote{As discussed above, we are, strictly speaking, writing the action for massive gravity in unitary gauge. Extending to a more general gauge by including St\"uckleberg fields by promoting $f_\mn$ to $f_{ab}\partial_\mu \phi^a \partial_\nu \phi^b$ is trivial.} is
\begin{equation}
   S_\mathrm{dRGT} = -\frac{\Mp^2}{2}\int d^4x\sdg R + m^2\Mp^2\int d^4x\sdg\sum_{n=0}^{4}\alpha_ne_n(\mathbb K) + \int d^4x\sdg\mathcal{L}_m\left(g, \Phi_i\right), \label{eq:sdrgtalpha}
\end{equation}
or, equivalently,
\begin{equation}
   S_\mathrm{dRGT} = -\frac{\Mp^2}{2}\int d^4x\sdg R + m^2\Mp^2\int d^4x\sdg\sum_{n=0}^{4}\beta_ne_n(\mathbb X) + \int d^4x\sdg\mathcal{L}_m\left(g, \Phi_i\right),\label{eq:sdrgtbeta}
\end{equation}
where we have defined the square-root matrix $\mathbb X$ as
\begin{equation}
 \mathbb X \equiv \sqrt{g^{-1}f},
\end{equation}
i.e., defined $\mathbb X$ so that (reintroducing explicit spacetime indices) $\mathbb X^\mu{}_\alpha \mathbb X^\alpha{}_\nu = g^{\mu\alpha}f_{\alpha\nu}$, and the related matrix $\mathbb K$ by
\begin{equation}
 \mathbb K \equiv \mathbb I - \mathbb X.
\end{equation}
Here, $\alpha_n$ and $\beta_n$ are dimensionless coupling constants, generally taken to be free parameters, and $e_n$ are the \emph{elementary symmetric polynomials} of the eigenvalues $\lambda_i$ of the matrix argument. In terms of the eigenvalues, assuming $i$ runs from 1 to 4, these are (taking as the argument a general matrix, $\mathbb A$, for concreteness)
\begin{align}
 e_0(\mathbb A) &= 1, \nonumber \\
 e_1(\mathbb A) &= \lambda_1 + \lambda_2 + \lambda_3 + \lambda_4, \nonumber \\
 e_2(\mathbb A) &= \lambda _1 \lambda _2 +\lambda _1 \lambda _3+\lambda _1 \lambda _4 +\lambda _2 \lambda _3+\lambda _2 \lambda_4+\lambda _3 \lambda_4, \nonumber \\
 e_3(\mathbb A) &= \lambda _1 \lambda _2 \lambda _3 +\lambda _1 \lambda _2 \lambda
   _4 +\lambda _1 \lambda _3 \lambda _4+\lambda _2 \lambda _3 \lambda _4, \nonumber \\
 e_4(\mathbb A) &= \lambda_1\lambda_2\lambda_3\lambda_4 = \det \mathbb A.
\end{align}
It is often more useful to write these polynomials directly in terms of the matrix,
\begin{align}
e_0(\mathbb{A}) &\equiv 1, \nonumber \\
e_1(\mathbb{A}) &\equiv [\mathbb{A}], \nonumber \\
e_2(\mathbb{A}) &\equiv \frac{1}{2}\left([\mathbb{A}]^2 - [\mathbb{A}^2]\right),\nonumber \\
e_3(\mathbb{A}) &\equiv \frac{1}{6}\left([\mathbb{A}]^ 3 - 3[\mathbb{A}][\mathbb{A}^2] + 2[\mathbb{A}^3]\right), \nonumber \\
e_4(\mathbb A) &\equiv \det\left(\mathbb{A}\right). \label{eq:eneigenvalues}
\end{align}
The formulations (\ref{eq:sdrgtalpha}) and (\ref{eq:sdrgtbeta}) in terms of $\mathbb K = \mathbb I - \sqrt{g^{-1}f}$ and $\mathbb X = \sqrt{g^{-1}f}$ are both very common in the literature. For reasons partly physical and partly historical, the $\mathbb K$ formulation is more common in massive gravity, while $\mathbb X$ is predominant in bigravity. The free parameters in the two formulations, $\alpha_n$ and $\beta_n$, are related by \cite{Hassan:2011vm}
\begin{equation}
 \beta_n = (4-n)!\displaystyle\sum_{i=n}^{4}\frac{(-1)^{i+n}}{(4-i)!(i-n)!}\alpha_i. \label{eq:alphabeta}
\end{equation}
Throughout this thesis we will use the $\mathbb X$ basis and $\beta_n$ parametrisation except when stated otherwise.

Notice that although these potential terms are very complicated, they have a significant amount of structure. This can be better understood by decomposing the metrics into their \emph{vielbeins}, defined by
\begin{equation}
g_\mn = \eta_{ab}E^a{}_\mu E^b{}_\nu, \qquad f_\mn = \eta_{ab}L^a{}_\mu L^b{}_\nu.
\end{equation}
Since vielbeins are, in a sense, the ``square roots'' of the metrics, and the dRGT potential terms are built out of a square root matrix, this is a natural language in which to formulate massive gravity. The interaction terms are, up to numerical constants, given by \cite{Hinterbichler:2012cn}
\begin{align}
e_0(\mathbb X) &\propto \tilde\epsilon_{abcd}\tilde\epsilon^{\mn\ab}E^a{}_\mu E^b{}_\nu E^c{}_\alpha E^d{}_\beta, \nonumber \\
e_1(\mathbb X) &\propto \tilde\epsilon_{abcd}\tilde\epsilon^{\mn\ab}E^a{}_\mu E^b{}_\nu E^c{}_\alpha L^d{}_\beta, \nonumber \\
e_2(\mathbb X) &\propto \tilde\epsilon_{abcd}\tilde\epsilon^{\mn\ab}E^a{}_\mu E^b{}_\nu L^c{}_\alpha L^d{}_\beta, \nonumber \\
e_3(\mathbb X) &\propto \tilde\epsilon_{abcd}\tilde\epsilon^{\mn\ab}E^a{}_\mu L^b{}_\nu L^c{}_\alpha L^d{}_\beta, \nonumber \\
e_0(\mathbb X) &\propto \tilde\epsilon_{abcd}\tilde\epsilon^{\mn\ab}L^a{}_\mu L^b{}_\nu L^c{}_\alpha L^d{}_\beta.
\end{align}
We can now understand the simplicity of the dRGT interaction potential: it is a linear combination of the only possible wedge products one can construct from $\mathbf{E}^a=E^a{}_\mu dx^\mu$ and $\mathbf{L}^a=L^a{}_\mu dx^\mu$.

Finally, notice that we have only coupled matter minimally to the dynamical metric, $g_\mn$. More general matter couplings are certainly possible, but the vast majority of attempts to couple the same matter sector to both metrics reintroduce the Boulware-Deser ghost \cite{Hassan:2012wr,Yamashita:2014fga,deRham:2014naa}. The question of formulating and studying more general matter couplings will form a significant part of this thesis, lying at the heart of \cref{chap:dc-finsler,chap:dc-background,chap:dc-drgt}.

\subsubsection{Hassan-Rosen Bigravity}
\label{sec:mg-bigravity}

In dRGT massive gravity, the reference metric, $f_\mn$, is fixed and must be put into the theory by hand. This leads to a multiplicity of theories of massive gravity: massive gravity around Minkowski, around de Sitter, around anti-de Sitter, and so on. As shown by Hassan and Rosen in \rcite{Hassan:2011zd}, the reference metric can be freely given dynamics without spoiling the ghost-free nature of the theory, as long as its kinetic term is also of the Einstein-Hilbert form. This leads to \emph{Hassan-Rosen bigravity} or \emph{massive bigravity},
\begin{align}
   S_\mathrm{HR} &= -\frac{M_g^2}{2}\int d^4x\sdg R(g) -\frac{M_f^2}{2}\int d^4x\sdf R(f) \nonumber \\
   &\hphantom{{}=}+ m^2M_g^2\int d^4x\sdg\sum_{n=0}^{4}\beta_ne_n(\mathbb X)+ \int d^4x\sdg\mathcal{L}_m\left(g, \Phi_i\right), \label{eq:actionHR}
\end{align}
where $R(g)$ and $R(f)$ are the Ricci scalars corresponding to each of the metrics. We have allowed for the two metrics to have different Planck masses, $M_g$ and $M_f$, although as long as both are finite, they can be set equal to each other by performing the constant rescalings \cite{Berg:2012kn}
\begin{equation}
 f_\mn \to M_\star^{-2}f_\mn, \qquad \beta_n \to M_\star^n\beta_n,
\end{equation}
where $M_\star \equiv M_f/M_g$. Therefore the $f$-metric Planck mass is a redundant parameter. We will generally perform this rescaling implicitly in later chapters, although for now we will leave both Planck masses in to help to elucidate some of the physical features of the theory.

In terms of free parameters bigravity is simpler than massive gravity: we have traded a constant matrix ($f_\mn$) for a constant number ($M_f$) which is not even physically relevant. We thus need to specify fewer theory ingredients to test its solutions.\footnote{Note however that the $f$-metric cosmological constant, $\beta_4$, is physically relevant in bigravity but not in massive gravity, as it is independent of $g_\mn$ and hence only contributes to the equation of motion for $f_\mn$.} On the other hand, it is less simple from the more theoretical point of view that it contains more degrees of freedom. The mass spectrum of bigravity contains two spin-2 fields, one massive and one massless. We can see this at the linear level by expanding each metric around the same background, $\bar g_\mn$, as
\begin{equation}
 g_\mn = \bar g_\mn + \frac{1}{M_g}h_\mn,\qquad f_\mn = \bar g_\mn + \frac{1}{M_f}l_\mn.
\end{equation}
For simplicity we assume the ``minimal model'' introduced in \rcite{Hassan:2011vm}, given by
\begin{equation}
 \beta_0 = 3,\qquad \beta_1 = -1, \qquad \beta_2=0,\qquad\beta_3=0,\qquad\beta_4=1.
\end{equation}
The quadratic Lagrangian is given by \cite{Hassan:2011zd}
\begin{align}
 \mathcal{L}_\mathrm{HR,linear} &= -\frac 1 4h^\mn \hat{\mathcal{E}}^\ab_\mn h_\ab -\frac 1 4l^\mn \hat{\mathcal{E}}^\ab_\mn l_\ab \nonumber \\
&\hphantom{{}=} - \frac{1}{8} m^2M_\eff^2 \left[ \left(\frac{h^\mu{}_\nu}{M_g} - \frac{l^\mu{}_\nu}{M_f}\right)^2 - \left(\frac{h}{M_g} - \frac{l}{M_f}\right)^2 \right],
\end{align}
where we have defined $M_\eff^{-2} \equiv M_g^{-2} + M_f^{-2}$. Indices are raised and lowered with the background metric. We notice the usual Einstein-Hilbert terms for each of the two metric perturbations, as well as two Fierz-Pauli terms with some additional mixing between $h_\mn$ and $l_\mn$. This can be easily diagonalised by performing the change of variables
\begin{equation}
 \frac{1}{M_\eff}u_\mn \equiv \frac{1}{M_f}h_\mn + \frac{1}{M_g}l_\mn, \qquad \frac{1}{M_\eff}v_\mn \equiv \frac{1}{M_f}h_\mn - \frac{1}{M_g}l_\mn.
\end{equation}
The resultant unmixed Lagrangian,
\begin{equation}
 \mathcal{L}_\mathrm{HR,linear} = -\frac 1 4u^\mn \hat{\mathcal{E}}^\ab_\mn u_\ab -\frac 1 4v^\mn \hat{\mathcal{E}}^\ab_\mn v_\ab - \frac{1}{8} m^2 \left(v_\mn v^\mn - v^2\right),
\end{equation}
contains a Fierz-Pauli term for $v_\mn$ and no interaction term for $u_\mn$. Therefore in the linearised theory $u_\mn$ corresponds to a massless graviton and $v_\mn$ to a ghost-free massive one with mass $m$. We can see that in the limit where one Planck mass is much larger than the other, the metric with the larger Planck mass corresponds mostly to the massless graviton: if $M_f \gg M_g$, then $(u_\mn,v_\mn)\to(l_\mn,h_\mn)$, and similarly if $M_g \gg M_f$, then $(u_\mn,v_\mn)\to(h_\mn,-l_\mn)$. This formalises the notion, which is intuitive from \cref{eq:actionHR}, that we can recover dRGT massive gravity by taking one of the Planck masses to infinity. In that case, the massless mode corresponds entirely to the metric with the infinite Planck mass, its dynamics freeze out so that it becomes fixed, and the massless mode decouples from the theory, leaving us with what we expect for massive gravity.\footnote{See, however, \rcite{Hassan:2014vja} for some caveats on taking the massive-gravity limit of bigravity.} Note, finally, that the notion of mass is only really well-defined around Minkowski space, as it follows from Poincar\'e invariance. More generally we can identify modes with a Fierz-Pauli term as massive, by analogy to the Minkowski case. As shown above, one can identify massive and massless linear fluctuations in this way around equal backgrounds for a special parameter choice, and indeed this can be done for general parameters as long as $g_\mn$ and $f_\mn$ are conformally related \cite{Hassan:2012wr}, but for general backgrounds there is no unambiguous splitting of the massive and massless modes in bigravity.

An interesting and useful property of Hassan-Rosen bigravity is that while $g_\mn$ and $f_\mn$ do not appear symmetrically in the action (\ref{eq:actionHR}), it nevertheless does treat does metrics on equal footing, ignoring the matter coupling. In particular, the mass term has the property
\begin{equation}
\sdg\sum_{n=0}^{4}\beta_ne_n\left(\sqrt{g^{-1}f}\right) = \sdf\sum_{n=0}^{4}\beta_{4-n}e_n\left(\sqrt{f^{-1}g}\right),
\end{equation}
which follows from the identity $\sdg e_n\left(\sqrt{g^{-1}f}\right) = \sdf e_{4-n}\left(\sqrt{f^{-1}g}\right)$ \cite{Hassan:2011zd}. This can easily be seen by formulating the $e_n$ polynomials in terms of the eigenvalues of $\mathbb X$ as in \cref{eq:eneigenvalues}. The result follows from using basic properties of the determinant and the fact that, because $\sqrt{g^{-1}f}$ and $\sqrt{f^{-1}g}$ are inverses of each other, their eigenvalues are inverses as well. As a consequence, the entire Hassan-Rosen action in vacuum is invariant under the exchange of the two metrics up to parameter redefinitions,
\begin{equation}
g_\mn \leftrightarrow f_\mn, \qquad M_g \leftrightarrow M_f, \qquad \beta_n \to \beta_{4-n}. \label{eq:bigravinterchanges}
\end{equation}
The fact that the matter coupling breaks this duality by coupling matter to only one metric will motivate the search for ``double couplings'' in later chapters.

By varying the action (\ref{eq:actionHR}) with respect to the $g$ and $f$ metrics we obtain the generalised Einstein equations for massive bigravity \cite{Hassan:2011vm},
\begin{align}
G_\mn(g) + m^2\sum_{n=0}^3\left(-1\right)^{n}\beta_ng_{\mu\alpha}Y^{\alpha}_{(n)\nu}\left(\sqrt{g^{-1}f}\right) &= \frac{1}{M_g^2}T_{\mu\nu}, \label{eq:Einsteingeng}\\
G_\mn(f) + \frac{m^2}{M^2_\star}\sum_{n=0}^3\left(-1\right)^{n}\beta_{4 - n}f_{\mu\alpha}Y^{\alpha}_{(n)\nu}\left(\sqrt{f^{-1} g}\right) &= 0, \label{eq:Einsteingenf}
\end{align}
where $G_\mn$ is the Einstein tensor computed for a given metric. The interaction matrices $Y_{(n)}(\mathbb{X})$ are defined as
\begin{align}
  Y_{(0)}(\mathbb X) &\equiv \mathbb{I}, \quad \nonumber \\
  Y_{(1)}(\mathbb X) &\equiv \mathbb{X} - \mathbb{I}[\mathbb{X}],\quad \nonumber \\
  Y_{(2)}(\mathbb X) &\equiv \mathbb{X}^2 - \mathbb{X}[\mathbb{X}] + \frac{1}{2}\mathbb{I}\left([\mathbb{X}]^2 - [\mathbb{X}^2]\right), \nonumber \\  
  Y_{(3)}(\mathbb X) &\equiv \mathbb{X}^3 - \mathbb{X}^2[\mathbb{X}] + \frac{1}{2}\mathbb{X}\left([\mathbb{X}]^2 - [\mathbb{X}^2]\right) \nonumber \\
  &\hphantom{{}\equiv} - \frac{1}{6}\mathbb{I}\left([\mathbb{X}]^ 3 - 3[\mathbb{X}][\mathbb{X}^2] + 2[\mathbb{X}^3]\right). \label{eq:ymatdef}
\end{align}
Notice that they satisfy the relation \cite{Hassan:2012wr}
\begin{equation}
Y_{(n)}(\mathbb X) = \displaystyle\sum_{i=0}^n(-1)^i\mathbb X^{n-i}e_i(\mathbb X).
\end{equation}
The tensors $g_{\mu\lambda}Y^\lambda_{(n)\nu}$ are symmetric and so do not need to be explicitly symmetrised \cite{Hassan:2012wr}, although this fact has gone unnoticed in much of the literature. Finally, $T_{\mu\nu}$ is the stress-energy tensor defined with respect to the matter metric, $g$,
\begin{equation}
T_{\mu\nu} \equiv - \frac{2}{\sqrt{-\det g}}\frac{\delta(\sqrt{-\det g}\mathcal{L}_m^g)}{\delta g^{\mu\nu}}.
\end{equation}
It is not difficult to check that when $T_\mn=0$, the Einstein equations are symmetric under the interchanges (\ref{eq:bigravinterchanges}).

General covariance of the matter sector implies conservation of the stress-energy tensor as in general relativity,
\begin{equation}\label{eq:ParticleEOM}
\nabla_g^\mu T_{\mu\nu} = 0.
\end{equation}
Furthermore, by combining the Bianchi identities for the $g$ and $f$ metrics with the field equations (\ref{eq:Einsteingeng}) and (\ref{eq:Einsteingenf}), we obtain the following two Bianchi constraints on the mass terms:
\begin{align}
 \nabla_g^{\mu}\frac{m^2}{2}\sum_{n=0}^3\left(-1\right)^{n}\beta_ng_{\mu\alpha}Y^{\alpha}_{(n)\nu}\left(\sqrt{g^{-1}f}\right) &= 0, \label{eq:Bianchig}\\
 \nabla_f^{\mu}\frac{m^2}{2M^2_\star}\sum_{n=0}^3\left(-1\right)^{n}\beta_{4 - n}f_{\mu\alpha}Y^{\alpha}_{(n)\nu}\left(\sqrt{f^{-1} g}\right) &= 0, \label{eq:Bianchif}
\end{align}
after using \cref{eq:ParticleEOM}. Only one of \cref{eq:Bianchig,eq:Bianchif} is independent: a linear combination of the two divergences can be formed which vanishes as an \emph{identity}, i.e., regardless of whether $g_\mn$ and $f_\mn$ satisfy the correct equations of motion \cite{Akrami:2013ffa}, so either of the Bianchi constraints implies the other.

\subsubsection{Field Equations for Massive Gravity}

Note that we can also easily obtain the equations of motion for dRGT massive gravity from the bimetric equations: the Einstein equation is simply \cref{eq:Einsteingeng} with $f_\mn$ fixed to the desired reference metric,
\begin{equation}
G_\mn + m^2\sum_{n=0}^3\left(-1\right)^{n}\beta_ng_{\mu\alpha}Y^{\alpha}_{(n)\nu}\left(\sqrt{g^{-1}f}\right) = \frac{1}{\Mp^2}T_{\mu\nu}, \label{eq:EinsteindRGT}
\end{equation}
and matter is conserved with respect to $g_\mn$ as usual. This quick ``derivation'' should be taken purely as a heuristic---i.e., if we had started off with the dRGT action (\ref{eq:sdrgtalpha}) and varied with respect to $g_\mn$, we would clearly obtain \cref{eq:EinsteindRGT} regardless of whether $f_\mn$ is dynamical---and \emph{not} as the outcome of a limiting procedure. Indeed, because $f_\mn$ lacks dynamics there is no analogue of its Einstein equation (\ref{eq:Einsteingenf}), and including that equation would lead to extra constraints. It turns out that massive gravity can be obtained as a limit of bigravity, but the process is more subtle than simply taking $M_f\to\infty$ (which freezes out the massless mode and equates it with $f_\mn$, as discussed above) \cite{Baccetti:2012bk,Hassan:2014vja}. Alternatively, the dRGT action can be obtained from the bigravity action by taking $M_f\to0$, but to obtain massive gravity we must throw away the $f_\mn$ Einstein equation (\ref{eq:Einsteingenf}) by hand. If we leave it in then it is determined algebraically in terms of $g_\mn$.\footnote{This is because we are effectively taking the dRGT action and varying with respect to $f_\mn$, treating it like a Lagrange multiplier. Hence $f_\mn$ cannot be picked freely in this case but is rather constrained in terms of $g_\mn$.} Plugging this into the mass term we simply obtain a cosmological constant; thus this is the general-relativity limit of massive gravity.\footnote{We thank Fawad Hassan for helpful discussions on these points.} This agrees with the linear analysis above, where we found that in the limit $M_f\to0$, the fluctuations of $g_\mn$ become massless.

\subsection{Cosmological Solutions in Massive Bigravity}
\label{sec:mg-bigrav-cosmo}

In this subsection we review the homogeneous and isotropic cosmology of massive bigravity. We will follow the framework derived in \rcite{vonStrauss:2011mq,Akrami:2012vf}, and use, with some generalisations, the notation and approach summarised in \rcite{Solomon:2014dua}. As discussed above, we will rescale $f_\mn$ and $\beta_n$ so that the two Planck masses are equal, $M_f=M_g$.

\subsubsection{Cosmological Equations of Motion}
\label{sec:mg-bigrav-cosmo-eq}

We assume that, at the background level, the Universe can be described by Friedmann-Lema\^{\i}tre-Robertson-Walker (FLRW) metrics for both $g_\mn$ and $f_\mn$. Specialising to spatially-flat metrics, we have
\begin{align}
ds_g^2 &= -N(t)^2dt^2 + a(t)^2d\vec{x}^2, \label{eq:frwg}\\
ds_f^2 &= -X(t)^2 dt^2 + Y(t)^2 d\vec{x}^2, \label{eq:frwf}
\end{align}
where $a(t)$ and $Y(t)$ are the spatial scale factors for $g_\mn$ and $f_\mn$, respectively, and $N(t)$ and $X(t)$ are their lapses. In the rest of this thesis we will leave the time dependences of these functions implicit. We will find it useful to define the ratios of the lapses and scale factors,
\begin{equation}
 x \equiv \frac XN,\qquad y \equiv \frac Ya.
\end{equation}
Notice that these quantities are \emph{coordinate-independent}: while we can freely choose either lapse or rescale either scale factor, their ratio is fixed. This is because bigravity is still invariant under general coordinate transformations as long as the same transformation is acted on each metric.\footnote{In group-theoretic terms, there are two diffeomorphism groups, one for each metric, and bigravity breaks the symmetry under each of them but maintains the symmetry under their diagonal subgroup. This is obvious from the fact that the mass term only depends on the metrics in the combination $g^{\mu\alpha}f_{\alpha\nu}$.}

With these choices of metrics, the generalised Einstein equations (\ref{eq:Einsteingeng}) and (\ref{eq:Einsteingenf}), assuming a perfect-fluid source with density $\rho = -{T^0}_0$, give rise to two Friedmann equations,
\begin{align}
3H^2 &= \frac{N^2}{M_g^2}\rho + m^2N^2\left(\beta_0 + 3\beta_1y + 3\beta_2y^2 + \beta_3y^3\right), \label{eq:friedmann_g}\\
3K^2 &= m^2X^2\left(\beta_1y^{-3} + 3\beta_2y^{-2} + 3\beta_3y^{-1} + \beta_4\right),\label{eq:friedmann_f}
\end{align}
where we have defined the Hubble rates as\footnote{In order to present the cosmological solutions for general lapses, we will define the $g$-metric Hubble rate differently here than in the rest of this thesis; in particular, $H$ is not necessarily the cosmic-time Hubble rate.}
\begin{equation}
H \equiv \dot{a}/{a}, \qquad K \equiv \dot{Y}/{Y},
\end{equation}
and overdots denote time derivatives. We will specialise in this thesis to pressureless dust, which obeys
\begin{equation}
\dot\rho + 3H\rho = 0. \label{eq:rhocons}
\end{equation}

The Bianchi constraint---either \cref{eq:Bianchig} or \cref{eq:Bianchif}---yields
\begin{equation}
m^{2}a^2P\left(X\dot a-N\dot Y\right)=0, \label{eq:bianchiconstraint-mg}
\end{equation}
where we have defined
\begin{equation}
P \equiv \beta_1+2\beta_2 y + \beta_3 y^2.
\end{equation}
The Bianchi constraint has two branches of solutions:
\begin{align}
\text{Algebraic branch:} && P&=0, \nonumber \\
\text{Dynamical branch:} && x &= \frac{\dot Y}{\dot a}. \nonumber
\end{align}
The algebraic branch is satisfied if $\beta_1 + 2\beta_2y + \beta_3y^2=0$, which seems to be nongeneric as it requires tuned initial conditions. Because the solutions on this branch have $y=\mathrm{const.}$, the mass term in \cref{eq:friedmann_g} clearly reduces to a cosmological constant. Thus the algebraic branch, at the background level, is equivalent to $\Lambda$CDM \cite{vonStrauss:2011mq,Comelli:2011zm}. At the level of linear perturbations, evidence has been found for several modes being strongly coupled \cite{Comelli:2012db}. Consequently we will focus our attention on the dynamical branch. In this case, the Bianchi constraint implies that the ratio of the lapses, $x$, can be written in terms of other background quantities as
\begin{align}
x &= \frac{Ky}{H}.\label{eq:xkyh}
\end{align}

The Friedmann equations, (\ref{eq:friedmann_g}) and (\ref{eq:friedmann_f}), and the Bianchi identity (\ref{eq:xkyh}) can be combined to find a purely algebraic, quartic evolution equation for $y$,
\begin{equation}
\beta_3y^4 + \left(3\beta_2 - \beta_4\right)y^3 + 3\left(\beta_1 - \beta_3\right)y^2 + \left(\frac{\rho}{M_g^2m^2} + \beta_0 - 3\beta_2\right)y - \beta_1 = 0.\label{eq:quartic-mg}
\end{equation}
The $g$-metric Friedmann equation (\ref{eq:friedmann_g}) and quartic equation (\ref{eq:quartic-mg}) completely determine the expansion history of the Universe. As in standard cosmology, we see that the cosmic expansion is governed by a Friedmann equation. It is sourced by a mass term that depends on $y$, the evolution of which is in turn determined by the quartic equation.

It will be useful to simplify the dynamics by expressing all background quantities solely in terms of $y(t)$ and then solving for $y(a)$. We can rearrange \cref{eq:quartic-mg} to solve for $\rho(y)$,
\begin{equation}
 \frac{\rho}{m^2M_g^2} = -\beta_3y^3 + (\beta_4 - 3\beta_2)y^2 + 3(\beta_3 - \beta_1)y + 3\beta_2 - \beta_0 + \beta_1y^{-1}. \label{eq:breprho}
\end{equation}
We can then substitute this into the Friedmann equation to find $H(y)$,
\begin{equation}
 3H^2 = m^2N^2\left(\beta_4y^2 + 3\beta_3 y + 3\beta_2 + \beta_1y^{-1}\right). \label{eq:brepH}
\end{equation}
By taking a derivative of the quartic equation (\ref{eq:quartic-mg}) and using the fluid conservation equation (\ref{eq:rhocons}) and our solution (\ref{eq:breprho}) for $\rho(y)$ we can find an evolution equation for $y(a)$,
\begin{align}
\frac{d\ln y}{d \ln a} = \frac{\dot y}{Hy} &= -3\frac{\beta_3y^4 + (3\beta_2 - \beta_4)y^3 + 3(\beta_1 - \beta_3)y^2 + (\beta_0 - 3\beta_2)y - \beta_1}{3\beta_3y^4 + 2(3\beta_2-\beta_4)y^3 + 3(\beta_1-\beta_3)y^2 + \beta_1}. \label{eq:brepyp}
\end{align}
Using the definition of $y$ to find $\dot y$, we can easily write $K(y)$,
\begin{align}
K = H + \frac{\dot y}{y}. \label{eq:brepK}
\end{align}
We can now write any background quantity in terms of $y$ alone, except for $y$ and $a$ themselves, and further we have two avenues for determining $y(a)$: integrating \cref{eq:brepyp}, or using the quartic equation (\ref{eq:quartic-mg}) with $\rho=\rho_0a^{-3}$. These expressions will be crucial throughout this thesis since they reduce the problem of finding any parameter---background or perturbation---to solving for $y(z)$, where $z=1/a-1$ is the redshift.\footnote{Note that while we have expressed all background quantities in terms of $y$ only, perturbations will in general depend on both $y$ and $a$.}

We note briefly that there has been some discussion in the literature over how to correctly take square roots in bigravity. There exist cosmological solutions in which $\det\sqrt{g^{-1}f}$ becomes zero at a finite point in time (and only at that time), and so it is important to determine whether to choose square roots to always be positive, per the usual mathematical definition, or to change sign on either side of the point where $\det\sqrt{g^{-1}f}=0$. This was discussed in some detail in \rcite{Gratia:2013gka} (see also \rcite{Gratia:2013uza}), where continuity of the vielbein corresponding to $\sqrt{g^{-1}f}$ demanded that the square root \emph{not} be positive definite. We will take a similar stance here, and make the only choice that renders the action differentiable at all times, i.e., such that the derivative of $\sqrt{g^{-1}f}$ with respect to $g_{\mu\nu}$ and $f_{\mu\nu}$ is continuous everywhere. In particular, for the FLRW backgrounds we are dealing with in this section, this choice implies that $\sqrt{-\det f}=XY^3$. This is important because, as we will see in \cref{chap:bigravity-stability}, it turns out that in the only cosmology with linearly-stable perturbations, the $f$ metric bounces, so $X=KY/H$ changes sign during cosmic evolution. With our square-root convention, the square roots will change sign as well, rather than develop cusps. Note that sufficiently small perturbations around the background will not lead to a different sign of this square root.

\subsubsection{Properties of Bimetric Cosmologies}
\label{sec:properties-bimetric-cosmo}

We can understand the qualitative behaviour of bimetric cosmologies by taking the early- and late-time limits, $\rho\to\infty$ and $\rho\to0$, respectively. We will use heuristic arguments to motivate results which were determined more rigorously in \rcite{Konnig:2013gxa} and can also be seen from a statistical comparison to observations of the expansion history \cite{Akrami:2012vf}. At early times, the quartic equation (\ref{eq:quartic-mg}) is solved either by $y\to0$ or $y\to\infty$. The former solution is quite easy to see: the quartic equation is of the form $\ldots + \rho y = 0$, where $\ldots$ contains only positive powers of $y$, so $y\to0$ will clearly be a solution. These are called \emph{finite-branch} solutions. The solutions with $y\to\infty$ at early times, or \emph{infinite-branch} solutions, occur when one of the higher powers of $y$ in the quartic equation scales at just the right rate to cancel out the $\rho y$ term. These solutions are rather less common; in order to enforce $\Omega_\mathrm m \to 1$ and $H^2>0$ at early times, viable infinite-branch solutions require $\beta_2=\beta_3=0$ and $\beta_4>0$ \cite{Konnig:2013gxa}. To see this, notice that $\Omega_\mathrm m = N^2\rho/(3M_g^2H^2)$ can be written in the limit $y\to\infty$, using \cref{eq:breprho,eq:brepH}, as
\begin{equation}
 \Omega_\mathrm m = 1 -\frac{\beta_3}{\beta_4}y + 3\frac{\beta_3^2}{\beta_4^2} - 3\frac{\beta_2}{\beta_4}.
\end{equation}
The condition $\beta_4>0$ (rather than just $\beta_4\neq0$) arises from demanding, per \cref{eq:brepH}, that $H^2$ be positive at all times.

On either branch, at late times $y$ will always flow to a constant, $y_c$, given by the quartic equation with $\rho=0$,
\begin{equation}
 \beta_3y_c^4 + \left(3\beta_2 - \beta_4\right)y_c^3 + 3\left(\beta_1 - \beta_3\right)y_c^2 + \left(\beta_0 - 3\beta_2\right)y_c - \beta_1 = 0. \label{eq:yc}
\end{equation}
Moreover, by taking a derivative of the quartic equation we see that $\dot y\neq0$ unless either $\dot\rho=0$ or $y=0$. Therefore $y$ evolves monotonically throughout cosmic history, flowing either from $0$ up to $y_c$ or from $\infty$ down to $y_c$.\footnote{Note that these are not necessarily the same $y_c$, as \cref{eq:yc} can have multiple roots.}

As long as $y_c>0$, the mass term asymptotes in the future to a cosmological constant. Hence bimetric cosmologies generally possess late-time acceleration with $H^2 \sim \mathcal{O}(m^2\beta_i)$. This is the case even if the $g$-metric cosmological constant, $\beta_0$, is turned off, hence these theories \emph{self-accelerate}. Because $y$ cannot be constant in these models,\footnote{Unless it is either $0$ at all times, which is trivial, or the special case $\beta_1=\beta_3=0$, in which case the Friedmann equation can be rewritten, with the help of the quartic equation, as a $\Lambda$CDM Friedmann equation with a rescaled gravitational constant \cite{vonStrauss:2011mq}.} the effective dark energy is dynamical. In particular, we do not have $w=-1$ except at the asymptotic future. Crucially, the parameters and the potential structure leading to the accelerated expansion are thought to be stable under quantum corrections \cite{deRham:2013qqa}, in stark contrast to a cosmological constant, which would need to be fine-tuned against the energy of the vacuum \cite{Weinberg:1988cp,Martin:2012bt,Burgess:2013ara}.\footnote{If matter couples to $g_\mn$ then matter loops will still contribute to $m^2\beta_0$ as usual. It is the rest of the dRGT potential which is stable under quantum corrections. Consequently we focus on self-accelerating models and assume---as is common in the literature---that some unknown mechanism removes the dangerous cosmological constant.} Thus we find that bigravity is an excellent candidate for technically-natural self-acceleration. Comparisons to background data---specifically the cosmic microwave background, baryon acoustic oscillations, and type Ia supernovae---show that these cosmological models can agree well with observations \cite{vonStrauss:2011mq,Akrami:2012vf,Konnig:2013gxa}.

Before ending this subsection, let us consider a worked example: the model with only $\beta_1$ nonzero. Because theoretical viability conditions require this term to be nonzero (ignoring the exact $\Lambda$CDM case with $\beta_1=\beta_3=0$), it is the simplest nontrivial one-parameter model which will lead to sensible cosmologies \cite{Konnig:2013gxa}. It has been shown in \rcite{Akrami:2012vf,Konnig:2013gxa} that this model provides a self-accelerating evolution which agrees with background cosmological observations and, as it possesses the same number of free parameters as the standard $\Lambda$CDM model, is a viable alternative to it. Indeed, it may be more viable than $\Lambda$CDM if the graviton mass turns out to be stable to quantum corrections, as mentioned above. The graviton mass in this case is given by $\sqrt{\beta_1}m$. Note that in order to give rise to acceleration at the present era, the graviton mass typically should be comparable to the present-day Hubble rate, $\beta_1m^2\sim H_0^2$.

In this simple case, the evolution equation (\ref{eq:brepyp}) for $y(a)$,
\begin{equation}
 \frac{d\ln y}{d\ln a} = -3\left(1-\frac{2}{1+3y^2}\right),
\end{equation}
can be integrated exactly to find
\begin{equation}
 y(a) = \frac 1 6 a^{-3}\left(-C \pm \sqrt{12a^6 + C^2} \right). \label{eq:MBMya}
\end{equation}
Assuming $y>0$ forces us to select the positive branch. Using the Friedmann and quartic equations, we can set the value of $C$ using initial conditions,
\begin{equation}
 C = -\frac{m^2\beta_1}{H_0^2} + 3\frac{H_0^2}{\beta_1m^2},
\end{equation}
where $H_0$ is the cosmic-time Hubble rate today. Equivalently, we can use the quartic equation to solve for $y$ and express the Friedmann equation as a modified expression for $H(\rho)$,
\begin{equation}
 H^2 = \frac{N^2}{6M_g^2}\left(\rho + \sqrt{12m^4M_g^4\beta_1^2 + \rho^2}\right).
\end{equation}
In either formulation, the late-time approach to a $\Lambda$-like behaviour is evident.

The minimal $\beta_1$-only model is also distinctive for having a phantom equation of state, $w(z)\approx-1.22_{-0.02}^{+0.02}-0.64_{-0.04}^{+0.05}z/(1+z)$ at small redshifts. Moreover, $w$ is related in a simple way to the matter density parameter \cite{Konnig:2013gxa}. This provides a concrete and testable prediction of the model that can be verified by future large-scale structure experiments, such as Euclid \cite{Laureijs:2011gra,Amendola:2012ys}, intensity mappings of neutral hydrogen \cite{Chang:2007xk,Bull:2014rha}, and combinations of structure and cosmic microwave background measurements \cite{Majerotto:2012mf}. The model has also been proven in \cite{Fasiello:2013woa} to satisfy an important stability bound at all times, avoiding the \emph{Higuchi ghost} which plagues theories of a massive graviton on expanding backgrounds \cite{Higuchi:1986py}. It is, however, worth noting that its linear cosmological perturbations are unstable until $z\sim0.5$, as shown in \cref{chap:bigravity-stability} of this thesis. We emphasise that this instability does not rule out the $\beta_1$-only model, but rather impedes our ability to use linear perturbation theory to describe perturbations at all times. This raises the interesting question of how to make predictions for structure formation during the unstable period, a question which is beyond the scope of this thesis.

The aforementioned studies have largely been restricted to the background cosmology of the theory. As the natural next step, in \cref{chap:bigravity-stability,chap:bigravity-subhorizon} we will extend the predictions of massive bigravity to the perturbative level, and study how consistent the models are with the observed growth of structures in the Universe.

\subsubsection{A No-Go Theorem for Massive Cosmology}
\label{sec:drgt-no-go}

As discussed in the previous section, we can easily obtain the equations of motion for massive gravity from those of bigravity; the $g$-metric equation is the same, and we lose the $f$-metric equation. Note that we also still have the Bianchi constraint (\ref{eq:bianchiconstraint-mg}). This fact will turn out to be crucial.

Let us assume that the reference metric is Minkowski, $f_\mn = \eta_\mn$. Under the assumption of homogeneity and isotropy for $g_\mn$ in unitary gauge, the Friedmann equation is simply \cref{eq:friedmann_g} with $y=a^{-1}$. This alone would define perfectly acceptable cosmologies. However, the Bianchi constraint (\ref{eq:bianchiconstraint-mg}) causes trouble. In the bigravity language we now have $X=Y=1$, so the constraint becomes
\begin{equation}
 m^2a^2P\dot a = 0.
\end{equation}
Because $\dot Y = 0$, we no longer have an interesting dynamical branch: it merely suggests $\dot a = 0$. Unfortunately, the algebraic branch, $P=0$, also only has $a=\mathrm{const.}$ as a solution; in bigravity it would fix $y=Y/a$ while allowing $a$ and $Y$ to change, but in massive gravity, it is $a$ which is fixed. Therefore $a$ is generally fixed to be constant, and this system has no dynamical solutions. This is the famous no-go theorem on cosmological solutions in dRGT massive gravity, and it is present for both flat and closed universes \cite{D'Amico:2011jj}. If we instead consider open universes or different reference metrics, such as FLRW or de Sitter, then FLRW solutions do exist, but they are unstable to the aforementioned Higuchi ghost and other linear and nonlinear instabilities
\cite{Gumrukcuoglu:2011ew,Gumrukcuoglu:2011zh,Vakili:2012tm,DeFelice:2012mx,Fasiello:2012rw,DeFelice:2013awa}.

The search for a viable cosmology with a massive graviton which avoids these conclusions has involved two routes. One is to extend dRGT by adding extra degrees of freedom. As discussed above, these problems are cured when the second metric is given dynamics. Other extensions of massive gravity, such as quasidilaton \cite{D'Amico:2012zv}, varying-mass \cite{D'Amico:2011jj,Huang:2012pe}, nonlocal \cite{Jaccard:2013gla,Foffa:2013vma,Dirian:2014ara}, and Lorentz-violating \cite{Comelli:2013paa,Comelli:2013tja} massive gravity, also seem to possess improved cosmological behaviour. The other approach is to give up on homogeneity and isotropy entirely. While FLRW solutions are mathematically simple, the Universe could in principle have anisotropies which have such low amplitude, are so much larger than our horizon, or both, that we cannot easily observe them. Remarkably, these cosmologies are much better behaved in massive gravity than is the standard FLRW case \cite{D'Amico:2011jj}. The general scenario of an FLRW metric with inhomogeneous St\"uckelberg fields has been derived in \rcite{Volkov:2012cf,Volkov:2012zb}. This includes, but is not limited to, the case in which the reference metric is still Minkowski space, but only has the canonical form $\eta_\mn = \operatorname{diag}(-1,1,1,1)$ in coordinates where $g_\mn$ is not of the FLRW form \cite{Gratia:2012wt}. The inhomogeneous and anisotropic solutions are reviewed thoroughly in \rcite{deRham:2014zqa}. See \rcite{DeFelice:2013bxa} for a review of cosmology in massive gravity and some of its extensions.

\section{Einstein-Aether Theory}
\label{sec:mg-aether}

In the final section of this chapter, we explore a different route to modifying gravity: allowing Lorentz symmetry to be violated. This is not a step taken lightly; Lorentz invariance is a cornerstone of modern physics. The two theories which have been separately successful at predicting nearly all experimental and observational data to date, general relativity to explain the structure of spacetime and gravity and the standard model of particle physics to describe particles and nongravitational forces in the language of quantum field theory, both contain Lorentz symmetry as a crucial underlying tenet.

What do we gain from exploring the breakdown of this fundamental symmetry? Given its foundational significance, the consequences of violating Lorentz invariance deserve to be fully explored and tested. Indeed, while experimental bounds strongly constrain possible Lorentz-violating extensions of the standard model \cite{Mattingly:2005re}, Lorentz violation confined to other areas of physics---such as the gravitational, dark, or inflationary sectors---is somewhat less constrained, provided that its effects are not communicated to the matter sector in a way that would violate the standard-model experimental bounds. Moreover, it is known that general relativity and the standard model should break down around the Planck scale and be replaced by a new, quantum theory of gravity. If Lorentz symmetry proves not to be fundamental at such high energies---for instance, because spacetime itself is discretised at very small scales---this may communicate Lorentz-violating effects to gravity at lower energies, which could potentially be testable. The study of possible consequences of its violation, and the extent to which they can be seen at energies probed by experiment and observation, may therefore help us to constrain theories with such behaviour at extremely high energies.

A pertinent recent example is Ho\v{r}ava-Lifschitz gravity, a potential ultraviolet completion of general relativity which achieves its remarkable results by explicitly treating space and time differently at higher energies \cite{Horava:2009uw}. The consistent nonprojectable extension \cite{Blas:2009qj,Blas:2009ck,Blas:2010hb} of Ho\v{r}ava-Lifschitz gravity is closely related to the model we will explore. Moreover, since we will be dealing with Lorentz violation in the gravitational sector, through a vector-tensor theory of gravity, the usual motivations for modifying gravity apply to this kind of Lorentz violation. Indeed, there are interesting models of cosmic acceleration, based on the low-energy limit of Ho\v{r}ava-Lifschitz gravity, in which the effective cosmological constant is technically natural \cite{Blas:2011en,Audren:2013dwa}. Generalised Lorentz-violating vector-tensor models have also been considered as candidates for both dark matter and dark energy \cite{Zlosnik:2006zu,Zuntz:2010jp}.

Lorentz violation need not have such dramatic, high-energy origins. Indeed, many theories with fundamental Lorentz violation may face fine-tuning problems in order to avoid low-energy Lorentz-violating effects that are several orders of magnitude greater than existing experimental constraints \cite{Collins:2004bp}. However, even a theory which possesses Lorentz invariance at high energies could spontaneously break it at low energies, and with safer experimental consequences.

Spontaneous violation of Lorentz invariance will generally result when a field that transforms nontrivially under the Lorentz group acquires a vacuum expectation value (VEV). A simple example is that of a vector field whose VEV is nonvanishing everywhere. As mentioned above, in order to avoid the experimental constraints such a vector field should not be coupled to the standard model fields, but in order to not be completely innocuous we will ask it to couple to gravity. Moreover, to model Lorentz violation in gravity without abandoning the successes of general relativity---in particular, without giving up general covariance---the (spontaneously) Lorentz-violating field must be a spacetime vector and must be dynamical.\footnote{The requirement that the field be dynamical stems from the fact that there is no nontrivial (i.e., nonzero) spacetime vector which is covariantly constant, i.e., if $\nabla_\mu u^\nu=0$ everywhere then necessarily $u^\nu=0$.}

A particularly simple, yet quite general, example of a model with these features is Einstein-aether theory (\ae-theory) \cite{Jacobson:2000xp,Jacobson:2008aj}. It adds to general relativity a dynamical, constant-length timelike vector field, called the aether and denoted by $u^{a}$, which spontaneously breaks Lorentz invariance by picking out a preferred frame at each point in spacetime while maintaining local rotational symmetry, thus breaking only the boost sector of Lorentz symmetry \cite{Eling:2004dk,Jacobson:2008aj}. The constant-length constraint plays two crucial roles. The first is phenomenological: it ensures that the aether picks a globally-nonzero VEV and so guarantees that Lorentz symmetry is in fact broken. The other role is to ensure that the theory is not sick: if the length is not fixed and the kinetic term is not gauge-invariant\footnote{Note that gauge invariance would uniquely pick out the Maxwell term, in which case we would simply have electromagnetism which clearly does not spontaneously break Lorentz symmetry.} then the length-stretching mode has a wrong-sign kinetic term and hence is ghostlike \cite{Elliott:2005va}. Note that \ae-theory is the most general effective field theory in which the rotation group is unbroken \cite{ArmendarizPicon:2010mz}, and hence it can be seen as the low-energy limit of any theory which violates boosts but maintains rotational symmetry.

\subsection{Pure Aether Theory}

Einstein-aether theory (which we will often refer to as ``pure'' Einstein-aether theory or \ae-theory) is a theory of the spacetime metric $g_{\mu\nu}$ and a vector field (the ``aether'') $u^{\mu}$. The action is \cite{Jacobson:2008aj,Carroll:2004ai}
\begin{equation}
S=\int d^{4}x\sqrt{-g}\left[  \frac{1}{16\pi G}R-K^{\mu\nu}{}_{\rho\sigma
}\nabla_{\mu}u^{\rho}\nabla_{\nu}u^{\sigma}+\lambda\left(  u^{\mu}u_{\mu
}+m^{2}\right)  \right]  ,\label{eq:aeaction}
\end{equation}
where we have defined
\begin{equation}
K^{\mu\nu}{}_{\rho\sigma}\equiv c_{1}g^{\mu\nu}g_{\rho\sigma}+c_{2}
\delta_{\rho}^{\mu}\delta_{\sigma}^{\nu}+c_{3}\delta_{\sigma}^{\mu}
\delta_{\rho}^{\nu}+c_{4}u^{\mu}u^{\nu}g_{\rho\sigma}.
\end{equation}
The action (\ref{eq:aeaction}) contains an Einstein-Hilbert term for the metric, a kinetic term
for the aether with four dimensionless coefficients $c_{i}$ (coupling the
aether to the metric through the covariant derivatives), and a nondynamical
Lagrange multiplier $\lambda$. Varying this action with
respect to $\lambda$ constrains the aether to be timelike with a constant norm,
$u^{\mu}u_{\mu}=-m^{2}$. The aether has units of mass; its length, $m$, has the same dimensions and corresponds to the Lorentz symmetry breaking scale.

The action (\ref{eq:aeaction}) is the most general diffeomorphism-invariant
action containing the metric, aether, and up to second derivatives of each. Higher derivatives are excluded because they would generically lead to ghostlike degrees of freedom. Most terms one can write down involving the aether are eliminated by the fixed norm condition. One could consider a term $R_\mn u^\mu u^\nu$, but this is equivalent under integration by parts to $(\nabla_\mu u^\mu)^2 + (1/2)F_\mn F^\mn - (\nabla_\mu u_\nu)(\nabla^\mu u^\nu)$, where $F_\mn = 2\nabla_{[\mu}u_{\nu]}$ is the field strength tensor, and so is already included in the \ae-theory action \cite{Jacobson:2000xp}. In what follows we will follow much of the literature on aether cosmology (e.g., \rcite{Carroll:2004ai,Lim:2004js}) and ignore the quartic self-interaction term by setting $c_{4}=0$.

It is generally assumed that (standard-model) matter fields couple to the metric only. Any coupling to the aether
would lead to Lorentz violation in the matter sector by inducing different maximum propagation speeds for different fields, an effect which is strongly constrained by experiments \cite{Mattingly:2005re}. These problematic standard-model couplings may, however, be forbidden by a supersymmetric extension of \ae-theory \cite{Pujolas:2011sk}. The work on \ae-theory which we detail in \cref{chap:aether} will be interested in exploring and constraining Lorentz violation in the gravitational sector and in a single non-standard-model scalar, hence we will not need to worry about such a coupling.

The gravitational constant $G$ that appears in \cref{eq:aeaction} is to be
distinguished from the gravitational constants which appear in the Newtonian
limit and in the Friedmann equations, both of which are modified by the
presence of the aether \cite{Carroll:2004ai}. The Newtonian gravitational
constant, $G_{N}$, and cosmological gravitational constant, $\gc$, are
related to the bare constant $G$ by
\begin{align}
G_{N} &  =\frac{G}{1+8\pi G\delta},\\
G_{c} &  =\frac{G}{1+8\pi G\alpha},
\end{align}
where
\begin{align}
\delta &  \equiv-c_{1}m^{2},\\
\alpha &  \equiv(c_{13}+3c_{2})m^{2}. \label{eq:alphadef}
\end{align}
We have introduced the notation $c_{13}\equiv c_{1}+c_{3}$, etc., which we
will use throughout.

\subsection{Coupling to a Scalar Inflaton}
\label{sec:aescalcoupling}

We now introduce to the theory a canonical scalar field $\phi$ which is
allowed to couple kinetically to the aether through its \emph{expansion}, $\theta\equiv\nabla_{\mu}u^{\mu}$ \cite{Donnelly:2010cr}. The full action reads
\begin{equation}
S=\int d^{4}x\sqrt{-g}\left[  \frac{1}{16\pi G}R - K^{\mu\nu}{}_{\rho\sigma
}\nabla_{\mu}u^{\rho}\nabla_{\nu}u^{\sigma}+\lambda\left(  u^{\mu}u_{\mu
}+m^{2}\right)  -\frac{1}{2}(\partial\phi)^{2}-V(\theta,\phi)\right]
.\label{eq:fullaction}
\end{equation}
Let us pause to motivate the generality of this model. Our aim in \cref{chap:aether} will be to constrain couplings between a Lorentz-violating field and a scalar, in particular a canonical, slowly-rolling scalar inflaton, in as general a way as possible. As mentioned above, Einstein-aether theory is the unique Lorentz-violating effective field theory in which rotational invariance is maintained \cite{ArmendarizPicon:2010mz},\footnote{However, we note that there is an allowed term, the quartic self-interaction parametrised by $c_4$, which we have turned off.} so any theory which spontaneously violates Lorentz symmetry without breaking rotational invariance will be described by the vector-tensor sector of our model at low energies. As for the scalar sector, the main restriction is that we have assumed a canonical kinetic term. While there are certainly coupling terms between the aether and the scalar which do not fall under the form $V(\theta,\phi)$, all of these terms have mass dimension greater than 4 and so are \emph{irrelevant} operators. Such terms are nonrenormalisable. While this is not necessarily disastrous from an effective field theory perspective, these terms are also nevertheless mostly important at short distances, and so should not factor into the cosmological considerations at the heart of this thesis. To see that all terms with dimension 4 or less fall into the framework (\ref{eq:fullaction}), notice that the aether, scalar, and derivative operators all have mass dimension 1, the aether norm is constant so $u_\mu u^\mu$ cannot be used in the coupling, and the aether and derivative operators carry spacetime indices which need to be contracted. Subject to these constraints, one can see that any terms which involve both $u^\mu$ and $\phi$ and have dimension 4 or less are either of the form $f(\theta,\phi)$ or can be recast into such a form under integration by parts. In particular, the only nontrivial interaction operators which are not irrelevant are $\phi\theta$ (dimension 3) and $\phi^2\theta$ (dimension 4).

This type of coupling was originally introduced with a more phenomenological motivation \cite{Donnelly:2010cr}. In a homogeneous and isotropic background, the aether aligns with the cosmic rest frame, so $\theta$ is essentially just the volume Hubble parameter, $\theta=3mH$. Hence the introduction of the aether allows a scalar inflaton to couple directly to the expansion rate. This is impossible in GR where $H$ is not proportional to any Lorentz scalar.

The aether equation of motion, obtained by varying the action with respect to
$u^{\mu}$, is
\begin{equation}
\lambda u^{\nu}= \nabla_{\mu}J^{\mu\nu}-\frac{1}{2}\nabla^{\nu}V_{\theta} \label{eq:aethereom}
\end{equation}
where the current tensor is defined by
\begin{equation}
J^{\mu}{}_{\sigma}\equiv -K^{\mu\nu}{}_{\sigma\rho}\nabla_{\nu}u^{\rho},
\end{equation}
and we are denoting partial derivatives of the potential by $V_\theta \equiv \partial V/\partial \theta$ and $V_\phi \equiv \partial V / \partial\phi$. Projecting this equation along $u^{\mu}$ allows us to obtain the Lagrange
multiplier $\lambda$,
\begin{equation}
\lambda= -\frac{1}{m^{2}} u_{\nu}\nabla_{\mu}J^{\mu\nu} + \frac{1}{2m^{2}
}u^{\mu}\nabla_{\mu}V_{\theta}.
\end{equation}
We will account for the modification to gravity by leaving the Einstein equations in the standard form (\ref{eq:einstein-intro}) and defining a stress-energy tensor for the combined aether-scalar system. Taking into
account the contribution from the Lagrange multiplier term, this is given by
\begin{equation}
T_{\mu\nu} = 2\frac{\delta\mathcal{L}}{\delta g^{\mu\nu}} + u^{\rho}
\frac{\delta\mathcal{L}}{\delta u^{\rho}}u_{\mu}u_{\nu}- \mathcal{L}g_{\mu\nu}
\end{equation}
where $\mathcal{L}$ is the Lagrangian for the aether and scalar. Using this
formula we find the stress-energy tensor,
\begin{align}
T_{\mu\nu} ={} &  2c_{1}(\nabla_{\mu}u^{\rho}\nabla_{\nu}u_{\rho}-
\nabla^{\rho}u_{\mu}\nabla_{\rho}u_{\nu})\nonumber\\
&  - 2[\nabla_{\rho}(u_{(\mu} J^{\rho}{}_{\nu)}) + \nabla_{\rho}(u^{\rho
}J_{(\mu\nu)}) - \nabla_{\rho}(u_{(\mu}J_{\nu)}{}^{\rho})]\nonumber\\
&  - 2m^{-2}u_{\sigma}\nabla_{\rho}J^{\sigma\rho} u_{\mu}u_{\nu}+ g_{\mu\nu
}\mathcal{L}_{u}\nonumber\\
&  + \nabla_{\mu}\phi\nabla_{\nu}\phi-\left( \frac{1}{2}\nabla_{\rho}
\phi\nabla^{\rho}\phi+V-\theta V_{\theta}\right) g_{\mu\nu}\nonumber\\
&  +(u^{\rho}\nabla_{\rho}V_{\theta})(g_{\mu\nu}+m^{-2}u_{\mu}u_{\nu
}),\label{eq:fullsetensor}
\end{align}	
where $\mathcal{L}_{u} \equiv K^{\mu\nu}{}_{\rho\sigma}\nabla_{\mu}u^{\rho
}\nabla_{\nu}u^{\sigma}$ is the Einstein-aether Lagrangian.
Finally, the inflaton obeys the usual Klein-Gordon equation,
\begin{equation}
\Box\phi=V_{\phi}.
\end{equation}
Notice that while this equation has the standard form, it couples the scalar to the aether since generally we will have $V_{\phi}=V_{\phi}(\theta,\phi)$.

Note that the equations of motion for the pure \ae-theory follow simply by setting $\phi=0$ and $V(\theta,\phi)=0$.

\subsection{Einstein-Aether Cosmology}
\label{sec:aecosmo}

In this section we examine the evolution of FLRW cosmological solutions in Einstein-aether theory. Consider a flat FLRW background geometry evolving
in conformal time, $\tau$,
\begin{equation}
ds^{2}=a^{2}(\tau)(-d\tau^{2}+d\vec{x}^{2}).\label{eq:ae-FRW}
\end{equation}
In pure \ae-theory, we take the $0$--$0$ and trace Einstein equations to obtain the Friedmann equations,
\begin{align}
\mathcal{H}^{2} &  =\frac{8\pi G_{c}}{3}a^{2}\rho_m,\\
\mathcal{H}^{\prime} &  =-\frac{4\pi G_{c}}{3}a^{2}\rho_{m}(1+3w),
\end{align}
where $\mathcal{H}\equiv a^{\prime}/a= d\ln a/d\tau$ is the
conformal time Hubble parameter. These are exactly the Friedmann equations of general relativity except that, as discussed above, the bare cosmological constant, $G$, is renormalised,
\begin{equation}
G_{c}=\frac{G}{1+8\pi G\alpha},
\end{equation}
with $\alpha=(c_{1}+3c_{2}+c_{3})m^{2}$.
The aether does not change the cosmological dynamics at all, but just modifies the gravitational constant. This arises because in a homogeneous and isotropic background the Einstein-aether terms for the vector field only contribute stress-energy that tracks the dominant matter fluid, so the associated energy density is proportional to $H^{2}$ \cite{Carroll:2004ai}.\footnote{Note, however, that perturbations of the aether do carry some dynamics \cite{Lim:2004js}.}

The aether does contribute dynamical stress-energy once we couple it to a scalar. In the theory (\ref{eq:fullaction}) the Friedmann equations are
\begin{align}
\mathcal{H}^{2} &  =\frac{8\pi G_{c}}{3}a^{2}\left(  V-\theta V_{\theta}
+\rho_{m}+\frac{1}{2}\phi^{\prime2}a^{-2}\right) \label{eq:friedmann} ,\\
\mathcal{H}^{\prime} &  =\frac{4\pi G_{c}}{3}a^{2}\left[  -3\frac{m}{a}\left(
3\frac{m}{a}V_{\theta\theta}(\mathcal{H}^{\prime}-\mathcal{H}^{2}
)+V_{\theta\phi}\phi^{\prime}\right)  -\rho_{m}(1+3w)+2(V-\theta V_{\theta
})-2\phi^{\prime2}a^{-2}\right]  , \label{eq:friedmann2}
\end{align}
For completeness we have included a matter component, but in the rest of this section and in \cref{chap:aether} we will assume that $\phi$ is gravitationally dominant and ignore any $\rho_m$. The scalar field obeys the usual cosmological Klein-Gordon equation,
\begin{equation}
\phi^{\prime\prime}+2{\mathcal{H}}\phi^{\prime} + a^2V_{\phi}=0.
\end{equation}
As discussed above, the coupling to $\theta$ is contained in the function $V_{\phi}$. In the background, $\theta=3mH$, with $H=\mathcal{H}/a$ the cosmic-time Hubble parameter, so this contributes either Hubble friction or a driving force \cite{Donnelly:2010cr}.

We need not write down the aether field equations in the background.
The vector field must be aligned with the cosmic rest frame due to homogeneity
and isotropy, and its value,
\begin{equation}
u^{\mu}=\frac ma\delta^{\mu}{}_{0},
\end{equation}
is
determined completely by the normalisation condition, $u_{\mu}u^{\mu}=-m^{2}$.
One can check that this solution satisfies the spatial component of the aether
equation, while the temporal component only determines the Lagrange
multiplier. In pure \ae-theory this solution is stable perturbatively
\cite{Lim:2004js,Kanno:2006ty,Li:2007vz,ArmendarizPicon:2010rs} and that
stability holds nonlinearly for most large perturbations \cite{Carruthers:2010ii}. This statement is subject to several constraints on the $c_i$ parameters which can be found in, e.g., \rcite{Lim:2004js,Jacobson:2008aj,ArmendarizPicon:2010rs}, and we will assume throughout this thesis that these constraints hold. One of the important results in \cref{chap:aether} is that the coupling between $u^\mu$ and $\phi$ can render cosmological solutions \emph{unstable} for large regions of parameter space that are allowed by other experimental, observational, and theoretical constraints.

When the scalar potential is $V(\theta,\phi)=V(\phi)$, the background aether is irrelevant apart from rescaling Newton's constant, and many choices for the potential can lead to periods
of slow-roll inflation \cite{Lyth:2009zz}. Adding a coupling to the aether
will change the dynamics but may still allow for inflation \cite{Donnelly:2010cr}. We will therefore aim to be as general about $V(\theta,\phi)$ as possible when discussing perturbation theory.

\cleardoublepage

\thispagestyle{empty}

		\vspace*{\fill}
		
		\begin{flushright}
		{\Huge{ \bf Part I}\\
		A Massive Graviton}
		\vspace{2cm}

\begin{quote}
\textit{It was therefore quite a shock when he said, ``But why should anybody be interested in getting exact solutions of such an ephemeral set of equations?''. I remember very well this word ``ephemeral.'' It meant that he did not consider his gravitational equations as the last word.}
\qauthor{Cornelius Lanczos, \textit{Einstein: The Man and His Achievement}}
\end{quote}

		\end{flushright}
\vspace*{\fill}

	\newpage
\thispagestyle{empty}

\newpage

\begin{savequote}[30pc]
There is nothing stable in the world; uproar's your only music.
\qauthor{John Keats, \textit{Letters}}
\end{savequote}

\chapter{Cosmological Stability of Massive Bigravity}
\label{chap:bigravity-stability}
\hrule
\vspace*{1cm}

In the previous chapter, and in particular in \cref{sec:massgrav}, we discussed an approach to modifying gravity in which its force-carrier particle, the graviton, is given a small mass. In particular, by specialising to the dRGT interaction potentials (\ref{eq:sdrgtbeta}) we ensure that the notorious Boulware-Deser ghost mode is absent, and by allowing both metrics to be dynamical and taking the graviton mass to be of the order of the present-day Hubble rate, we can obtain cosmological solutions which agree well with observations of the cosmic expansion history. These solutions are self-accelerating: the Hubble parameter goes to a constant at late times even in the absence of a cosmological constant. The action of this bimetric theory, or \emph{bigravity}, is given by \cref{eq:actionHR}, and the associated modified gravitational field equations were presented as \cref{eq:Einsteingeng,eq:Einsteingenf}.

We have discussed work comparing these FLRW solutions to tracers of the expansion history, most notably in \rcite{vonStrauss:2011mq,Akrami:2012vf,Konnig:2013gxa}. The natural next step is to move beyond the assumption of homogeneity and isotropy and allow for linear perturbations to FLRW which could describe the formation of large-scale structure. In particular, by employing the frequently-used subhorizon and quasistatic approximations we can dramatically simplify the complicated system of perturbed field equations while still capturing most observable linear modes. Indeed, this is precisely what we will do in \cref{chap:bigravity-subhorizon}.

The quasistatic limit is, however, a valid approximation only if the full system is stable for large wavenumbers. Previous work \cite{Comelli:2012db,DeFelice:2014nja,Comelli:2014bqa} has identified a region of instability in the past.\footnote{This should not be confused with the Higuchi ghost instability, which affects most massive gravity cosmologies and some in bigravity, but is, however, absent from the simplest bimetric models which produce $\Lambda$CDM-like backgrounds \cite{Fasiello:2013woa}.} The aim of this chapter is to investigate this problem in detail.

The rest of this chapter is organised as follows. In \cref{sec:stability-perts}, we present the linearised Einstein and fluid conservation equations in massive bigravity, which are derived in \cref{app:perteqs}, present a scheme for counting the number of independent, dynamical degrees of freedom, and then use that counting to pick a useful gauge. Putting this all together, we discuss how to reduce the system of ten Einstein and conservation equations to two equations for the two independent fields. In \cref{sec:bigrav-instabilities}, we solve these equations when the background can be assumed to be slowly varying, which is valid on small scales, and obtain the eigenfrequencies for the various bimetric models. With these in hand, we analytically determine the epochs of stability and instability for all the models with up to two free parameters which have been shown to produce viable cosmological background evolution. The behaviour of more complicated models can be reduced to these simpler ones at early and late times.

We show that several models which yield sensible background cosmologies in close agreement with the data are in fact plagued by an instability that only turns off at recent times. This does not necessarily rule out these regions of the bimetric parameter space, but rather presents a question of how to interpret and test these models, as linear perturbation theory is quickly invalidated. Remarkably, we find that only a particular bimetric model---in which only the $\beta_{1}$ and $\beta_{4}$ parameters are nonzero (that is, the linear interaction and the cosmological constant for the reference metric are turned on)---is stable at all times when the evolution is on a particular branch. This shows that a cosmologically-viable bimetric model without an explicit cosmological constant does indeed exist, and raises the question of how to nonlinearly probe other corners of bigravity. We summarise and discuss our results in \cref{sec:bigravity-stability-summary}.

\section{Linear Cosmological Perturbations}
\label{sec:stability-perts}

In this section we set up the formalism for cosmological perturbation theory in massive bigravity. We define the scalar perturbations to the FLRW metrics by extending \cref{eq:frwg,eq:frwf} to\footnote{By leaving the lapse $N$ in the $g$ metric general, we retain the freedom to later work in cosmic or conformal time. There is a further practical benefit: since this choice makes the symmetry between the two metrics manifest, and the action is symmetric between $g$ and $f$ as described in \cref{sec:mg-bigravity}, this means the $f$ field equations can easily be derived from the $g$ equations by judicious use of \texttt{ctrl-f}.}
\begin{align}
ds_g^2 &= -N^2(1+E_g)dt^2 + 2Na\partial_iF_gdt dx^i + a^2\left[(1+A_g)\delta_{ij} + \partial_i\partial_jB_g\right]dx^idx^j, \\
ds_f^2 &= -X^2(1+E_f)dt^2 + 2XY\partial_iF_fdt dx^i + Y^2\left[(1+A_f)\delta_{ij} + \partial_i\partial_jB_f\right]dx^idx^j,
\end{align}
where the perturbations $\{E_{g,f},A_{g,f},F_{g,f},B_{g,f}\}$ are allowed to depend on both time and space. Spatial indices are raised and lowered with the Kronecker delta. The stress-energy tensor is defined up to linear order by
\begin{align}
T{}^0{}_0 &= -\bar\rho(1+\delta), \nonumber \\
T{}^i{}_0 &= -\left(\bar\rho + \bar P\right)v^i, \nonumber \\
T{}^0{}_i &= \left(\bar\rho + \bar P\right)\left(v_i + \partial_iF_g\right), \nonumber \\
T{}^i{}_j &= \left(\bar P + \delta P\right)\delta^i{}_j + \Sigma{}^i{}_j, \label{eq:pertstresstens}
\end{align}
where $\delta$ is the density contrast, $v^i \equiv dx^i/dt$ is the 3-velocity, and $\Sigma^i{}_j$ is the anisotropic stress, with $\Sigma^i{}_i=0$. We specialise immediately to dust ($P=\delta P = \Sigma^i{}_j = 0$) and define the velocity divergence, $\theta \equiv \partial_i v^i$.

\subsection{Linearised Field Equations}
\label{sec:linearisedeqs}

The linearised Einstein and conservation equations are arrived at by a fairly lengthy computation. We summarise the results here; details on the derivation can be found in \cref{app:perteqs}. The $g$-metric Einstein equations are
\begin{itemize}

\item $0$--$0$:
\begin{equation}
\frac{3H}{N^2}\left(HE_g - \dot{A}_g\right) + \nabla^2 \left[\frac{A_g}{a^2} + H \left(\frac{2F_g}{Na} - \frac{\dot{B}_g}{N^2}\right) \right] + \frac{m^2}{2}yP\left(3\Delta A + \nabla^2\Delta B\right) = \frac{1}{M_g^2}\delta T{}^0{}_0,
\end{equation}

\item $0$--$i$:
\begin{equation}
\frac{1}{N^2}\partial_i\left(\dot A_g - HE_g\right) + m^2\frac{P}{x + y}\frac{Y}{N} \partial_i\left(xF_f - yF_g\right) = \frac{1}{M_g^2}\delta T{}^0{}_i,
\end{equation}

\item $i$--$i$:
\begin{align}
\frac{1}{N^2}\left[\left(2\dot H + 3H^2 - 2\frac{\dot N}{N}H\right)E_g + H\dot E_g - \ddot{A}_g - 3H\dot{A}_g + \frac{\dot N}{N}\dot A_g \right] + \frac{1}{2}\left(\partial_j^2 + \partial_k^2\right)D_g  \nonumber \\
+ m^2\left[\frac{1}{2}xP\Delta E + yQ\left(\Delta A+ \frac{1}{2}\left(\partial_j^2 + \partial_k^2\right)\Delta B\right)\right] = \frac{1}{M_g^2}\delta T{}^i{}_i, \label{eq:fullgii}
\end{align}

\item Off-diagonal $i$--$j$:
\begin{equation}
-\frac{1}{2}\partial^i\partial_j D_g - \frac{m^2}{2}yQ\partial^i\partial_j \Delta B = \frac{1}{M_g^2}\delta T{}^i{}_j,
\end{equation}

\end{itemize}
where $H\equiv\dot{a}/a$ is the usual $g$-metric Hubble parameter (in cosmic or conformal time, depending on $N$), $\partial_j^2 + \partial_k^2$ in the $i$--$i$ spatial Einstein equation refers to derivatives with respect to the other two Cartesian coordinates, i.e., $\nabla^2 - \partial^i \partial_i$ where the $i$ indices are not summed over, and we have defined
\begin{align}
P &\equiv \beta_1 + 2\beta_2y + \beta_3y^2, \\
Q &\equiv \beta_1 + \left(x + y\right)\beta_2 + xy\beta_3, \label{eq:pertQdef} \\
x &\equiv X/N, \\
y &\equiv Y/a, \\
\Delta A &\equiv A_f - A_g, \\
\Delta B &\equiv B_f - B_g, \\
\Delta E &\equiv E_f - E_g,
\end{align}
as well as
\begin{equation}
 D_g \equiv \frac{A_g + E_g}{a^2} + \frac{H}{N}\left(\frac{4F_g}{a} - \frac{3\dot{B}_g}{N}\right) + \frac{2\dot{F}_g}{Na} - \frac{1}{N^2}\left(\ddot{B}_g - \frac{\dot N}{N}\dot{B}_g\right).
\end{equation}
The linearised Einstein equations for the $f$ metric are
\begin{itemize}

\item $0$--$0$:
\begin{equation}
\frac{3K}{X^2}\left(KE_f - \dot{A}_f\right) + \nabla^2 \left[\frac{A_f}{Y^2} + K \left(\frac{2F_f}{XY} - \frac{\dot{B}_f}{X^2}\right) \right] - \frac{m^2}{2M_\star^2}\frac{P}{y^3}\left(3\Delta A + \nabla^2\Delta B\right) = 0,
\end{equation}

\item $0$--$i$:
\begin{equation}
\frac{1}{X^2}\partial_i\left(\dot A_f - KE_f\right) + \frac{m^2}{M_\star^2}\frac{P}{y^2}\frac{1}{x + y}\frac{a}{X} \partial_i\left(yF_g - xF_f\right) = 0,
\end{equation}

\item $i$--$i$:
\begin{align}
\frac{1}{X^2}\left[\left(2\dot K + 3K^2 - 2\frac{\dot X}{X}K\right)E_f + K\dot E_f - \ddot{A}_f - 3K\dot{A}_f + \frac{\dot X}{X}\dot A_f \right] + \frac{1}{2}\left(\partial_j^2 + \partial_k^2\right)D_f  \nonumber \\
- \frac{m^2}{M_\star^2}\frac{1}{xy^2}\left[\frac{1}{2}P\Delta E + Q\left(\Delta A+ \frac{1}{2}\left(\partial_j^2 + \partial_k^2\right)\Delta B\right)\right] = 0,
\end{align}

\item Off-diagonal $i$--$j$:
\begin{equation}
-\frac{1}{2}\partial^i\partial_j D_f + \frac{m^2}{2M_\star^2}\frac{Q}{xy^2}\partial^i\partial_j \Delta B = 0,
\end{equation}

\end{itemize}
where $K \equiv\dot{Y}/Y$ is the $f$-metric Hubble parameter and we have again defined
\begin{equation}
 D_f \equiv \frac{A_f + E_f}{Y^2} + \frac{K}{X}\left(\frac{4F_f}{Y} - \frac{3\dot{B}_f}{X}\right) + \frac{2\dot{F}_f}{XY} - \frac{1}{X^2}\left(\ddot{B}_f - \frac{\dot X}{X}\dot{B}_f\right).
\end{equation}
Finally the fluid conservation equation, $\nabla_\mu^g T^\mu{}_\nu=0$, can be split into time and space parts, neither of which is changed from general relativity,
\begin{itemize}
 \item Energy conservation ($\nu = 0$):
\begin{equation}
 \nabla_{\mu}^gT^\mu{}_0 = -\left(\dot{\bar\rho} + 3H\bar\rho\right)(1+\delta) -\bar\rho\left[\dot\delta + \theta + \frac{3}{2}\dot{A}_g + \frac{1}{2}\nabla^2\dot{B}_g\right],
\end{equation}

\item Momentum conservation ($\nu = i$):
\begin{equation}
 \nabla_{\mu}^gT^\mu{}_i = \left(\dot{\bar\rho} + 4H\bar\rho\right)\left(v_i + \partial_iF_g\right) + \bar\rho\left(\dot{v}_i + \partial_i\dot{F}_g\right) + \frac{1}{2}\bar\rho\partial_iE_g.
\end{equation}

\end{itemize}
We have assumed for simplicity that the fluid comprises only pressureless dust. These are in agreement with the results found elsewhere in the literature using various choices of gauge-invariant variables \cite{Berg:2012kn,Comelli:2012db,Konnig:2014xva}.

It is worth mentioning that the $g$-metric $i$--$i$ equation, (\ref{eq:fullgii}), is identically zero in GR after taking into account the $0$--$i$, off-diagonal $i$--$j$, and momentum conservation equations and hence gives no information; in massive (bi)gravity, however, it is crucial, and is manifestly only important when $m\neq0$. In a gauge with $F_g=F_f=0$ it has the simple form
\begin{equation}
m^2\left[P\left(xE_f-yE_g\right)+2yQ\Delta A\right]=0. \label{eq:simpleii}
\end{equation}
Performing the same steps on the $f$-metric $i$--$i$ equation, we arrive again at \cref{eq:simpleii}. Hence both $i$--$i$ equations carry the same information. We see there is an extra algebraic constraint hidden in the system of perturbation equations; this is closely related to the nontrivial constraint which eliminates the Boulware-Deser ghost,\footnote{We thank Shinji Mukhoyama for discussions on this point.} and will become important shortly when discussing the degrees-of-freedom counting in bigravity.

Herein we will decompose the perturbations into Fourier modes without writing mode subscripts: every variable will implicitly refer to the Fourier mode of that variable with wavenumber $k$.

\subsection{Counting the Degrees of Freedom}

While we have ten equations for ten variables, there are only two independent degrees of freedom.\footnote{The discussion in this section is indebted to useful conversations with Macarena Lagos and Pedro Ferreira.} These can be seen as corresponding, for example, to the scalar modes of the two gravitons or to the matter perturbation and the scalar mode of the massive graviton. To understand the degrees-of-freedom counting, we will start with the simpler waters of general relativity, using the language we have employed for bigravity and following the spirit of the discussion in \rcite{Lagos:2013aua}. The time-time and time-space perturbations, $E_g$ and $F_g$, as well as the velocity perturbation, $\theta$, are \emph{auxiliary} in that they appear in the second-order action without derivatives.\footnote{Specifically, they appear without time derivatives. Recall that we are working in Fourier space where spatial derivatives effectively amount to multiplicative factors of $ik$.} Therefore their equations of motion are algebraic constraints which relate them to other perturbation variables, and they can be removed from the system trivially. This leaves us with three dynamical variables, $A_g$, $B_g$, and $\delta$, two of which can be \emph{gauge fixed}, or set to a desired value (such as zero) by a coordinate transformation. Hence at linear order general relativity only has one dynamical (scalar) degree of freedom propagating on FLRW backgrounds. In inflation, for instance, this is often taken to be the comoving curvature perturbation, $\zeta$.

This is relatively straightforward to extend to massive bigravity; we point the reader to \rcite{Lagos:2013aua,Lagos:2014lca} for in-depth discussions. Few complications are introduced because the only new components of the theory are an Einstein-Hilbert term for $f_\mn$, which has exactly the same derivative structure as in general relativity, and a mass term, which has no derivatives. Therefore we can see immediately that five of the perturbations---$E_g$, $E_f$, $F_g$, $F_f$, and $\theta$---are nondynamical and can be integrated out in terms of the dynamical variables and their derivatives. As discussed in \cref{sec:mg-bigrav-cosmo-eq}, the coordinate invariance in massive bigravity is effectively the same as in general relativity, as long as we view the gauge transformations as acting on the coordinates, rather than on the fields. This can be seen by the fact that the Einstein-Hilbert terms are clearly invariant under separate diffeomorphisms for $g_\mn$ and $f_\mn$, and the mass term is invariant as long as $g^{-1}f$ is, which is the case if we act the same coordinate transformation on each of them. We can therefore gauge fix two of the dynamical variables.

Because we have started with ten perturbation variables, five of which were auxiliary and two of which can be gauge fixed, we are now left with three dynamical variables. However, they are not all independent. After the auxiliary variables are integrated out, one of the initially-dynamical variables \emph{becomes} auxiliary, i.e., its derivatives drop out of the second-order action, and it can itself be integrated out. This leaves us, as promised, with two independent, dynamical degrees of freedom.

It is therefore possible to reduce the ten linearised Einstein equations to a much simpler system of two coupled second-order differential equations. As we will see, with the right choice of gauge this process is fairly simple. This will allow us, in \cref{sec:bigrav-instabilities}, to check whether the solutions to that system are stable.

\subsection{Gauge Choice and Reducing the Einstein Equations}
\label{sec:reduce}

In \rcite{Lagos:2013aua} a method for identifying the gauge-invariant degrees of freedom was presented in which Noether identities are used to identify ``good'' gauges, i.e., gauges in which the equations of motions for the gauge-fixed variables are contained in the remaining equations of motion. This methodology was applied to massive bigravity in \rcite{Lagos:2014lca}. While the method was developed with a focus on deriving the second-order action for the perturbations, rather than starting with the equations of motion as we do here, we will find that this method of picking a gauge will be convenient.

Many common gauges choose to fix auxiliary variables, but this makes the job of reducing the perturbation equations to the minimal number of degrees of freedom difficult. By contrast, in the Noether-identity method only dynamical variables are fixed. Specifically, one chooses to eliminate those variables whose equations of motion are redundant, i.e., are contained within the equations of motion for fields which we leave in, so that no physical information is lost. Because the equations of motion for the perturbation variables are the same as the Einstein equations we are using, such a gauge choice works well for our purposes. The end result is that we should choose to eliminate one of $\{A_g,A_f\}$ and one of $\{B_g,B_f,\chi\}$ \cite{Lagos:2014lca}, where $\chi\equiv k^{-2}\delta + (3/2)k^{-2}A_g - (1/2)B_g$, which characterises the fluid flow, is the basic scalar dynamical degree of freedom for a perturbed fluid \cite{Mukhanov:1990me}. This uses up all of the available gauge freedom.

In this chapter we will work in a gauge in which $A_f = \chi = 0$. This has the advantage of treating the two metrics symmetrically: the remaining independent, dynamical fields are $B_g$ and $B_f$, with $A_g$ having become auxiliary in the process. Our goal is to derive the reduced system of equations of motion for $B_g$ and $B_f$, as well as expressions relating all of the rest of the perturbation variables to these. Because the resultant equations are extraordinarily lengthy, we will not present them but will simply summarise the steps.

Five equations---the $0$--$0$ and $0$--$i$ Einstein equations for each of the metrics and the energy-conservation equation---correspond directly to the equations of motion for the five auxiliary variables. In these equations the auxiliary variables only appear linearly and without derivatives. Therefore we can easily ``integrate them out'' by solving the system of those five equations to obtain $\{E_g,E_f,F_g,F_f,\theta\}$ in terms of the remaining degrees of freedom, $\{A_g,B_g,B_f\}$, and their derivatives.

We are left with $A_g$, $B_g$, and $B_f$, and their equations of motion are the $g$-metric $i$--$i$, $g$-metric $i$--$j$, and $f$-metric $i$--$j$ equations, respectively. As discussed above, the $i$--$i$ equations effectively become constraints after manipulation with the other equations of motion. We demonstrated this in a gauge where $F_g=F_f=0$; while mathematically simple, this gauge is not very helpful as it only eliminates auxiliary variables. In the more convenient gauge we are now using there is an equivalent statement: after integrating out the five auxiliary variables, all derivatives of $A_g$ vanish from the $g$-metric $i$--$i$ equation, so that it can be used to solve algebraically for $A_g$ in terms of $B_g$, $B_f$, and their first derivatives. This is the result, mentioned above, that after integrating out the auxiliary variables, one of the dynamical variables becomes auxiliary. We note that $A_g$ only depends on $B_g$ and $B_f$ up to first derivatives. This fact is crucial because $\dot A_g$ (though not $\ddot A_g$) appears in the remaining equations of motion. If $A_g$ depended on second derivatives, then higher derivatives would appear upon integrating it out, and we would be in danger of a ghost instability. Indeed, as mentioned above, the fact that $A_g$ loses its dynamics is nothing other than the Boulware-Deser ghost being rendered nondynamical by the specific potential structure of massive bigravity \cite{Langlois:2014jba}.

\section{Stability Analysis}
\label{sec:bigrav-instabilities}

Having reduced the system of linearised Einstein and conservation equations using the steps outlined in \cref{sec:reduce}, we can write our original ten equations as just two coupled second-order differential equations. Defining the vector
\begin{equation}
\underline{\mathbf x} \equiv \begin{pmatrix}B_g\\B_f\end{pmatrix},
\end{equation}
this system takes the simple form
\begin{equation}
\ddot{\underline{\mathbf x}} + \bm A \dot{\underline{\mathbf x}} + \bm B \underline{\mathbf x}=0,\label{eq:2nd_order_DEQ}
\end{equation}
where $\bm A$ and $\bm B$ are matrices with extremely unwieldy forms which depend only on background quantities and on $k$. Equivalently we can write \cref{eq:2nd_order_DEQ} in the same form in terms of $N=\ln a$ (not to be confused with the $g$-metric lapse, $N$),
\begin{equation}
\underline{\mathbf x}'' + \tilde{\bm A} \underline{\mathbf x}' + \tilde{\bm B} \underline{\mathbf x}=0,\label{eq:2nd_order_DEQ-N}
\end{equation}
where we use primes to denote derivatives with respect to $N$. We will choose this formulation to simplify the analysis and better understand its physical consequences.

We are now in a good position to analytically probe the stability of linear cosmological perturbations in bigravity. If we neglect the dependence of $\bm A$ and $\bm B$ and treat them as constants, then \cref{eq:2nd_order_DEQ-N} is clearly solved by a linear superposition of exponentials
\begin{equation}
\underline{\mathbf x}=\displaystyle\sum_i \underline{\mathbf x}_ie^{i\omega_i N}. \label{eq:eigenfreqsol}
\end{equation}
We refer to $\omega_i$ as \emph{eigenfrequencies} because they can be determined by solving for the eigenvalues of $i\omega_i\bm A + \bm B$. While $\bm A$ and $\bm B$ are not truly constant, they do in fact vary slowly enough for the \emph{WKB approximation} to hold, in which case \cref{eq:eigenfreqsol} is the correct first-order solution for $\underline{\mathbf x}$. The criterion for the WKB approximation to hold is $|\omega'/\omega^{2}|\ll1$; this will be satisfied in the cases in which we are interested.

The criterion for stability is that all eigenfrequencies be real. This is necessary to obtain purely oscillating solutions for the perturbation variables; if any eigenfrequency had an imaginary piece, then there would be exponentially growing modes. The eigenfrequencies we obtain from \cref{eq:2nd_order_DEQ-N} are inordinately complicated.\footnote{We recognise that the number of unused synonyms for ``these equations are very long'' is growing short as this chapter progresses.} To simplify the analysis and focus on subhorizon scales, which account for most of the modes we can observe, we will take the limit $k \gg \sh$, where, as elsewhere in this thesis, we denote the conformal-time Hubble rate by $\sh$.

As discussed in \cref{sec:properties-bimetric-cosmo}, the only theory with one nonzero $\beta_i$ parameter which allows for viable cosmological background expansion is the $\beta_1$ model, providing an excellent fit to background data including Type Ia supernovae, the cosmic microwave background, and baryon-acoustic oscillations \cite{Akrami:2012vf,Konnig:2013gxa}. However, the analysis in this chapter shows that it suffers from instabilities throughout most of cosmic history. In the limit of large $k/\sh$ we find the eigenfrequencies for this model are 
\begin{equation}
\omega_{\beta_{1}}=\pm\frac{k}{\mathcal{H}}\frac{\sqrt{-1+12y^{2}+9y^{4}}}{1+3y^{2}}.\label{eq:eigenfreq_b1}
\end{equation}
The condition for stability, i.e., for the object inside the square root to be positive, is
\begin{equation}
y>\sqrt{\frac{1}{3} \left(\sqrt{5}-2\right)} \approx 0.28.
\end{equation}
This suggests that there is an instability problem at early times; recall from \cref{sec:properties-bimetric-cosmo} that the $\beta_1$ model has only finite-branch solutions, meaning that $y$ evolves monotonically from $0$ at early times to, at late times, a positive constant. Therefore the $\beta_1$ model always suffers from an instability at early times, with a turnover from unstable to stable occurring when $y\approx0.28$. This instability is quite dangerous. Consider scales of $k\sim100\sh$, which is a typical mode size for structure observations. Those modes would then grow, assuming $\operatorname{Im}(\omega)\sim\mathcal{O}(1)$, as roughly $e^{100N}$, which is far too rapid for linear theory to be applicable for more than a fraction of an $e$-fold.

During what time period is this instability present? If $y$ reached $0.28$ at sufficiently early times, one might expect that the presence of radiation or other new fields, which we have ignored in favour of dust, could ensure that modes are stable. However, the time at which the instability turns off is generically close to the time at which the expansion begins to accelerate, so this is clearly a modern problem. We can find the exact region of instability recalling that, in this model, we can solve for $y(a)$ exactly, c.f. \cref{eq:MBMya},
\begin{equation}
 y(a) = \frac 1 6 a^{-3}\left(-C + \sqrt{12a^6 + C^2} \right).
\end{equation}
where
\begin{equation}
 C = -B_1 + \frac{3}{B_1},
\end{equation}
where $B_1 \equiv m^2\beta_1/H_0^2$ and $H_0$ is the cosmic-time Hubble rate today. The best-fit value for $B_1$ using a combination of SNe, CMB, and BAO data is $B_1 = 1.448\pm0.0168$ \cite{Akrami:2012vf,Solomon:2014dua}. For $B_1=1.448$ exactly, $\omega_{\beta_1}$ switches from imaginary to real at $N=-0.49$, corresponding to a relatively recent redshift, $z=0.63435$. This number is fairly sensitive to the choice of datasets. The CMB and BAO data are taken from observations which assume general relativity in their analysis. Restricting the analysis to supernovae alone, the best fit is $B_1 = 1.3527 \pm 0.0497$. In this case, the instability ends at $N=-0.38$ or $z=0.47$. At any epoch before this, the perturbation equations are unstable for large $k$. This behaviour invalidates linear perturbation theory on subhorizon scales and may rule out the model if the instability is not cured at higher orders. This is not necessarily out of the realm of possibility. As discussed earlier, massive bigravity possesses the Vainshtein mechanism, in which nonlinear effects suppress the helicity-0 mode of the massive graviton in dense environments, thereby recovering general relativity \cite{Vainshtein:1972sx,Babichev:2013usa}. It may be the case that such a mechanism will also impose general-relativistic behaviour on nonlinear cosmological perturbations.

Now let us move on to more general models. As we mentioned in \cref{sec:properties-bimetric-cosmo}, the other one-parameter models are not viable in the background,\footnote{With the exception of the $\beta_{0}$ model, which is simply $\Lambda$CDM.} i.e., none of them has a matter-dominated epoch in the asymptotic past and produces a positive Hubble rate \cite{Konnig:2013gxa}.\footnote{We frequently discuss the viability of various models in this section; all such results were derived in \rcite{Konnig:2013gxa}.} Nevertheless it is worthwhile to calculate the eigenfreqencies in these cases in order to study the asymptotic behaviour of the viable multiple-parameter models. For simplicity, from now on we refer
to a model in which, e.g., only $\beta_{1}$ and $\beta_{2}$ are nonzero as the $\beta_{1}\beta_{2}$ model, and so on.

At early times, every viable, finite-branch, multiple-parameter model is approximately described by the single-parameter model with the lowest-order interaction. For instance, the $\beta_{1}\beta_{2}$, $\beta_{1}\beta_{3}$, and $\beta_{1}\beta_{2}\beta_{3}$ models all reduce to $\beta_{1}$, the $\beta_{2}\beta_{3}$ model reduces to $\beta_{2}$, and so on. Similarly, in the early Universe, the viable, infinite-branch models reduce to single-parameter models with the highest-order interaction. This is clear from the structures of the terms in the Friedmann equation and the $P$ and $Q$ parameters introduced above. It is only through these terms that the $\beta_i$ parameters enter the perturbation equations. Therefore, in order to determine the early-time stability, we need only look at the eigenfrequencies of single-parameter models. In addition to $\omega_{\beta_1}$ presented in \cref{eq:eigenfreq_b1} above, we have
\begin{align}
\omega_{\beta_{2}} & =\pm\frac{k}{\mathcal{H}}\frac 1y,\label{eq:eigenfreq_b2}\\
\omega_{\beta_{3}} & =\pm\frac k\sh\frac{\sqrt{-3+8y^2-y^{4}}}{\sqrt{3}\left(1-y^{2}\right)},\label{eq:eigenfreq_b3}\\
\omega_{\beta_{4}} & =\pm\frac{k}{\mathcal{H}}\frac{1}{\sqrt{2}}.\label{eq:eigenfreq_b4}
\end{align}
We see that the $\beta_{2}$ and $\beta_{4}$ models are stable at all times, while the $\beta_3$ model suffers from an early-time instability just like the $\beta_1$ model. We can now extend these results to the rest of the bigravity parameter space by using the single-parameter models to test the early-time stability.

Since much of the power of massive bigravity lies in its potential to address the dark energy problem in a technically-natural way, let us first consider models without an explicit $g$-metric cosmological constant, i.e., $\beta_{0}=0$. On the finite branch, all such models with $\beta_1 \neq0$ reduce, at early times, to the $\beta_{1}$ model. As we have seen, this possesses an imaginary sound speed for large $k$, cf. \cref{eq:eigenfreq_b1}, and is therefore unstable in the early Universe. Hence the finite-branch $\beta_{1}\beta_{2}\beta_{3}\beta_{4}$ model and all its subsets with $\beta_{1}\neq0$ are all plagued by instabilities. This is particularly significant because all of these models otherwise have viable background evolutions \cite{Konnig:2013gxa}. This leaves the $\beta_{2}\beta_{3}\beta_{4}$ model; this is stable on the finite branch as long as $\beta_{2}\neq0$, but its background is not viable. We conclude that there are no models with $\beta_{0}=0$ which live on a finite branch, have a viable background evolution, and predict stable linear perturbations at all times.

This conclusion has two obvious loopholes: we can either include a cosmological constant, $\beta_{0}$, or turn to an infinite-branch model. We first consider including a nonzero cosmological constant, bearing in mind this may not be as interesting theoretically as the models which self-accelerate. Adding a cosmological constant can change the stability properties, although it turns out not to do so in the finite-branch models with viable backgrounds. In the $\beta_{0}\beta_{1}$ model, the eigenfrequencies, 
\begin{equation}
\omega_{\beta_{0}\beta_{1}}=\pm\frac k\sh\frac{\sqrt{-1+2\left(\beta_{0}/\beta_{1}\right)y+12y^2+9y^{4}}}{1+3y^{2}},
\end{equation}
are unaffected by $\beta_{0}$ at early times and therefore still imply exponential mode growth in the asymptotic past. This result extends (at early times) to the rest of the bigravity parameter space with $\beta_0,\beta_1\neq0$. No other finite-branch models yield viable backgrounds. In conclusion, all of the solutions on a finite branch, for any combination of parameters, are either unviable (in the background) or linearly unstable in the past.

Let us now turn to the infinite-branch models. There are two candidates with viable background histories. The first is the the $\beta_0\beta_2\beta_4$ model. The reality of $\omega_{\beta_2}$ and $\omega_{\beta_4}$, cf. \cref{eq:eigenfreq_b2}, suggests that this model is linearly stable. At the background level, this case is something of an exception as it is the only bimetric cosmology with exact $\Lambda$CDM evolution: the structure of the Friedmann equation and quartic equation conspire to allow the dynamics to be rewritten with a modified gravitational constant and an effective cosmological constant \cite{vonStrauss:2011mq},
\begin{equation}
3H^2 = \frac{\beta_4}{\beta_4-3\beta_2}\frac{\rho}{M_g^2} + m^2\frac{\beta_0\beta_4-9\beta_2^2}{\beta_4-3\beta_2}.
\end{equation}
The quartic equation for $y$ can be solved to find
\begin{equation}
y^2 = \frac{\frac{\rho}{m^2M_g^2}+\beta_0-3\beta_2}{\beta_4-3\beta_2}.
\end{equation}
This implies that all solutions to this model live on the infinite branch. In order for $y$ to be real at all times, we are required to have $\beta_0 - 3\beta_2 >0$ and $\beta_4 - 3\beta_2 > 0$. Unfortunately, for these reasons we cannot have a viable self-accelerating solution; if $\beta_0$ were set to zero (or were much smaller than $\beta_2$ and $\beta_4$), then the effective cosmological constant would be negative. The modified gravitational constant would also be negative if $\beta_4$ were positive. From a cosmological point of view, these models are therefore not altogether interesting.

Finally there is a small class of viable and interesting models which have stable cosmological evolution: the self-accelerating $\beta_{1}\beta_{4}$ model and its generalisation to include $\beta_0$.\footnote{We do not have the freedom to include nonzero $\beta_2$ or $\beta_3$; in either case the background evolution would not be viable \cite{Konnig:2013gxa}. We can see this from the expressions (\ref{eq:breprho}) and (\ref{eq:brepH}) for $\rho(y)$ and $H^2(y)$. If $\beta_3$ were nonzero, then $\Omega_{\mathrm m} = \rho/(3M_g^2H^2)$ would diverge as $y$ at early times. Setting $\beta_3=0$, we find $\Omega_\mathrm{m} \to 1-3\beta_2/\beta_4$ as $y\to\infty$. If we demand a matter-dominated history, then $\beta_2$ must at the very least be small compared to $\beta_4$.} Here, $y$ evolves from infinity in the past and asymptotes to a finite de Sitter value in the future. For these $\beta_{0}\beta_{1}\beta_{4}$ models we perform a similar eigenfrequency analysis and obtain
\begin{align}
\omega_{\beta_{0}\beta_{1}\beta_{4}} & =\pm\frac k\sh\frac{\sqrt{-1+2\left(\beta_{0}/\beta_{1}\right)y+12y^{2}+\left(9+2\beta_{0}\beta_{4}/\beta_{1}^{2}\right)y^{4}-2\left(\beta_{4}/\beta_{1}\right)\left[4y^{3}+3y^{5}-(\beta_{4}/\beta_{1})y^{6}\right]}}{1+3y^{2}-2\left(\beta_{4}/\beta_{1}\right)y^{3}}.\label{eq:eqigenfreq_b0b1b4}
\end{align}
Restricting ourselves to the self-accelerating models (i.e., $\beta_{\text{0}}=0$), we obtain 
\begin{align}
\omega_{\beta_{1}\beta_{4}} & =\pm\frac k\sh\frac{\sqrt{-1+12y^2+9y^{4}-2\left(\beta_{4}/\beta_{1}\right)\left[4y^3+3y^5-(\beta_{4}/\beta_{1})y^{6}\right]}}{1+3y^{2}-2\left(\beta_{4}/\beta_{1}\right)y^{3}}.\label{eq:eigenfreq_b1b4}
\end{align}
Notice that at early times, i.e., for large $y$, the eigenvalues (\ref{eq:eqigenfreq_b0b1b4}) and (\ref{eq:eigenfreq_b1b4}) reduce to the expression (\ref{eq:eigenfreq_b4}) for $\omega_{\beta_{4}}$. This frequency is real, and therefore the $\beta_{1}\beta_{4}$ model, as well as its generalisation to include a cosmological constant, is stable on the infinite branch at early times.

\begin{figure}
\centering
\includegraphics[width=.7\textwidth]{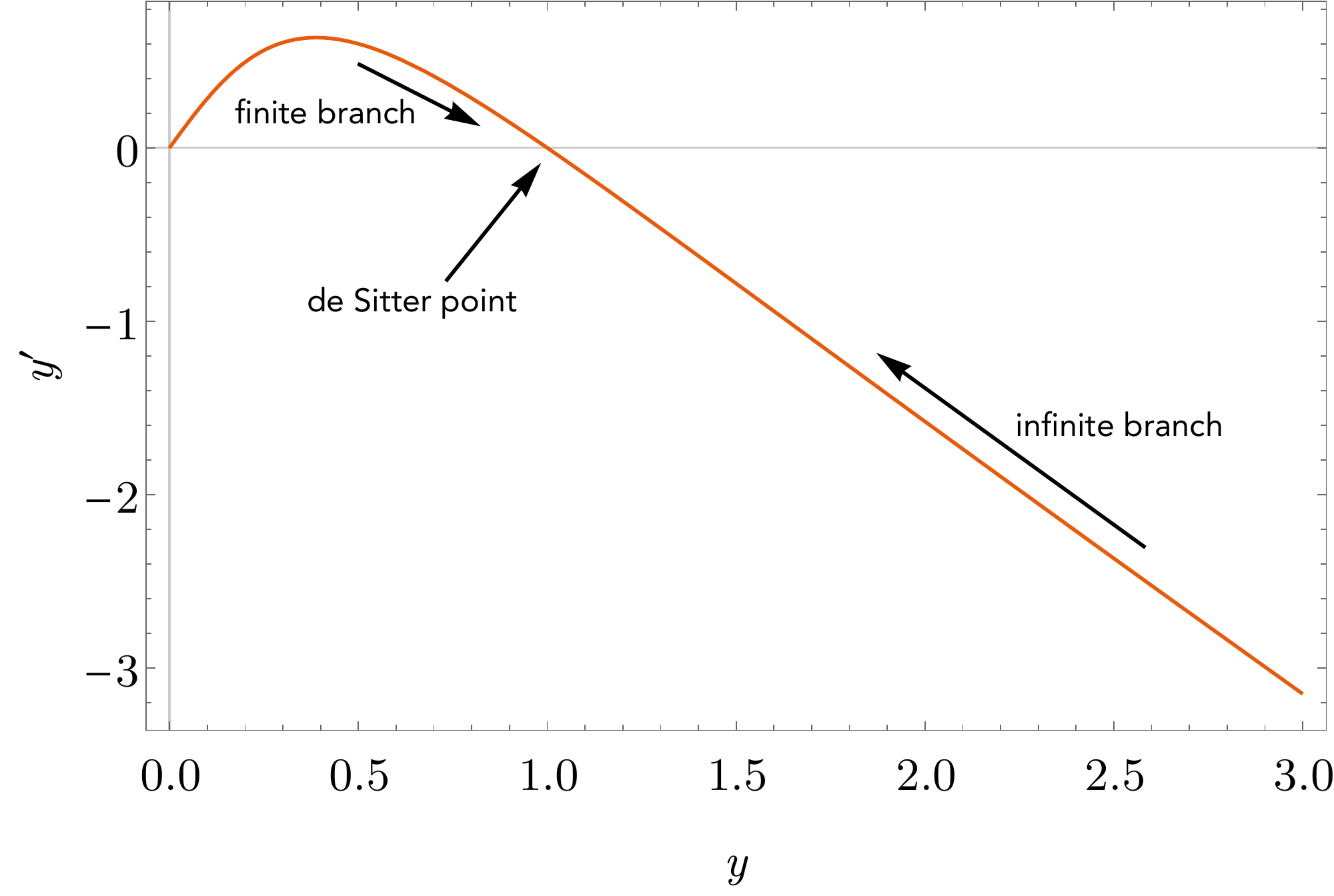}
\caption[Plot of the function $y'(y)$ for the $\beta_{1}\beta_{4}$ model for $\beta_4=2\beta_1$.]{Plot of the function $y'(y)$ for the $\beta_{1}\beta_{4}$ model for $\beta_4=2\beta_1$. For both the finite and infinite branches, the final state is the de Sitter point. The arrows show the direction of movement of $r$.}
\label{fig:branches}
\end{figure}

Interestingly, the eigenfrequencies for this particular model can also be written as 
\begin{equation}
\omega_{\beta_{0}\beta_{1}\beta_{4}} =\pm\frac{k}{\mathcal{H}}\sqrt{-\frac{y''}{3y'}} = \pm \frac{k}{\sh}\sqrt{-\frac13 \frac{dy'}{dy}}.\label{eq:eqigenfreq_b0b1b4-1}
\end{equation}
Therefore, the condition for the stability of this model in the infinite branch, where $y'<0$, is simply $y''>0$. One might wonder whether this expression for $\omega$ is general or model specific. While it does not hold for the $\beta_{2}$ and $\beta_{3}$ models, c.f. \cref{eq:eigenfreq_b2,eq:eigenfreq_b3}, it is valid for all of the submodels of $\beta_{0}\beta_{1}\beta_{4}$, including the single-parameter models presented in \cref{eq:eigenfreq_b1,eq:eigenfreq_b4}. We can see from this, for example, that the finite-branch ($y'>0$) $\beta_{1}$ model is unstable at early times because initially $y''$ is positive. In \cref{fig:branches} we show schematically the evolution of the $\beta_{1}\beta_{4}$ model on the finite and infinite branches. The stability condition on either branch is $y''/y'=dy'/dy<0$. For the parameters plotted, $\beta_4=2\beta_1$, one can see graphically that this condition is met, and hence the model is stable, only at late times on the finite branch but for all times on the infinite branch. Our remaining task is to extend this to other parameters.

Let us now prove that the infinite-branch $\beta_{1}\beta_{4}$ model is stable in the subhorizon limit at all times as long as the background expansion is viable, which restricts us to the parameter range $0 < \beta_4 < 2 \beta_1$ \cite{Konnig:2013gxa}. From \cref{eq:eigenfreq_b1b4} we can see that the subhorizon perturbations are clearly stable if and only if
\begin{equation}
-1+12y^2+9y^{4}-2\left(\beta_{4}/\beta_{1}\right)\left[4y^3+3y^5-(\beta_{4}/\beta_{1})y^{6}\right]>0 \label{eq:stabilitycondition}
\end{equation}
At early times, $y\to\infty$, this is dominated by a manifestly positive term. Indeed we have already seen that the eigenfrequencies match those in the $\beta_{4}$ model (\ref{eq:eigenfreq_b4}) which are purely real. At later times, \cref{eq:stabilitycondition} is satisfied for all $y>1$ as long as we restrict to the viable parameter range, $0<\beta_4/\beta_1<2$. We can therefore rephrase the question of stability as a question about the background evolution: do the infinite-branch models in this region of the parameter space always have $y>1$?

The answer is \emph{yes}. Recall from \cref{sec:properties-bimetric-cosmo} that, on the infinite branch, $y$ evolves monotonically from $y=\infty$ to $y=y_c$, where $y_c$ is defined by \cref{eq:yc},
\begin{equation}
\beta_{4}y_{c}^{3}-3\beta_{1}y_c^{2}+\beta_{1}=0.\label{eq:rcubic}
\end{equation}
Because the evolution of $y$ is monotonic, $y>1$ at all times if $y_c\geq1$. Moreover, because $y=y_c$ corresponds, through the quartic equation (\ref{eq:quartic-mg}), to $\rho=0$, we are only interested in the largest real root of \cref{eq:rcubic}. For the largest allowed value of $\beta_4$, $\beta_{4}=2\beta_{1}$ exactly, we find $y_c=1$. We must then ask whether for $0<\beta_{4}<2\beta_{1}$, $y_c$ remains greater than 1. Writing $p\equiv y_c-1$, using Descartes' rule of signs, and restricting ourselves to $0<\beta_{4}<2\beta_{1}$, we can see that $p$ has one positive root, i.e., there is always exactly one solution with $y_c>1$ in that parameter range. Therefore, in all infinite-branch solutions with $0<\beta_{4}<2\beta_{1}$, $y$ evolves to some $y_c>1$ in the asymptotic future. We conclude that all of the infinite-branch $\beta_{1}\beta_{4}$ cosmologies which are viable at the background level are also linearly stable at all times in the subhorizon limit, providing a clear example of a bimetric cosmology which is a viable competitor to $\Lambda$CDM. This stability has been confirmed and extended to the superhorizon limit in a complementary analysis in \rcite{Lagos:2014lca}.

As a side remark, we note that in this model the asymptotic past corresponds to the limit $y\rightarrow\infty$ and $y'\rightarrow-\frac{3}{2}y$, i.e., $y\rightarrow a^{-3/2}$. This implies that $Y\sim a^{-1/2}$, i.e., the second metric initially collapses while ``our'' metric expands. On the approach to the final de Sitter stage, $y$ approaches a constant $y_c$, so the scale factors $a$ and $Y$ both expand exponentially. This infinite-branch model therefore contains a bouncing cosmology for the $f$ metric.

This bounce has an unusual consequence. Recall from \cref{eq:xkyh} that, after imposing the Bianchi identity, we have $f_{00} = -\dot{Y}^2/\sh^2$. Therefore, when $y$ bounces, $f_{00}$ becomes zero: at that one point, the lapse function of the $f$ metric vanishes.\footnote{Moreover, the square root of this, $\dot Y/\mathcal{H}$, appears in the mass terms. We choose branches of the square root such that this quantity starts off negative at early times and then becomes positive.} Nevertheless, this does not render the solution unphysical, for the following reasons. First, the $f$ metric does not couple to matter and so, unlike the $g$ metric, it does not have a geometric interpretation. A singularity in $f_\mn$ therefore does not necessarily imply a singularity in observable quantities. In fact, we find no singularity in any of our background or perturbed variables. Second, although the Riemann tensor for the $f$ metric is singular when $f_{00}=0$, the Lagrangian density $\sdf R(f)$ remains finite and nonzero at all times, so the equations of motion can be derived at any points in time.

\section{Summary of Results}
\label{sec:bigravity-stability-summary}

In this chapter, we introduced the tools for perturbation theory in massive bigravity and used them to test the stability of the theory.

We began by presenting the cosmological perturbation equations; these are derived in \cref{app:perteqs}. We went on to detail the way in which the physical degrees of freedom are counted and described how to pick a good gauge and integrate out nondynamical variables. By doing so we reduced the ten linearised Einstein and fluid conservation equations to a system of two coupled, second-order differential equations. These describe the evolution of the two independent, dynamical degrees of freedom present at linear order around FLRW solutions in massive bigravity.

We then identified the stable and unstable models by employing a WKB approximation and calculating the eigenfrequencies of the perturbation equations. This analysis revealed that many models with viable background cosmologies exhibit an instability on small scales until fairly recently in cosmic history. However, we also found a class of viable models which are stable at all times. These are defined by giving nonzero, positive values to the interaction parameters $\beta_1$ and $\beta_4$, setting $\beta_2=\beta_3=0$, and choosing solutions in which the ratio $y=Y/a$ of the two scale factors decreases from infinity to a finite late-time value. A cosmological constant can be added without spoiling the stability, although it is not necessary; the theory is able to self-accelerate.

On the surface, these results would seem to place in jeopardy a large swath of bigravity's parameter space, such as the ``minimal'' $\beta_{1}$-only model which is the only single-parameter model that is viable at the background level \cite{Konnig:2013gxa}. It is important to emphasise that the existence of such an instability does \emph{not} automatically rule these models out. It merely impedes our ability to use linear theory on deep subhorizon scales (recall that the instability is problematic specifically for large $k$). Models that are not linearly stable can still be realistic if only the gravitational potentials become nonlinear, or even if the matter fluctuations also become nonlinear but in such a way that their properties do not contradict observations. The theory can be saved if, for instance, the instability is softened or vanishes entirely when nonlinear effects are taken into account. We might even expect such behaviour: bigravity models exhibit a Vainshtein mechanism \cite{Vainshtein:1972sx,Babichev:2013usa} which restores general relativity in environments where the new degrees of freedom are highly nonlinear. Consequently two very important questions remain: can these unstable models still accurately describe the real Universe, and if so, how can we perform calculations for structure formation?

Until these questions are answered, the infinite-branch $\beta_{1}\beta_{4}$ model seems to be the most promising target at the moment for studying massive bigravity. In the next chapter, we will calculate its predictions for structure formation, confront them with data, and discuss the potential of near-future probes like {\sc Euclid} to test this model against $\Lambda$CDM.

\begin{savequote}[30pc]
The wonder is, not that the field of the stars is so vast, but that man has measured it.
\qauthor{Anatole France, \textit{The Garden of Epicurus}}
\end{savequote}

\chapter{Linear Structure Growth in Massive Bigravity}
\label{chap:bigravity-subhorizon}
\hrule
\vspace*{1cm}

We have, to this point, reviewed the background FLRW solutions of massive bigravity in \cref{chap:mg} and begun to analyse linear perturbations around these solutions in \cref{chap:bigravity-stability}. In addition to introducing the formalism for cosmological perturbation theory in bigravity, the specific aim of the previous chapter was to identify which models are stable at the linear level and which are not. The natural next step is to use perturbation theory to derive observable predictions for the stable models.

In this chapter we undertake a study of the cosmological large-scale structure (LSS) in massive bigravity with the aim of understanding the ways in which bigravity deviates from general relativity and its potentially-testable cosmological signatures. This is motivated in particular by anticipation of the forthcoming Euclid mission which is expected to improve the accuracy of the present large-scale structure data by nearly an order of magnitude \cite{Laureijs:2011gra,Amendola:2012ys}.

At the background level, careful statistical analyses show that several bimetric models can provide as good a fit as the standard $\Lambda$CDM model, including one case, the $\beta_1$-only model, which has the same number of free parameters as $\Lambda$CDM \cite{Akrami:2012vf,Akrami:2013pna,Konnig:2013gxa}.\footnote{As discussed in \cref{chap:bigravity-stability}, this model is linearly unstable. However, if the instability is cured at higher orders in perturbation theory before the background solution is spoiled, then at the background level this is a perfectly viable model.} This is a blessing and a curse; while it is encouraging that massive bigravity can produce realistic cosmologies in the absence of a cosmological constant, the quality of background observations is not good enough to distinguish these models from $\Lambda$CDM or from each other. In particular, the parameter constraints obtained from the expansion history have strong degeneracies within the theory itself.

To efficiently test the theory and distinguish its cosmology from others one needs to move beyond the FLRW metric and study how consistent bimetric models are with the observed growth of cosmic structure. We restrict ourselves to the linear, subhorizon r\'egime and examine whether there are any deviations from the standard model predictions which may be observable by future LSS experiments.

Note that due to the aforementioned instability, some (though not all) bimetric models cannot be treated with linear perturbation theory at all times, as any individual mode would quickly grow large during the period of instability (from early times until some recent redshift). Structure formation must be studied nonlinearly in these cases. In order to demonstrate our methodology, we choose to study every model with a sensible background evolution and one or two free $\beta_i$ parameters, so some unstable cases will be included. For the most part these should be seen as toy models, useful for illustrative purposes, although our results about quantities which do not depend on initial conditions, such as the anisotropic stress, will be observationally relevant at later times. One specific model which we study, the infinite-branch $\beta_1\beta_4$ model, is stable at all times, and so we will pay extra attention to its comparison to observations.

This chapter is organised as follows. Taking the subhorizon and quasistatic limit of the full perturbation equations, we arrive at a convenient closed-form evolution equation in \cref{sec:subhorizon-eqs} that captures the modifications to the general relativistic growth rate of linear structure, as well as the leading-order scale-dependence which modifies the shape of the spectrum at near-horizon scales. The coefficients in the closed-form equation are given in \cref{app:hcoeff}. The results are analysed numerically in \cref{sec:subhorizon-results}, where we discuss the general features of the models and confront them with the observational data. We conclude in \cref{sec:bigravity-subhorizon-summary}.

\section{Perturbations in the Subhorizon Limit}
\label{sec:subhorizon-eqs}

We define the perturbed metrics in conformal time ($N=a$) as
\begin{align}
ds_g^2 &= a^2\left\{-(1+E_g)d\tau^2 + 2\partial_iF_gd\tau dx^i + \left[(1+A_g)\delta_{ij} + \partial_i\partial_jB_g\right]dx^idx^j\right\}, \\
ds_f^2 &= -X^2(1+E_f)d\tau^2 + 2XY\partial_iF_fd\tau dx^i + Y^2\left[(1+A_f)\delta_{ij} + \partial_i\partial_jB_f\right]dx^idx^j,
\end{align}
and immediately specialise to dust ($P=\delta P=0$) and work in Fourier space. The linearised Einstein and fluid conservation equations have been presented in \cref{sec:stability-perts} and are derived in \cref{app:perteqs}. These equations are quite complicated; in order to isolate the physics of interest, namely that of linear structure in the subhorizon r\'egime, we focus on the subhorizon, quasistatic limit of the field equations. This limit is defined by taking $k^2\Phi\gg H^2\Phi\sim H\dot\Phi \sim \ddot\Phi$ for any variable $\Phi$, where we have expanded in Fourier modes with wavenumber $k$, and assuming $K \sim H$ so we can take the same limit in the $f$-metric equations. Moreover, we will take $A_{g,f}$ and $E_{g,f}$ to be of the same order as $kF_{g,f}$ and $k^2B_{g,f}$, as these terms all appear this way in the linearised metric. Finally, we will work in Newtonian gauge for the $g$ metric, defined by setting $F_g = B_g = 0$. Since we are working in the singly-coupled theory where $g_\mn$ is the ``physical'' metric, i.e., only $g_\mn$ couples (minimally) to matter, this is as sensible a gauge choice as it is in GR.\footnote{Given two separate diffeomorphisms for the $g$ and $f$ metrics, only the diagonal subgroup of the two preserves the mass term. In practice, this means that we have a single coordinate system which we may transform by infinitesimal diffeomorphisms, exactly as in GR.}

With these definitions, we can write down the subhorizon evolution equations. The energy constraint (the $0$--$0$ Einstein equation) for the $g$ metric is
\begin{equation}
\left(\frac{k}{a}\right)^2 \left(A_g + \frac{m^2}{2}yPa^2B_f\right) + \frac{3}{2}m^2yP\left(A_g-A_f\right) = \frac{\bar\rho}{M_g^2}\delta. \label{eq:00g}
\end{equation}
The trace of the $i$--$j$ equation yields
\begin{equation}
\left(\dot H - H^2 + \frac{a^2\bar\rho}{2M_g^2}\right)E_g + m^2a^2\left[\frac{1}{2}xP\left(E_f - E_g\right) + yQ\left(A_f - A_g\right)\right] = 0. \label{eq:iig}
\end{equation}
The off-diagonal piece of the $i$--$j$ equation tells us
\begin{equation}
A_g + E_g + m^2a^2yQB_f = 0. \label{eq:ijg}
\end{equation}
We have not presented the momentum constraint (the $0$--$i$ Einstein equation) as it has already been used, along with the momentum conservation equation (\ref{eq:thetacons}), to simplify the trace $i$--$j$ equation (\ref{eq:iig}).\footnote{In the approach of Ref. \cite{Konnig:2014dna}, where this limit is taken by dropping all derivative terms, this step is crucial for the results to be consistent; in our case it is simply useful for rewriting derivatives of $\dot A_g$ and $\ddot A_g$ in terms of $E_g$, so that the equation is manifestly algebraic in the perturbations.} Note also that we have used the off-diagonal piece, \cref{eq:ijg}, to eliminate some redundant terms in the trace equation. For the $f$ metric, the corresponding equations are
\begin{align}
\left(\frac{k}{a}\right)^2 \left(A_f - \frac{m^2}{2}\frac{Pa^2}{y}B_f\right) + \frac{3m^2}{2}\frac{P}{y}\left(A_f - A_g\right) &= 0, \label{eq:00f} \\
\left[-\dot K + \left(H + \frac{\dot x}{x}\right)K\right]E_f + m^2\frac{a^2x}{y^2}\left[\frac{1}{2}P\left(E_f - E_g\right) + Q\left(A_f - A_g\right)\right] &= 0,
 \label{eq:iif} \\
A_f + E_f - m^2\frac{Qa^2}{x}B_f &= 0. \label{eq:ijf}
\end{align}
Finally, due to the minimal coupling between matter and $g_\mn$, the fluid conservation equations are unchanged from GR,
\begin{align}
\dot\delta + \theta &=0, \label{eq:deltacons} \\
\dot\theta + H\theta - \frac{1}{2}k^2 E_g &= 0. \label{eq:thetacons}
\end{align}

In GR the trace equation, (\ref{eq:iig}), adds no new information: it becomes an identity after using the Friedmann equation. In massive bigravity this equation does carry information and it is crucial that we use it. However, we can still simplify it by using the background equations, obtaining\footnote{This equation holds beyond the subhorizon limit, in a particular gauge; see \cref{app:perteqs}.}
\begin{equation}
m^2\left[P\left(xE_f-yE_g\right)+2yQ\Delta A\right]=0. \label{eq:iig-simplified}
\end{equation}
Note that the $m\to0$ limit yields an identity, as expected. There is a further interesting feature: if we substitute the background equations into the $f$-metric trace equation, (\ref{eq:iif}), we obtain \cref{eq:iig-simplified} again. Hence one of the two trace equations is redundant and can be discarded, so what looks like a system of six equations is actually a system of five.

With these relations, the system of equations presented in this section is closed.

\section{Structure Growth and Cosmological Observables}
\label{sec:subhorizon-results}

In this section we study the linearised growth of structure in the quasistatic and subhorizon limit, first solving the field equations to obtain predictions and then comparing to data. Deviations from the predictions of general relativity can be summarised by a few parameters which are observable by large-scale structure surveys such as Euclid \cite{Laureijs:2011gra,Amendola:2012ys}. The main aim of this section is to see under what circumstances these parameters are modified by observable amounts in the linear r\'{e}gime by massive bigravity.

\subsection{Modified Gravity Parameters}

We will focus on three modified growth parameters, defined in the Euclid Theory Working Group review \cite{Amendola:2012ys} as $f$ (and its parametrisation $\gamma$), $Q$, and $\eta$. They are:
\begin{description}
\item[Growth rate ($f$) and index ($\gamma$):] These parameters measure the growth of structure, and are defined by
\begin{equation}
f(a,k) \equiv \frac{d\log \delta}{d\log a} \approx \Omega_m^\gamma, \label{eq:gammadef}
\end{equation}
where $\Omega_m \equiv a^2\bar\rho/(3M_g^2H^2)$ is the usual matter density parameter.
\item[Modification of Newton's Constant ($Q$):] The function $Q(a,k)$\footnote{Not to be confused with the background quantity defined in \cref{eq:pertQdef}, $Q \equiv \beta_1 + \left(x + y\right)\beta_2 + xy\beta_3$.} parametrises modifications to Newton's constant in the Poisson equation,
\begin{equation}
\frac{k^2}{a^2}A_g \equiv \frac{Q(a,k)\bar\rho}{M_g^2}\delta. \label{eq:mgQdef}
\end{equation}
\item[Anisotropic Stress ($\eta$):] Effective anisotropic stress leads the quantity $A_g + E_g$ to deviate from its GR value of zero, which we can parametrise by the parameter $\eta(a,k)$,
\begin{equation}
\eta(a,k) \equiv -\frac{A_g}{E_g}. \label{eq:etadef}
\end{equation}
\end{description}
In GR, these parameters have the values $\gamma \approx 0.545$ and $Q = \eta = 1$.

We have five independent Einstein equations [\cref{eq:00g,eq:ijg,eq:00f,eq:ijf,eq:iig-simplified}] for five metric perturbations\footnote{After gauge fixing there are six metric perturbations, but once we substitute the $0$--$i$ equations into the trace $i$--$j$ equations, $F_f$ drops out of our system. In a gauge where $F_g=F_f=0$, as was used in \rcite{Konnig:2014dna}, the equivalent statement is that the $B_g$ and $B_f$ parameters are only determined up to their difference, $B_f - B_g$, which is gauge invariant.} and $\delta$. Crucially, this system is algebraic. There are five equations for six variables, so we can only solve for any five of the perturbations in terms of the sixth. Of the modified growth parameters, $Q$ and $\eta$ are ratios of perturbations so are insensitive to how we solve the system. However, to find $\gamma$ we need to solve a differential equation for $\delta$. It is therefore simplest to solve for the perturbations $\{A_{g,f},E_{g,f},B_f\}$ in terms of $\delta$.

Solving the system, we find each perturbation can be written in the form $f(\tau,k)\delta$, for some function $f(t)$. We do not display the solutions here as they are quite unwieldy, although we do note that in the limit with only $\beta_1 \neq 0$ studied in \rcite{Konnig:2014dna}, and taking into account differences in notation and gauge, our expressions for the perturbations match theirs.

With these solutions for $\{A_{g,f},E_{g,f},B_f\}$ in hand, we can immediately read off $Q$ and $\eta$. To calculate the growth index, $\gamma$, we need to solve a conservation equation for the density contrast, $\delta$. The fluid conservation equations, (\ref{eq:deltacons}) and (\ref{eq:thetacons}), are unchanged from GR, so as in GR we can manipulate them to find the usual evolution equation for $\delta$ sourced by the gravitational potential,
\begin{equation}
\ddot \delta + H\dot\delta + \frac{1}{2}k^2E_g(\delta) = 0. \label{eq:deltaevol}
\end{equation}
At this point we diverge from the usual story. In GR, there is no anisotropic stress and the Poisson equation holds; combining the two, we find $k^2E_g = -(a^2\bar\rho/M_g^2)\delta$. Both of these facts are changed in massive bigravity, so there is a modified (and rather more complicated) relation between $k^2E_g$ and $\delta$. However, since we do have such a relation, $\delta$ still obeys a closed second-order equation which we can solve numerically.

Finally, we note that the three modified gravity parameters are encapsulated by five time-dependent parameters. The expressions for $\eta$ and $Q$ can be written in the forms
\begin{align}
\eta &= h_2\left(\frac{1 + k^2h_4}{1+k^2h_5}\right), \label{eq:expreta} \\
Q &= h_1\left(\frac{1 + k^2h_4}{1+k^2h_3}\right), \label{eq:exprQ}
\end{align}
where the $h_i$ are functions of time only and depend on $m^2\beta_i$. We present their explicit forms in \cref{app:hcoeff}. The same result has been obtained for Horndeski gravity \cite{DeFelice:2011hq,Amendola:2012ky}, which is the most general scalar-tensor theory with second-order equations of motion \cite{Horndeski:1974wa}. The similarity is a consequence of the fact that massive bigravity introduces only a single new spin-0 degree of freedom, its equations of motion are second-order, and the new mass scale it introduces (the graviton mass) is comparable to the Hubble scale \cite{Konnig:2014dna,Baker:2014zva}.

Furthermore, the structure growth equation, (\ref{eq:deltaevol}), can be written in terms of $Q$ and $\eta$ and hence the $h_i$ coefficients as
\begin{equation}
\ddot \delta + H\dot\delta - \frac{1}{2}\frac{Q}{\eta}\frac{a^2\bar\rho}{M_g^2}\delta = 0. \label{eq:exprdelta}
\end{equation}
The quantity $Q/\eta$, sometimes called $Y$ in the literature \cite{Amendola:2012ky,Amendola:2013qna,Konnig:2014dna}, represents deviations from Newton's constant in structure growth, and is effectively given in the subhorizon r\'{e}gime by $(h_1h_5)/(h_2h_3)$.

\subsection{Numerical Solutions}

In this section we numerically solve for the background quantities and modified gravity parameters for one- and two-parameter bigravity models.\footnote{We focus on these simpler models to illustrate bigravity effects on growth. Current growth data are not able to significantly constrain these models, so we would not gain anything by adding more free parameters.} We look in particular for potential observable signatures, as the growth data are currently not competitive with background data for these theories, although we expect future LSS experiments such as Euclid \cite{Laureijs:2011gra,Amendola:2012ys} to change this. The recipe is straightforward: using \cref{eq:brepyp} we can solve directly for $y(z)$, which is all we need to find solutions for $\eta(z,k)$ and $Q(z,k)$ using \cref{eq:expreta,eq:exprQ}. Finally these can be used, along with \cref{eq:brepH,eq:breprho}, to solve \cref{eq:exprdelta} numerically for $\delta(z,k)$ and hence for $f(z,k)$. We fit $f(z,k)$ to the parametrisation $\Omega_m^\gamma$ in the redshift range $0<z<5$ unless stated otherwise.

The likelihoods for these models were analysed in detail in Ref. \cite{Akrami:2012vf}, using the Union2.1 compilation of Type Ia supernovae \cite{Suzuki:2011hu}, Wilkinson Microwave Anisotropy Probe (WMAP) seven-year observations of the CMB \cite{Komatsu:2010fb}, and baryon acoustic oscillation (BAO) measurements from the galaxy surveys 2dFGRS, 6dFGS, SDSS and
WiggleZ \cite{Beutler:2011hx,Percival:2009xn,Blake:2011en}. We compute likelihoods based on growth data compiled in Ref. \cite{Macaulay:2013swa}, including growth histories from the 6dFGS \cite{Beutler:2012px}, LRG$_{200}$, LRG$_{60}$ \cite{Samushia:2011cs}, BOSS \cite{Tojeiro:2012rp}, WiggleZ \cite{Blake:2012pj}, and VIPERS \cite{delaTorre:2013rpa} surveys.

Both the numerical solutions of background quantities and the likelihood computations are performed as in Ref. \cite{Akrami:2012vf}, where they are described in detail. Following Ref. \cite{Akrami:2012vf}, we will normalise the $\beta_i$ parameters to present-day Hubble rate, $H_0$, by defining
\begin{equation}
 B_i \equiv \frac{m^2}{H_0^2}\beta_i.
\end{equation}
Throughout, we will assume that the $g$-metric cosmological constant, $B_0$, vanishes, as we are interested in the solutions which accelerate due to modified-gravity effects.

\subsubsection{The Minimal Model}
\label{sec:b1}

We begin with the ``minimal'' model in which only $B_1$ is nonzero. This is the only single-$B_i$ theory which is in agreement with background observations \cite{Akrami:2012vf}; the other models also have theoretical viability issues \cite{Konnig:2013gxa}. Note, however, that the linear perturbations are unstable at early times until relatively recently, $z\sim0.5$, as discussed in \cref{chap:bigravity-stability}. This restricts the real-world applicability of the results presented herein, as the quasistatic approximation we employ will not be viable. Our results will hold for observations within the stable period. Specifically, our results for $Q$ and $\eta$ will certainly hold, while the growth rate, $f$, may vary if the initial conditions for $\delta$ are significantly changed from what we assume herein. Otherwise this should be seen as an illustrative example.

\begin{figure}
\centering
\includegraphics{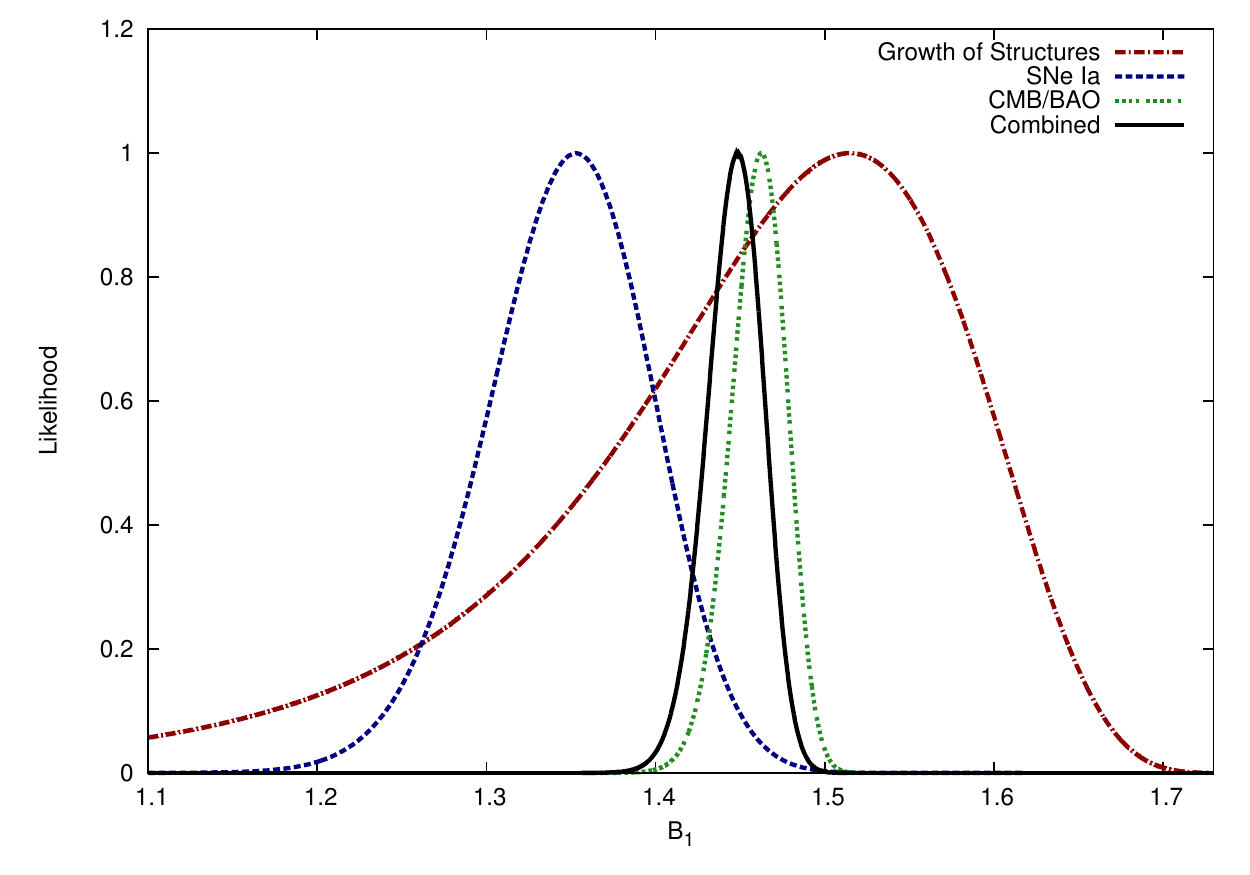}
\caption[Likelihoods for the $B_1$ model.]{The likelihood for $B_1$ in the $B_1$-only model from growth data (red), as well as background likelihoods for comparison. The fits for $B_1$ effectively depend only on the background data; the combined likelihood (black) is not noticeably changed by the addition of the growth data.}
\label{fig:b1-likelihood}
\end{figure}

The likelihoods for $B_1$ are plotted in \cref{fig:b1-likelihood} based on supernovae, BAO/CMB, growth data, and all three combined, although the growth likelihood is so wide that it has a negligible effect on the combined likelihood. The point was raised in Ref. \cite{Konnig:2014dna} that the WMAP analysis is performed assuming a $\Lambda$CDM model and hence may not apply perfectly to these data. We will take an agnostic point of view on this and consider both the best-fit value of $B_1$ from supernovae alone ($B_1=1.3527\pm0.0497$) and from the combination of supernovae and CMB/BAO ($B_1=1.448\pm0.0168$).\footnote{These differ slightly from the best-fit $B_1=1.38\pm0.03$ reported by Ref. \cite{Konnig:2013gxa}, also based on the Union2.1 supernovae compilation.} The results do not change qualitatively with either choice.

\begin{figure}
\centering
\includegraphics{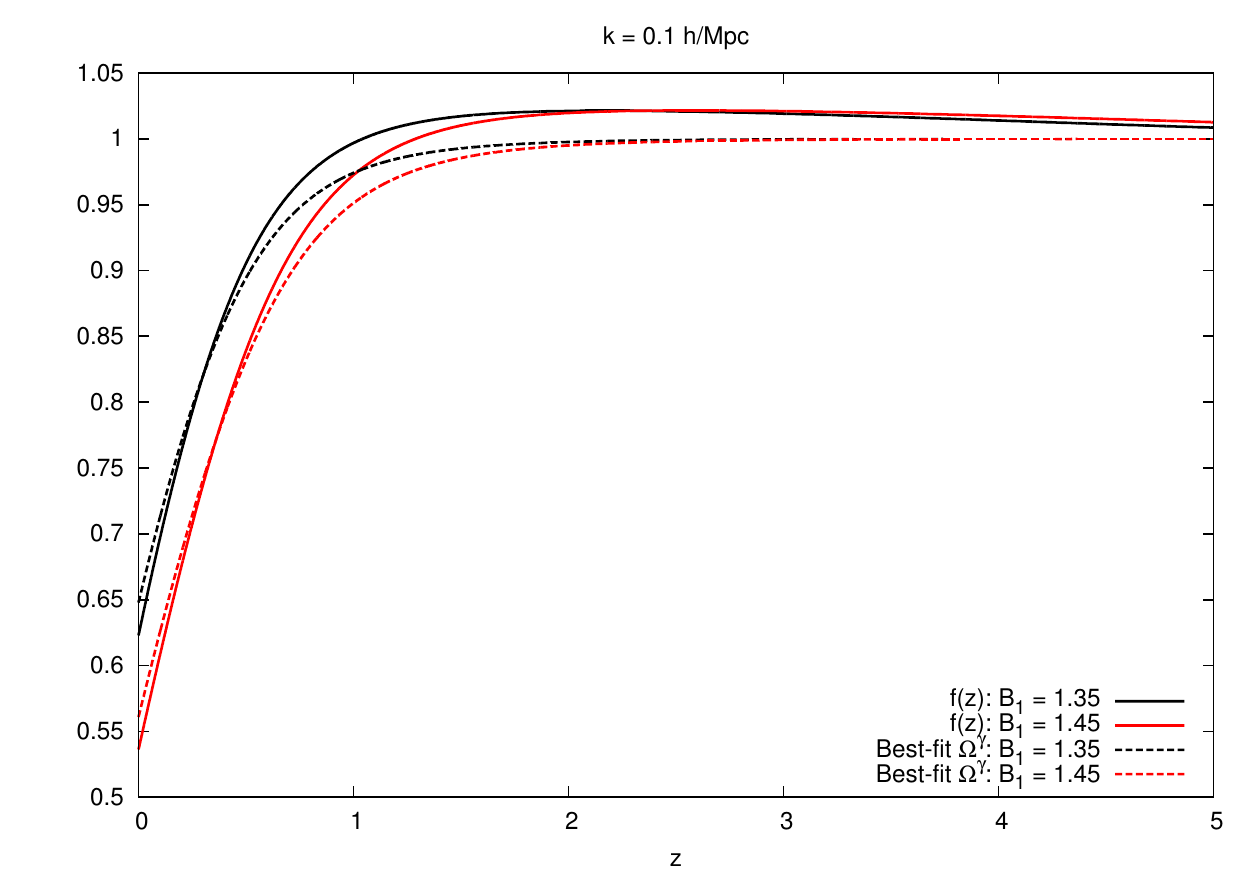}
\includegraphics{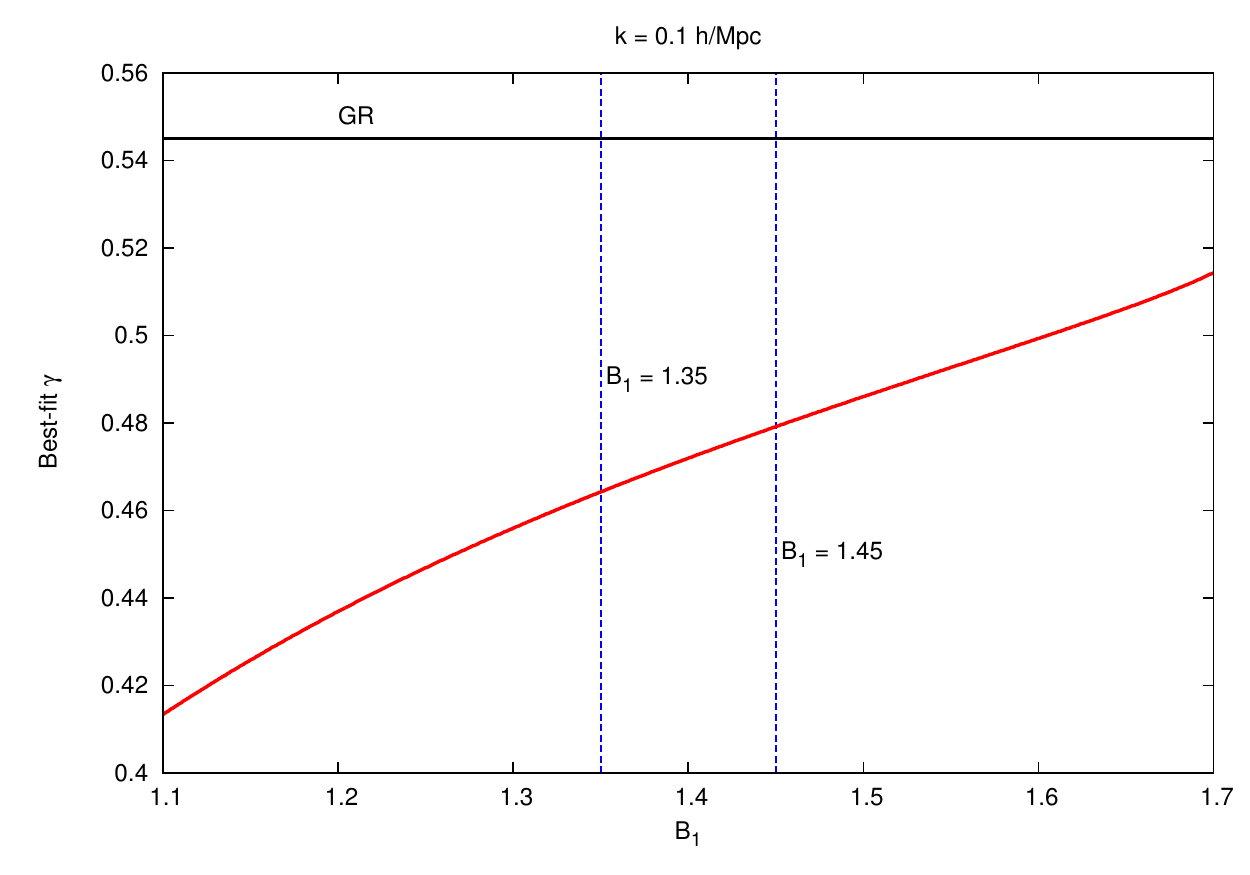}
\caption[Growth rate and growth index for the $B_1$ model.]{\textbf{First panel:} The growth rate, $f = d\ln\delta/d\ln a$, for the SNe best-fit parameter, $B_1=1.35$ (in black), and for the SNe/BAO/CMB combined best-fit parameter, $B_1=1.45$ (in red). The full growth rate (solid line) is plotted alongside the $\Omega^\gamma$ parametrisation (dotted line) with best fits $\gamma=0.46$ and 0.48 for $B_1=1.35$ and 1.45, respectively. \textbf{Second panel:} The best-fit value of $\gamma$ as a function of $B_1$. For comparison, the GR prediction ($\gamma\approx0.545$) is plotted as a black horizontal line. The blue lines correspond to the best-fit values of $B_1$ from different background data sets. This is a prediction of a clear deviation from GR.}
\label{fig:b1-gamma}
\end{figure}

The growth rate, $f$, at $k=0.1$~$h$/Mpc is plotted in the first panel of \cref{fig:b1-gamma}, along with the parametrisation $\Omega_m^\gamma$ with best fits $\gamma=0.46$ for $B_1=1.35$ and $\gamma=0.48$ for $B_1=1.45$. This is in agreement with the results of Ref. \cite{Konnig:2014dna}, who additionally found that $f(z)$ is fit much more closely by a redshift-dependent parametrisation, $f \approx \Omega_m^{\gamma_0}\left(1+\frac{\gamma_1}{1+z}\right)$. In the second panel we plot the best-fit value of $\gamma$ as a function of $B_1$. All values of $B_1$ consistent with background observations give a value of $\gamma$ that is far from the GR value (including $\Lambda$CDM and minimally-coupled quintessence models) of $\gamma\approx0.545$. While present observations of LSS are unable to easily distinguish this model from $\Lambda$CDM (cf. \cref{fig:b1-likelihood}), the Euclid satellite expects to measure $\gamma$ within $0.02$ \cite{Laureijs:2011gra,Amendola:2012ys} and should easily be able to rule out either the minimal massive bigravity model or GR. Note that there is a caveat, in that we have calculated $\gamma$ by fitting over a redshift range ($0<z<5$) which includes the unstable period of this model's history ($z\gtrsim0.5$) during which linear theory breaks down. As emphasised above, these predictions should only be compared to data during the stable period. Therefore if this model does describe reality, Euclid may measure a different growth rate at higher redshifts; a nonlinear analysis is required to answer this with certainty.

\begin{figure}
\centering
\includegraphics{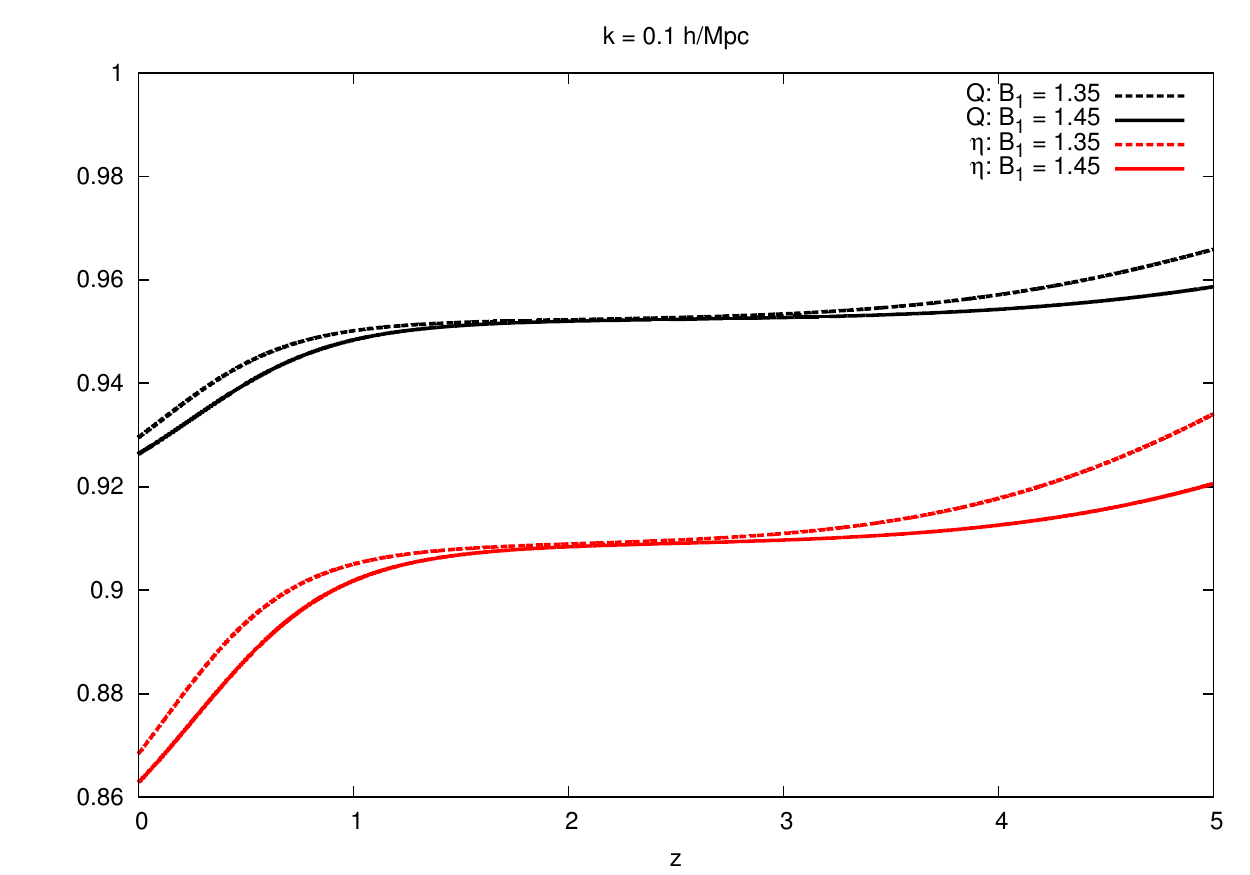}
\includegraphics{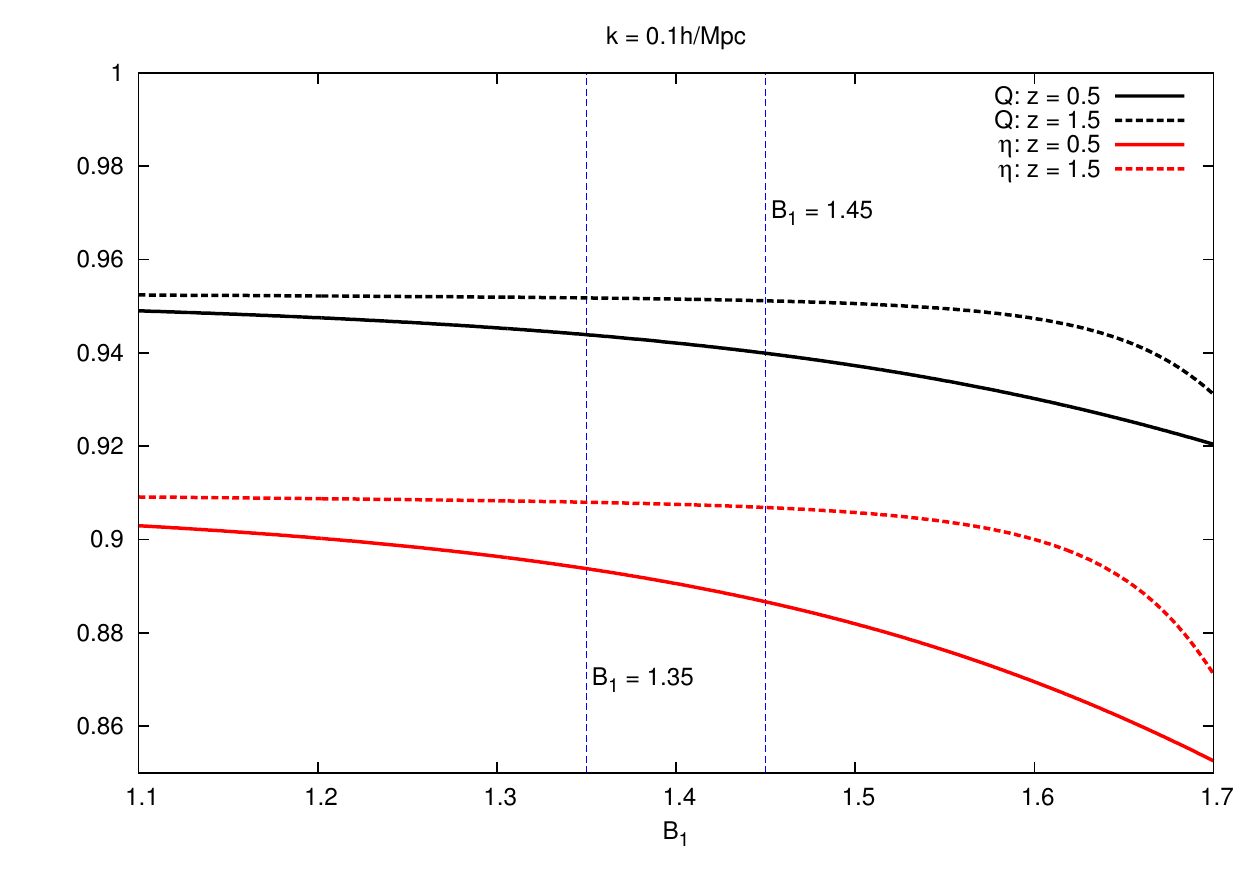}
\caption[Modified gravity parameters $Q$ and $\eta$ for the $B_1$ model as functions of redshift and $B_1$.]{The modified gravity parameters $Q$ (modification of Newton's constant) and $\eta$ (anisotropic stress) in the $B_1$-only model as functions of $z$ and $B_1$. They exhibit $\mathcal{O}(10^{-2})$--$\mathcal{O}(10^{-1})$ deviations from the GR prediction, which will be around the range of observability of a Euclid-like mission.}
\label{fig:b1-qe}
\end{figure}

\begin{figure}
\centering
\includegraphics{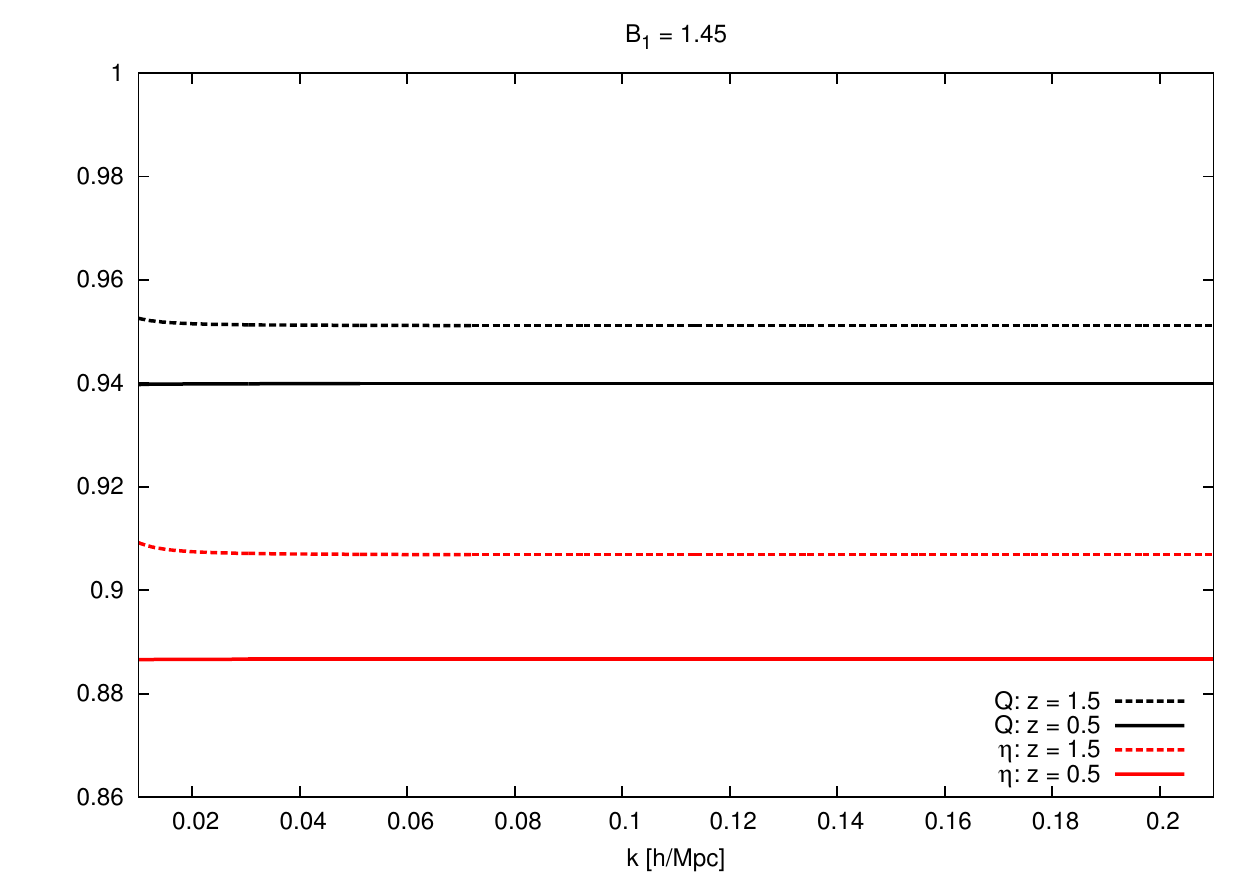}
\caption[Modified gravity parameters $Q$ and $\eta$ for the $B_1$ model as functions of scale.]{The modified gravity parameters $Q$ (modification of Newton's constant) and $\eta$ (anisotropic stress) in the $B_1$-only model as functions of $scale$. They depend only very weakly on scale; consequently, more stringent constraints can be placed on them in a model-independent way by future surveys.}
\label{fig:b1-qe-k}
\end{figure}

We next look at the modified gravity parameters $\eta(z,k)$ and $Q(z,k)$. In \cref{fig:b1-qe,fig:b1-qe-k} they are plotted with respect to $z$, $B_1$, and $k$, respectively, with the other two quantities fixed. $Q$ deviates from the GR value $Q=1$ by $\sim0.05$, while $\eta$ deviates from GR by up to $\sim0.15$. From the first panel of \cref{fig:b1-qe} we notice that $Q$ and $\eta$ lose their dependence on $B_1$ momentarily around $z\sim2.5$. This feature persists to other values of $B_1$ as well. Additionally, we can see from the third panel that $Q$ and $\eta$ only depend extremely weakly on $k$ in the linear subhorizon r\'egime. Future structure experiments like Euclid will be able to constrain $Q$ and $\eta$ more tightly in a model-independent way because they are effectively scale-independent; in particular, because of scale independence they are expected to be able to measure $\eta$ within 10\% \cite{Amendola:2013qna}, which would bring this minimal model to the cusp of observability.

\subsubsection{Two-Parameter Models}

At the background level, four models with two nonzero $B_i$ parameters provide good fits to the data: $B_1B_2$, $B_1B_3$, $B_1B_4$, and $B_2B_3$ \cite{Akrami:2012vf}. Even though these models all possess a two-dimensional parameter space, only an effectively one-dimensional subspace matches the background data (cf. figures 7 and 8 of Ref. \cite{Akrami:2012vf} and figure 4 of Ref. \cite{Konnig:2013gxa}). We will restrict ourselves to those subspaces by fixing one $B_i$ parameter in terms of another, usually $B_1$. We do this by identifying the effective present-day dark energy density, $\Omega_\Lambda^\mathrm{eff}$,\footnote{This does not need to coincide with the value of $\Omega_{\Lambda}$ derived in the context of $\Lambda$CDM models. For the $B_1$-only model and hence all the two-parameter finite-branch models, they happen to be similar in value, although this was not \textit{a priori} guaranteed, while in the infinite-branch $B_1B_4$ model, the best-fit value to the background data is $\Omega_\Lambda^\mathrm{eff}=0.84_{-0.02}^{+0.03}$ \cite{Konnig:2013gxa}.} from the Friedmann equation (\ref{eq:friedmann_g}):
\begin{equation}
\Omega_\Lambda^\mathrm{eff} \equiv B_1y_0 + B_2y_0^2 + \frac{1}{3}B_3y_0^3, \label{eq:omegalambdadef}
\end{equation}
and plugging that into the quartic equation for $y$ (\ref{eq:quartic-mg}), evaluated at the present day ($y=y_0$) and using $\Omega_{m,0} + \Omega_\Lambda^\mathrm{eff}=1$. This procedure fixes one $B_i$ parameter in terms of the other and $\Omega_\Lambda^\mathrm{eff}$. The value of $\Omega_\Lambda^\mathrm{eff}$ can be determined by fitting to the data, as was done in Ref. \cite{Akrami:2012vf}. However, for finite-branch solutions of the $B_1B_i$ models we can also simply take the limit in which only $B_1$ is nonzero, recovering the single-parameter model discussed above; by then using the SNe/CMB/BAO combined best-fit value $B_1=1.448$ we find $\Omega_\Lambda^\mathrm{eff} = 0.699$.

A detailed study of the conditions for a viable background was undertaken in Ref. \cite{Konnig:2013gxa}. There are two results that are particularly relevant for the present study and bear mentioning. First, a viable model requires $B_1\geq0$. Second, the two-parameter models are all (with one exception, which we discuss below) finite-branch models, in which, as discussed in \cref{sec:properties-bimetric-cosmo}, $y$ evolves from 0 at $z=\infty$ to a finite value $y_c$ at late times, which can be determined from \cref{eq:breprho} by setting $\rho=0$ at $y=y_c$. Consequently, the present-day value of $y_0$, which is generally simple to calculate, must always be smaller than $y_c$ (or above $y_c$ in a model with an infinite branch). In the $B_1B_3$ and $B_1B_4$ models this will rule out certain regions of the parameter space \textit{a priori}.

There is one two-parameter model which was shown in \rcite{Akrami:2012vf} to fit the background data well but was ruled out on theoretical grounds in \rcite{Konnig:2013gxa}: the $B_2B_3$ model. The theoretical issue is that $y$ becomes negative in finite time going towards the past. This itself may not render the model observationally unviable, as long as any issues occur outside of the redshift range of observations. However, we have solved \cref{eq:brepyp} numerically for $y(z)$, and found that $y$ generically goes to $-\infty$ at finite $z$, which means that at higher redshifts there is not a sensible background cosmology at all. This problem can be avoided by introducing new physics at those higher redshifts to modify the evolution of $y$, or by increasing $B_2$ enough that the pole in $y$ occurs at an unobservably high redshift.\footnote{For $B_2=(5,50,500)$, and $B_3$ chosen to give an effective $\Omega_\Lambda^\mathrm{eff}\approx0.7$ today, the pole occurs at $z\approx(1.99,8.19,27.95)$.} However, these are nonminimal solutions, and so we do not study the $B_2B_3$ model.

Recall from \cref{chap:bigravity-stability} that all of the two-parameter models except for the infinite-branch $B_1B_4$ model suffer from an early-time instability. Consequently, caution should be used when applying the results for any of the other models in this section to real-world data. We emphasise again that our quasistatic approximation is only valid at low redshifts, and that moreover the growth rate should be recalculated using whatever initial conditions the earlier period ends with. (The predictions for $Q$ and $\eta$ do not depend on solving any differential equation and therefore apply without change.) Modulo this caveat, we present the quasistatic results for the unstable models as proofs of concept, as examples of how to apply our methods. The infinite-branch $B_1B_4$ predictions, presented at the end of this section, can be straightforwardly applied to data.

We evaluate all quantities at $k=0.1$~$h$/Mpc. The modified gravity parameters in all of the two-parameter models we study depend extremely weakly on $k$, as in the $B_1$-only model.

As we have already mentioned, in these models the two $B_i$ parameters are highly degenerate at the background level. One of the main goals of this section is to see whether observations of LSS have the potential to break this degeneracy.

\paragraph{$\mathbf{B_1}\mathbf{B_2}$:} 

The $B_1B_2$ models which fit the background data \cite{Akrami:2012vf} live in the parameter subspace
\begin{equation}
B_2=\frac{-B_1^2 + 9 \Omega_\Lambda^\mathrm{eff} - \sqrt{B_1^4 + 9 B_1^2 \Omega_\Lambda^\mathrm{eff}}}{9\Omega_\Lambda^\mathrm{eff}}, \label{eq:b2b1rel}
\end{equation}
with $\Omega_\Lambda^\mathrm{eff} \approx 0.7$. This line has a slight thickness because we must rely on observations to fit $\Omega_\Lambda^\mathrm{eff}$. We can subsequently determine $y_0$ from \cref{eq:omegalambdadef}.

\begin{figure}
\centering
\includegraphics{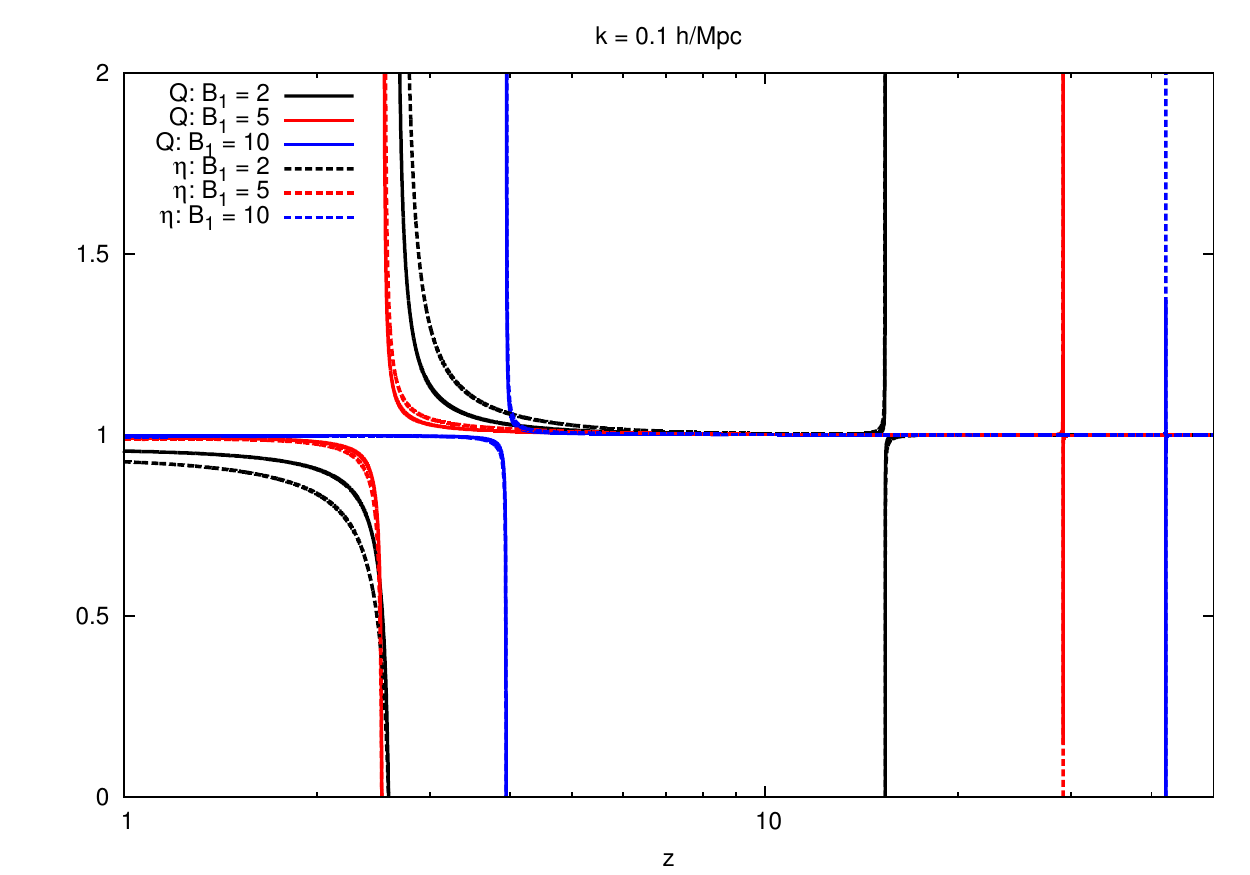}
\includegraphics{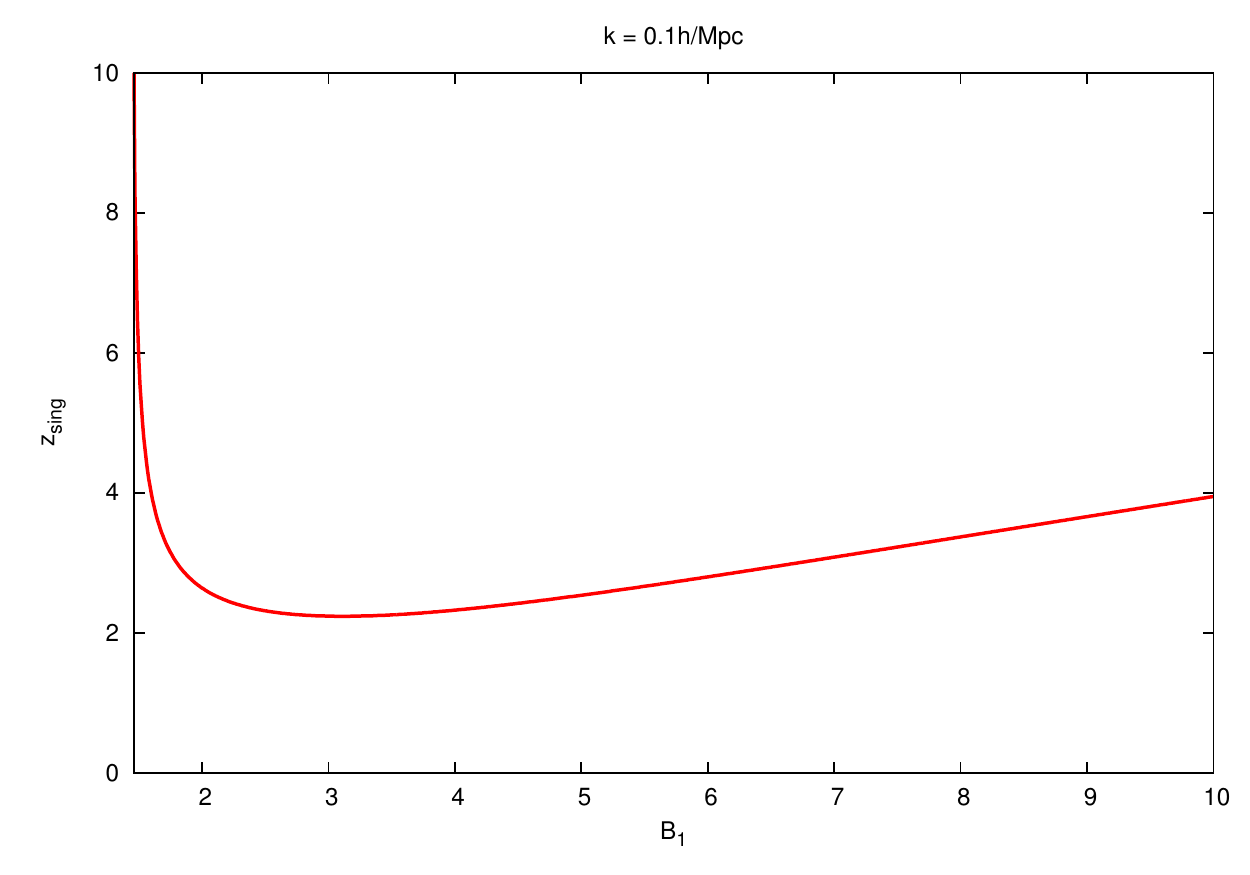}
\caption[Poles in $Q$ and $\eta$ for the $B_1B_2$ model.]{\textbf{First panel:} $Q$ and $\eta$ for a few parameter values in the instability range of the $B_1B_2$ model. Generally there are two poles, each of short duration. Note that the redshift at which the pole occurs depends only on $B_1$ and $B_2$ and not on the initial conditions for the perturbations. \textbf{Second panel:} The redshift of the most recent pole in $Q$ (the pole in $\eta$ occurs nearly simultaneously) in terms of $B_1>1.448$. This parameter range corresponds to $B_2<0$, cf. \cref{eq:b2b1rel}. The minimum is at $(B_1,B_2,z)=(3.11,-2.51,2.24)$.}
\label{fig:zsing}
\end{figure}

This model possesses an instability when $B_2<0$.\footnote{A similar singular evolution of linear perturbations in a smooth background has been observed in the cosmology of Gauss-Bonnet gravity \cite{Kawai:1999pw,Koivisto:2006xf}. This instability is different from the early-time instabilities discussed in \cref{chap:bigravity-stability}, as those do not arise in the quasistatic limit which we are now taking.} This is not entirely unexpected: the $B_2$ term is the coefficient of the quadratic interaction, and so a negative $B_2$ might lead to a tachyonic instability. However, the instability of the $B_1B_2$ model is somewhat unusual: in the subhorizon limit, $Q$ and $\eta$ develop poles, but they only diverge during a brief period around a fixed redshift, as shown in the first panel of \cref{fig:zsing}, regardless of wavenumber or initial conditions. We can find these poles using the expressions in \cref{app:hcoeff}. The exact solutions are unwieldy and not enlightening, but there are three notable features. First, as mentioned, the instability only occurs for $B_2<0$. (When $B_2>0$, these poles occur when $y$ is negative, which is not physical.) Second, the instability develops at high redshifts, $z>2$. The redshift of the latest pole (which can be solved for by taking the limit $k/H\to\infty$) is plotted as a function of $B_1$ in the second panel of \cref{fig:zsing}. As a result, measurements of $Q$ and $\eta$ at $z\lesssim2$ would generally not see divergent values. However, such measurements would see the main instabilities at much lower redshifts. Finally, the most recent pole occurs at $y=0$ for $B_2=0$ ($B_1=\sqrt{3\Omega_\Lambda^\mathrm{eff}} \approx 1.45$), and approaches $y=y_0/2$ as $B_2\to-\infty$ ($B_1\to\infty$).

\begin{figure}
\centering
\includegraphics{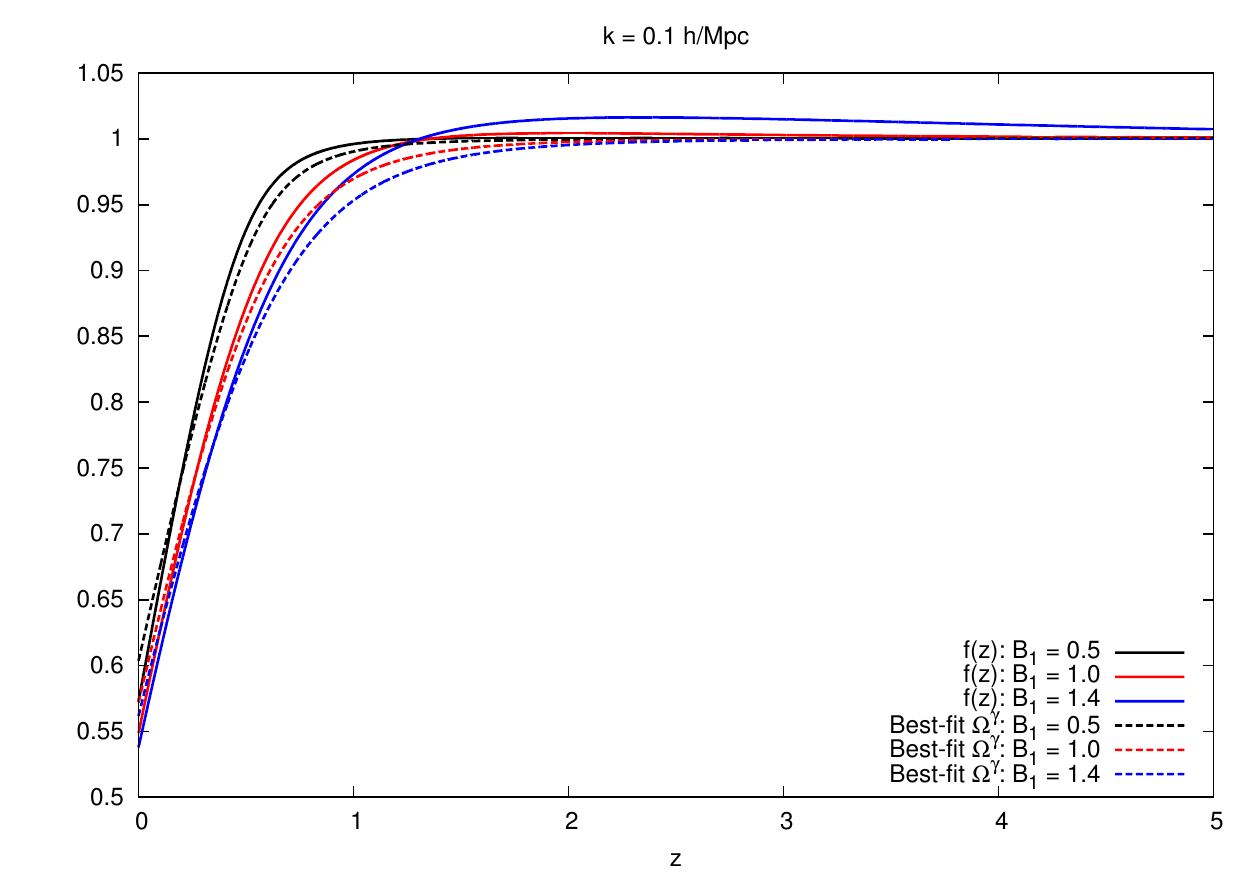}
\includegraphics{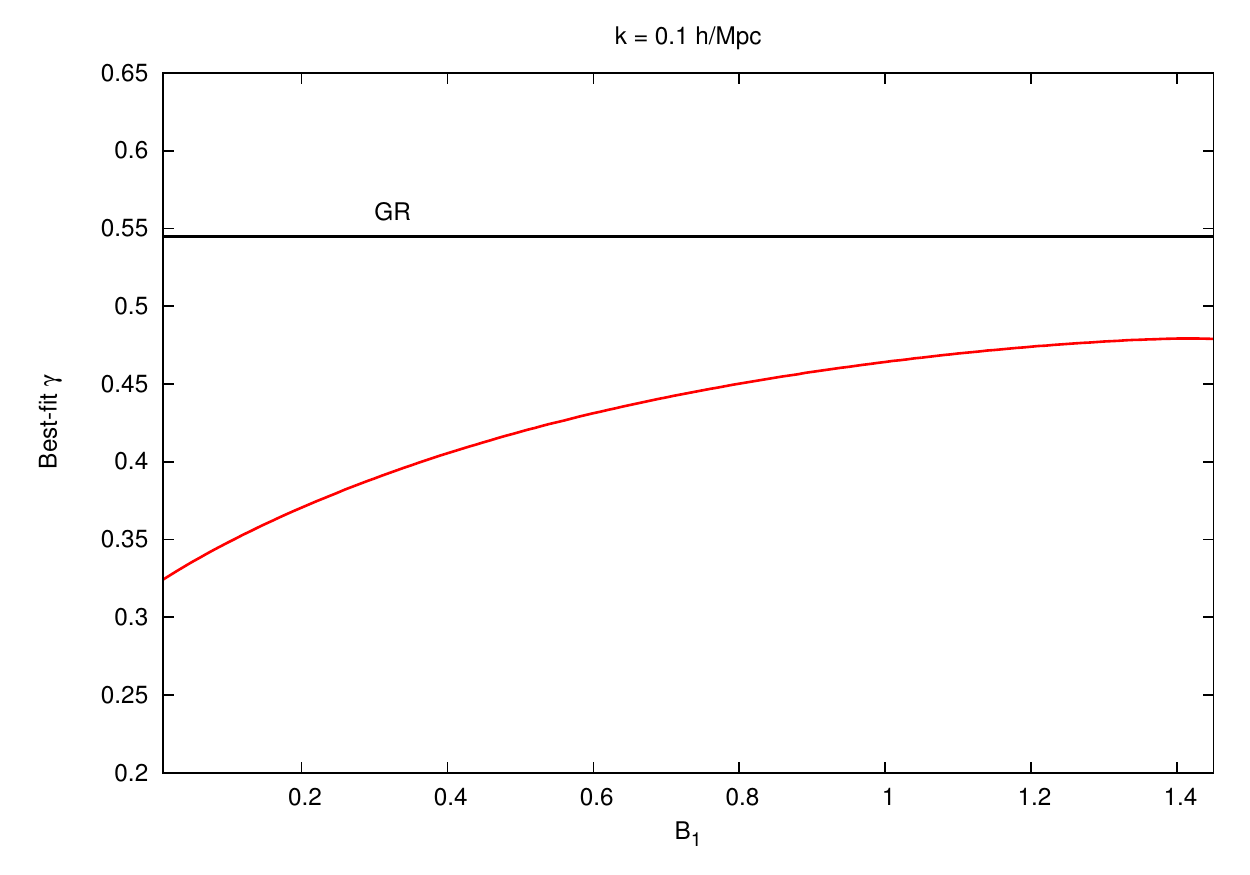}
\caption[Modified growth rate for the $B_1B_2$ model.]{Growth-rate results for the $B_1B_2$ model, with $\Omega_\Lambda^\mathrm{eff}=0.699$ and $B_1\leq1.448$. The significant deviation from GR in $\gamma$ should be observable by a Euclid-like experiment. Moreover, it has the potential to break the degeneracy between $B_1$ and $B_2$ when fitting to background observations.}
\label{fig:b1-b2-f}
\end{figure}

\begin{figure}
\centering
\includegraphics{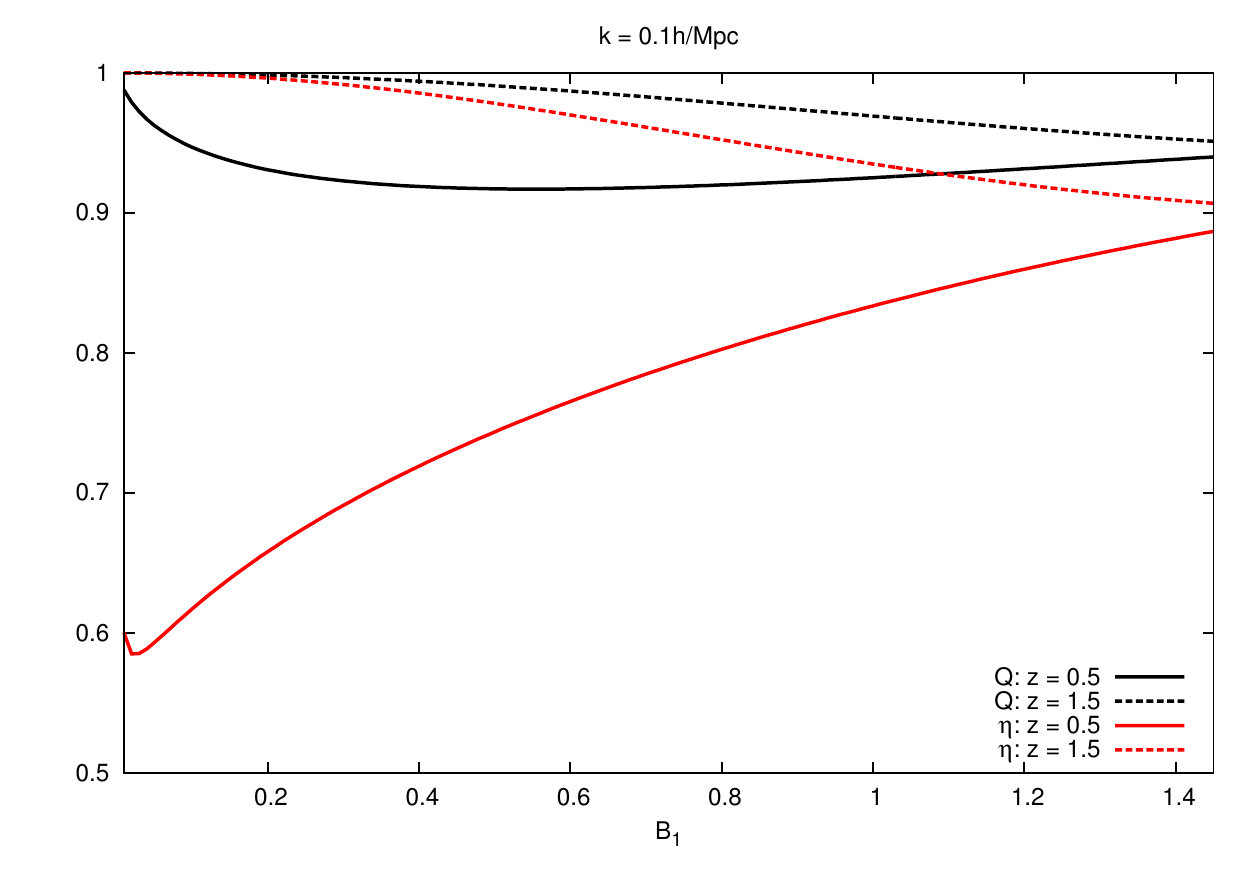}
\includegraphics{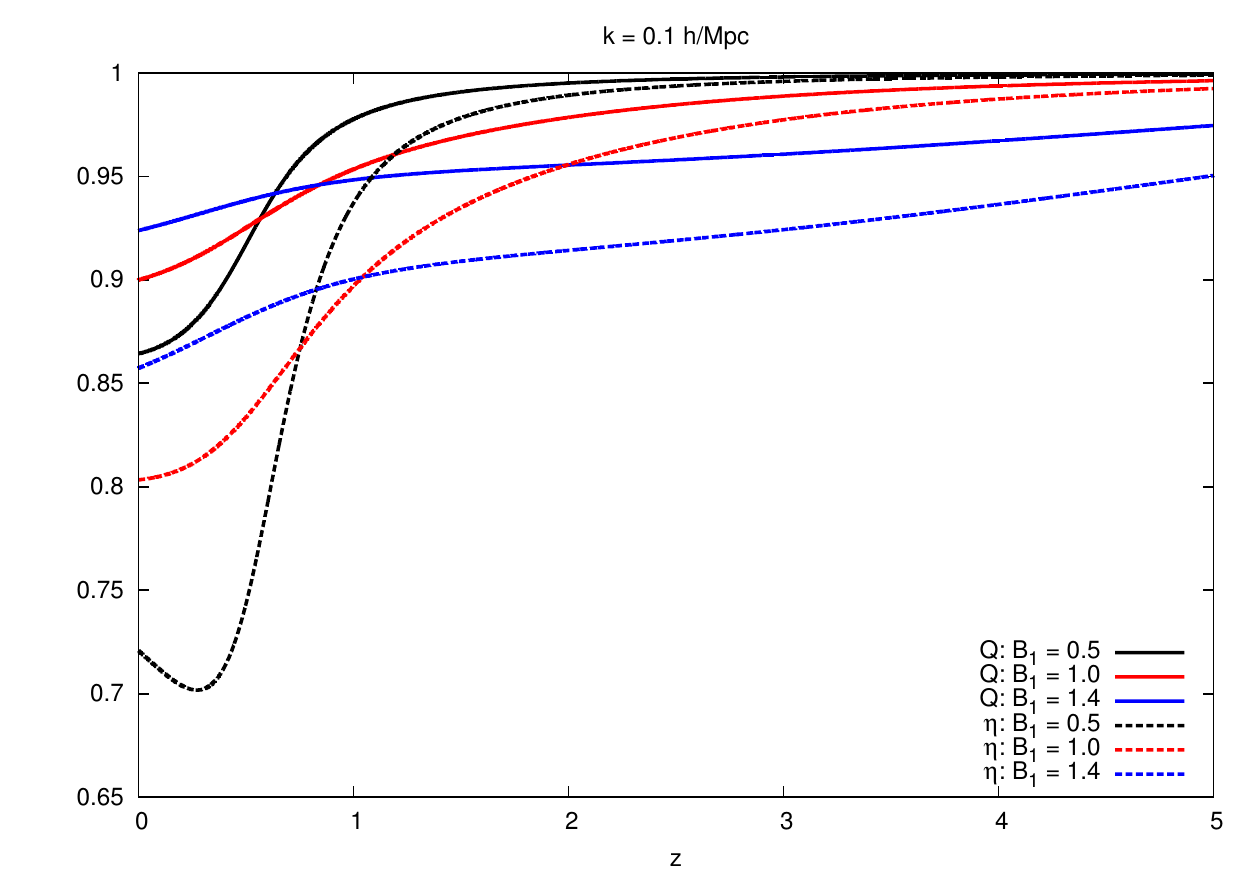}
\caption[Modified gravity parameters $Q$ and $\eta$ for the $B_1B_2$ model.]{The modified gravity parameters $Q$ and $\eta$ for the $B_1B_2$ model, with $\Omega_\Lambda^\mathrm{eff}=0.699$ and $B_1\leq1.448$. As with the growth rate in \cref{fig:b1-b2-f}, we notice significant deviations from GR. The anisotropic stress, $\eta(z\sim0.5)$, is a particularly good target for observations.}
\label{fig:b1-b2-qeta}
\end{figure}

This particular instability is avoided if we restrict ourselves to the range $0<B_1\leq1.448$, for which $B_2>0$. Some typical results for this region of parameter space are plotted in \cref{fig:b1-b2-f,fig:b1-b2-qeta}. The first panel of \cref{fig:b1-b2-f} plots $f(z)$ and the best-fit $\Omega_m^\gamma$ parametrisation for selected values of $B_1$ [with $B_2$ given by \cref{eq:b2b1rel}], while the second panel shows the best-fit value of $\gamma$ as a function of $B_1$. For smaller values of $B_1$ this parametrisation fits $f$ well, more so than in the $B_1$-only model discussed in \cref{sec:b1} (which is the $B_1=1.448$ limit of this model). We find that $\gamma$ is always well below the GR value of $\gamma\approx0.545$, especially at low $B_1$.

In \cref{fig:b1-b2-qeta} we plot $Q$ and $\eta$, both in terms of $B_1$ at fixed $z$ and in terms of $z$ at fixed $B_1$. In comparison to the $B_1$-only model (at the far-right edge in the first panel), lowering $B_1$ tends to make these parameters more GR-like, except for $\eta$ evaluated at late times ($z\sim0.5$), which dips as low as $\eta\sim0.6$. Because these quantities are all $k$-independent in the linear, subhorizon r\'egime, future LSS experiments like Euclid would be able to measure $\eta$ at these redshifts to within about 10\% \cite{Amendola:2013qna} and thus effectively distinguish between $\Lambda$CDM and significant portions of the parameter space of the $B_1B_2$ model, testing the theory and breaking the background-level degeneracy.

\paragraph{$\mathbf{B_1}\mathbf{B_3}$:} 

The $B_1B_3$ models which are consistent with the background data \cite{Akrami:2012vf} lie along a one-dimensional parameter space (up to a slight thickness) given by
\begin{equation}
B_3 = \frac{-32 B_1^3 + 81 B_1 \Omega_\Lambda^\mathrm{eff} \pm \sqrt{\left(8 B_1^2-27 \Omega_\Lambda^\mathrm{eff}\right)^2 \left(16 B_1^2+27 \Omega_\Lambda^\mathrm{eff}\right)}}{243 (\Omega_\Lambda^\mathrm{eff})^2}, \label{eq:b3b1rel}
\end{equation}
with $\Omega_\Lambda^\mathrm{eff} \approx 0.7$. There is a subtlety here: the physical branch is constructed in a piecewise fashion, taking the $+$ root for $B_1<(3/2)^{3/2}\sqrt{\Omega_\Lambda^\mathrm{eff}}= 1.536$ and the $-$ root otherwise \cite{Konnig:2013gxa}. We solve for $y(z)$ using the initial condition,\footnote{There is also a positive root, but this is not physical. When $B_3<0$, that root yields $y_0<0$. When $B_3>0$, which is only the case for a small range of parameters, then the positive root of $y_0$ is greater than the far-future value $y_c$ and hence is also not physical.} derived from \cref{eq:quartic-mg},
\begin{equation}
y_0 = \frac{1-\sqrt{1-\frac{4}{3}B_1B_3}}{2B_3}.
\end{equation}
As discussed at the beginning of this section, $y_0$ should not be larger than the value of $y$ in the far future, $y_c$. Demanding this, we find a maximum allowed value for $B_3$,\footnote{Note, per \cref{eq:b3b1rel}, that this is equivalent to simply imposing $\Omega_\Lambda^\mathrm{eff}<1$, which must be true since we have chosen a spatially-flat universe \textit{a priori}.}
\begin{equation}
B_3 < \frac{1}{243}\left(-32B_1^3 + 81B_1+\sqrt{\left(16B_1^2+27\right)\left(8B_1^2-27\right)^2}\right).
\end{equation}
For $\Omega_\Lambda^\mathrm{eff}=0.699$ this implies we need to restrict ourselves to $B_1>1.055$. This sort of bound is to be expected: we know that the $B_3$-only model is a poor fit to the data \cite{Akrami:2012vf}, so we cannot continue to get viable cosmologies the entire way through the $B_1\to0$ limit of the $B_1B_3$ model.

\begin{figure}
\centering
\includegraphics{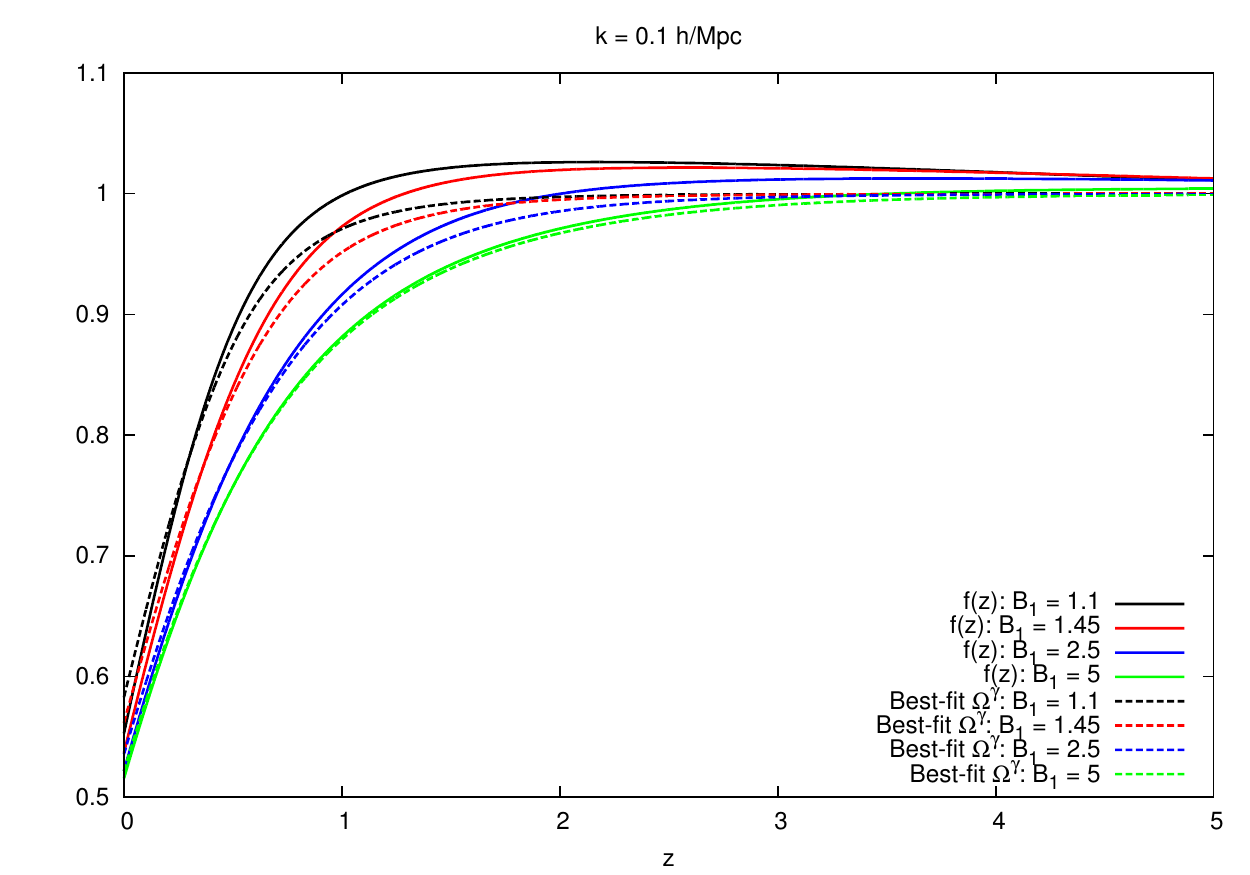}
\includegraphics{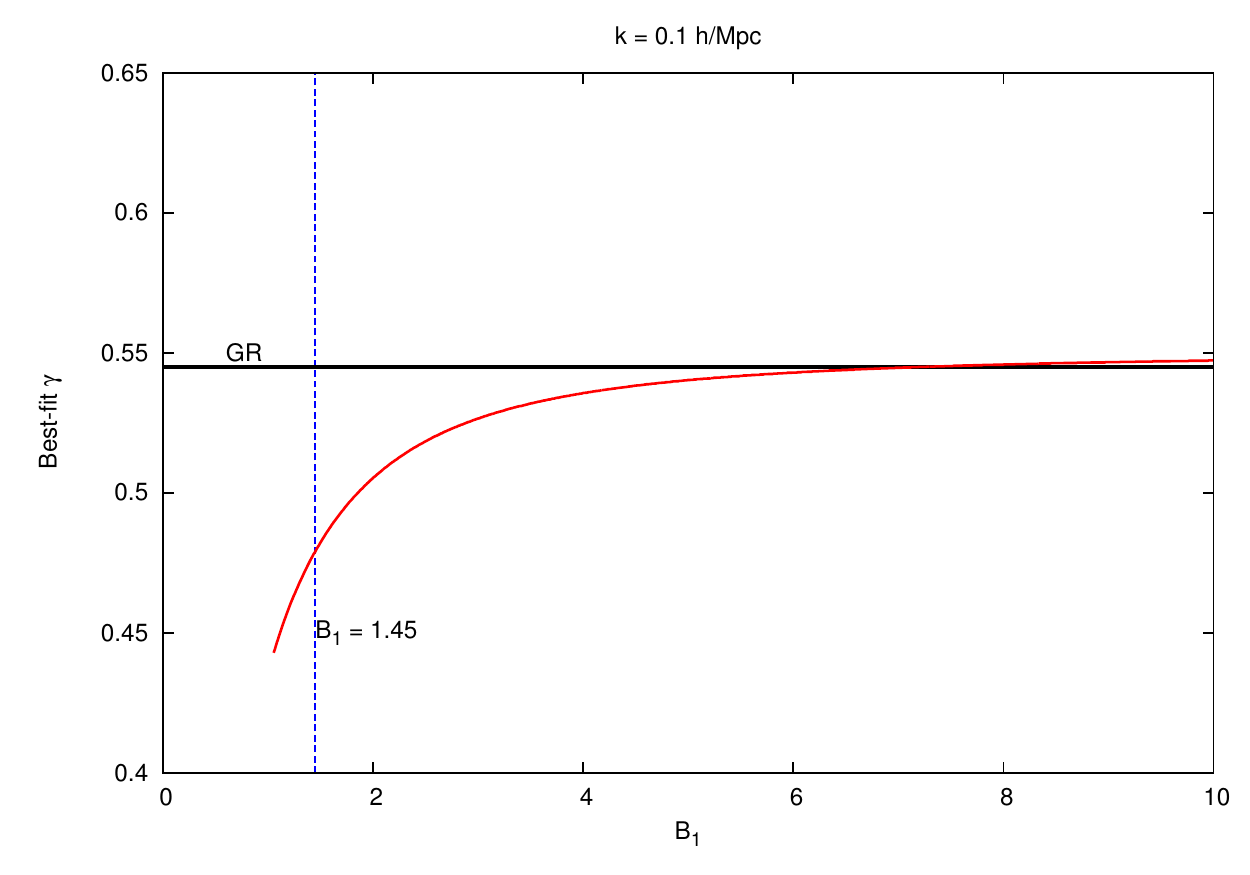}
\caption[Modified growth rate for the $B_1B_3$ model.]{Growth-rate results for the $B_1B_3$ model, with $\Omega_\Lambda^\mathrm{eff}=0.699$ and $B_1>1.055$. While $B_1$ and $B_3$ are degenerate in the background (along a line given by \cref{eq:b3b1rel}), perturbations clearly can break this degeneracy, with significant deviations at low values of $B_1$ (small $|B_3|$), and GR-like behaviour at large values of $B_1$ (large, negative $B_3$).}
\label{fig:b1-b3-f}
\end{figure}

\begin{figure}
\centering
\includegraphics{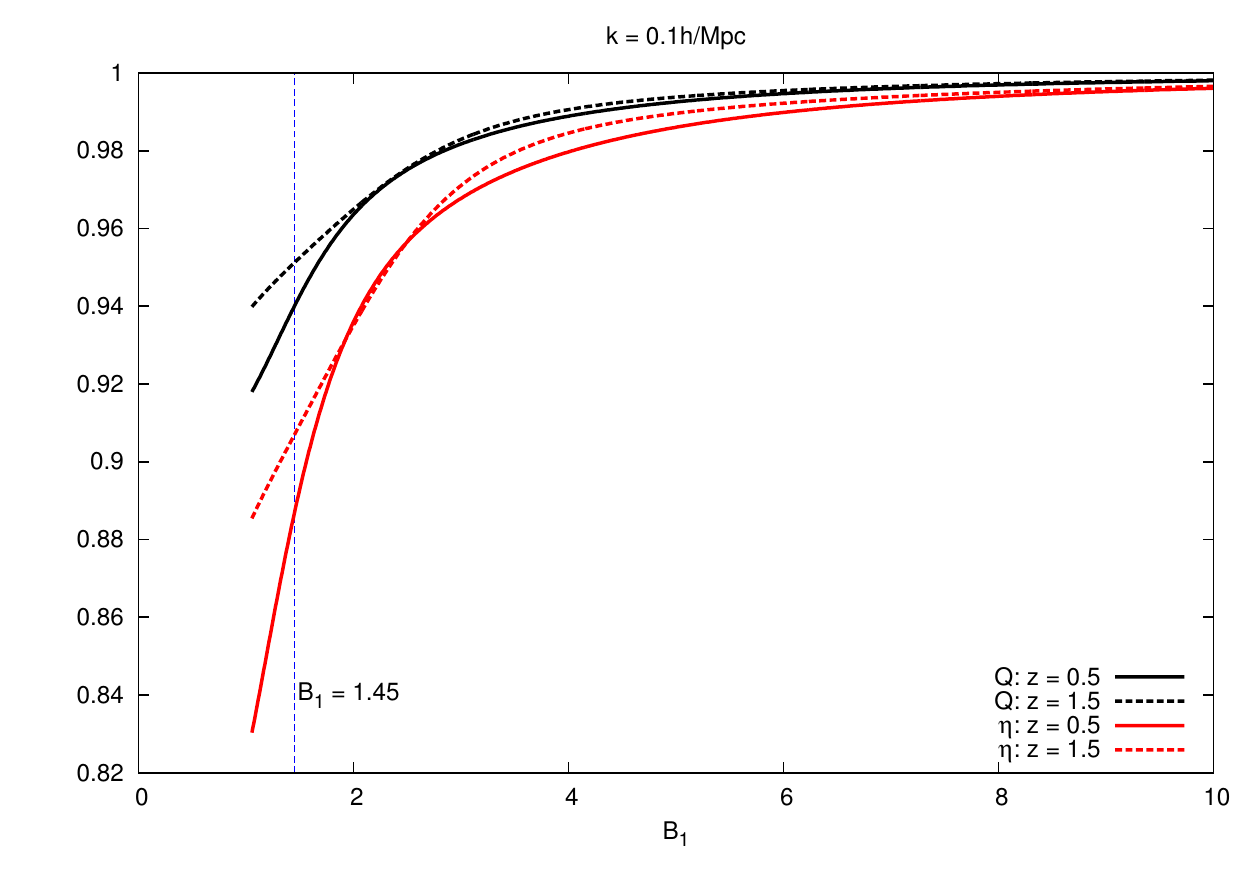}
\includegraphics{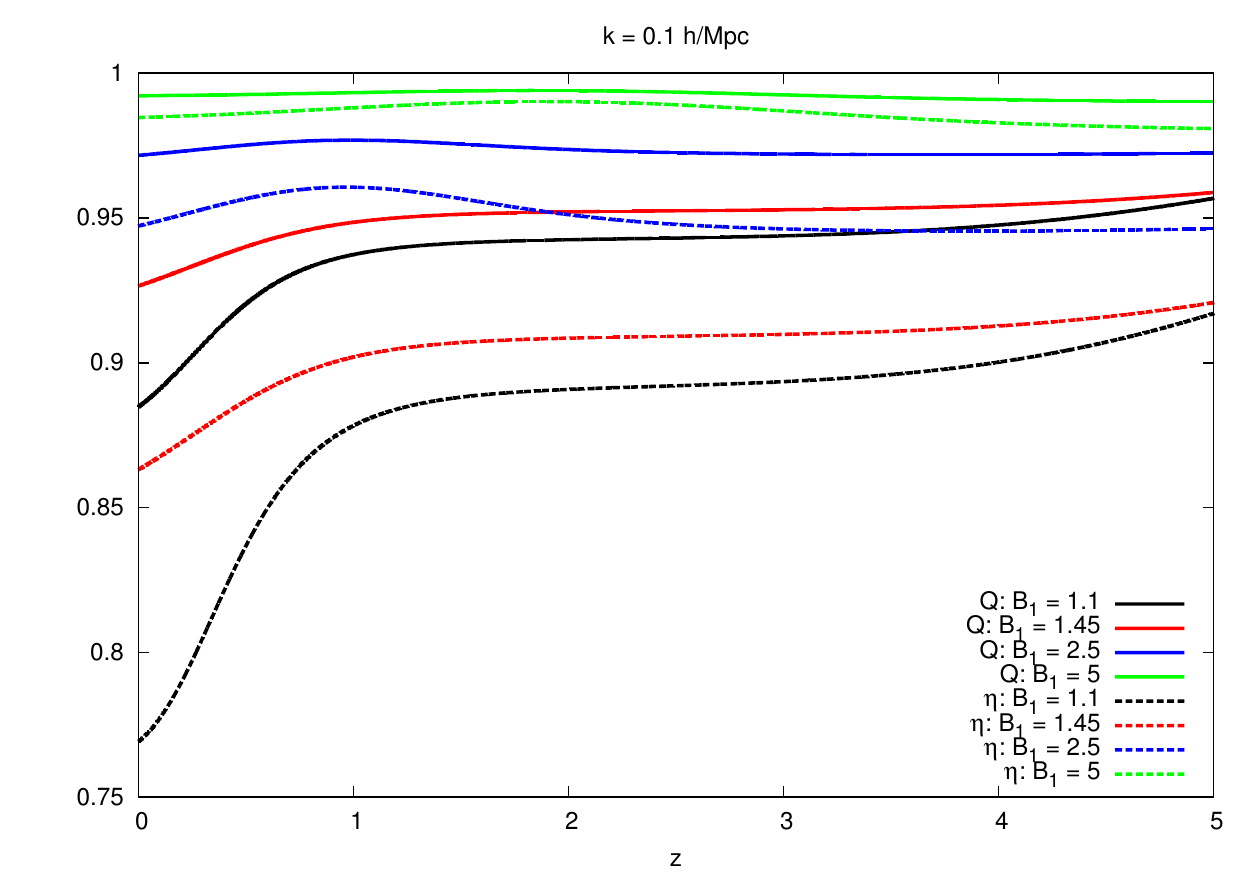}
\caption[Modified gravity parameters $Q$ and $\eta$ for the $B_1B_3$ model.]{The modified gravity parameters $Q$ and $\eta$ for the $B_1B_3$ model, with $\Omega_\Lambda^\mathrm{eff}=0.699$ and $B_1>1.055$. As with the growth rate, we notice deviations from GR for small $B_1$ (small $|B_3|$), with parameters approaching their GR values for large $B_1$ (large, negative $B_3$).}
\label{fig:b1-b3-qeta}
\end{figure}

We plot the results for the $B_1B_3$ model in \cref{fig:b1-b3-f,fig:b1-b3-qeta}. These display the tendency, which we will also see in the $B_1B_4$ model, that large $|B_i|$ values lead to modified gravity parameters that are closer to GR. For example, $\gamma$ can be as low as $\gamma\approx0.45$ for the lowest allowed value of $B_1$, but by $B_1\sim3$ it is practically indistinguishable from the GR value, assuming a Euclid-like precision of $\sim0.02$ on $\gamma$ \cite{Laureijs:2011gra}. Again we note that this value of $\gamma$ has been obtained assuming $0<z<5$, which is not a valid range for observations because of the early-time instability. For lower values of $B_1$, current growth data (see, e.g., \rcite{Macaulay:2013swa}) are not sufficient to significantly constrain the parameter space, but these non-GR values of $\gamma$ and $\eta$ should be well within Euclid's window.

\paragraph{$\mathbf{B_1}\mathbf{B_4}$:}

This model comprises the lowest-order $B_1$ term in conjunction with a cosmological constant for the $f$ metric, $B_4$.\footnote{In the singly-coupled version of massive bigravity we are studying, matter loops only contribute to the $g$-metric cosmological constant, $B_0$.} Note that $B_4$ does not contribute directly to the Friedmann equation (\ref{eq:friedmann_g}), but only affects the dynamics through its effect on the evolution of $y$.

The $B_1B_4$ model has two viable solutions for $y(z)$: a finite branch with $0<y<y_c$, and an infinite branch with $y_c<y<\infty$. The infinite-branch model is the only two-parameter bimetric model which is linearly stable at all times, as shown in \cref{chap:bigravity-stability}. Therefore this should be considered the most viable bimetric massive gravity theory to date. In this section we will elucidate its predictions for subhorizon structure formation.

As discussed in the beginning of this section, $y_c$ is the value of $y$ in the asymptotic future and can be calculated by setting $\rho=0$ in \cref{eq:breprho}. We will consider the two branches separately.

\begin{figure}
 \centering
 \includegraphics[width=0.75\textwidth]{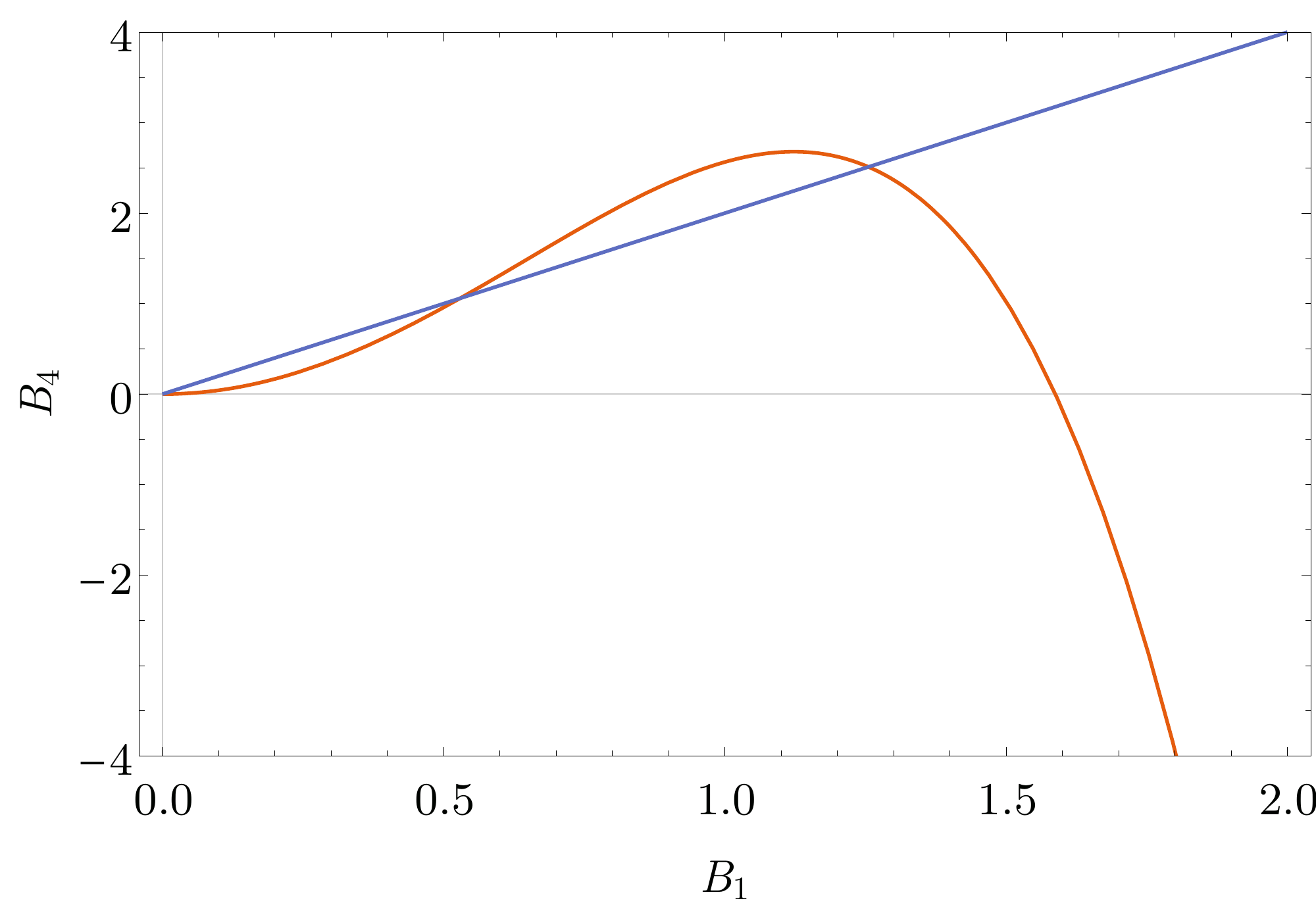}
  \includegraphics[width=0.75\textwidth]{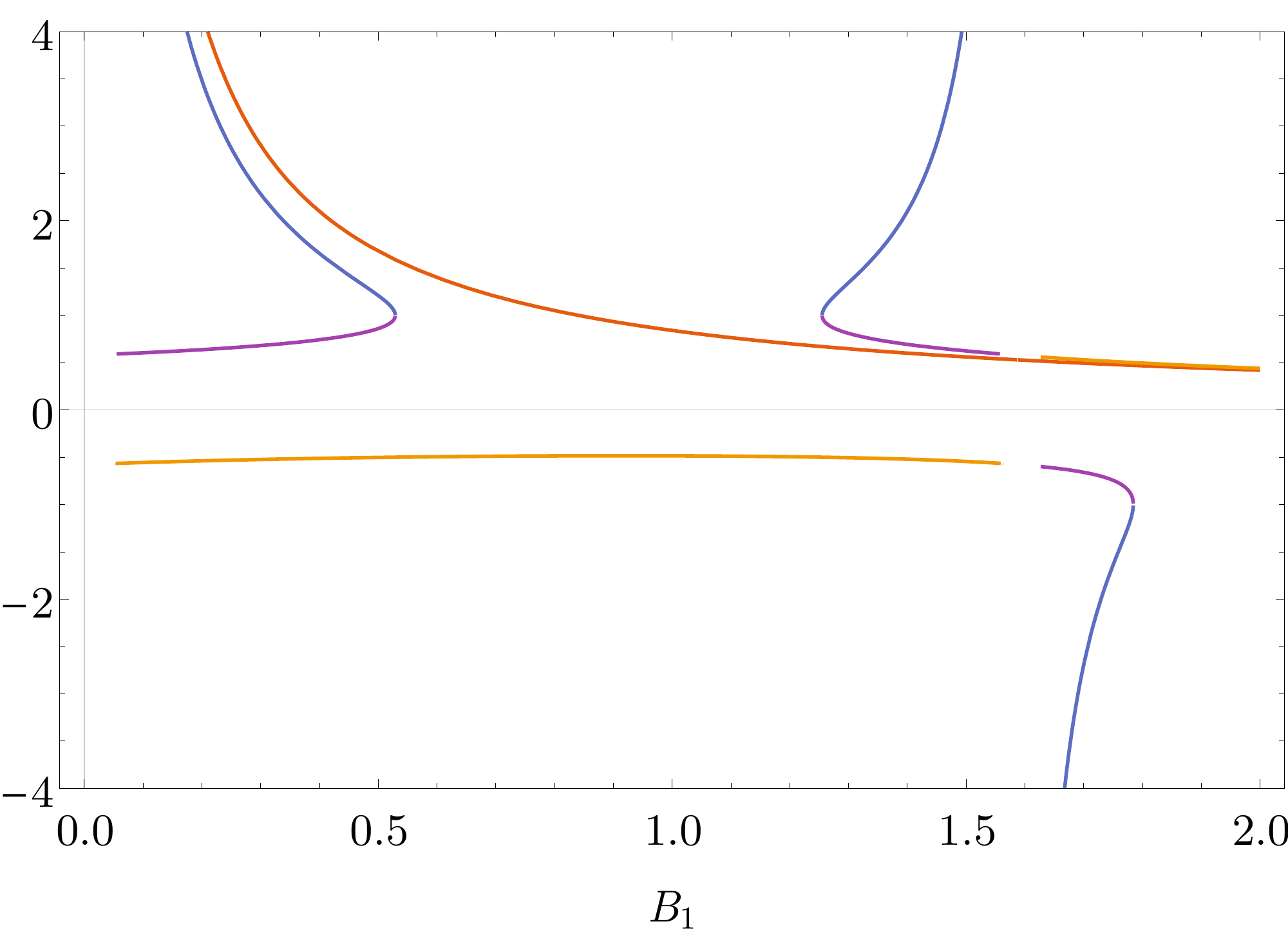}
 \caption[Allowed regions in the $B_1$--$B_4$ parameter space.]{Allowed regions in the $B_1$--$B_4$ parameter space. \textbf{First panel:} Regions in parameter space with $B_4<B_1$, which is required for the viability of the background. The orange line fixes $B_4$ as a function of $B_1$ for $\Omega_\Lambda^\eff=0.84$, per \cref{eq:b1-b4}. The blue line is $B_4=2B_1$. \textbf{Second panel:} Plotted are $y_0$ (orange line) and the three roots of $y_c$, again with $\Omega_\Lambda^\eff=0.84$. Of the two regions with $B_4<B_1$, we can see that the region with smaller $B_1$ has $y_0>y_c$ and therefore corresponds to the infinite branch, while the region with larger $B_1$ has $y_0<y_c$ and therefore possesses finite-branch solutions.}
 \label{fig:b1b4regions}
\end{figure}

For a given $\Omega_\Lambda^\mathrm{eff}$, $B_4$ is related to $B_1$ by
\begin{equation}
B_4 = \frac{3\Omega_\Lambda^\mathrm{eff}B_1^2 - B_1^4}{(\Omega_\Lambda^\mathrm{eff})^3}, \label{eq:b1-b4}
\end{equation}
while $y_0$ is given by
\begin{equation}
 y_0 = \frac{\Omega_\Lambda^\mathrm{eff}}{B_1}.
\end{equation}
Background viability conditions impose $B_1>0$ for both branches and $B_4>0$ on the infinite branch \cite{Konnig:2013gxa}. The late-time value of $y$, $y_c$, is determined by
\begin{equation}
 B_4y_c^3 - 3B_1y_c^2 + B_1 = 0, \label{eq:b1b4yc}
\end{equation}
from which we can determine that real, positive solutions for $y_c$ only exist if $B_4 < 2B_1$. Combined with \cref{eq:b1-b4} and the requirements that $B_1,\Omega_\Lambda^\mathrm{eff}>0$, we find two possible regions for $B_1$, as can be seen from the example plotted in the first panel of \cref{fig:b1b4regions}.
We can identify each of these two regions with the two solution branches by comparing the solutions of \cref{eq:b1b4yc} for $y_c$ to $y_0 = \Omega_\Lambda^\eff/B_1$. Restricting to positive, real roots of $y_c$ one can see (graphically, for example) that the first region, with the smaller values of $B_1$, has $y_0>y_c$ for all roots of $y_c$, and hence can only comprise infinite-branch solutions, while $y_0 < y_c$ in the second region, as is plotted in the second panel of \cref{fig:b1b4regions}. This identification is supported by observational data \cite{Konnig:2013gxa} and also makes intuitive sense. Consider the limit $B_4=0$. In the second region this has $B_1>0$ and corresponds to the $B_1$-only model, a finite-branch model, which we discussed in \cref{sec:b1}. In the first region the point $B_4=0$ coincides with $B_1=0$, which is simply a CDM model with no modification to gravity, in agreement with the fact that there should not be an infinite-branch $B_1$-only model.

These considerations place constraints on the allowed ranges of $B_1$, as in the $B_1B_3$ model, which depend on the best-fit value of $\Omega_\Lambda^\mathrm{eff}$. The $B_1$-only model ($B_4=0$, $B_1>0$) is on the finite branch, so that on that branch we can use $\Omega_\Lambda^\mathrm{eff}=0.699$ as we did in the other models. This implies $B_1>1.244$ for the finite branch. On the infinite branch, SNe observations are best fit by $\Omega_\Lambda^\mathrm{eff}=0.84$ \cite{Konnig:2013gxa}; consequently we restrict ourselves to $B_1<0.529$ for the infinite branch. Note that the infinite-branch model therefore predicts an unusually low matter density, $\Omega_{m,0}\approx 0.18$.

\begin{figure}
\centering
\includegraphics{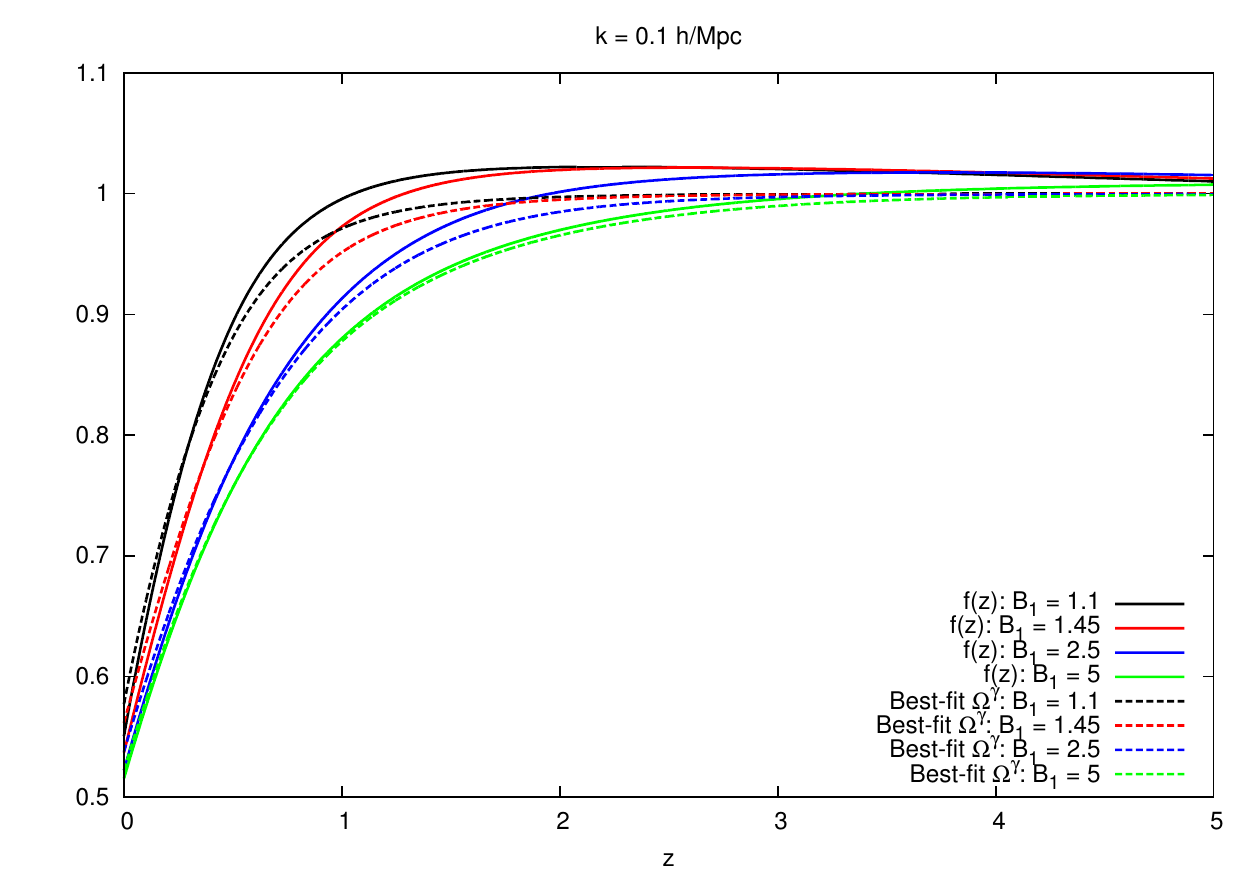}
\includegraphics{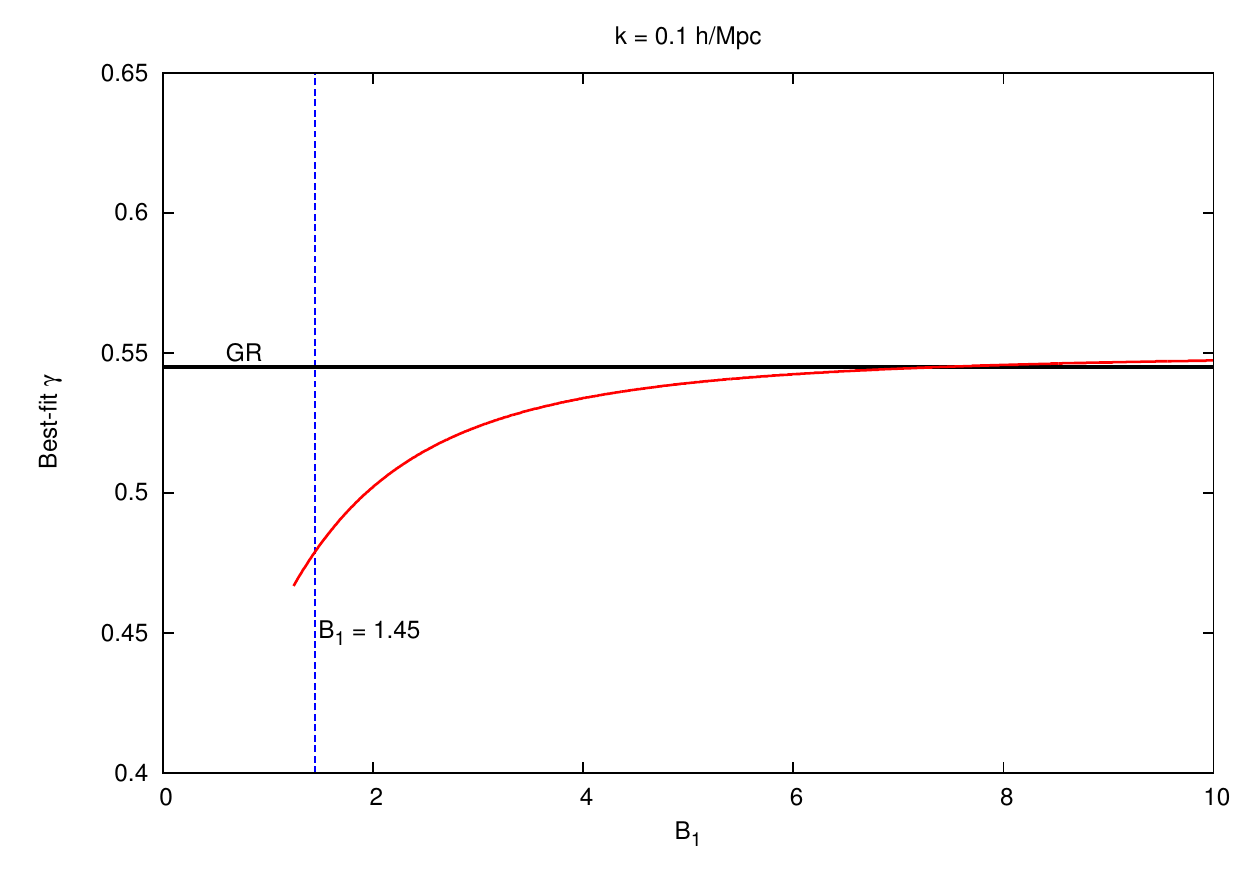}
\caption[Modified growth rate for the finite-branch $B_1B_4$ model.]{Growth-rate results for the $B_1B_4$ model on the finite branch, with $\Omega_\Lambda^\mathrm{eff}=0.699$ and $B_1>1.244$. The behaviour is quite similar to that of the $B_1B_3$ model, plotted in \cref{fig:b1-b3-f,fig:b1-b3-qeta}.}
\label{fig:b1-b4-finite-f}
\end{figure}

\begin{figure}
\centering
\includegraphics{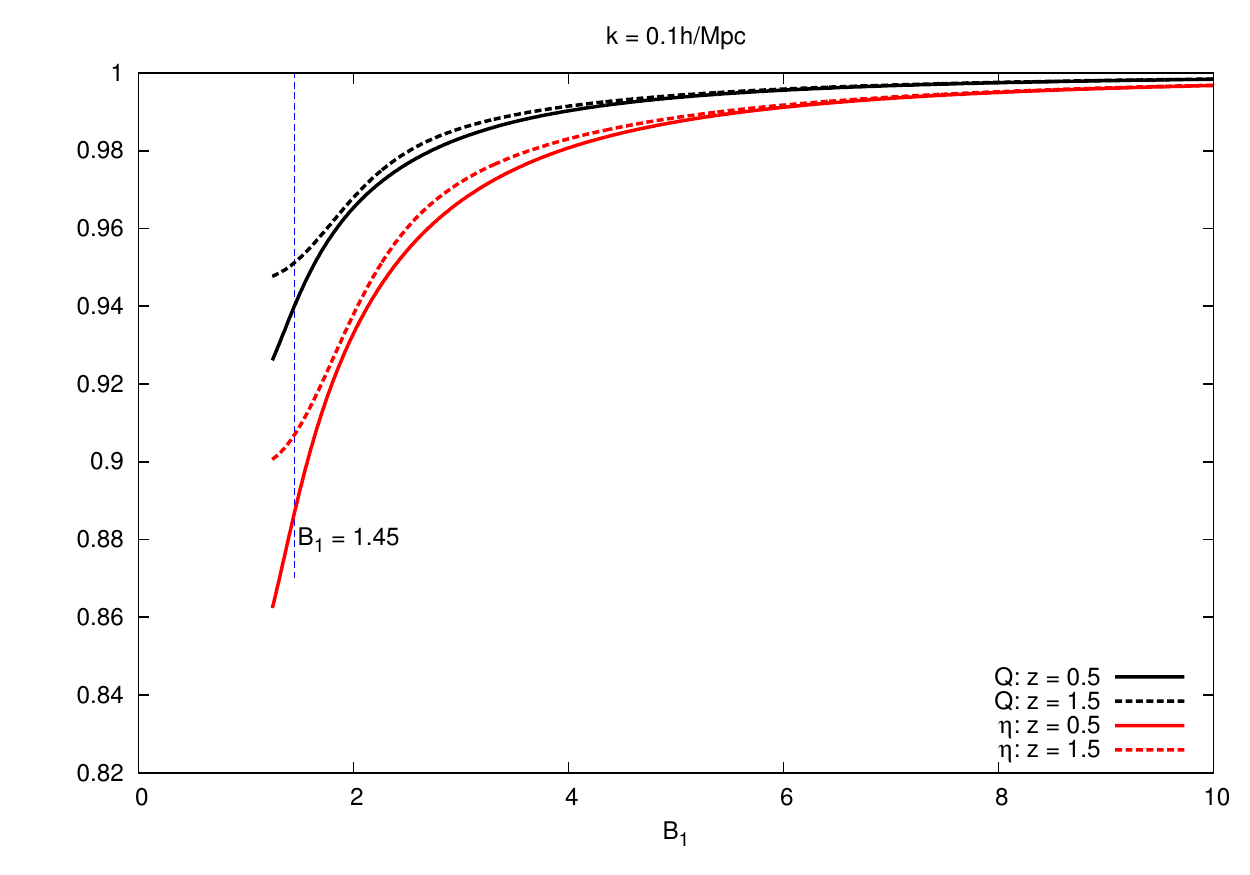}
\includegraphics{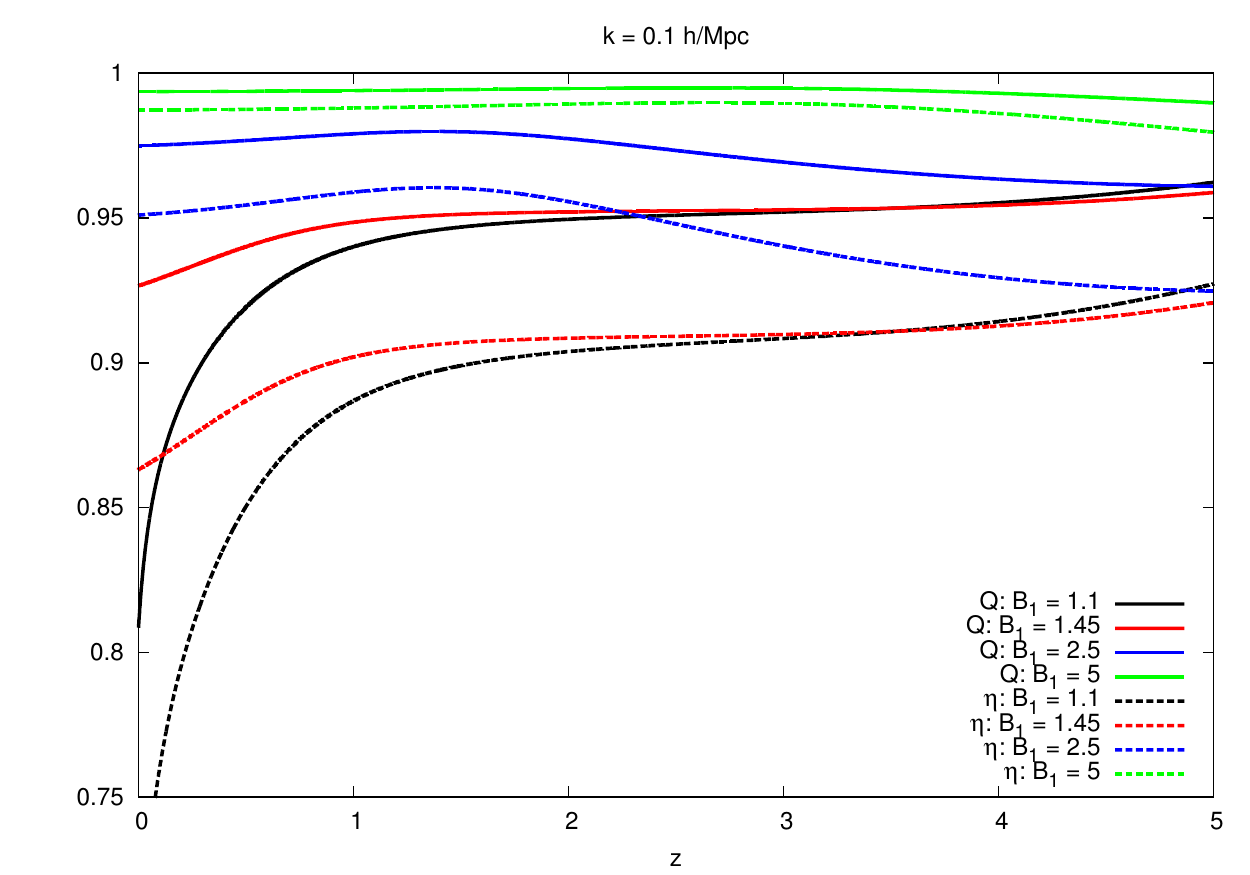}
\caption[Modified gravity parameters $Q$ and $\eta$ for the finite-branch $B_1B_4$ model.]{The modified gravity parameters $Q$ and $\eta$ for the $B_1B_4$ model on the finite branch. Results are displayed for $\Omega_\Lambda^\mathrm{eff}=0.699$ and $B_1>1.244$.}
\label{fig:b1-b4-finite-qeta}
\end{figure}

We plot the results for the finite branch in \cref{fig:b1-b4-finite-f,fig:b1-b4-finite-qeta}. Qualitatively, this model predicts subhorizon behaviour quite similar to that of the $B_1B_3$ model, discussed above and plotted in \cref{fig:b1-b3-f,fig:b1-b3-qeta}.

\begin{figure}
\centering
\includegraphics{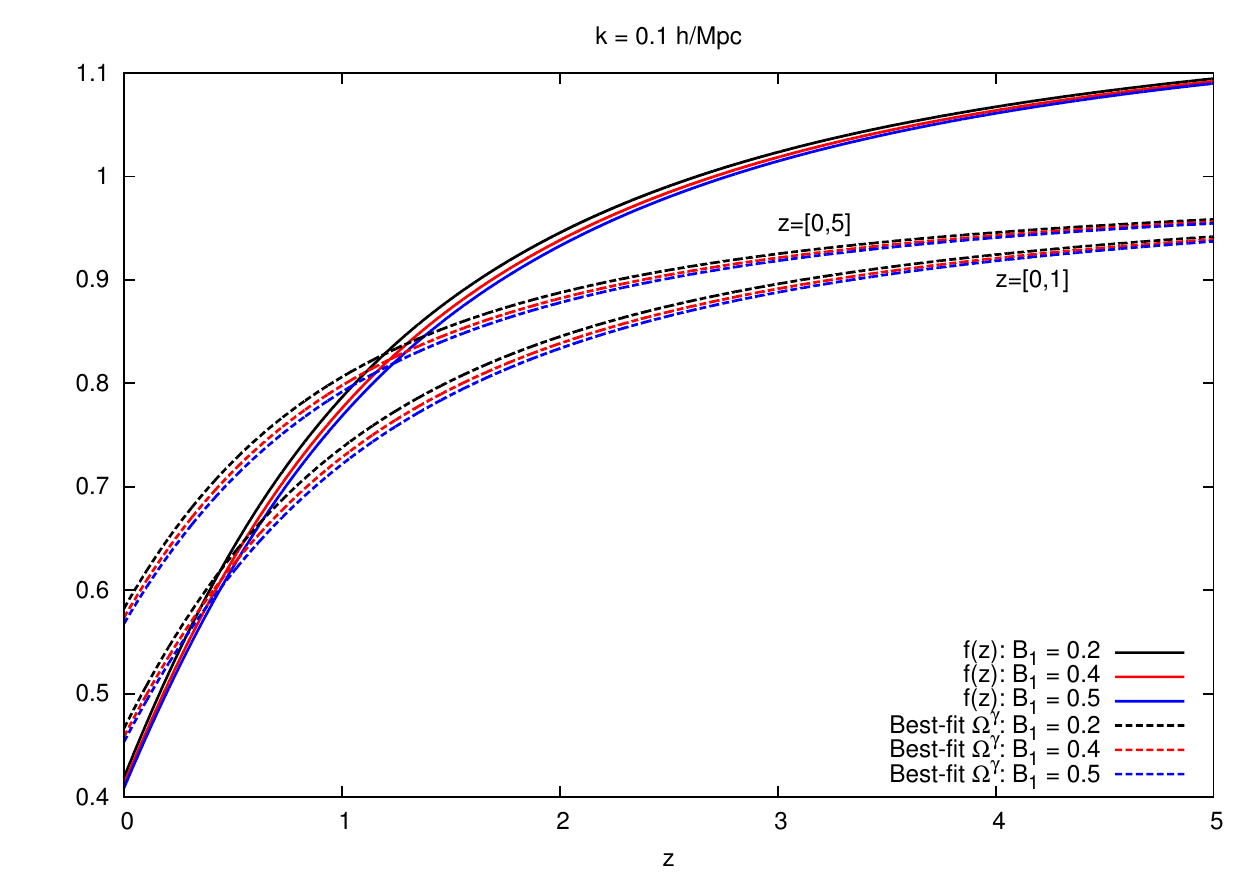}
\includegraphics{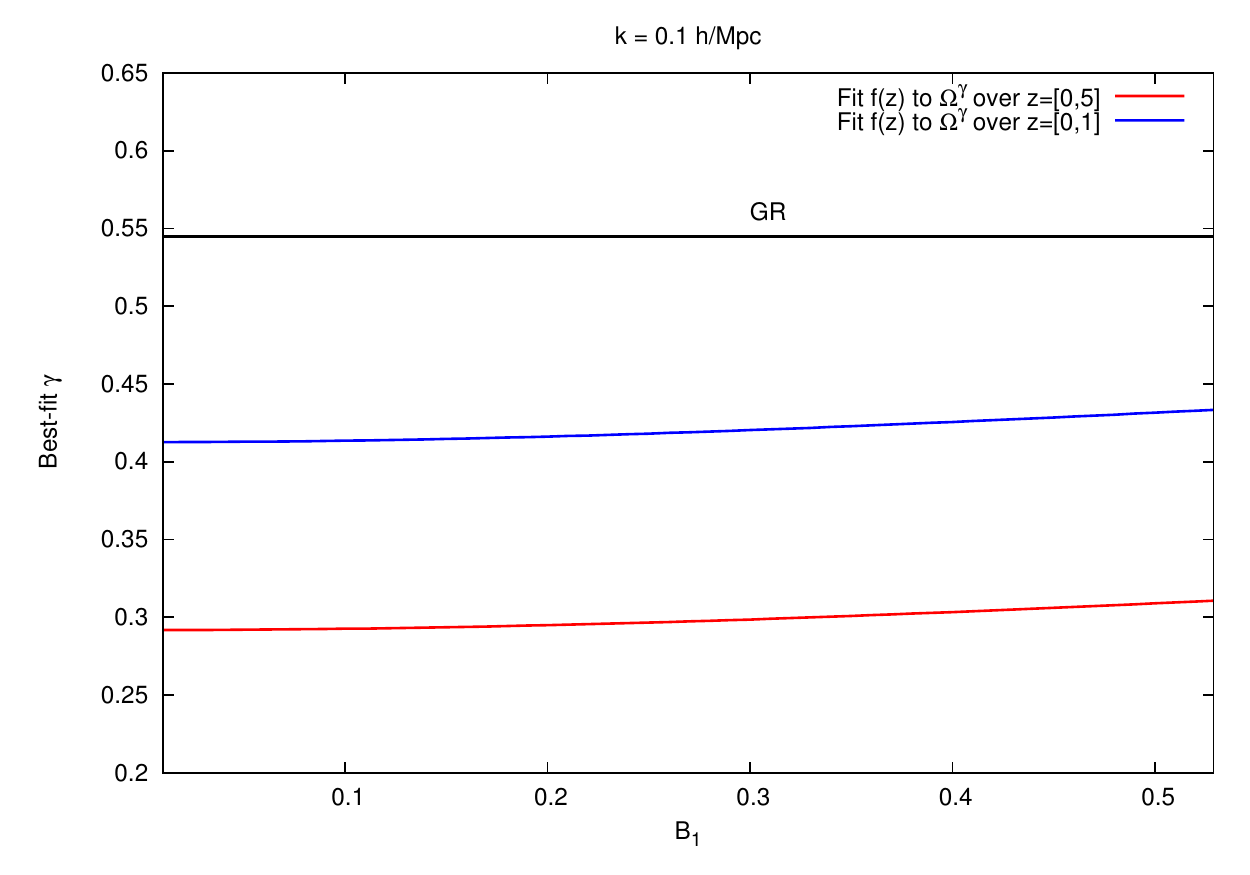}
\caption[Modified growth rate for the infinite-branch $B_1B_4$ model.]{Growth-rate results for the $B_1B_4$ model on the infinite branch, with $\Omega_\Lambda^\mathrm{eff}=0.84$ and $B_1<0.529$. This model is qualitatively different from the others we study, as it does not possess a limit to the minimal $B_1$-only model, and has the strongest deviations from $\Lambda$CDM among all of the models presented in this work. Here we plot $f(z)$ as well as the best-fit parametrisation $\Omega_m^\gamma$ with the fitting done over the redshift ranges $0<z<1$ and $0<z<5$.}
\label{fig:b1-b4-infinite-f}
\end{figure}

\begin{figure}
\centering
\includegraphics{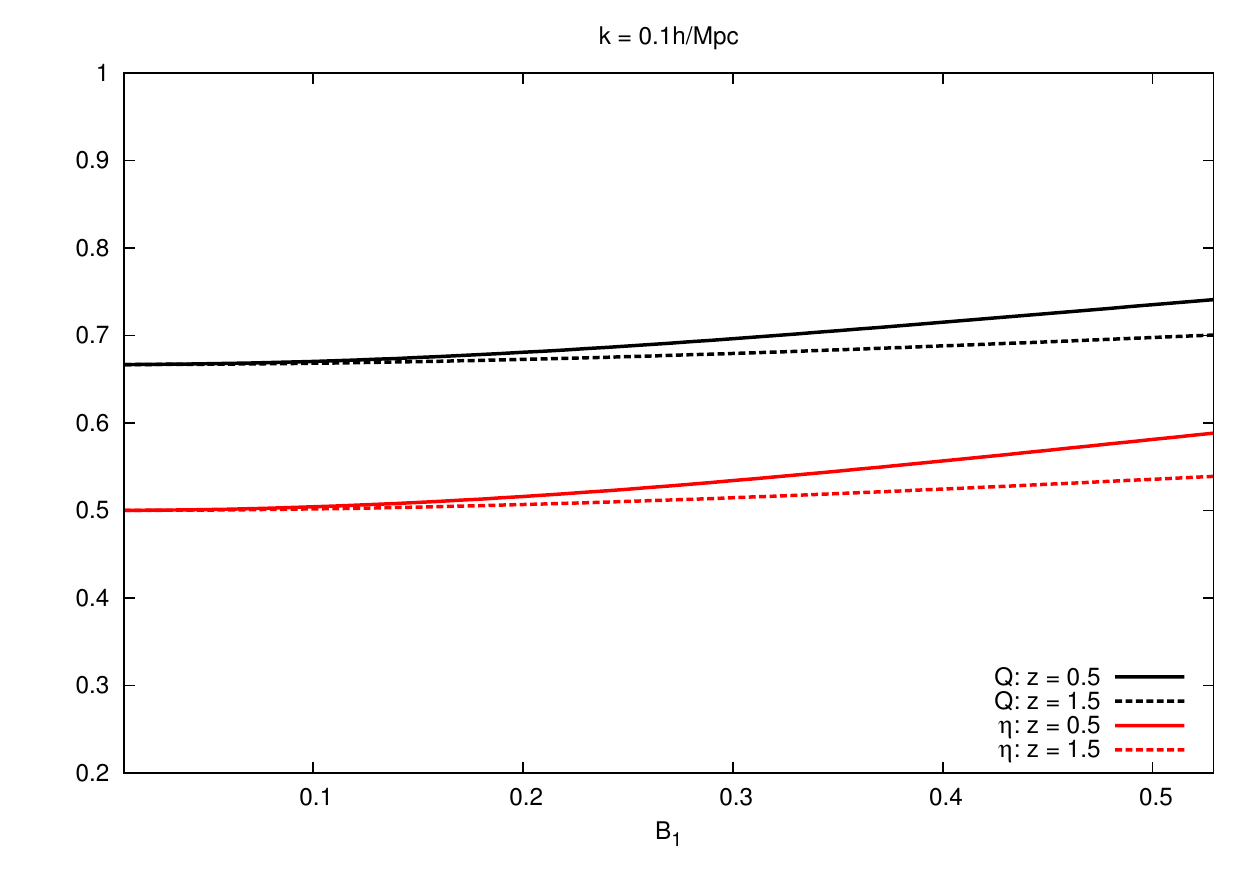}
\includegraphics{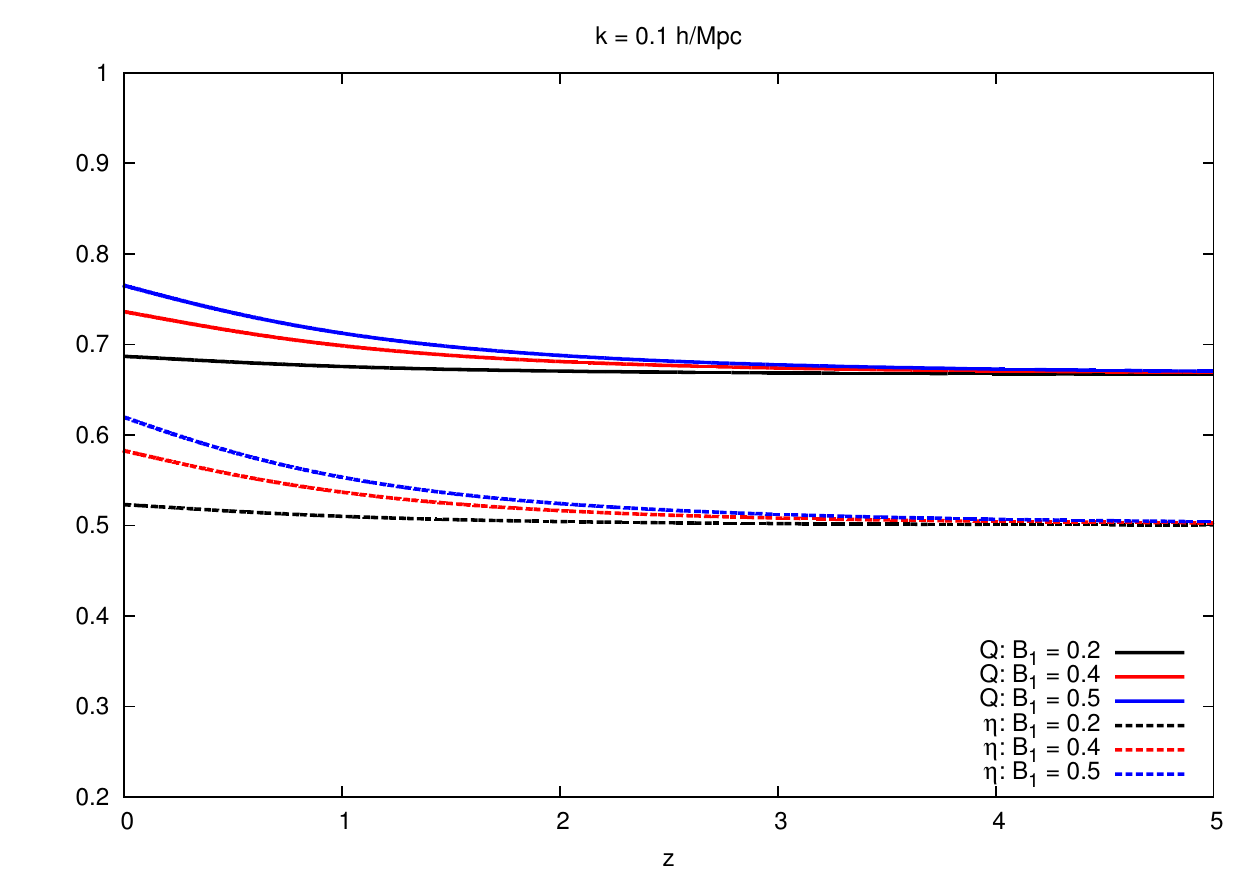}
\caption[Modified gravity parameters $Q$ and $\eta$ for the infinite-branch $B_1B_4$ model.]{The modified gravity parameters $Q$ and $\eta$ for the $B_1B_4$ model on the infinite branch, with $\Omega_\Lambda^\mathrm{eff}=0.84$ and $B_1<0.529$. We find strong deviations from GR which will be easily visible to near-future LSS experiments.}
\label{fig:b1-b4-infinite-qeta}
\end{figure}

Now let us move on to the stable infinite-branch model, the modified gravity parameters of which are plotted in \cref{fig:b1-b4-infinite-f,fig:b1-b4-infinite-qeta}.  This is the only model we study which does not possess a limit to the minimal $B_1$-only model, and it predicts significant deviations from $\Lambda$CDM. In this model, $\eta$ deviates from 1 by nearly a factor of 2 at all observable epochs and for all allowed values of $B_1$, providing a clear observable signal of modified gravity. This is a significant feature of this model; it has no free parameters which can be tuned to make its predictions arbitrarily close to $\Lambda$CDM and therefore is unambiguously testable.

We can understand this behaviour analytically as follows. In the asymptotic past, we can take the limits of our expressions for $\eta$ and $Q$ to find
\begin{equation}
\lim_{N\rightarrow-\infty}\eta=\frac{1}{2}\qquad\text{and}\qquad\lim_{N\rightarrow-\infty}Q=\frac{2}{3}.
\end{equation}
Unusually, structure growth does not reduce to that of $\Lambda$CDM even though at early times the graviton mass is very small compared to the Hubble scale. In the future one finds $\eta\to1$ if $k$ is kept finite, but this is somewhat unphysical: for any finite $k$ there will be an epoch of horizon exit in the future after which the subhorizon QS approximation breaks down. We can see both the asymptotic past and asymptotic future behaviours in the second panel of \cref{fig:b1-b4-infinite-qeta}, although the late-time approach of $\eta$ to unity is not entirely visible.

The growth rate and index, $f(z)$ and $\gamma$, also deviate strongly from the $\Lambda$CDM predictions. Using the $\Omega_m^\gamma$ parametrisation, we find that $\gamma$ is even lower than the range $\sim0.45$--$0.5$ which we found in the $B_1$-only model and in the other two-parameter models. However, the $\Omega_m^\gamma$ parametrisation is an especially bad fit to $f(z)$ in this case; we fit $\gamma$ to $f(z)$ in the redshift range $0<z<5$ (as in the rest of this chapter) and $0<z<1$ (which is the redshift range of present observations \cite{Macaulay:2013swa}) in \cref{fig:b1-b4-infinite-f}, obtaining significantly different results and still never agreeing well with the data.

\begin{figure}
\centering
\includegraphics[width=0.8\textwidth]{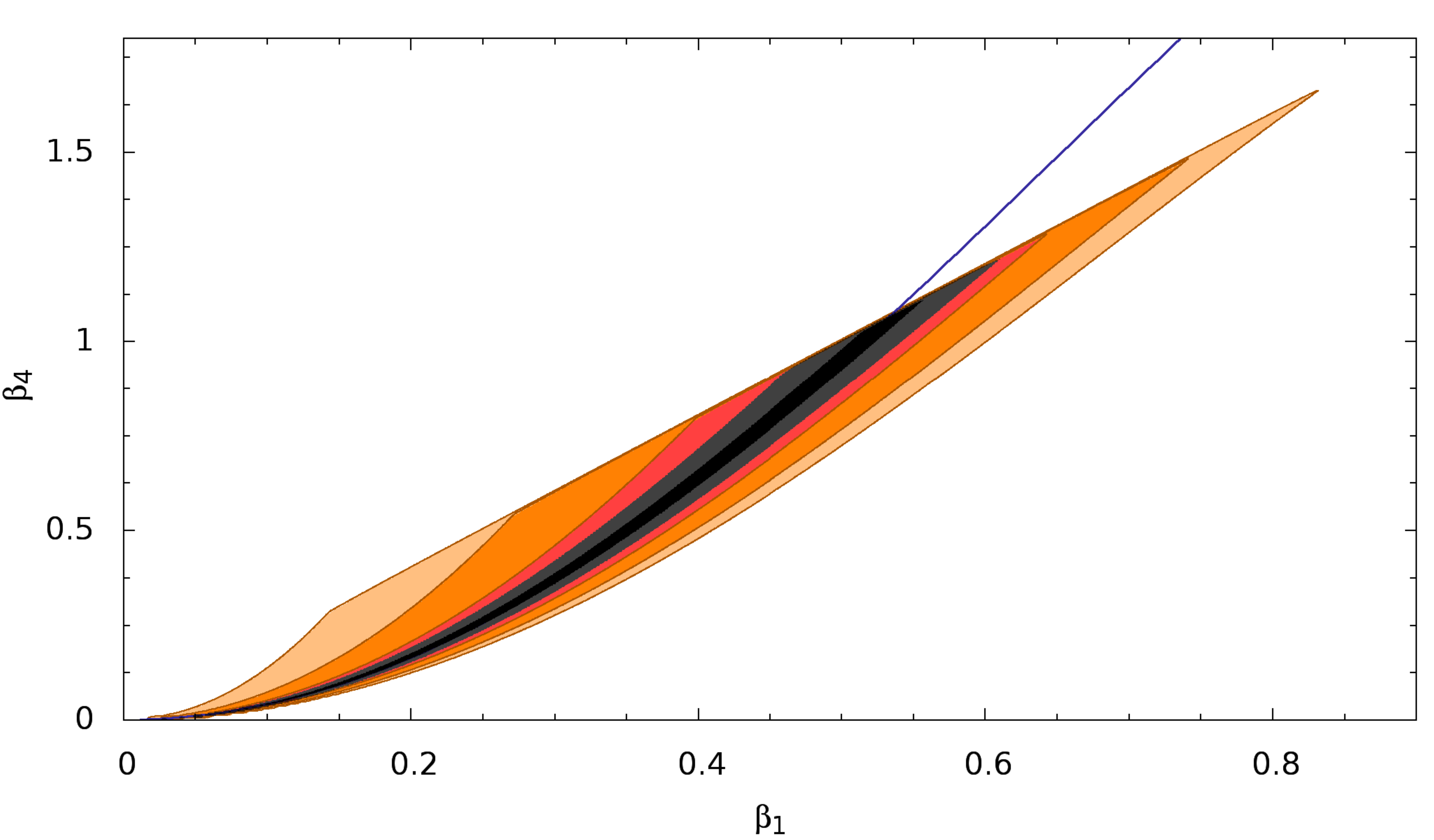}
\caption[Likelihoods for $B_1$ and $B_4$ using background and growth-rate data.]{Likelihoods for $B_1$ and $B_4$. The red, orange, and light-orange filled regions correspond to 68\%, 95\% and 99.7\% confidence levels using growth data alone. The black (68\%) and gray (99.7\%) regions illustrate the combination of the likelihoods from measured growth data and type Ia supernovae. The blue line indicates the degeneracy curve given by \cref{eq:b1-b4} with the best-fit value of $\Omega_\Lambda^\eff$. The viability condition enforces the likelihood to vanish when $\beta_{4}>2\beta_{1}$.}
\label{fig:likelihood}
\end{figure}

As shown in \cref{fig:likelihood}, the confidence region obtained from the growth data is in agreement with type Ia supernovae (SNe) data (see Ref.~\cite{Konnig:2013gxa} for the likelihood from the SCP Union 2.1 Compilation of SNe Ia data \cite{Suzuki:2011hu}). The growth data alone provide $\beta_{1}=0.40_{-0.15}^{+0.14}$ and $\beta_{4}=0.67_{-0.38}^{+0.31}$ with a $\chi_{\mathrm{min}}^{2}=9.72$ (with 9 degrees of freedom) for the best-fit value and is in agreement with the SNe Ia likelihood. The likelihood from growth data is, however, a much weaker constraint than the likelihood from background observations. Thus, the combination of both likelihoods, providing $\beta_{1}=0.48_{-0.16}^{+0.05}$ and $\beta_{4}=0.94_{-0.51}^{+0.11}$, is similar to the SNe Ia result alone.

\begin{figure}
\centering
\includegraphics[width=0.7\textwidth]{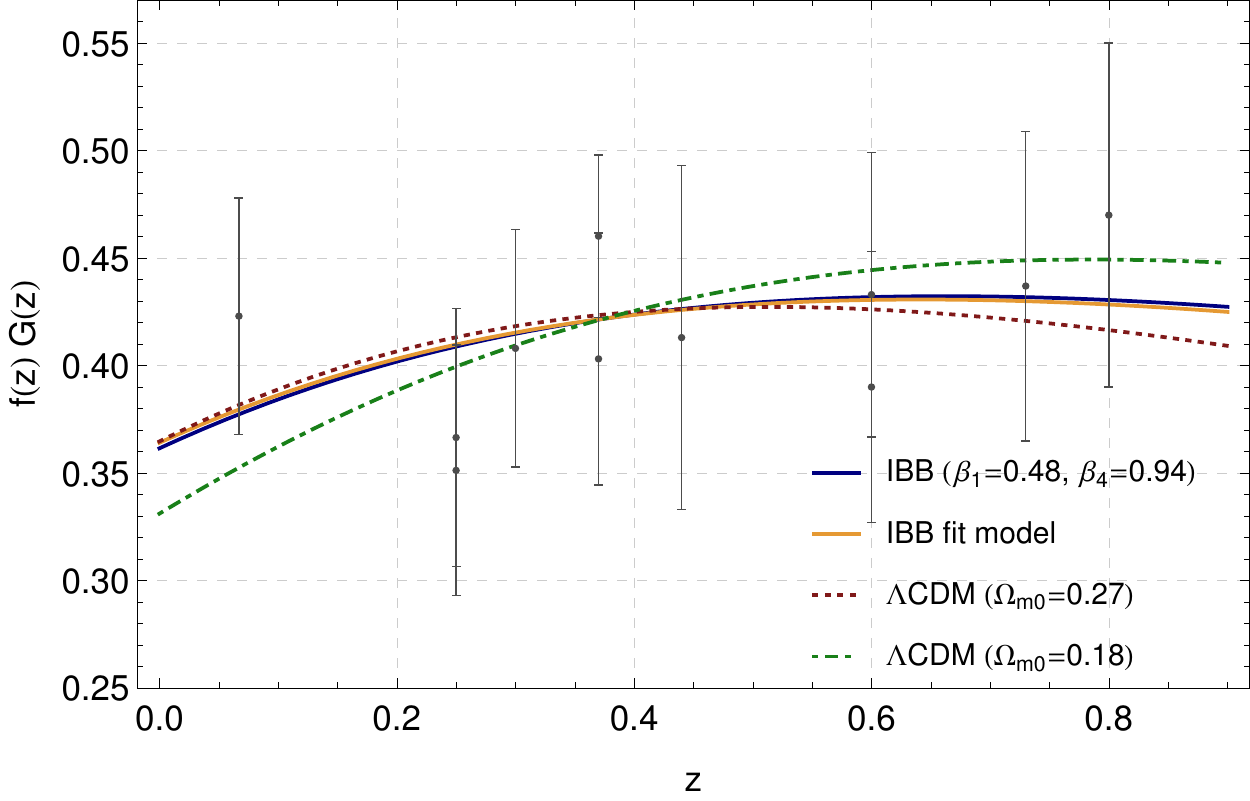}
\caption[Growth history for the best-fit infinite-branch model.]{Growth history for the best-fit infinite-branch $\beta_1\beta_4$ model (solid blue) with $\beta_{1}=0.48$ and $\beta_{4}=0.94$ compared to the result obtained from the best fit (\ref{eq:fitmodel_f}) (solid orange) with $\gamma_{0}=0.47$ and $\alpha=0.21$ and the $\Lambda$CDM predictions for $\Omega_{m,0}=0.27$ (dotted red) and $\Omega_{m,0}=0.18$ (dotted-dashed green). The latter value for the matter density is the same as is predicted by the infinite-branch model. Note that a vertical shift of each single curve is possible due to the marginalisation over $\sigma_{8}$. Here we choose $\sigma_{8}$ for each curve individually such that it fits the data best. The growth histories are compared to observed data compiled by Ref.~\cite{Macaulay:2013swa}.
\label{fig:growthhistory}}
\end{figure}

In \cref{fig:growthhistory} we compare the growth rate directly to the observational data compiled in \rcite{Macaulay:2013swa}, using the best-fit values determined above. The available growth data are unable to distinguish between the infinite-branch model and $\Lambda$CDM. We also find that an alternative parametrisation,
\begin{equation}
f(z)\approx\Omega_{m}^{\gamma_{0}}\left(1+\alpha\frac{z}{1+z}\right),\label{eq:fitmodel_f}
\end{equation}
is able to provide a much better fit to $f(z)$ than the usual $\Omega_m^\gamma$. The best-fit values for this parametrisation are $\gamma_{0}=0.47$ and $\alpha=0.21$.

\section{Summary of Results}
\label{sec:bigravity-subhorizon-summary}

In this chapter we have examined the evolution of cosmological perturbations on subhorizon scales in massive bigravity, describing linear structure formation during the matter-dominated era. We solved the linearised Einstein and conservation equations in the quasistatic, subhorizon limit in Newtonian gauge for $g_\mn$. In this limit, we found that the perturbations are described by a system of six algebraic equations for five variables. We obtained a consistent solution to this system relating each of the metric perturbations appearing in the subhorizon Einstein equations to the matter density contrast. This allowed us to derive a modified evolution equation for the density contrast, which differs from its GR counterpart by a varying effective Newton's constant. We also obtained algebraic expressions for the anisotropic stress, $\eta(z,k)$, and the parameter measuring modifications to Newton's constant in the Poisson equation, $Q(z,k)$, purely in terms of background quantities. We then solved for the background numerically to obtain these parameters, and finally integrated the structure growth equation to derive the growth rate, $f(z,k)$, and its best-fit parametrisation, $f\sim\Omega_m^\gamma$.

We have studied every subset of the theory which is viable at the background level and contains one or two free parameters, excluding the $g$-metric cosmological constant as we are interested in self-accelerating theories. Among the single-parameter models, only the case with the lowest-order interaction term, $\beta_1\neq0$, is in agreement with the background data. As emphasised in \rcite{Akrami:2012vf,Akrami:2013pna,Konnig:2013gxa}, this ``minimal'' bigravity model is especially appealing because it possesses late-time acceleration and fits the background data well with the same number of free parameters as $\Lambda$CDM. We have found that it predicts modified gravity parameters that differ significantly from GR: $\gamma\sim0.46$--$0.48$ (in agreement with Ref. \cite{Konnig:2014dna}), $Q\sim0.94$--$0.95$, and $\eta\sim0.88$--$0.90$. For reference, the $\Lambda$CDM predictions are $\gamma\approx0.545$ and $Q=\eta=1$. Future large-scale structure experiments such as Euclid \cite{Laureijs:2011gra,Amendola:2012ys} will easily be able to distinguish this simple model from GR, if we can trust its predictions. However, we have shown in \cref{chap:bigravity-stability} that this model suffers from an early-time instability. If this can be overcome, or if observations are restricted to late times ($z\lesssim0.5$), our results demonstrate that by going to the level of linear perturbations, this theory can be probed in the near future by multiple observables which deviate significantly from general relativity.

We additionally examined the four two-parameter models which are viable in the background, all of which keep $\beta_1>0$ while turning on a second, higher-order interaction term. Two of these models, in which either the cubic interaction, $\beta_3$, or the $f$-metric cosmological constant, $\beta_4$, is nonzero (the latter specifically in the ``finite branch,'' which reduces to the minimal model in the $\beta_4\to0$ limit), have similar behaviour to each other. They predict GR-like values for all three modified gravity parameters in the limit where $m^2\beta_1/H_0^2$ is large, becoming indistinguishable from $\Lambda$CDM (given a Euclid-like experiment) for $m^2\beta_1/H_0^2\gtrsim3$. These reduce to the predictions of the minimal model in the limit $m^2\beta_1/H_0^2 \approx 1.45$ (the best-fit value for $\beta_1$ in the minimal model). For lower values of $\beta_1$ (corresponding to positive $\beta_3$ or $\beta_4$) these models predict even more dramatic deviations from GR: $\gamma$ can dip to $0.45$, and $\eta$ at recent times can be as low as $\sim0.75$. Euclid is expected to measure these parameters to within about $0.02$ and $0.1$, respectively \cite{Laureijs:2011gra,Amendola:2012ys,Amendola:2013qna}, and therefore has the potential to break the degeneracies between $\beta_1$ and $\beta_3$ or $\beta_1$ and $\beta_4$ which is present at the level of background observations.

The $\beta_1\beta_2$ model has an instability when $\beta_2$, the coefficient of the quadratic term, is negative. This instability does not necessarily rule out the theory, but it might signal the breakdown of linear perturbation theory, in which case nonlinear studies are required in order to understand the formation of structure. This is different from the early-time instability discussed in \cref{chap:bigravity-stability} which does not show up in the subhorizon, quasistatic r\'egime. It is possible that these perturbations at some point become GR-like due to the Vainshtein mechanism. Moreover, because the instability occurs at a characteristic redshift (which depends on the $\beta_i$ parameters), there may be an observable excess of cosmic structure around that redshift.

The parameter range of the $\beta_1\beta_2$ model over which this instability is absent, $0<\beta_1\lesssim1.45H_0^2/m^2$ (corresponding to $H_0^2/m^2>\beta_2>0$), is quite small; near the large $\beta_1$ end the predictions recover those of the minimal $\beta_1$-only model, while at the small $\beta_1$ end the perturbations can differ quite significantly from GR, with $\gamma$ as low as $\sim0.35$ and $\eta$ as low as 0.6. However, the exact $\beta_1=0$ limit of this theory is already ruled out by background observations \cite{Akrami:2012vf}, so one should take care when comparing the model to observations in the very low $\beta_1$ region of this parameter space.

Finally, we examined the ``infinite branch'' of the $\beta_1\beta_4$ model, which is the only bimetric model\footnote{Up to the addition of a cosmological constant, which is uninteresting.} that avoids the early-time instabilities uncovered in \cref{chap:bigravity-stability}. This is called an infinite-branch model because the ratio of the $f$ metric scale factor to the $g$ metric scale factor, $y$, starts at infinity and monotonically decreases to a finite value. In the rest of the models we study, $y$ starts at zero and then increases; consequently, in the $\beta_4\to0$ limit this theory reduces to pure CDM, rather than to the $\beta_1$-only model.

The predictions of the infinite-branch $\beta_1\beta_4$ theory deviate strongly from GR. The model predicts a growth rate $f(z)$ which is not well-parametrised by a $\Omega_m^\gamma$ fit, but has best-fit values of $\gamma$ on the low side ($0.3$--$0.4$, depending on the fitting range). The anisotropic stress $\eta$ is almost always below $0.7$ and can even be as low as $0.5$, a factor of two away from the GR prediction. Across its entire parameter space, this model has the most significantly non-GR values of any we study. Its predictions should be well within Euclid's window.

As the only sector of the massive bigravity parameter space which both self-accelerates and is always linearly stable around cosmological backgrounds, it is a prime target for observational study, especially because its observables are far from those of $\Lambda$CDM for any choices of its parameters.

\begin{savequote}[30pc]
O, that way madness lies; let me shun that; \\
No more of that.
\qauthor{Lear, \textit{King Lear}, 3.4}
\end{savequote}

\chapter{The Geometry of Doubly-Coupled Bigravity}
\label{chap:dc-finsler}
\hrule
\vspace*{1cm}

The existence of a consistent bimetric theory raises an intriguing question: which is the physical metric? In \cref{chap:bigravity-stability,chap:bigravity-subhorizon} we chose to couple only one of the two metric, $g_\mn$, directly to matter, while the other dynamical metric, $f_\mn$, only interacts with matter fields indirectly through its interactions with $g_\mn$. It is specifically in this \emph{singly-coupled} context that the absence of the Boulware-Deser ghost was originally proven \cite{Hassan:2011hr,Hassan:2011tf,Hassan:2011ea}, and extending that analysis to other matter couplings is decidedly nontrivial \cite{Yamashita:2014fga,deRham:2014naa,Hassan:2014gta,deRham:2014fha}. It is therefore natural to interpret $g_\mn$ in this case as the usual ``metric'' of spacetime, while $f_\mn$ is an extra spin-2 field that is required in order to give mass to the graviton. Since the fields $g_\mn$ and $f_\mn$ have metric properties, we have called both $g_\mn$ and $f_\mn$ ``metrics,'' even though strictly speaking, the singly-coupled theory could more accurately be called a theory of ``gravity coupled to matter and a symmetric 2-tensor'' \cite{Berg:2012kn}.

In this chapter we aim to explore the consequences of coupling matter to both metrics in massive bigravity. As discussed in \cref{sec:mg-bigravity}, the bimetric action (\ref{eq:actionHR}) in vacuum places both metrics on equal footing: it is invariant under the interchanges (\ref{eq:bigravinterchanges})
\begin{equation}
g_\mn \leftrightarrow f_\mn, \qquad M_g \leftrightarrow M_f, \qquad \beta_n \to \beta_{4-n}. \label{eq:bigravinterchanges2}
\end{equation}
This metric-interchange duality is broken by the addition of matter in the singly-coupled version of bigravity. The structure of the vacuum theory might hint that any fundamental theory which gives rise to massive bigravity does not discriminate between the two metrics. Consequently it is important to explore \emph{doubly-coupled} bigravity, in which matter couples to both metrics in a way that maintains the interchange duality (\ref{eq:bigravinterchanges2}).

We are specifically interested in the possibility of double-coupling schemes in which there is no ``effective metric'' whose geodesics describe the motion of matter. In these cases, we must introduce new tools for understanding the physical geometry of spacetime. We will focus on one of the most straightforward possibilities for double coupling, in which matter is minimally coupled to each metric. This theory was introduced and its cosmological solutions were studied in \rcite{Akrami:2013ffa}. Subsequent work showed this type of coupling revives the Boulware-Deser ghost at arbitrarily low energy scales and therefore cannot be a fundamental theory of bigravity \cite{Yamashita:2014fga,deRham:2014naa}. Indeed, there is only one known double coupling which avoids the ghost at low energies, and even in this theory it re-emerges at or above the strong-coupling scale \cite{deRham:2014naa,Noller:2014sta}.\footnote{Using the same principles, further candidate double couplings have been constructed in \rcite{Heisenberg:2014rka}, but it is not yet known which, if any, of these are ghost-free.} This coupling is the phenomenologically interesting one and will be investigated in depth in \cref{chap:dc-background,chap:dc-drgt}. These theories couple matter to a single effective metric built out of $g_\mn$ and $f_\mn$ and so avoid the problem of determining the physical spacetime. It is, however, not yet clear whether such theories are immune from other types of pathologies, and it may well be the case that the unknown, healthy doubly-coupled theory of bigravity will not admit an effective-metric formulation at all. It is our goal in this chapter to demonstrate the difficulties such theories would have with regards to defining observables by studying arguably the simplest example of doubly-coupled bigravity without an effective-metric description. In this context, the traditional notion of a ``physical metric'' may have to be discarded, leaving us faced with entirely new conceptual challenges in interpreting even the observables of the theory. 

This chapter is organised as follows. In \cref{sec:physmet} we argue that there is no effective metric to which matter minimally couples, and that such a metric does not even exist for most individual fields. In \cref{sec:lightprop} we explore light propagation in this theory in the geometric optics limit, and discuss the problem of relating cosmological observables to the underlying theory when we can no longer describe photon trajectories as null geodesics in a metric. In \cref{sec:ppfinsler} we examine the dynamics of point particles, finding that they effectively live in a Finsler spacetime, a geometry which depends nontrivially on the coordinate intervals. Finally, we conclude in \cref{sec:dc-finsler-summary}.

\section{The Lack of a Physical Metric}
\label{sec:physmet}

We consider a doubly-coupled bimetric theory in which the action (\ref{eq:actionHR}) is extended by the addition of a minimal-coupling term between matter fields, $\Psi_i$, and $f_\mn$,
\begin{align}
   S_\mathrm{HR} &= -\frac{M_g^2}{2}\int d^4x\sdg R(g) -\frac{M_f^2}{2}\int d^4x\sdf R(f) + m^2M_g^2\int d^4x\sdg\sum_{n=0}^{4}\beta_ne_n(\mathbb X) \nonumber \\
   &\hphantom{{}=}+ \alpha_g\int d^4x\sdg\mathcal{L}_m\left(g, \Phi_i\right) + \alpha_f\int d^4x\sdf\mathcal{L}_m\left(f, \Phi_i\right).
\end{align}
This extends the symmetry (\ref{eq:bigravinterchanges2}) to the entire action, as long as we also exchange $\al_g$ and $\al_f$,
\begin{equation}
g_\mn \leftrightarrow f_\mn, \qquad M_g \leftrightarrow M_f, \qquad \beta_n \to \beta_{4-n}, \qquad \al_g \leftrightarrow \al_f.
\end{equation}
The presence of the interaction term is crucial; if one were to couple two pure, noninteracting GR sectors to the same matter, the Bianchi identities would constrain that matter to be entirely nondynamical \cite{Akrami:2013ffa}. Note that $\ag$ and $\af$ are not both necessary to fully specify the theory; only their ratio is physical, as can be seen by rescaling the action by $\ag^{-1}$. For the purposes of this chapter, we will find it useful to leave both in so as to keep the symmetry between the two metrics explicit.

An immediate concern is the violation of the equivalence principle. However, because the Vainshtein mechanism screens massive-gravity effects \cite{Vainshtein:1972sx}, it is not obvious how stringent the constraints from tests of GR in the solar system would be: the modifications might be hidden from local experiments while showing up at cosmological scales. The cosmology of this doubly-coupled theory has been studied and shown to produce viable late-time accelerating background expansion without an explicit cosmological constant term \cite{Akrami:2013ffa}, and with a phenomenology which can be interestingly different from that of the singly-coupled theory \cite{Akrami:2012vf}. We emphasise again that this model itself possesses the Boulware-Deser ghost and hence we cannot trust its cosmological solutions, but a ghost-free doubly-coupled theory may well have similar properties. Indeed, the cosmological phenomenology of this theory is quite similar to that of the healthier doubly-coupled theory introduced in \rcite{deRham:2014naa}, as we will show in \cref{chap:dc-background}.

We can readily confirm that no physical Riemannian metric exists in the sense that all matter species would minimally couple to it and thus follow its geodesics. Indeed, for some matter fields such a metric does not exist at all. Consider a massive scalar field. Its action is given by
\begin{equation}
S_\phi = -\ag \int d^4x \sdg \left(-\frac 12 g^\mn \partial_\mu\phi \partial_\nu\phi - V(\phi)\right) - \af \int d^4x \sdf \left(-\frac 12 f^\mn \partial_\mu\phi \partial_\nu\phi - V(\phi)\right).
\end{equation}
Let us assume that $\phi$ is minimally coupled to an effective metric, $h_{\mu\nu}[g_{\mu\nu},f_{\mu\nu}]$. This metric is defined through the relation
\begin{equation}
S_\phi \equiv \int d^4x \sdh \mathcal{L}_\phi^h \equiv \ag \int d^4x \sdg \mathcal{L}_\phi^g + \af \int d^4x \sdf \mathcal{L}_\phi^f.
\end{equation}
where we have defined the various scalar Lagrangians as
\begin{equation}
\mathcal{L}_\phi^{(g,f,h)} = -\frac{1}{2}(g,f,h)^{\mn}\partial_\mu\phi\partial_\nu\phi - V(\phi).
\end{equation}
The kinetic and potential terms, respectively, yield the conditions
\begin{align}
\ag\sqrt{-g}g^{\mn} + \af\sqrt{-f}f^{\mn} &= \sdh h^\mn \label{eq:scalkin} \\
\ag\sqrt{-g} + \af\sqrt{-f} &= \sqrt{-h}. \label{eq:scalpot}
\end{align}
These are not necessarily consistent with each other: taking the determinant of \cref{eq:scalkin}, we find
\begin{equation}
\operatorname{det}\left(\ag\sqrt{-g}g^{\mn} + \af\sqrt{-f}f^{\mn}\right) = h,
\end{equation}
so that \cref{eq:scalkin,eq:scalpot} overdetermine $h_\mn$ unless $g_\mn$ and $f_\mn$ satisfy the nontrivial relation
\begin{equation}
\left(\ag\sdg + \af\sdf\right)^2 = -\operatorname{det}\left(\ag\sqrt{-g}g^{\mn} + \af\sqrt{-f}f^{\mn}\right).
\end{equation}
However, even the simplest case, $g_\mn=f_\mn=\eta_\mn$, fails this test, as well as more complicated but physically-relevant cases like FLRW.

Therefore, any choice of $h_\mn$ will generally result in either a noncanonical kinetic term or a spacetime-varying mass for the scalar. For some special choices of $g_\mn$ and $f_\mn$, particularly if they are related by a constant conformal factor, then this will only rescale the mass or kinetic term by a constant amount. In this particular theory, however, such a relation between $g_\mn$ and $f_\mn$ is far from general \cite{Akrami:2013ffa}.

If the scalar field is massless, $V(\phi)=0$, then we lose the constraint (\ref{eq:scalpot}). Massless scalars therefore do have physical metrics defined by \cref{eq:scalkin}. As a consistency check, we can confirm that the Klein-Gordon equation in $h_\mn$,
\begin{equation}
\Box_h \phi = 0,
\end{equation}
yields the correct massless Klein-Gordon equation \cite{Akrami:2013ffa}
\begin{equation}
\left(\ag\sdg \Box_g + \af \sdf \Box_f \right)\phi = 0.
\end{equation}
This is straightforward to show using the identity $g^\mn \Gamma^\rho_\mn = -\frac{1}{\sdg}\partial_\mu(\sdg g^{\mu\rho})$. This is the only example we have found of a field for which we can construct an effective metric.

Consider the electromagnetic field $A_\mu$. This is of paramount importance for cosmology, since we make observations by tracking photons. Its action is
\begin{equation}
S_A = \frac{1}{4}\ag \int d^4x \sdg g^{\mu\alpha}g^{\nu\beta}F_\mn F_\ab - \frac{1}{4}\af \int d^4x \sdf f^{\mu\alpha}f^{\nu\beta}F_\mn F_\ab, \label{eq:maxwellaction}
\end{equation}
where $F_\mn = \partial_\mu A_\nu - \partial_\nu A_\mu$ is the usual field-strength tensor and does not depend on a metric. If $A_\mu$ is minimally coupled to an effective metric, $h_\mn$, then we can write \cref{eq:maxwellaction} as
\begin{equation}
S_A = -\frac{1}{4} \int d^4x \sdh h^{\mu\alpha}h^{\nu\beta}F_\mn F_\ab.
\end{equation}
This implies that $h_\mn$ obeys
\begin{equation}
\ag \sdg g^\mn g^\ab + \af \sdf f^\mn f^\ab = \sdh h^\mn h^\ab.
\end{equation}
However, this equation overconstrains $h_\mn$. Consider the $00$--$00$, $00$--$ii$, and $ii$--$ii$ components,
\begin{align}
\ag \sdg \left(g^{00}\right)^2 + \af \sdf \left(f^{00}\right)^2 &= \sdh \left(h^{00}\right)^2, \label{eq:max00} \\
\ag \sdg g^{00} g^{ii} + \af \sdf f^{00} f^{ii} &= \sdh h^{00} h^{ii}, \label{eq:max0i} \\
\ag \sdg \left(g^{ii}\right)^2 + \af \sdf \left(f^{ii}\right)^2 &= \sdh \left(h^{ii}\right)^2. \label{eq:maxii}
\end{align}
where repeated indices are not summed over. Solving for $h^{00}$ and $h^{ii}$ using \cref{eq:max00,eq:max0i}, \cref{eq:maxii} becomes a constraint on $g_\mn$ and $f_\mn$,
\begin{equation}
g^{00}f^{ii} = f^{00}g^{ii}. \label{eq:maxconstraint}
\end{equation}
Note that we have chosen an arbitrary spatial index, $i$, in an arbitrary coordinate system; \cref{eq:maxconstraint} therefore applies to any diagonal spatial component in any coordinates. This equation is not satisfied by general choices of $g_\mn$ and $f_\mn$. An FLRW universe is a simple example where this condition fails to be satisfied. Thus, except in special circumstances, there is no physical metric for the electromagnetic field.

Similar arguments should hold for other fields. The massless scalar appears to be a special case because it lacks a potential term to constrain $h_\mn$ and because it has no indices, so its kinetic term only includes one appearance of the metric.

\section{Light Propagation and the Problem of Observables}
\label{sec:lightprop}

We have shown that the electromagnetic field is not minimally coupled to any effective metric. This case is of particular physical relevance because we make observations by tracking photons. For cosmological observations, especially, it is crucial to know how light propagates.

Even in this simplified case, photons turn out not to travel on null geodesics of any metric. To see this, we will consider the plane-wave approximation for the Maxwell field,
\begin{equation}
A_\mu = \operatorname{Re}\left[\left(a_\mu + \epsilon b_\mu\right)e^{i\psi/\epsilon}\right], \label{eq:Aansatz}
\end{equation}
and take the \emph{geometric optics limit} in which the wavelength is tiny compared to the characteristic curvature scale, $\epsilon\equiv\lambda/R\ll1$. This provides a rigorous approach to describing light propagation in curved spacetime. In this ansatz, $a_\mu$ is the leading-order polarisation vector and $\psi$ is the phase. Herein we will drop the real evaluation for compactness. Because this is a ``pregeometric'' approach, we can utilise it to tackle the propagation of light rays in bigravity. The stress-energy tensor for the electromagnetic field, which can be derived from the action (\ref{eq:maxwellaction}), is
\begin{equation}
T^\mu{}_\nu = \frac{1}{4\pi}\left(F^{\mu\alpha}F_{\nu\alpha} - \frac{1}{4}\delta^\mu{}_\nu F_{\alpha\beta}F^{\alpha\beta}\right).
\end{equation}
We must be careful about which quantities depend on a metric and which don't. The field tensor is defined as usual in terms of the electromagnetic 4-potential $A_\mu$, which is itself defined with lower indices just in terms of the fields, so $A^\mu$ is the same in both metrics. Similarly, because of the symmetries of the Christoffel symbols, $F_{\mu\nu}$ can be defined equivalently in terms of covariant or partial derivatives; because of the latter, we see that $F_{\mu\nu}$ with lower indices is also independent of the metric. 

Stress-energy conservation is given by \cite{Akrami:2013ffa}
\begin{equation}
\ag\sdg\nabla^g_\mu T^\mu_g{}_\nu + \af\sdf\nabla^f_\mu T^\mu_f{}_\nu=0, \label{eq:cons}
\end{equation}
where $\nabla^g_\mu$ and $\nabla^f_\mu$ are the covariant derivatives defined with respect to $g_\mn$ and $f_\mn$, respectively. To apply this to the electromagnetic field, we first need to know (in terms of $g_\mn$, for concreteness) the divergence of the stress-energy tensor. The identity $\nabla_{[\alpha}F_{\mu\nu]}=0$ holds independently of the theory of gravity and in either metric, because it relies only on the usual expression for the commutation of covariant derivatives and the symmetries of the Riemann tensor. Using this, we find
\begin{equation}
\nabla_\mu T^\mu{}_\nu = F_{\nu\alpha}\nabla_\mu F^{\mu\alpha}.
\end{equation}
Plugging this into \cref{eq:cons} and using the fact that it should apply for arbitrary $F_{\nu\alpha}$ (because, as mentioned above, this is independent of the metrics), we find a straightforward generalisation of the Maxwell equations,
\begin{equation}
\ag\sdg \nabla^g_\mu F^{\mu\nu}_g + \af\sdf\nabla^f_\mu F^{\mu\nu}_f = 0, \label{eq:maxwell}
\end{equation}
where $g$ and $f$ subscripts on $F^\mn$ tell us which metric is being used to raise indices.

We have yet to use our gauge freedom. We will choose to work in a Lorenz gauge, where $\nabla_\mu A^\mu=0$. Since this cannot be simultaneously satisfied in both metrics, we will choose to apply this gauge with respect to $g_\mn$. As we will see shortly, this choice does not make a difference at leading order in the geometric optics approximation. After specialising to this gauge and commuting some covariant derivatives, the Maxwell equation reduces to
\begin{equation}
\ag\sdg \left(g^\mn\Box_gA_\mu - R_g^\mn A_\mu\right) + \af\sdf \left(f^\mn\Box_fA_\mu - \nabla_f^\nu\nabla_f^\mu A_\mu - R_f^\mn A_\mu\right) = 0. \label{eq:max}
\end{equation}
Plugging the ansatz (\ref{eq:Aansatz}) into \cref{eq:max} and keeping only the leading-order terms in $\epsilon$---i.e., those obtained by acting the covariant derivatives on the exponential term twice---we obtain
\begin{equation}
a_\mu\left(\ag\sdg g^\mn g^\ab + \af\sdf f^\mn f^\ab - \af\sdf f^{\mu\alpha}f^{\nu\beta}\right)k_\alpha k_\beta = 0, \label{eq:geooptprop}
\end{equation}
where $k_\mu \equiv \partial_\mu \psi$ is the wavevector. Note that in the singly-coupled limit, this gives us the standard result that $k_\alpha$ is null in $g_{\mu\nu}$.

As discussed above, we cannot use this to define a metric, $h^\mn$, in which $k_\alpha$ is lightlike. This creates problems when applying the standard methods of relativistic cosmology to compare observable quantities to the underlying theory. In a bimetric cosmology, as we have seen, there are two scale factors and two Hubble rates. When matter couples to both metrics, neither of these quantities plays the role that they play in general relativity. Had we been able to identify an effective metric from \cref{eq:geooptprop}, then the scale factor of that metric would have been the geometrical quantity that entered the expression for the redshift, and its Hubble rate (computed using the effective metric's lapse) would be the ``physical'' Hubble rate. The next step, relating the theoretical redshift to the shift in wavelengths observed by a telescope, would involve understanding the proper time of a massive observer, which we tackle in the next section. However, \cref{eq:geooptprop} defies the usual, simple categorisation. While we can, in principle, use this to compute light propagation, this approach does not shed light on the identification of a physical scale factor to compare to observations.

\section{Point Particles and Non-Riemannian Geometry}
\label{sec:ppfinsler}

The situation we have described in bimetric theories is radically different from the extensively-studied nonminimally coupled theories where the behaviour of matter can be described in terms of a single metric. In the context of scalar-tensor theories, for example, it is well known that there are conformally-equivalent descriptions of the theory where either the gravity sector is general relativity whilst matter has a nonminimal coupling (the Einstein frame), or matter is minimally coupled whilst the gravity sector is modified (the Jordan frame). All physical predictions are completely independent of the frame in which they are calculated after properly taking into account the rescaling of units in the Einstein frame \cite{Brans:1961sx}. One can generalise to nonuniversal couplings, allowing different Jordan frame metrics for different matter species, or to couplings to multiple fields. These bring about new technical but not fundamental difficulties. However, the doubly-coupled bimetric theories we are studying do not admit a Jordan frame at all for most types of matter. They possess mathematically two metrics but physically none, and to understand them we need to step beyond the confines of metric geometry. 

For concreteness, let us look at the simplest possible type of matter: a point particle of mass $m$. Its action is given by
\begin{equation}
S_\mathrm{pp} = -m\ag \int d\lambda \sqrt{-g_\mn \dot x^\mu \dot x^\nu} - m \af \int d\lambda \sqrt{-f_\mn \dot x^\mu \dot x^\nu}, \label{eq:ppaction-lambda}
\end{equation}
where overdots denote derivatives with respect to a parameter $\lambda$ along the particle's trajectory, $x^\mu(\lambda)$. Varying with respect to $\lambda$, we obtain the ``geodesic'' equation \cite{Akrami:2013ffa}
\begin{equation}
\ag g_\ab\left(\frac{du_g^\alpha}{ds_g} +\overset g \Gamma \vphantom{\Gamma}^\alpha_{\mu\nu}u_g^\mu u_g^\nu\right) + \af f_\ab\frac{ds_f}{ds_g}\left(\frac{du_f^\alpha}{ds_f} +\overset f\Gamma\vphantom{\Gamma}^\alpha_{\mu\nu}u_f^\mu u_f^\nu\right) = 0,  \label{eq:ppeom}
\end{equation}
where $u_g^\mu \equiv dx^\mu/ds_g$ is the four-velocity properly normalised with respect to $g_\mn$, such that $g_{\mu\nu}u_g^\mu u_g^\nu=1$, and $u_f^\mu$ is defined analogously for the $f_\mn$ geometry. In defining $u_g^\mu$ and $u_f^\mu$ we have introduced the line elements for the two metrics, $ds_g^2 \equiv g_\mn dx^\mu dx^\nu$ and $ds_f ^2 \equiv f_\mn dx^\mu dx^\nu$.

Is \cref{eq:ppeom} the geodesic equation for a Riemannian metric? In other words, can the motion of point particles in this bimetric theory be described as geodesic motion of an effective metric? We can gain insight on this question by writing \cref{eq:ppaction-lambda} in the form
\begin{equation}
S_\mathrm{pp} = -im \ag \int ds_g - im \ag \int ds_f. \label{eq:ppaction-s}
\end{equation}
To see that this is equivalent to \cref{eq:ppaction-lambda}, note that we can write
\begin{equation}
d\lambda \sqrt{-g_\mn \dot x^\mu \dot x^\nu} = ids_g \sqrt{g_\mn u_g^\mu u_g^\nu} = ids_g,
\end{equation}
where we have used the fact that (by definition) $g_\mn u_g^\mu u_g^\nu = 1$. Similar logic holds for $f_\mn$. The form (\ref{eq:ppaction-s}) is less useful calculationally, particularly for deriving the geodesic equation (\ref{eq:ppeom}), but it opens up a helpful rephrasing of the question of an effective metric: we want to find a line element, $ds$, for which $S_\mathrm{pp} = -im\int ds$. This would imply
\begin{equation}
ds \equiv \ag ds_g + \af ds_f.
\end{equation}
Squaring this and plugging back in the definitions of $ds_g$ and $ds_f$, we find that the cross-term introduces a non-Riemannian piece,
\begin{equation}
ds^2 = \left(\ag^2g_\mn + \af^2f_\mn\right)dx^\mu dx^\nu + 2\alpha_g\alpha_f\sqrt{g_\mn f_\ab dx^\mu dx^\nu dx^\alpha dx^\beta}, \label{eq:ppline}
\end{equation}
and so point particles do not move on geodesics of an effective metric.

In fact, \cref{eq:ppline} is the line element of a \emph{Finsler geometry} \cite{Cartan:1934,Bekenstein:1992pj}. A Finsler spacetime can be defined by the most general line element that is homogeneous of degree 2 in the coordinate intervals $dx^\mu$, i.e.,
\begin{align}
ds^2 &= f(x^\mu,dx^\nu),\\
f(x^\mu,\lambda dx^\nu) &= \lambda^2f(x^\mu,dx^\nu). \label{eq:finslerhomog}
\end{align}
The homogeneity property (\ref{eq:finslerhomog}) conveniently allows us to write the Finsler line element in a pseudometric form. Following Bekenstein \cite{Bekenstein:1992pj}, let us write \cref{eq:finslerhomog} taking $\lambda = 1+\epsilon$ and $\epsilon\ll1$,
\begin{equation}
f + \epsilon \frac{\partial f}{\partial dx^\mu}dx^\mu + \frac 12 \epsilon^2 \frac{\partial^2 f}{\partial dx^\mu \partial dx^\nu}dx^\mu dx^\nu + \mathcal{O}(\epsilon^3) = \left(1 + 2\epsilon + \epsilon^2\right)f, \label{eq:finsler-exp}
\end{equation}
where $f$ here is shorthand for $f(x^\mu,dx^\nu)$. Taking the $\mathcal{O}(\epsilon^2)$ piece, we find the line element can be written in terms of a \emph{quasimetric} $\mathcal{G}_\mn$,
\begin{equation}
f = ds^2 = \mathcal{G}_{\mu\nu}dx^\mu dx^\nu, \label{eq:f-qm}
\end{equation}
where the quasimetric is defined by
\begin{equation}
\mathcal{G}_{\mu\nu} \equiv \frac{1}{2}\frac{\partial^2 f}{\partial dx^\mu \partial dx^\nu}. \label{eq:qm}
\end{equation}
It is worth noting that this is an exact relation, as we have not thrown away any information by expanding in $\epsilon$. We have simply used the fact that the condition (\ref{eq:finslerhomog}) has to be satisfied at every single order in $\epsilon$. Indeed, this result equivalently follows from Euler's theorem for homogeneous functions. Euler's theorem for a multivariate function (in this case, a function of each component of $dx^\nu$, holding $x^\mu$ fixed) can be written\footnote{Note that this is the $\mathcal O (\epsilon)$ piece of \cref{eq:finsler-exp}. This form of the line element, $f = p_\mu dx^\mu$, defines the \emph{Finsler one-form}, $p_\mu$ \cite{Pfeifer:2013gha}.}
\begin{equation}
f = \frac 12 \frac{\partial f}{\partial dx^\nu}dx^\nu. \label{eq:euler-thm}
\end{equation}
Differentiating this with respect to $dx^\mu$, we obtain
\begin{equation}
\frac{\partial f}{\partial dx^\mu} = \frac 12 \left[\frac{\partial f}{\partial dx^\nu} \delta^\nu{}_\mu + \frac{\partial^2 f}{\partial dx^\mu \partial dx^\nu} dx^\nu\right] \Longrightarrow \frac{\partial f}{\partial dx^\mu} = \frac{\partial^2 f}{\partial dx^\mu \partial dx^\nu} dx^\nu.
\end{equation}
Plugging this back into \cref{eq:euler-thm}, we recover the result (\ref{eq:f-qm}, \ref{eq:qm}).

Note that the quasimetric can depend on the coordinate intervals, $dx^\mu$, which is how it differs from the metric of a usual Riemannian spacetime. We can see this by applying the \cref{eq:qm} to the definition (\ref{eq:ppline}) of $f$ to explicitly calculate $\mathcal{G}_{\mu\nu}$,
\begin{equation}
\mathcal{G}_{\mu\nu} =  \ag^2g_\ab + \af^2f_\ab  + \ag\af\left[\frac{ds_f}{ds_g}\left(g_{\mu\nu} - u^g_\mu u^g_\nu\right) + \frac{ds_g}{ds_f}\left(f_\ab - u^f_\mu u^f_\nu \right) + 2u^g_{(\mu}u^f_{\nu)}\right].
\end{equation}
In addition to constant conformal relations to the original metrics, this quasimetric is \emph{disformally} related to the particle's four-velocity. The link between Finsler geometries and disformal relations is not new; certain Finsler geometries can be described as Riemannian spacetime with matter disformaly coupled to a scalar or vector field, such as a disformal scalar-tensor theory where matter couples to the effective metric $g_\mn + \partial_\mu \phi \partial_\nu \phi$ \cite{Bekenstein:1992pj}. Disformal theories of gravity have attracted attention recently \cite{Koivisto:2012za,Zumalacarregui:2012us}; it would be interesting if these theories turned out to be related to bigravity.

This formulation in terms of Finsler geometry opens up our understanding of point-particle dynamics. For massive particles, we can define the proper time, $\tau$, from the line element in the usual way,
\begin{equation}
d\tau^2 = -ds^2.
\end{equation}
It follows trivially that, in terms of this proper time, massive point particles travel on unit-norm timelike geodesics with respect to the quasimetric,
\begin{equation}
\mathcal{G}_\ab \frac{dx^a}{d\tau}\frac{dx^b}{d\tau} = -1.
\end{equation}

We are also now in a position to extend the action (\ref{eq:ppaction-lambda}) to massless particles. This action vanishes in the limit $m\to0$ and so is technically only defined for massive particles. In general relativity, the geodesic equation does hold for massless particles. This can be seen from the fact that $m$ drops out of the geodesic equation, but to show it rigorously, it is common to introduce a Lagrange multiplier, often called the \emph{einbein}. The same logic carries over to our bimetric theory uninterrupted. Let us write the action (\ref{eq:ppaction-lambda}) in terms of a parameter $\lambda$ and introduce the einbein, $e(\lambda)$, as
\begin{equation}
S = -\frac{1}{2}\int d\lambda \left[e^{-1}(\lambda)\left(\frac{ds}{d\lambda}\right)^2 + m^2e(\lambda)\right]. \label{eq:einbeinaction}
\end{equation}
For $m\neq0$, varying this with respect to $e$ we find
\begin{equation}
e= \frac{1}{m}\frac{ds}{d\lambda}.
\end{equation}
Plugging this into the action (\ref{eq:einbeinaction}), we obtain the original action, \cref{eq:ppaction-lambda}. But we can now extend the treatment to $m=0$. In this case, varying with respect to $e$ yields
\begin{equation}
ds^2=0.
\end{equation}
Then, varying with respect to $x^a$, we find the same geodesic equation as for the massive point particles. In other words, we have found that massless point particles travel on null geodesics of $\mathcal G_\mn$. We may want to use a different form than (\ref{eq:ppeom}) for the geodesic equation when dealing with massless point particles, since in general $ds_g$ and $ds_f$ may vanish for a massless particle.\footnote{This will be the case in particular if $g_\mn$ and $f_\mn$ are conformally related, as then a lightlike path in one metric is also lightlike in the other.} We can write the geodesic equation (for a massive or massless point particle) in terms of the quasimetric as
\begin{equation}
\mathcal{G}_\mn \ddot x^\nu + \left(\mathcal{G}_{\mn,\alpha} - \frac{1}{2}\mathcal{G}_{\nu\alpha,\mu}\right)\dot x^\nu \dot x^\alpha = 0.
\end{equation}
Note that we do not write this in terms of Christoffel symbols because we do not have to; if we had, we would need to calculate the inverse quasimetric, which is both difficult and unnecessary.

This result is straightforward to extend to theories with $N$ interacting metrics $g^i_\mn$, corresponding to a massless graviton with a tower of $N-1$ massive gravitons \cite{Hinterbichler:2012cn,Tamanini:2013xia}. In this case, the line element is defined by
\begin{equation}
ds^2 = \lp\sum_{i=1}^N \alpha_i \sqrt{g^i_\mn dx^\mu dx^\nu}\rp^2,
\end{equation}
which is clearly Finslerian.

\section{Summary of Results}
\label{sec:dc-finsler-summary}

We have examined an example of a bimetric theory in which, due to their minimal coupling to both metrics, matter fields do not feel a universal physical metric. We have found that when coupling matter to multiple metrics, the massless scalar might be unique in minimally coupling to a Riemannian effective metric. The massive scalar, Maxwell field, and point particle all provide counter-examples. We examined in detail light propagation in the geometric optics limit and showed that there is a distinct problem in relating observations to the underlying theory. We can make progress by generalising the line element beyond a Riemannian form. In particular, we showed that point particles follow geodesics of a Finsler spacetime, which is nonmetric. This geometry that emerges for a pointlike observer depends quite nontrivially upon, in addition to the two metric structures, the observer's own four-velocity through a disformal coupling.

These considerations in this chapter may force us to rethink the geometric nature of spacetime, even in a metric theory of gravity. Consider the fundamental question of how to relate bigravity to observations, such as cosmological measurements. The textbook methods lean heavily on the existence of a ``Jordan-frame'' metric to which matter is minimally coupled. Here, however, such a metric does not exist universally, and may not exist at all for certain species of matter. How, then, should one calculate the redshift and the luminosity distance of a cosmological source in terms of the underlying FLRW geometries? Even the proper time along a timelike path is no longer trivial, as we cannot use the assumption $d\tau = -ds$. Indeed, because of the different effective metrics (or lack thereof), the notion of proper time is likely no longer even unique, depending instead on which matter fields an observer uses to construct her clock.

Perhaps the best approach to solving physical problems in bimetric spacetimes without an effective metric is to go back to ``primitive,'' pre-geometric constructions. In the absence of a single spacetime on which to formulate physics, we may need to simply consider particle motion coupled to two (or more) spin-2 fields in a way that only looks geometric because it is the nature of the spin-2 particle to invoke geometry \cite{Gupta:1954zz,Weinberg:1965rz}.

Paradoxically, once we have doubled geometry, we lose the ability to use its familiar methods. This is a call to go back to the basics, and rediscover the justification for results which we have taken for granted for the better part of the last century.

\begin{savequote}[30pc]
The universe is full of magical things patiently waiting for our wits to grow sharper.
\qauthor{Eden Phillpotts, \textit{A Shadow Passes}}
\end{savequote}

\chapter{Cosmological Implications of Doubly-Coupled Massive Bigravity}
\label{chap:dc-background}
\hrule
\vspace*{1cm}

So far we have studied the cosmological solutions of massive bigravity in \cref{chap:bigravity-stability,chap:bigravity-subhorizon} with matter coupled only to one metric, and discussed some of the theoretical issues with extending to a bimetric matter coupling in \cref{chap:dc-finsler}. As emphasised in the introduction of \cref{chap:dc-finsler}, the singly-coupled theory spoils the metric interchange symmetry present in vacuum; the kinetic and mass terms treat the metrics on equal footing, but this is broken when one couples matter to only one metric. It is therefore compelling to investigate other types of matter coupling that extend this metric-interchange symmetry to the entire theory. Moreover, as demonstrated in \cref{chap:bigravity-stability}, cosmological background viability and linear stability rule out all but a small handful of the parameter space of the singly-coupled theory. By extending the matter coupling, we may be able to open up the space of observationally-allowed bimetric theories.

The most significant obstacle to the construction of such a theory is that almost all attempts to couple matter to both metrics, such as the double minimal coupling discussed in \cref{chap:dc-finsler}, reintroduce the Boulware-Deser ghost (cf. \cref{sec:bd}) at arbitrarily low energies \cite{Hassan:2012wr,Yamashita:2014fga,deRham:2014naa}. One of the papers demonstrating this, \rcite{deRham:2014naa}, also proposed a double coupling which is significantly better-behaved with respect to the Boulware-Deser ghost. While a ghost does appear in this theory, it appears at a scale at least as high as the strong coupling scale and possibly parametrically larger, in which case it is outside the domain of the validity of the effective theory. While this may present a problem for highly anisotropic solutions, the absence of the ghost around FLRW solutions was demonstrated explicitly \cite{deRham:2014naa}. The status of the ghost in this specific coupling has also been investigated in \rcite{Hassan:2014gta,deRham:2014fha,Noller:2014ioa}.

In this theory, matter couples minimally to an effective metric constructed out of the two metrics appearing in the gravitational sector of the theory, regardless of whether the second metric is dynamical. This would alleviate the problem of constructing physical observables discussed in \cref{chap:dc-finsler}, as matter would move on geodesics of the effective metric. This proposal has been derived using complementary methods and extended to a multi-metric framework in \rcite{Noller:2014sta}, while the cosmology of this new coupling has been investigated in the dRGT context in \rcite{Gumrukcuoglu:2014xba} and will be discussed in \cref{chap:dc-drgt}.

In this chapter, we study the background cosmology of massive bigravity when matter couples to the effective metric proposed in \rcite{deRham:2014naa}. We show that the background expansion can asymptotically approach $\Lambda$CDM at both early and late times, and for certain parameter values is identical to $\Lambda$CDM always. At the background level, this type of coupling is therefore consistent with observations. In a future study, we will investigate whether this holds true for cosmological perturbations.

This chapter is organised as follows. In \cref{sec:dcbg-action} we present the effective metric and the symmetries that are present in the action. In \cref{sec:dcbg-flrw} we derive the cosmological equations of motion and discuss their main features. A parameter scan of the minimal models, where only one of the interaction terms is nonvanishing, is performed in \cref{sec:minimod}. In \cref{sec:specialparams} we discuss some special parameter choices. We conclude in \cref{sec:bg-summary}.

\section{Doubly-Coupled Bigravity}
\label{sec:dcbg-action}

In this chapter we will extend the bigravity action (\ref{eq:actionHR}) to a doubly-coupled version with an effective metric, $g^\eff_\mn$,\footnote{We will denote the effective metric with ``eff'' written as a superscript or subscript interchangeably.} given by
\begin{align}
   S_\mathrm{HR,dc} &= -\frac{M_g^2}{2}\int d^4x\sdg R(g) -\frac{M_f^2}{2}\int d^4x\sdf R(f) \nonumber \\
   &\hphantom{{}=}+ m^2M_g^2\int d^4x\sdg\sum_{n=0}^{4}\beta_ne_n(\mathbb X)+ \int d^4x\sdg\mathcal{L}_m\left(g_\eff, \Phi_i\right). \label{eq:actionunscaled}
\end{align}
The effective metric, first introduced in \rcite{deRham:2014naa}, is defined by\footnote{In \rcite{deRham:2014naa} the effective metric is given in an explicitly symmetric form, but this is not needed since $g_{\mu\alpha}\mathbb X^{\alpha}{}_{\nu} = g_{\nu\alpha}\mathbb X^{\alpha}{}_{\mu}$, as first shown in \rcite{Hassan:2012wr}; see also \cref{app:sym}.}
\begin{equation}
\label{eq:effectiveg}
g^\eff_{\mu\nu} = \alpha^2 g_{\mu\nu} + 2\alpha\beta g_{\mu\alpha}\mathbb X^{\alpha}{}_{\nu}+\beta^2 f_{\mu\nu},\qquad \mathbb X^{\mu}{}_{\nu} = (\sqrt{g^{-1}f})^\mu{}_\nu.
\end{equation}
As shown in \cref{app:sym}, the effective metric is symmetric under the interchange $g_\mn\leftrightarrow f_\mn$ and $\alpha\leftrightarrow \beta$. This makes the entire action symmetric under the transformations
\begin{equation}
g_\mn\leftrightarrow f_\mn,\qquad M_g \leftrightarrow M_f,\qquad \beta_{n}\to \beta_{4-n},\qquad\alpha\leftrightarrow\beta. \label{eq:bg-dualitytrans}
\end{equation}
There is thus a duality between the two metrics present in the action which is spoiled when matter couples to only one of the metrics (taken by setting either $\alpha=0$ or $\beta=0$).

The effective metric has the convenient property that its determinant is in fact in the form of the ghost-free interaction potential in \cref{eq:actionunscaled}. In particular, the determinant can be written as \cite{deRham:2014naa}
\begin{equation}
\sqrt{-g_\eff}=\sqrt{-g}\det\left(\alpha+\beta \mathbb X\right). \label{eq:detgeff}
\end{equation}
The right-hand side is a \emph{deformed determinant}, and it appears naturally when constructing a ghost-free potential \cite{Hassan:2011vm}. Indeed, this deformed determinant is nothing other than a subset of the ghost-free dRGT potential with specific choices for $\beta_n$,
\begin{equation}
\det\left(\alpha+\beta \mathbb X\right)=\sum_{n=0}^{4}\alpha^{4-n}\beta^n e_{n}\left(\mathbb X\right). \label{eq:defdet}
\end{equation}
Therefore matter loops, which will generate a term of the form $\sqrt{-g_\eff}\Lambda_v$, will by construction not lead to a Boulware-Deser ghost. This simple criterion in fact dooms many other forms of double coupling and is, in part, what motivated \rcite{deRham:2014naa,Noller:2014sta} to construct this specific form of $g_\mn^\eff$.

The action (\ref{eq:actionunscaled}) contains two Planck masses ($M_g$ and $M_f$), five interaction parameters ($\beta_n$, of which $\beta_0$ and $\beta_4$ are the cosmological constants for $g$ and $f$, respectively), and two parameters describing how matter couples to each metric ($\alpha$ and $\beta$). The Planck masses and the coupling parameters $\alpha$ and $\beta$ only enter observable quantities through their ratios. Moreover, one of those ratios is redundant: as described in \cref{app:sym}, the action can be freely rescaled so that either $M_f/M_g$ or $\beta/\alpha$ is set to unity.\footnote{See also \cref{sec:mg-bigravity} for the redundancy of the Planck masses in the singly-coupled theory.} Therefore the physically-relevant parameters are $\beta_n$ and either $M_f/M_g$ or $\beta/\alpha$. In this chapter we will rescale the Planck masses so that there is one effective gravitational coupling strength, $M_\eff$. We will also keep $\alpha$ and $\beta$ explicit to make the $\alpha \leftrightarrow \beta$ symmetry manifest, but the reader should bear in mind that only their ratio matters physically. All observational constraints will be given solely in terms of $\beta/\alpha$, from which it is straightforward to take the singly-coupled limit, $\beta/\alpha\to0$.

The Einstein equations have been derived in \rcite{Schmidt-May:2014xla} and can be written in the form
\begin{align}
(\mathbb X^{-1})^{(\mu}{}_\alpha G^{\nu)\alpha}_g + m^2\displaystyle\sum_{n=0}^{3}(-1)^n\beta_ng^{\alpha\beta}(\mathbb X^{-1})^{(\mu}{}_\alpha Y^{\nu)}_{(n)\beta} = \frac{\alpha}{M_g^2}\sqrt{\frac{g_\eff}{g}}\left(\alpha (\mathbb X^{-1})^{(\mu}{}_\alpha T^{\nu)\alpha}+\beta T^{\mu\nu}\right),\label{eq:eomg}
\end{align}
\begin{align}
\mathbb X^{(\mu}{}_\alpha G^{\nu)\alpha}_f + m^2\frac{M_g^2}{M_f^2}\displaystyle\sum_{n=0}^{3}(-1)^n\beta_{4-n}f^{\alpha\beta}\mathbb X^{(\mu}{}_\alpha \hat{Y}^{\nu)}_{(n)\beta} = \frac{\beta}{M_f^2}\sqrt{\frac{g_\eff}{f}}\left(\alpha T^{\mu\nu}+\beta \mathbb X^{(\mu}{}_\alpha T^{\nu)\alpha}\right).\label{eq:eomf}
\end{align}
The matrices $Y$ and $\hat{Y}$ depend on $\sqrt{g^{-1}f}$ and $\sqrt{f^{-1}g}$, respectively, and are the same as were defined in \cref{eq:ymatdef}. The Einstein tensors $G_g^\mn$ and $G_f^\mn$ have their indices raised with $g_\mn$ and $f_\mn$, respectively. Note that the terms with $g_\eff$ can be simplified using \cref{eq:detgeff}. The stress-energy tensor $T^\mn$ is defined with respect to the effective metric $g_\eff$ as
\begin{equation}
 \delta\left[\sdgb\calL\left(g_\eff,\Phi\right)\right] = \frac{1}{2}\sdgb T^\mn\delta g^\eff_\mn,
\end{equation}
and obeys the usual conservation equation
\begin{equation}
\nabla^\eff_\mu T^\mn = 0,
\end{equation}
where $\nabla^\eff_\mu$ is the covariant derivative for $g^\eff_\mn$.

\section{Cosmological Equations and Their Solutions}
\label{sec:dcbg-flrw}

To describe homogeneous and isotropic cosmologies, we specialise to the Friedmann-Lema\^itre-Robertson-Walker (FLRW) ans{\"a}tze for both $g_\mn$ and $f_\mn$,
\begin{align}
ds_{g}^{2}&=-N_{g}^{2}dt^{2}+a_{g}^{2}d\vec{x}^{2}, \\
ds_{f}^{2}&=-N_{f}^{2}dt^{2}+a_{f}^{2}d\vec{x}^{2},
\end{align}
where $N_{g,f}$ and $a_{g,f}$ are the lapses and scale factors, respectively, of the two metrics. Because both metrics are on equal footing, we have changed the notation slightly from \cref{chap:mg,chap:bigravity-stability,chap:bigravity-subhorizon} to be more symmetric between the two metrics. As in general relativity, we can freely rescale the time coordinate to fix either $N_g$ or $N_f$; however, their ratio is gauge-invariant and will remain unchanged. The effective metric becomes
\begin{equation}
ds_\eff^{2}=-N^{2}dt^{2}+a^{2}d\vec{x}^{2},
\end{equation}
where the effective lapse and scale factor are related to those of $g_\mn$ and $f_\mn$ by
\begin{align}
N &= \alpha N_{g} + \beta N_{f}, \\
a &= \alpha a_{g} + \beta a_{f}.
\end{align}
The equations of motion can be derived either directly from \cref{eq:eomg,eq:eomf}, or by plugging the FLRW ans{\"a}tze into the action and varying with respect to the scale factors and lapses, as was done in \rcite{deRham:2014naa}. We have checked that both approaches yield the same result. Defining
\begin{align}
\label{eq:B0}
B_{0}(y)&\equiv\beta_{0}+3\beta_{1}y+3\beta_{2}y^2+\beta_{3}y^3, \\
\label{eq:B1}
B_{1}(y)&\equiv\beta_{1}y^{-3}+3\beta_{2}y^{-2}+3\beta_{3}y^{-1}+\beta_{4},
\end{align}
where, as before,
\begin{equation}
y \equiv \frac {a_{f}}{a_{g}},
\end{equation}
the Friedmann equations for $g_\mn$ and $f_\mn$ are
\begin{align}
\label{eq:friedeqg}
3H_{g}^2 &= \frac{\alpha\rho}{M_\eff^2}\frac{a^3}{a_{g}^3} + m^2B_0, \\
\label{eq:friedeqf}
3H_{f}^2 &= \frac{\beta\rho}{M_\eff^2}\frac{a^3}{a_{f}^3} + m^2B_1.
\end{align}
Here the energy density, $\rho$, is a function of the effective scale factor, $a$, and we have defined the $g$- and $f$-metric Hubble rates as
\begin{equation}
H_{g} \equiv \frac{\dot a_{g}}{N_{g}a_{g}}, \qquad H_{f} \equiv \frac{\dot a_{f}}{N_{f}a_{f}}.
\end{equation}
Notice that the two Friedmann equations for $H_g$ and $H_f$ map into one another under the interchange $\beta_n \to \beta_{4-n}$, $\alpha \leftrightarrow \beta$, and $g_\mn \leftrightarrow f_\mn$ (which sends $H_g \leftrightarrow H_f$, $y\to y^{-1}$, and $B_0\leftrightarrow B_1$), as expected from the properties of the action described in \cref{app:sym}.

The stress-energy tensor is conserved with respect to the effective metric, so we immediately have
\begin{equation}
\dot{\rho}+3\frac{\dot{a}}{a}\left(\rho+p\right)=0,
\end{equation}
where the density, $\rho$, and pressure, $p$, are defined in the usual way from the stress-energy tensor. By taking the divergence of either Einstein equation with respect to the associated metric (e.g., taking the $g$-metric divergence of \cref{eq:eomg}) and using the Bianchi identity and stress-energy conservation, we obtain the ``Bianchi constraint,''
\begin{equation}
\label{eq:bg-bianchi}
\left[m^{2}\left(\beta_1 a_g^2+2\beta_2 a_g a_f + \beta_3 a_f^2\right)-\frac{\alpha\beta a^{2}p}{M_\eff^2}\right]\left(N_{f}\dot{a}_{g}-N_{g}\dot{a}_{f}\right)=0.
\end{equation}
In complete analogy with the singly-coupled case discussed in \cref{sec:mg-bigrav-cosmo-eq}, which can be obtained by setting $\alpha$ or $\beta$ to zero, \cref{eq:bg-bianchi} gives rise to two possible branches of solutions, one algebraic and one dynamical \cite{Comelli:2011zm,vonStrauss:2011mq,Volkov:2011an}.\footnote{In the singly-coupled theory, \cref{eq:bg-bianchi} would be a constraint equation arising from the Bianchi identity and stress-energy conservation. When using the effective coupling, the stress-energy conservation holds with respect to the effective metric, rather than $g_\mn$ or $f_\mn$. This gives rise to the pressure-dependent term in the left bracket. Due to this term, both branches---obtained by setting either bracket to zero---can be regarded as dynamical. We choose to adopt the terminology from the singly-coupled case here, however.}

\subsection{Algebraic Branch of the Bianchi Constraint}

As discussed in \cref{sec:mg-bigrav-cosmo-eq}, in the singly-coupled case, setting the first bracket of \cref{eq:bg-bianchi} to zero gives an algebraic constraint on $y$ that can be shown to give solutions that are indistinguishable from general relativity at all scales \cite{vonStrauss:2011mq}. In the doubly-coupled theory, the presence of the pressure term makes the phenomenology of the algebraic branch richer.

In this section, without any ambition to examine all possible solutions, we briefly outline some of the properties of a few specific solutions on the algebraic branch of the Bianchi constraint (\ref{eq:bg-bianchi}). In this branch we have
\begin{equation}
m^{2}\left(\beta_1 a_g^2+2\beta_2 a_g a_f + \beta_3 a_f^2\right)=\frac{\alpha\beta a^{2}p}{M_\eff^2}.
\label{eq:firstbrancheq}
\end{equation}
If the Universe is dominated by dust ($p=0$), then as in the singly-coupled theory this is a polynomial equation for $y$,
\begin{equation}
\beta_1+2\beta_2 y+\beta_3 y^2=0,
\end{equation}
which is solved by a constant $y=y_c$. Notice that when $y$ is constant, the mass terms in the two Friedmann equations become constant, so $H_g$ and $H_f$ are determined by Friedmann equations containing effective cosmological constants.\footnote{These are not, however, $\Lambda$CDM cosmologies for the effective metric due to the nontrivial coupling to $\rho$.} Using the fact that $a=\left(\alpha+\beta y_c\right)a_g=\left(\alpha/y_c+\beta\right)a_f$, we can show that the observed Hubble rate, $H = \dot a/(aN)$, for a constant $y$ is given by
\begin{equation}
H = H_g\left(\alpha + \beta\frac{N_f}{N_g}\right)^{-1} = H_f\left(\alpha\frac{N_g}{N_f} + \beta\right)^{-1}.
\end{equation}
If the ratio $N_g/N_f$ is constant, the solutions on this branch contain an exact cosmological constant (at least at the background level) given by a combination of the metric interaction terms.

Since for a constant $y$, the two Hubble rates are related by
\begin{equation}
H_f = H_g \frac{N_g}{N_f},
\end{equation}
the bimetric interactions mimic a cosmological constant when $H_g/H_f=\mathrm{const}$. This is only possible if the parameters satisfy $\alpha y_c^3 B_1=\beta B_0$. For more general parameter values, we have
\begin{equation}
\left(\frac{N_f}{N_g}\right)^2 = \frac{3\alpha H_g^2 y_c^3}{3\beta H_g^2+m^2\left(\alpha B_1 y_c^3 - \beta B_0\right)},
\end{equation}
which is dynamical, so these cosmologies are not exactly $\Lambda$CDM.

For nonzero pressure, $p\neq0$, we can rewrite the constraint (\ref{eq:firstbrancheq}) as
\begin{equation}
m^{2}\beta_{1}\left(1+2\frac{\beta_{2}}{\beta_{1}}y+\frac{\beta_{3}}{\beta_{1}}y^{2}\right)=\frac{\alpha^{3}\beta\left(1+\frac{2\beta}{\alpha}y+\frac{\beta^{2}}{\alpha^{2}}y^{2}\right)p}{M_\eff^{2}}.
\end{equation}
We will not attempt to classify the solutions in this more complicated scenario. However, we note that for the special parameter choice
\begin{equation}
\beta_{2}=\beta_{1}\frac{\beta}{\alpha},\qquad\beta_{3}=\beta_{1}\frac{\beta^{2}}{\alpha^{2}},
\end{equation}
we obtain
\begin{equation}
\frac{p}{M_\eff^{2}}=\frac{m^{2}\beta_{1}}{\alpha^{3}\beta},
\end{equation}
i.e., we are required to have a constant $p$, corresponding to a vacuum equation of state, $w=-1$. 

\subsection{Dynamical Branch of the Bianchi Constraint}

As is most often done in singly-coupled bigravity models---see, for example, \rcite{vonStrauss:2011mq,Berg:2012kn,Akrami:2012vf,Akrami:2013pna,Konnig:2013gxa} and \cref{chap:mg,chap:bigravity-stability,chap:bigravity-subhorizon}---in the remainder of this chapter we will restrict our study to solutions where the second bracket in \cref{eq:bg-bianchi} vanishes, as these will turn out to be consistent with observational data. In this branch we have a dynamical constraint on the ratio between $N_f$ and $N_g$,
\begin{equation}
\label{eq:NfNg}
\frac{N_{f}}{N_{g}}=\frac{\dot{a}_{f}}{\dot{a}_{g}}=\frac{da_{f}}{da_{g}}.
\end{equation}
This implies the simple relation $H_{f}y=H_{g}$. Furthermore, the physical Hubble rate $H$, defined as
\begin{equation}
\label{eq:H}
H \equiv \frac{\dot a}{Na},
\end{equation}
becomes
\begin{equation}
H = \frac{H_g}{\alpha+\beta y} = \frac{y H_f}{\alpha+\beta y}.
\end{equation}
Combining the two Friedmann equations, we obtain the equations for $H$ and $y$,
\begin{align}
H^{2}&=\frac{\rho}{6M_{\eff}^{2}}\left(\alpha+\beta y\right)\left(\alpha+\beta y^{-1}\right)+\frac{m^{2}\left(B_{0}+y^{2}B_{1}\right)}{6\left(\alpha+\beta y\right)^{2}}, \label{eq:Heff}\\
0 &= \frac{\rho}{M_\eff^{2}}\left(\alpha+\beta y\right)^{3}\left(\alpha -\beta y^{-1}\right)+m^{2}\left(B_0-y^2B_1\right). \label{eq:bg-quartic}
\end{align}
\Cref{eq:Heff,eq:bg-quartic} determine the expansion history completely and are invariant under the combination of $\beta_n \to \beta_{4-n}$, $\alpha\leftrightarrow \beta$, and $y \to y^{-1}$. They have the same structure as in singly-coupled bigravity (cf. \cref{sec:mg-bigrav-cosmo-eq}): there is a single Friedmann equation sourced by $\rho$ and $y$, while $y$ evolves according to an \emph{algebraic} equation whose only time dependence comes from $\rho$. Notice that due to \cref{eq:bg-quartic} one can write many different, equivalent forms of the Friedmann equation for $H^2$. It is therefore dangerous to directly identify the factors in front of $\rho$ in \cref{eq:Heff} as a time-varying gravitational constant and the term proportional to $m^2$ as a dynamical dark energy component: both of these effects are present, but they cannot be straightforwardly separated from each other.

From \cref{eq:bg-quartic}, we see that as $\rho\to\infty$ in the far past, either $y\to \beta/\alpha$ or $y\to -\alpha/\beta$. One can show that if $\rho\sim a^{-p}$ then $H^2\sim a^{-2p/3}$ as $y\to -\alpha/\beta$. Since this scenario is observationally excluded, we will not consider this limit. Recall from \cref{sec:properties-bimetric-cosmo} that in the singly-coupled theory there are also infinite-branch solutions where $y\to\infty$ at early times \cite{Konnig:2013gxa}. Indeed, as we saw in \cref{chap:bigravity-stability}, these infinite-branch solutions are crucial in order to avoid linear instabilities. However, in the doubly-coupled theory, there are no solutions to \cref{eq:bg-quartic} in which $y\to\infty$ as $\rho\to\infty$. This is because of the new term proportional to $\alpha\beta y^3\rho$; none of the terms in $B_0-y^2B_1$, which grows at most as $y^3$, can possibly cancel off this term as $\rho\to\infty$.

An interesting feature is that in the early Universe the mass term drops away but we are left with a modification to the gravitational constant,
\begin{equation}
H^{2}\to\frac{(\alpha^2+\beta^2)\rho}{3M_{\eff}^{2}}.
\end{equation}
Since the coefficient in front of $\rho$ in the Friedmann equation during radiation domination can be probed by big bang nucleosynthesis, this could in principle be used to constrain the parameters of the theory. However, this will only work if the corresponding factor in front of $\rho$ in local gravity measurements has a different dependence on $\alpha$ and $\beta$. The solar-system predictions for this theory have not, to date, been worked out.

In the far future, as $\rho\to 0$, we have two possibilities. The first is that $y$ goes to a constant $y_c$, determined by
\begin{equation}\label{eq:bg-quarticfuture}
\beta_{3}y_c^{4}+\left(3\beta_{2}-\beta_{4}\right)y_c^{3}+3\left(\beta_{1}-\beta_{3}\right)y_c^{2}+\left(\beta_{0}-3\beta_{2}\right)y_c-\beta_{1}=0.
\end{equation}
These models approach a de Sitter phase at late times (whether they \emph{self}-accelerate is a subtle question which we address below), with a cosmological constant given by
\begin{equation}\label{eq:effCC}
\Lambda=\frac{m^2\left[\beta_1+\left(\beta_0+3\beta_2\right)y_c+3\left(\beta_1+\beta_3\right)y_c^2+\left(3\beta_2+\beta_4\right)y_c^3+\beta_3 y_c^4\right]}{2y_c\left(\alpha+\beta y_c\right)^2}.
\end{equation}
The second possibility is that, for some parameter choices, $|y| \to\infty$ such that the leading-order $\beta_n$ term in \cref{eq:bg-quartic} exactly cancels the leading density term, $y^4\rho$. It is unclear whether these solutions are viable; in this chapter, we will restrict ourselves to solutions where $y$ is asymptotically constant in the past and future, starting at $y=\beta/\alpha$ and ending with $y=y_c$. This implies that $a_g$ and $a_f$ are proportional to one another in both the early and late Universe. As long as $y$ does not exhibit any singular behaviour, the evolution between $y=\beta/\alpha$ and $y=y_c$ is monotonic. This can be seen by taking a time derivative of \cref{eq:bg-quartic} and setting $\dot y=0$. 

The monotonicity of the evolution of $y$ implies that in the special case where $y_c=\beta/\alpha$, then we will have $y=\beta/\alpha$ at all times, and the expansion history is identical to $\Lambda$CDM. This is a new feature of the doubly-coupled theory: in the singly-coupled case, $y_c$ becomes zero in the presence of matter, which makes such a case trivially identical to general relativity. A constant $y$ occurs in any model where the $\beta_n$ parameters and $\beta/\alpha$ are chosen to satisfy
\begin{equation}\label{eq:betaalpha}
\beta_{3}\left(\frac{\beta}{\alpha}\right)^{4}+\left(3\beta_{2}-\beta_{4}\right)\left(\frac{\beta}{\alpha}\right)^{3}+3\left(\beta_{1}-\beta_{3}\right)\left(\frac{\beta}{\alpha}\right)^{2}+\left(\beta_{0}-3\beta_{2}\right)\left(\frac{\beta}{\alpha}\right)-\beta_{1}=0,
\end{equation}
which is simply \cref{eq:bg-quarticfuture} with $y_c=\beta/\alpha$. An interesting implication of solutions with constant $y$ is that, since \cref{eq:NfNg} implies $N_f/N_g = da_f/da_g=y$, the two metrics are proportional, $f_\mn = y^2 g_\mn$.\footnote{It is not difficult to see that there are no cases in which the two metrics are related by a \emph{dynamical} conformal factor; from \cref{eq:NfNg} any conformal relation means that $da_f/da_g=a_f/a_g$, but this implies $a_f/a_g=\mathrm{const.}$}

\section{Comparison to Data: Minimal Models}
\label{sec:minimod}

In this section, we compare the background expansion derived above to observations and perform a parameter scan of the \emph{minimal models}, in which only one of the $\beta_n$ is nonzero. Due to the duality property of the solutions, we only have to look at the $\beta_0$, $\beta_1$, and $\beta_2$ cases. We will restrict ourselves to positive $\beta/\alpha$; in principle negative values could also be allowed, but we have not yet investigated the physical implications of these values.\footnote{Note that $\beta<0$ leads to instabilities in the case of doubly-coupled dRGT massive gravity, in which one of the metrics is nondynamical \cite{Gumrukcuoglu:2014xba}.} The minimal models admit exact $\Lambda$CDM solutions when $\beta/\alpha=\left\{0,\frac{1}{\sqrt3},1\right\}$ for the $\beta_0$, $\beta_1$, and $\beta_2$ cases, respectively, as is evident from \cref{eq:betaalpha}. 

Since we have so far calculated the equations of motion only for homogeneous backgrounds, we will limit this study to purely geometrical tests of the background expansion, 
including the redshift-luminosity relation of Type Ia supernovae (SNe) \cite{Suzuki:2011hu}, the observed angular scales of cosmic microwave background (CMB) 
anisotropies \cite{Ade:2013zuv}, and baryon-acoustic oscillations (BAO) \cite{Anderson:2012sa,Beutler:2011hx,Blake:2011en}. 
Since the latter two depend on the physical size of the sound horizon scale around the time when the CMB photons decoupled from the
baryon plasma, we can cancel out this dependence by using only the ratio of the observed angular scales in the CMB and BAO \cite{Sollerman:2009yu,vonStrauss:2011mq}. In this way, we obtain a cosmological probe that is highly insensitive to the physics of the early Universe, and almost exclusively sensitive to the expansion history of the Universe between $z\sim 1000$ and today.

We can calculate the effective equation of state for the background model described in eqs.~(\ref{eq:Heff}) and (\ref{eq:bg-quartic}) using
\begin{equation}
w_\eff=-1-\frac{1}{3}\frac{d \log H^2}{d \log a}.
\end{equation}
Since in this chapter we restrict ourselves to solutions where $y$ approaches constant values in the 
infinite past and future, for matter-dominated models we are guaranteed to have an 
effective equation of state where $w_\eff\to 0$ as $a\to 0$ (ignoring radiation) and $w_\eff\to -1$ as $a\to\infty$, mimicking the asymptotic behaviour of the $\Lambda$CDM model. Except for some special parameter choices which are exactly $\Lambda$CDM (see the discussion above, as well as \cref{sec:specialparams}), we expect the model to deviate from the concordance model at all finite times.

It is well-known that $\Lambda$CDM is able to provide an excellent fit to background expansion data, so we expect the success of the bimetric model to depend on how close the effective equation of state is to that of $\Lambda$CDM. All solutions that look exactly like $\Lambda$CDM will trivially be able to fit existing background expansion data. Note, however, that this does not mean that these models are equivalent to $\Lambda$CDM, since they may give different predictions for perturbations, i.e., when studying structure formation.

\begin{figure}
\centering
\includegraphics[scale=0.4]{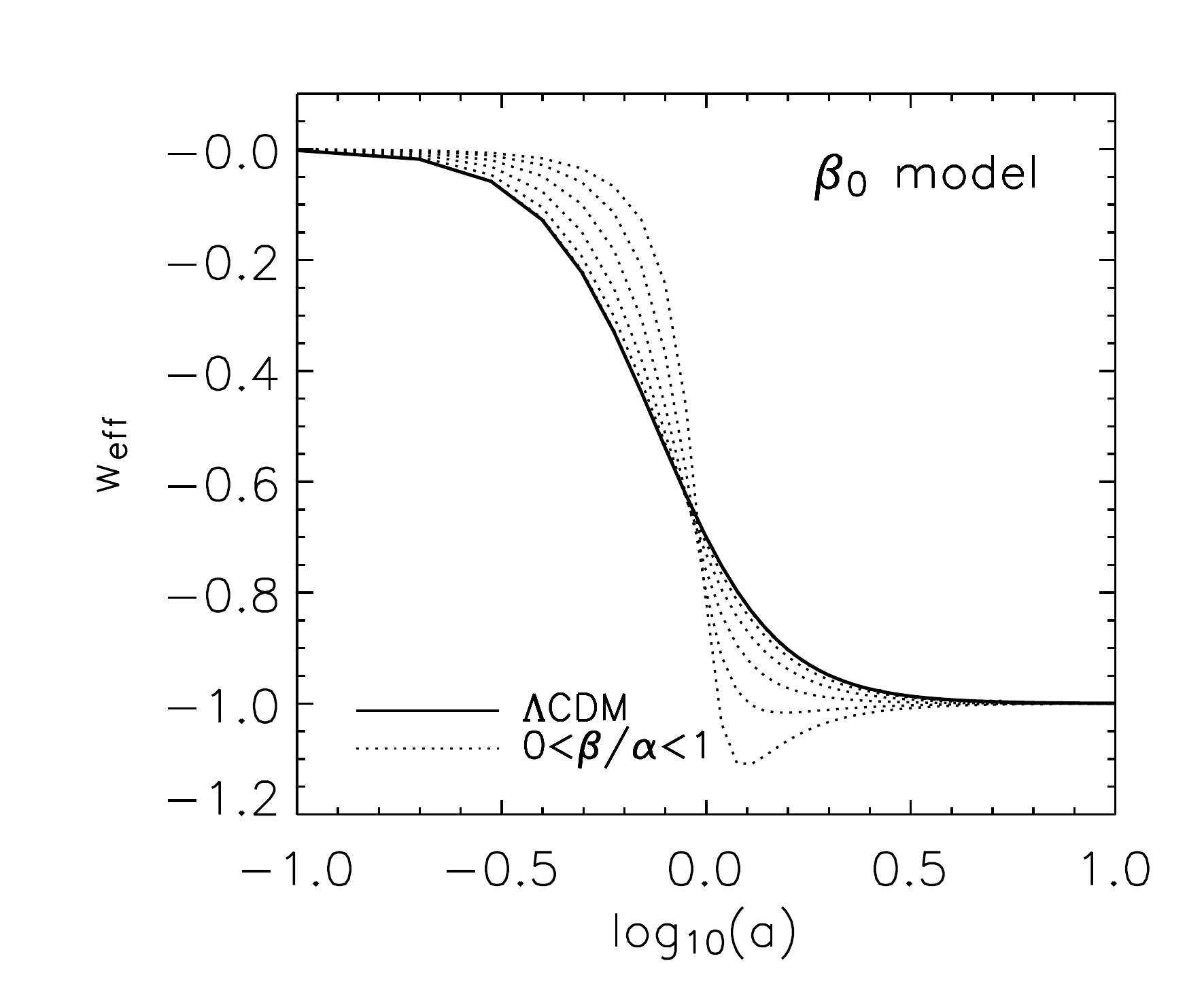}
\includegraphics[scale=0.4]{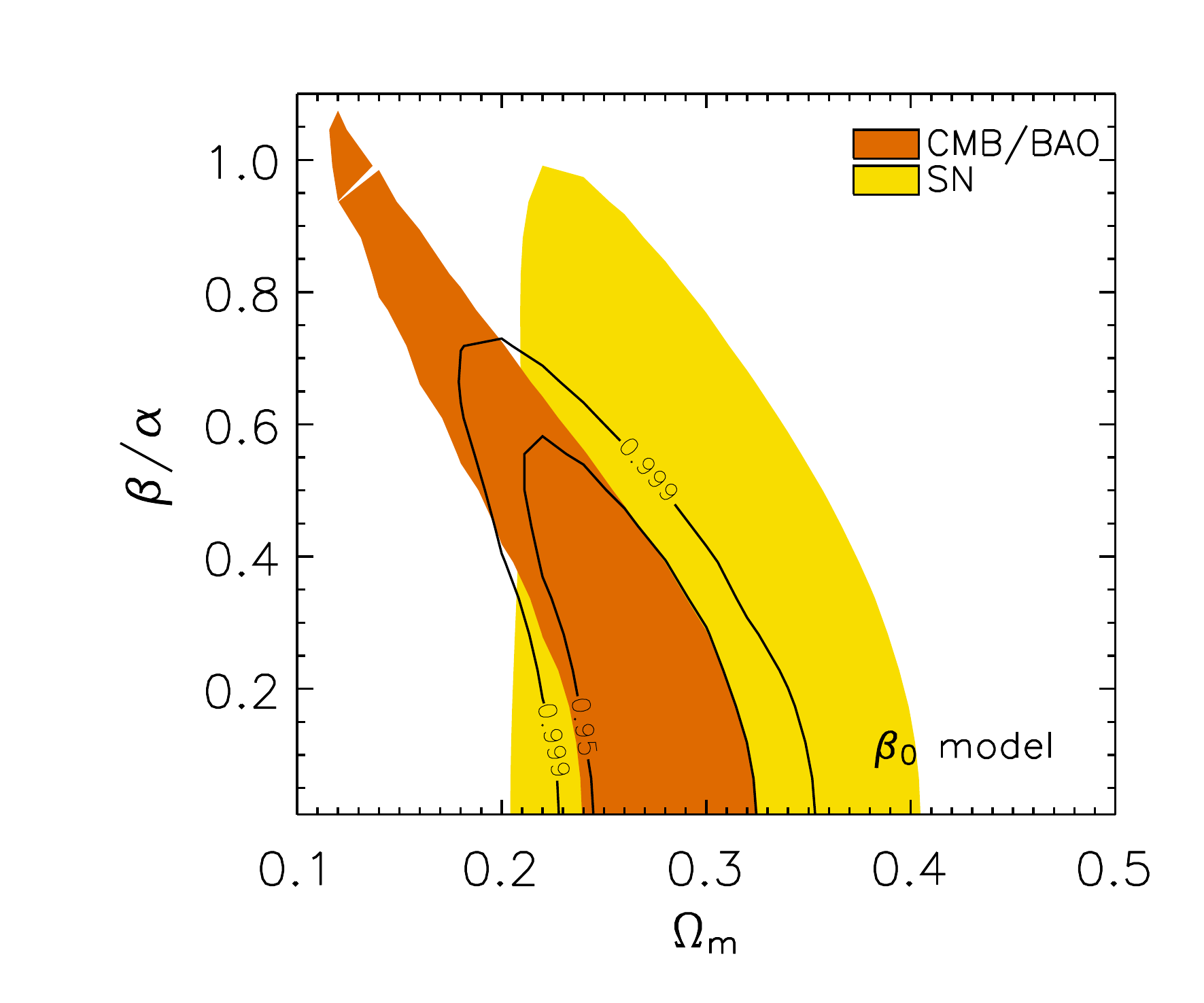}
\caption[Effective equation of state and confidence contours for the doubly-coupled $\beta_0$ model.]{\textbf{Left panel:} The effective equation of state, $w_\eff$, for the $\beta_0$ model with $0<\beta/\alpha<1$  (dotted lines) compared to $w_\eff$ of the $\Lambda$CDM model (solid line). When $\beta/\alpha\to 0$, the effective equation of state for the $\beta_0$ model approaches that of the $\Lambda$CDM model. In all cases, $\Omega_m=0.3$. \textbf{Right panel:} Confidence contours for $\Omega_m$ and $\beta/\alpha$ for the $\beta_0$ model as fitted to SNe, CMB, and BAO data.}
\label{fig:weffd0}
\end{figure}

In \cref{fig:weffd0}, we study the $\beta_0$ model, i.e., when only $\beta_0$ is turned on. Notice, cf. \cref{eq:Heff}, that this model has no nontrivial interactions between the two metrics, so it deviates from $\Lambda$CDM only through the novel matter coupling. In the left panel of \cref{fig:weffd0}, we compare the effective equation of state for different values of $\beta/\alpha$ with that of $\Lambda$CDM. We fix $\Omega_m=0.3$, where
\begin{equation}
\Omega_m \equiv \frac{\alpha^2\rho_0}{3M_\eff^2 H_0^2},
\end{equation}
and the subscript 0 indicates a value today. In the right panel of \cref{fig:weffd0}, we plot background constraints on $\Omega_m$ and $\beta/\alpha$. Note that the value of $\beta_0$ is set by the requirement that we have a flat geometry. Shaded contours show constraints from SNe and CMB/BAO data, respectively, corresponding to a 95\% confidence level for two parameters. Combined constraints are shown with solid lines corresponding to 95\% and 99.9\% confidence levels for two parameters.  As expected, when $\beta/\alpha \to 0$, the effective equation of state coincides with $\Lambda$CDM since this limit corresponds to the singly-coupled case where $\beta_0$ acts as a cosmological constant.  Note also that as $\beta/\alpha$ is increased, so is the factor multiplying the matter density in the Friedmann equation, and therefore the preferred matter density, $\Omega_m$, becomes smaller. 

\begin{figure}
\begin{centering}
\includegraphics[scale=0.4]{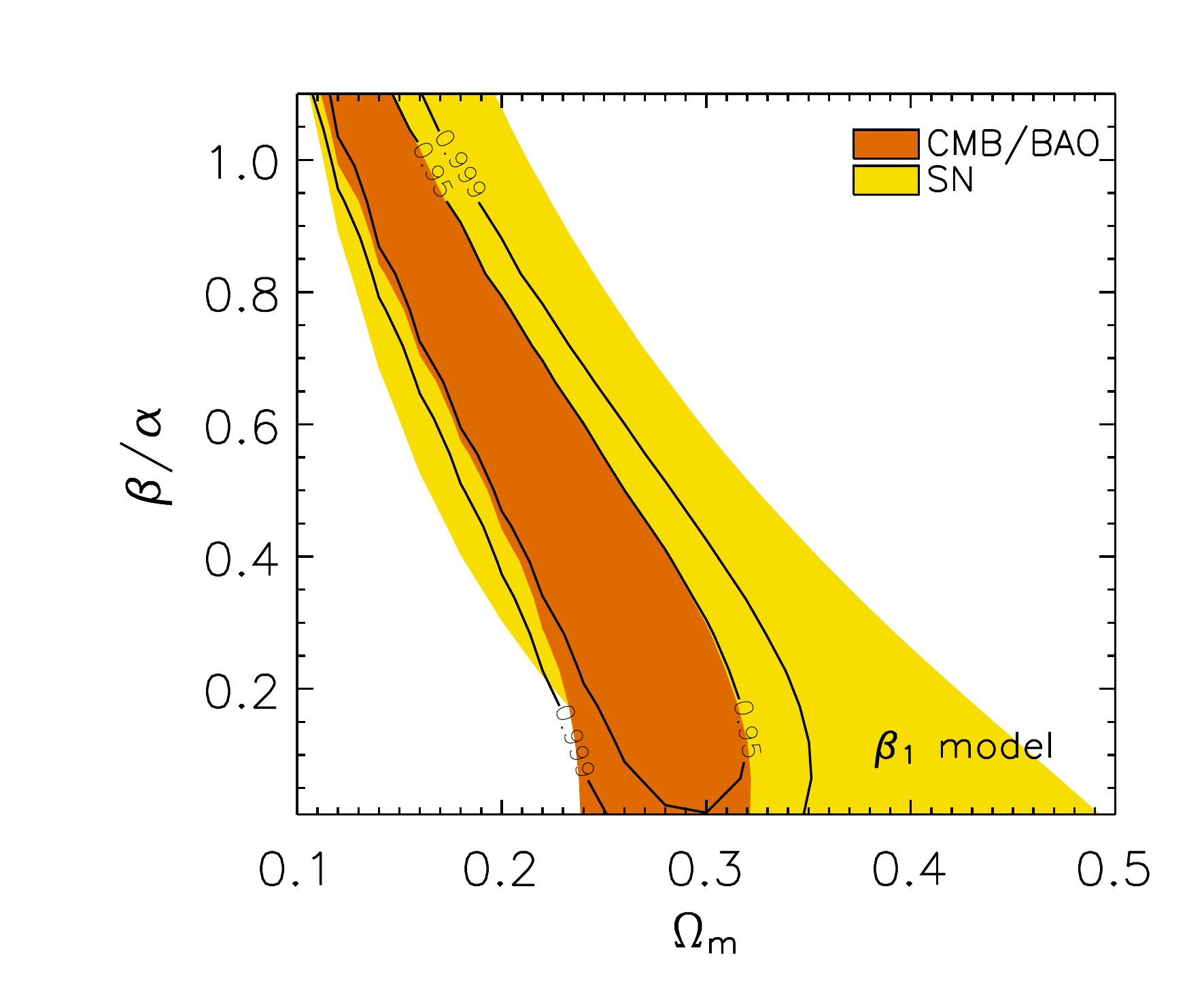}
\includegraphics[scale=0.4]{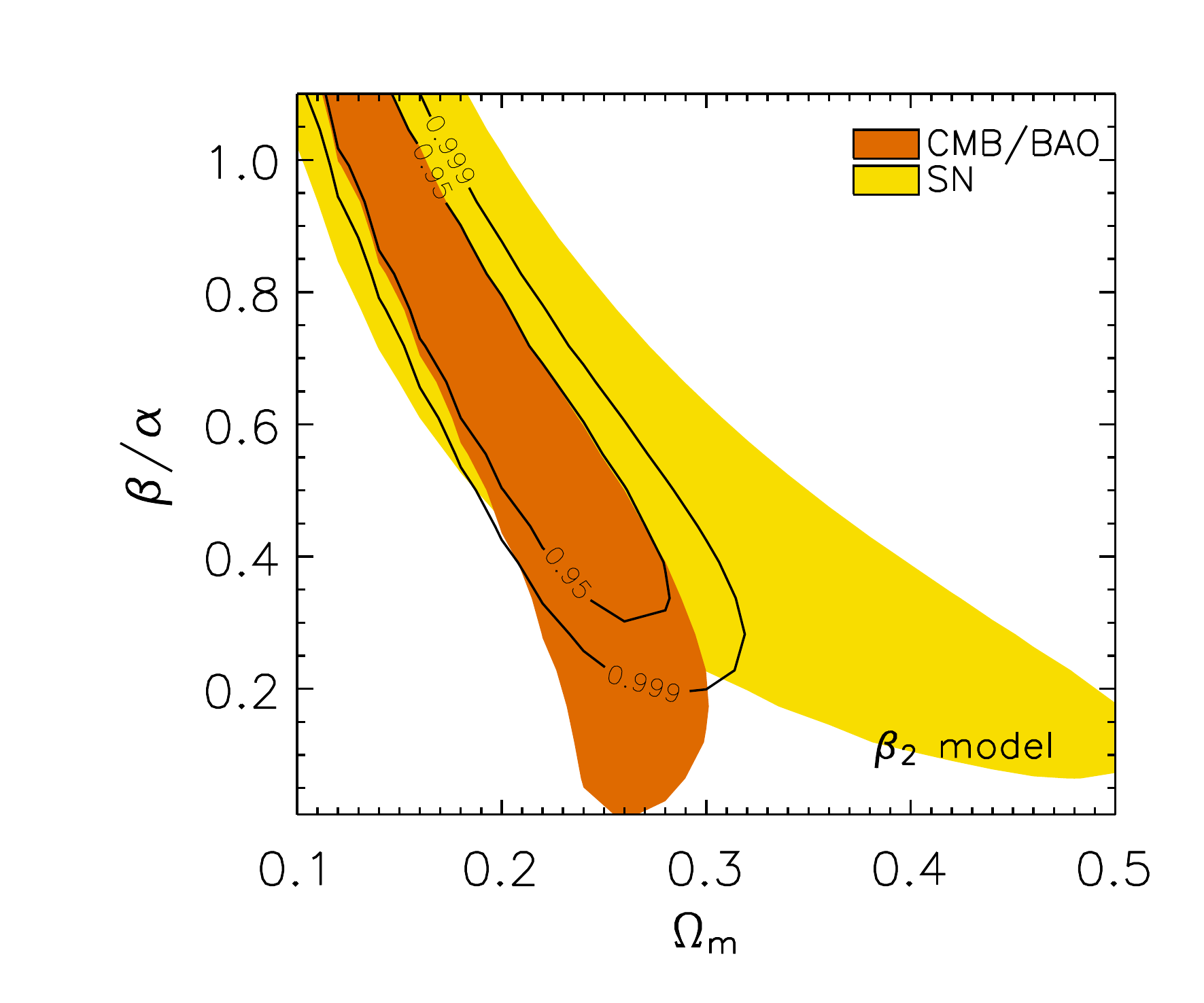}
\par\end{centering}
\caption[Confidence contours for the doubly-coupled $\beta_1$ and $\beta_2$ models.]{Confidence contours for $\Omega_m$ and $\beta/\alpha$ for the $\beta_1$ and $\beta_2$ minimal models as fitted to SNe, CMB, and BAO data. In each case, we are able to obtain as good a fit as the concordance $\Lambda$CDM model.}
\label{fig:comb}
\end{figure}

In \cref{fig:comb} we plot background constraints on the $\beta_1$ and $\beta_2$ models. Since we know that the values $\beta/\alpha=\frac{1}{\sqrt{3}}$ and $\beta/\alpha=1$ give exact $\Lambda$CDM solutions for the $\beta_1$- and $\beta_2$-only models, respectively, we expect these values to provide good fits to the data. This is indeed the case, as can be seen in the plots.
The $\beta_2$ model is especially interesting in this regard, as $\beta/\alpha=1$ corresponds to the case where the two metrics $g_\mn$ and $f_\mn$ give equal contributions to the effective metric (or $M_g=M_f$ when using the equal coupling strength framework described in appendix \ref{app:sym}). Notice that the $\beta_2$ model favours $\beta>0$, as we would expect since the $\beta_2$-only singly-coupled model is not in agreement with background data \cite{Akrami:2012vf} and is ruled out by theoretical viability conditions \cite{Konnig:2013gxa}.

One of the attractive features of the double coupling is that it allows sensible cosmological solutions with only one of the $\beta_n$  turned on.
For more general combinations of the $\beta_n$ parameters, we expect the data to favour values that cluster around the value of $\beta/\alpha$ given by solving \cref{eq:betaalpha}, since this value yields an exact $\Lambda$CDM background expansion. We do not find it meaningful to do such a parameter scan at this moment, since it is only by including other probes, such as spherically symmetric solutions and cosmological perturbations, that we can exclude a larger part of the parameter space. However, in the next section, we discuss a few special cases that may turn out to be of particular interest for further investigations.

\section{Special Parameter Cases}
\label{sec:specialparams}

\subsection{Partially-Massless Gravity}
\label{sec:pm}

Partial masslessness arises when a new gauge symmetry is present that eliminates the helicity-0 mode of the massive graviton,\footnote{So that a partially-massless graviton has four polarisations rather than the five of a massive graviton, hence the name.} removing two of the problems with massive gravity discussed in \cref{sec:massgrav}: the vDVZ discontinuity in the $m\to0$ limit of linearised massive gravity \cite{vanDam:1970vg,Zakharov:1970cc} and the need for Vainshtein screening to reconcile the theory with solar system tests \cite{Vainshtein:1972sx}. This is because both of these aspects of massive gravity are direct results of the fifth force mediated by the helicity-0 mode. Moreover, this new gauge symmetry would both determine the cosmological constant in terms of the graviton mass and protect a small cosmological constant against quantum corrections. Thus it is potentially a solution to both the old and new cosmological-constant problems: why the cosmological constant is not huge, and why it is not exactly zero, respectively.

Massive gravity and bigravity contain a candidate partially-massless theory \cite{deRham:2012kf,Hassan:2012gz}, obtained by making the parameter choices
\begin{equation}
\beta_0=3\beta_2=\beta_4, \qquad \label{eq:pmparams}
\beta_1=\beta_3=0.
\end{equation}
For more on partially-massless gravity and its connection to massive (bi)gravity, we refer the reader to \rcite{Hassan:2013pca}, as well as \rcite{Hassan:2012gz,Hassan:2014vja} and references therein. In singly-coupled bigravity, the partially-massless parameter choices could only be imposed in vacuum; including matter forces $y$ to be zero, which trivially reduces to general relativity. The nontrivial implications of the partially-massless scenario have been demonstrated for other doubly-coupled bigravity theories (see \rcite{Akrami:2013ffa}, though note that the theory discussed therein appears to have a ghost \cite{Yamashita:2014fga,deRham:2014naa}). Here we discuss this class in the context of the present doubly-coupled theory.

For the partially-massless parameter choices, \cref{eq:bg-quartic} implies that $y=\beta/\alpha$ at all times, and the Friedmann equation becomes
\begin{align}
H^{2}=\frac{\alpha^2+\beta^2}{3M_\eff^{2}}\rho+\frac{m^{2}\beta_{0}}{3(\alpha^2+\beta^2)}.
\end{align}
The cosmology of the candidate partially-massless theory is therefore equivalent to standard $\Lambda$CDM with an effective cosmological constant, $m^{2}\beta_0/(\alpha^2+\beta^2)$, and a rescaled gravitational coupling for matter. Consequently, the background expansion is identical to that of general relativity, albeit with shifted constants. Notice that this is a qualitatively new feature as compared to the singly-coupled theory.

Doubly-coupled bigravity with the parameters (\ref{eq:pmparams}) is thus a strong candidate partially-massless theory of gravity. In the context of single-metric (dRGT) massive gravity, with matter coupled only to the dynamical metric, this parameter choice leads to a theory which is not partially massless and in fact suffers from an infinitely strongly-coupled helicity-0 mode \cite{deRham:2013wv}. If doubly-coupled bigravity is shown to possess the partially-massless gauge symmetry nonlinearly and around all backgrounds, it should automatically become one of the most interesting available theories of gravity beyond general relativity.

\subsection{Vacuum Energy and the Question of Self-Acceleration}
\label{sec:vac}

As discussed in \cref{chap:intro}, one of the primary motivations for modifying general relativity is the possibility of having self-accelerating solutions, i.e., cosmologies which accelerate at late times even in the absence of a cosmological constant or vacuum energy contribution. In general relativity, as well as in singly-coupled bigravity, these two are degenerate: the vacuum energy and a cosmological constant may have different origins, but they are mathematically indistinguishable. In bigravity with matter coupled to the effective metric, however, this question becomes rather subtle, as the vacuum energy from the matter sector produces more than just the cosmological constant terms for $g_\mn$ and $f_\mn$, which are equivalent to $\beta_0$ and $\beta_4$.

We have shown in \cref{sec:dcbg-action} that quantum corrections to matter coupled to $g^\eff_\mn$ will generate \emph{all} of the ghost-free bimetric interaction terms. If we take the matter loops to generate a cosmological constant term $\sqrt{-g_\eff}\Lambda_v$, then we can see from \cref{eq:detgeff,eq:defdet} a pure vacuum-energy contribution can be written in the form of the bigravity interaction potential with parameters
\begin{equation}\label{eq:lambdabeta}
\beta_n = \frac{\Lambda_v \alpha^{4-n}\beta^n}{m^{2}}.
\end{equation}

Let us assume that the $\beta_n$ parameters take this particular form, i.e., the only metric interactions arise from matter loops. The quartic equation~(\ref{eq:bg-quartic}) can then be solved only if $y=\beta/\alpha$ (or $\rho = -M_\eff^2\Lambda_v$), and the Friedmann equation becomes
\begin{equation}\label{eq:lambdahubble}
H^{2}=\frac{\left(\alpha^{2}+\beta^{2}\right)\rho}{M_{\eff}^{2}}+\frac{\left(\alpha^{2}+\beta^{2}\right)\Lambda_v}{3}.
\end{equation}
\Cref{eq:lambdabeta,eq:lambdahubble} reduce to the known expression for the $\Lambda$CDM solutions with a cosmological constant proportional to either $\beta_0$ or $\beta_4$ in the singly-coupled limit (where either $\beta\to0$ or $\alpha\to0$).

It is, of course, not surprising that matter loops lead to an accelerating expansion. However, the appearance of the vacuum energy in \emph{all} the bigravity interaction terms has novel implications. First, because the vacuum energy contributes to all the interaction terms, the mass scale $m$ is not protected against quantum corrections from matter loops \cite{deRham:2014naa}. Therefore, any values we obtain for these parameters from comparison of the theory to observations must be highly fine-tuned.\footnote{If the case described in \cref{sec:pm} is truly partially massless, this may be an exception, as there is a new gauge symmetry to protect against quantum corrections.} This is in contrast to singly-coupled bigravity, in which the only parameter that receives contributions from quantum loops is $\beta_0$ (if one couples matter to $g_\mn$), just as in general relativity where the cosmological constant is unstable in the presence of matter fields. In the singly-coupled theory, both the scale $m$ and the structure of the interaction potential are stable to quantum corrections \cite{deRham:2012ew,deRham:2013qqa}, a very useful fact which is lost once we couple matter to $g^\eff_\mn$.\footnote{Indeed, the fact that a small graviton mass is stable against quantum corrections is one of the main motivations for studying massive (bi)gravity, particularly as a candidate to explain the accelerating Universe.} This is not a problem in the double coupling studied in \cref{chap:dc-finsler}, as loops would only induce $g$- and $f$-metric cosmological constants, $\beta_0$ and $\beta_4$, although that theory is not ghost-free. Candidate expressions for $g_\mn^\eff$ where the matter sector would only contribute quantum corrections to $\beta_0$ and $\beta_4$ have been studied in \rcite{Heisenberg:2014rka}, although it is not yet known whether any of these are free of the Boulware-Deser ghost at low energies.

The other implication is that self-accelerating solutions are no longer straightforward to define in this theory. Typically, self-acceleration refers to cosmologies which accelerate at late times even when the vacuum energy is set to zero. Since in general relativity and singly-coupled bigravity, there is a single parameter which is degenerate with the vacuum energy ($\Lambda$ in the former and $\beta_0$ or $\beta_4$ in the latter), one can simply set its value to zero and look for other accelerating solutions. In the present doubly-coupled theory, however, \emph{all} interaction terms are degenerate with the vacuum energy: given an interaction potential, there is no way to unambiguously determine the value of $\Lambda_v$. In that respect, we cannot set some of the parameters to zero in order to restrict ourselves to accelerating solutions arising from nonvacuum, massive-gravity interaction terms (unless we set all the parameters to zero, which will give uninteresting solutions). Therefore, from a particle physics point of view this theory lacks, or at the very least cannot unambiguously define, self-accelerating solutions.

\subsection{Maximally-Symmetric Bigravity}
\label{sec:maxsymbigrav}

The parameter choice
\begin{equation}
\beta_0=\beta_4,\qquad \beta_1=\beta_3, \qquad \alpha=\beta,
\end{equation}
is special in the sense that the duality transformation (\ref{eq:bg-dualitytrans}) maps solutions to themselves.\footnote{Vacuum solutions for this model were previously studied in \rcite{Hassan:2014vja}.} Thus this theory is \emph{maximally symmetric} between the two metrics: they appear in the theory in completely equal ways. In this case, the quartic equation (\ref{eq:bg-quartic}) becomes
\begin{equation}
\left(y^{2}-1\right)\left[\beta_{1}\left(y^{2}+1\right)+3\beta_{2}y-\beta_{0}y+\frac{\alpha^{4}\rho}{m^{2}M_{\eff}^{2}}\left(1+y\right)^{2}\right]=0.
\end{equation}
As expected, there is an exact $\Lambda$CDM solution given by $y=1$. Indeed, the two metrics are completely equal, $g_\mn = f_\mn$, because the Bianchi constraint imposes $N_f/N_g=da_f/da_g=1$. The second-order polynomial for $y$ in brackets gives two solutions which are inverses of one another. This is not surprising, since when $g_\mn\leftrightarrow f_\mn$ we have $y\to y^{-1}$.

\section{Summary of Results}
\label{sec:bg-summary}

In this chapter we have presented the main features of the background expansion for massive bigravity with matter ``doubly coupled'' to both metrics through an effective metric, given by
\begin{equation}
g^\eff_{\mu\nu} = \alpha^2 g_{\mu\nu} + 2\alpha\beta g_{\mu\alpha}\mathbb X^{\alpha}{}_{\nu}+\beta^2 f_{\mu\nu},\qquad \mathbb X^{\mu}{}_{\nu} = (\sqrt{g^{-1}f})^\mu{}_\nu.
\end{equation}
This coupling was introduced in \rcite{deRham:2014naa,Noller:2014sta}, and has been further discussed in \rcite{deRham:2014fha,Hassan:2014gta,Noller:2014ioa,Gumrukcuoglu:2014xba}. This matter coupling has several advantages: it retains the metric-interchange symmetry in the presence of matter, leads to sensible cosmological solutions, and has a straightforward physical interpretation.

The expansion history is described by a Friedmann equation for the effective metric (\ref{eq:Heff}) and a quartic equation (\ref{eq:bg-quartic}) which algebraically describes the evolution of $y=a_f/a_g$, the ratio of the $f$- and $g$-metric scale factors. One can always choose the parameters of the theory such that the background expansion is exactly that of $\Lambda$CDM; any parameter choice which leads to $y=\beta/\alpha$ in \cref{eq:bg-quartic} will have this behaviour. For more general parameter values, the background expansion will deviate from $\Lambda$CDM but may still be consistent with observational data. To explore this, we confronted the models with only $\beta_0$, $\beta_1$, or $\beta_2$ nonzero with observational data. The other single-parameter models---with $\beta_3$ or $\beta_4$ nonzero---are then automatically included in this analysis due to the duality between solutions under $g_\mn\leftrightarrow f_\mn$, $\beta_n \to \beta_{4-n}$, and $\alpha\leftrightarrow\beta$, as described in \cref{app:sym}.

A novel feature of the effective coupling studied here is that $g_\mn$ and $f_\mn$ can be conformally related to each other at the background level in the presence of matter. In the singly-coupled case, this is only possible in vacuum, where the solutions are de Sitter. A special example of this is the parameter choice leading to a candidate partially-massless theory. This potentially has a novel gauge symmetry which would eliminate the problematic fifth force and protect a small vacuum energy against quantum corrections. In this case the background is identical to $\Lambda$CDM in the presence of matter. This suggests that doubly-coupled bigravity is a promising candidate for a theory of partially-massless gravity.

This matter coupling has a problematic feature, namely that loop corrections for any matter coupled to $g^\eff_\mn$ will generate all five dRGT interaction terms. Therefore the structure of the potential and the mass scale $m$ lose their stability against quantum corrections, which had been one of the most impressive features of the singly-coupled theory. We have discussed an important consequence of this: while many solutions to the theory accelerate at late times, it is no longer possible to unambiguously identify solutions that \emph{self-accelerate}, as the effective cosmological constant at late times can always be identified at least in part with a vacuum energy contribution.

We end with a brief comment concerning our expectations for perturbations around these cosmological solutions. We have shown in \cref{chap:bigravity-stability} that the singly-coupled models are often unstable for small $y$. One might hope that these doubly-coupled models will have better stability properties: $y$ is always nonzero and can be made to have a large minimum value by tuning $\beta/\alpha$. Moreover, we found in \cref{chap:bigravity-stability} that the $\beta_2$-only model did have stable perturbations in the singly-coupled case, but that model is not viable in the background. As we have shown, this model is in excellent agreement with background data if $\beta/\alpha$ is not too small, so it may provide another avenue for stable cosmological solutions in massive bigravity.

\begin{savequote}[30pc]
If the Lord Almighty had consulted me before embarking upon his creation, I should have recommended something simpler.
\qauthor{Alfonso X of Castile}
\end{savequote}

\chapter{Cosmological Implications of Doubly-Coupled Massive Gravity}
\label{chap:dc-drgt}
\hrule
\vspace*{1cm}

In \cref{sec:drgt-no-go} we described a no-go theorem for cosmological solutions in dRGT massive gravity, i.e., in the theory where the only gravitational degree of freedom is a massive graviton. If the reference metric is taken to be that of Minkowski space, then dynamical flat and closed FLRW solutions do not exist; the Bianchi constraint (\ref{eq:bianchiconstraint-mg}) restricts the scale factor to be constant. This can be avoided by either choosing open solutions or changing the reference metric, but the resultant solutions are unstable. Therefore, the search for a viable cosmology with a massive graviton has necessarily involved extending dRGT by adding extra degrees of freedom (as in the bimetric theory which we have studied in \cref{chap:bigravity-stability,chap:bigravity-subhorizon,chap:dc-finsler,chap:dc-background}) or by breaking the assumptions of homogeneity and isotropy, either in the metric or in the St\"uckelberg sector.

The double coupling discussed in \cref{chap:dc-background} has been shown to avoid both of these no-go theorems, opening up the intriguing possibility of obtaining sensible cosmological solutions with only a single massive graviton \cite{deRham:2014naa,Gumrukcuoglu:2014xba}. In this scenario, matter is coupled to an effective or Jordan-frame metric,
\begin{equation}
g_\mn^\eff \equiv \alpha^2g_\mn + 2\alpha\beta g_{\mu\alpha}\mathbb X^\alpha{}_\nu + \beta^2\eta_\mn, \label{eq:geffdef}
\end{equation}
where $g_\mn$ is the dynamical metric, $\eta_\mn$ is the Minkowski reference metric, and $\mathbb X^\mu{}_\nu\equiv(\sqrt{g^{-1}\eta})^\mu{}_\nu$. The properties of this effective metric were discussed in some detail in \cref{sec:dcbg-action}. However, we remind the reader that the theory with this matter coupling is believed to be ghost-free at least within the effective theory's r\'egime of validity and that the Boulware-Deser ghost is absent about FLRW backgrounds \cite{deRham:2014naa,deRham:2014fha,Hassan:2014gta}.

In this chapter we explore the basic properties of these newly-allowed massive gravity cosmologies. Unusually, the proof in \rcite{deRham:2014naa} that the no-go theorem is avoided turns out to rely crucially on coupling a fundamental field (in this case, a scalar field) to the effective metric. In a standard late-Universe setup where matter is described by a perfect fluid with a constant equation of state (or even more generally when $w$ only depends on the scale factor), this result does not hold, and FLRW solutions are constrained to be nondynamical, just as in standard dRGT. More generally, the pressure of at least one component in the Universe must depend on something besides the scale factor---such as the lapse or the time derivative of the scale factor---for massive gravity cosmologies to be consistent. This is why fields, which have kinetic terms where the lapse appears naturally, are required in order to obtain sensible cosmological solutions. Consequently the standard techniques of late-time cosmology cannot be applied to this theory.

While we do not aim to rule out these models, the inability to obtain cosmological solutions with just, e.g., dust or radiation is an unusual feature which makes it difficult to derive precise predictions for cosmology, as the nature of the ``extra matter'' is not presently known. These solutions exhibit pathologies in the early- and late-time limits if all matter couples to the effective metric, and the scalar field physics would need to be highly contrived to avoid these issues. Moreover, the reliance on extra matter, such as a scalar field, which may well be gravitationally subdominant and high-energy implies a violation of the decoupling principle, in which the low-energy expansion of the Universe should not be overly sensitive to high-energy physics.

The rest of this chapter is organised as follows. In \cref{sec:cosmo-back-drgt} we derive and discuss the cosmological evolution equations in this theory. In \cref{sec:dyn-sol} we elucidate the conditions under which the no-go theorem is violated and dynamical cosmological solutions exist. We discuss in \cref{sec:einvsjord} some of the nonintuitive features of the Einstein-frame formulation of the theory, and how these are resolved in a Jordan-frame description. In \cref{sec:drgt-scalar} we study cosmologies containing only a scalar field, and generalise this to include a perfect fluid coupled to the effective metric in \cref{sec:scalandfluid}. In \cref{sec:mixedcouplings} we consider an alternative setup in which the scalar field couples to the effective metric while the perfect fluid couples to the dynamical metric. We conclude in \cref{sec:dc-drgt-summary}.

\section{Cosmological Backgrounds}
\label{sec:cosmo-back-drgt}

The Einstein equation with all matter fields coupled to $g^\eff_\mn$ was derived in \rcite{Schmidt-May:2014xla} (see also \cref{sec:dcbg-action}) and can be written in the form\footnote{Our convention is that indices on the Einstein tensor $G^\mn$ are raised with $g^\mn$.}
\begin{equation}
 (\mathbb X^{-1})^{(\mu}{}_\alpha G^{\nu)\alpha} + m^2\displaystyle\sum_{n=0}^{3}(-1)^n\beta_ng^{\alpha\beta}(\mathbb X^{-1})^{(\mu}{}_\alpha Y^{\nu)}_{(n)\beta} = \frac{\alpha}{\Mp^2}\det\left(\alpha+\beta \mathbb X\right)\left(\alpha (\mathbb X^{-1})^{(\mu}{}_\alpha T^{\nu)\alpha}+\beta T^{\mu\nu}\right), \label{eq:einstein-drgt}
\end{equation}
where the stress-energy tensor is defined as usual with respect to the effective metric,
\begin{equation}
T^\mn = \frac{2}{\sdgb}\frac{\delta\left[\sdgb\calL\left(g^\eff_\mn,\Phi\right)\right]}{\delta g^\eff_\mn},
\end{equation}
and the matrices $Y_{(n)}$ are defined in \cref{eq:ymatdef}. Let us assume a flat FLRW ansatz for $g_\mn$ of the form\footnote{Note the differences in notation between this chapter and \cref{chap:dc-background}, such as our use of $a$ for the scale factor of $g_\mn$ rather than of $g_\mn^\eff$.}
\begin{equation}
 g_\mn dx^\mu dx^\nu = -N^2(t)dt^2 + a^2(t)\delta_{ij}dx^idx^j, \label{eq:FRW}
\end{equation}
and choose unitary gauge for the St\"uckelberg fields, $\eta_\mn=\operatorname{diag}(-1,1,1,1)$, so the effective metric is given by
\begin{equation}
 g^\eff_\mn dx^\mu dx^\nu = -N_\eff^2(t)dt^2 + a_\eff^2(t)\delta_{ij}dx^idx^j,
\end{equation}
where the effective lapse and scale factor are related to $N$ and $a$ by
\begin{equation}
 N_\eff = \alpha N + \beta, \qquad a_\eff = \alpha a + \beta. \label{eq:Naeff}
\end{equation}

We will define the Hubble rates for $g_\mn$ and $g_\mn^\eff$ by
\begin{equation}
 H \equiv \frac{\dot a}{aN},\qquad H_\eff \equiv \frac{\dot a_\eff}{a_\eff N_\eff}. \label{eq:Hdef}
\end{equation}
Notice that, because of the inclusion of the lapses in these definitions, these quantities correspond to what would be the cosmic-time Hubble rates in general relativity, obtained by setting $N=1$ or $N_\eff=1$. While we need not include the lapse in the definition of $H$ when working with diffeomorphism-invariant theories like general relativity or massive bigravity, instead choosing to set $N$ to a convenient value and thereby pick a physically-meaningful time coordinate like cosmic time or conformal time, the lack of diffeomorphism invariance in massive gravity means that neither the lapse nor the time coordinate has any meaning on its own, but will only appear through the combination $Ndt$. The time component of \cref{eq:einstein-drgt} yields the Friedmann equation,
\begin{equation}
3H^2 = \frac{\alpha\rho}{\Mp^2}\frac{a_\eff^3}{a^3} + m^2\left(\beta_0 + \frac{3\beta_1}{a} + \frac{3\beta_2}{a^{2}} + \frac{\beta_3}{a^{3}}\right), \label{eq:drgt-fried}
\end{equation}
where $\rho \equiv -g^\eff_{00}T^{00}$ is the density of the matter source. The spatial component of \cref{eq:einstein-drgt} gives us the acceleration equation,
\begin{equation}\label{eq:acc}
3H^{2} + \frac{2\dot{H}}{N}+ \frac{\alpha p}{\Mp^{2}}\frac{N_{\eff}a_{\eff}^{2}}{Na^{2}} =m^{2}\left[\beta_{0}+\beta_{1}\left(\frac{1}{N}+\frac{2}{a}\right)+\beta_{2}\left(\frac{2}{aN}+\frac{1}{a^{2}}\right)+\frac{\beta_{3}}{Na^{2}}\right],
\end{equation}
where the pressure is defined by $p\equiv(1/3)g^\eff_{ij}T^{ij}$. Notice that the double coupling leads to a time-dependent coefficient multiplying the density and pressure terms in \cref{eq:drgt-fried,eq:acc}. The Friedmann equation for the effective Hubble rate, $H_\eff$, can be determined from \cref{eq:drgt-fried} by the relation
\begin{equation}
H_\eff = \alpha\frac{Na}{N_\eff a_\eff}H, \label{eq:Heffdef}
\end{equation}
which follows from \cref{eq:Naeff}.

Matter is covariantly conserved with respect to $g^\eff_\mn$,
\begin{equation}
\nabla^\eff_\mu T^\mn=0, 
\end{equation}
from which we can obtain the usual energy conservation equation written in terms of the effective scale factor,
\begin{equation}\label{eq:continuity}
\dot\rho + 3\frac{\dot a_\eff}{a_\eff}\left(\rho + p\right) = 0.
\end{equation}
As in general relativity, this holds independently for each species of matter as long as we assume that interactions between species are negligible. Finally, we can take the divergence of the Einstein equation (\ref{eq:einstein-drgt}) with respect to $g_\mn$ and specialise to the FLRW background to find, after imposing stress-energy conservation, the ``Bianchi constraint,''
\begin{equation}
m^2M_\mathrm{Pl}^2a^2P(a) \dot a = \alpha\beta a_\eff^2p \dot a, \label{eq:bianchi-drgt}
\end{equation}
where we have defined
\begin{equation}
P(a) \equiv \beta_1 + \frac{2\beta_2}{a} + \frac{\beta_3}{a^2}.
\end{equation}
This can equivalently be derived using \cref{eq:drgt-fried,eq:acc,eq:continuity}. The pressure, $p$, appearing in \cref{eq:bianchi-drgt} is the total pressure of the Universe, or, if different species couple to different metrics, the total pressure of all matter coupling to $g^\eff_\mn$.

Let us pause to count the number of equations and variables in this system. We have four free functions---the scale factor, the lapse, the density, and the pressure---and four equations---Friedmann, acceleration, conservation, and Bianchi constraint. Of the four equations, only three are independent, much like in general relativity. The remaining freedom is fixed by specifying an equation of state. The acceleration equation can usually be derived from the other three, but unlike in general relativity it is \emph{not} always redundant: if the Bianchi constraint yields $\dot a=0$, then the acceleration equation does give new information, and in fact is what would be used to determine the lapse \cite{D'Amico:2011jj}. This situation is similar to general relativistic cosmology, but with one new variable and one new equation: because we have broken diffeomorphism invariance, the lapse cannot be fixed by a coordinate transformation, and furthermore the divergence of the Einstein equations leads to a nontrivial constraint. This is in contrast to general relativity, where the same procedure results in an identity.\footnote{This is because the Bianchi identity and stress-energy conservation are related to the diffeomorphism invariance of the Einstein-Hilbert and matter actions, respectively, but we have now added a mass term which does \emph{not} obey this gauge symmetry, after fixing the St\"uckelberg fields.}

We emphasise that if all matter couples to the same $g^\eff_\mn$ then the expansion history inferred from observations is given by $a_\eff$ and $H_\eff$, for the simple reason that all observations are observations of matter (including light). In deriving any cosmological observables, the ``proper time,'' $d\tau \equiv N_\eff dt$, will play the same role as the cosmic time coordinate in general relativity. In particular, $\tau$ corresponds to the time measured by point-particle clocks, while the distance light travels is given by $dr = d\tau / a_\eff(\tau)$. Therefore in principle we need only know $H_\eff(a_\eff)$ in order to connect to standard background observables. The coordinate time, $t$, is just the coordinate in which the reference metric, $\eta_\mn$, has the standard Minkowski form, and has no other physical significance.

Since $g_\mn$ and $g^\eff_\mn$ play the exact same roles as the Einstein-frame and Jordan-frame metrics, respectively, in other modified gravity theories, we will use these terms freely.

\section{Do Dynamical Solutions Exist?}
\label{sec:dyn-sol}

In the original, singly-coupled formulation of massive gravity, $\beta=0$ and so the right-hand side of \cref{eq:bianchi-drgt} vanishes, with the result that $a$ is constrained to be constant. This is nothing other than the no-go theorem on flat FLRW solutions in massive gravity. A nondynamical cosmology is, of course, still a solution when $\alpha$ and $\beta$ are nonzero, in which case the values of $a$ and $N$ are determined from \cref{eq:drgt-fried,eq:acc}. The question is now under which circumstances the theory also allows for dynamical $a$. 

To begin with, let us assume that $p=w\rho$, where $w$ can depend on the effective scale factor but nothing else. Assuming that $\dot{a}\neq0$, \cref{eq:bianchi-drgt} becomes
\begin{equation}
m^2M_\mathrm{Pl}^2a^2P(a) = \alpha\beta w a_\eff^2\rho, \label{eq:bianchi-drgtpf}
\end{equation}
and $\rho$ is a function only of $a$ (or equivalently $a_\eff$). To see this, consider \cref{eq:continuity} in the form
\begin{equation}
 \frac{d\ln\rho}{d\ln a} + 3\left[1+w(a)\right] = 0.
\end{equation}
Integrating this will clearly yield $\rho = \rho(a)$. Unless the left-hand side of \cref{eq:bianchi-drgtpf} has the exact same functional form for $a$ as the right hand side (which is, e.g., the case when $w=-1/3$ and $\beta_2=\beta_3=0$), this equation is not consistent with a time-varying $a$. The theory does therefore not give viable cosmologies using the standard equation of state $p=w\rho$, where $w$ is constant or depends on the scale factor.

This conclusion is avoided if the pressure also depends on the lapse. In this case, \cref{eq:bianchi-drgt} becomes a constraint on the lapse, unlocking dynamical solutions.\footnote{Another possibility is that the pressure depends on $\dot a$. The dynamics would be determined by \cref{eq:bianchi-drgt}, while the lapse would be constrained by the Friedmann equation. It is unclear whether these would give rise to Friedmann-like evolution, and we do not discuss this case any further.} The most obvious way to obtain a lapse-dependent pressure is to source the Einstein equations with a fundamental field rather than an effective fluid. This was exploited by \rcite{deRham:2014naa} to find dynamical cosmologies with a scalar field coupled to $g^\eff_\mn$. We discuss this case in more detail below. Therefore, while physical dust-dominated solutions may exist, we must either include additional degrees of freedom or treat the dust in terms of fundamental fields. The standard methods of late-time cosmology cannot be applied to doubly-coupled massive gravity.

\section{Einstein Frame vs. Jordan Frame}
\label{sec:einvsjord}

Before examining the cosmological solutions when the pressure depends on the lapse, it behoves us to further clarify the somewhat unusual differences between this theory's Einstein and Jordan frames. It turns out that the Friedmann equation in the Einstein frame is completely independent of the matter content of the Universe (up to an integration constant which behaves like pressureless dust): $H(a)$ always has a predetermined form [see \cref{eq:FEgen}]. In the Einstein-frame description, matter components with nonzero pressure affect the cosmological dynamics through the lapse, $N$. Because the lapse is involved in the transformation from the Einstein frame, $H$, to the Jordan frame, $H_\eff$, cf. \cref{eq:Heffdef}, the Jordan-frame Friedmann equation (corresponding to the observable Hubble rate) does depend on matter.

We proceed to demonstrate this explicitly. Regardless of the functional form of $p$, and whether or not it depends on the lapse, for $\dot a\neq 0$ the pressure is constrained by \cref{eq:bianchi-drgt} to have an implicit dependence on $a$ given by
\begin{equation}
p(a) = \frac{m^2M_\mathrm{Pl}^2a^2P(a)}{\alpha\beta a_\eff^2}. \label{eq:pa}
\end{equation}
The continuity equation (\ref{eq:continuity}) can then be integrated to obtain
\begin{equation}
\rho(a) = \frac{C}{a_\eff^3} - \frac{3m^2M_\mathrm{Pl}^2}{\beta a_\eff^3}\left(\frac{\beta_1}{3} a^3+\beta_2 a^2 + \beta_3 a \right), \label{eq:rhoa}
\end{equation}
where $C$ is a constant of integration that includes any pressureless dust. Inserting this into \cref{eq:drgt-fried} we find a generic form for the Einstein-frame Friedmann equation,
\begin{equation}
3H^2=m^2 \left(c_0+3 c_1a^{-1}+ 3c_2a^{-2}+c_3a^{-3}\right),
\label{eq:FEgen}
\end{equation}
where we have defined the coefficients
\begin{align}
c_0 &\equiv  \beta _0 - \frac{\alpha}{\beta}  \beta_1, \nonumber \\
c_1 &\equiv \beta _1 - \frac{\alpha}{\beta}  \beta_2, \nonumber \\
c_2 &\equiv \beta _2 - \frac{\alpha}{\beta}  \beta_3, \nonumber \\
c_3 &\equiv \beta _3 + \frac{\alpha C}{m^2\Mp^2}. \label{eq:cidef}
\end{align}
Notice that the functional forms of $p(a)$, $\rho (a)$, and $H^2(a)$ are completely independent of the energy content of the Universe, except for an integration constant scaling like pressureless matter. It is interesting to note that in the vacuum energy case studied in \cref{sec:vac} with $\beta_n = (\alpha/\beta)\beta_{n+1}$, all of the $c_i$ coefficients apart from $c_3$ vanish. In other words, if the metric interactions took the form of a cosmological constant for $g_\mn^\eff$, then the Einstein-frame Friedmann equation would scale as $a^{-3}$.

\section{Massive Cosmologies with a Scalar Field}
\label{sec:drgt-scalar}

If we include matter whose pressure does not only depend on the scale factor, $a_\eff$, then the Bianchi constraint (\ref{eq:bianchi-drgt}) may not rule out dynamical cosmological solutions. For a pressure that also depends on the lapse, \cref{eq:bianchi-drgt,eq:FEgen} determine $H$ and $N$, which in turn can be used to derive the Jordan-frame Friedmann equation. Because the lapse enters into the frame transformation (\ref{eq:Heffdef}), the Jordan frame can be sensitive to matter even though, as discussed above, the Einstein frame is not. The lapse thus plays an important and novel role in massive gravity compared to general relativity.

As discussed above, lapse-dependent pressures are not difficult to obtain: they enter whenever considering a fundamental field with a kinetic term. Consider a universe dominated by a scalar field, $\chi$, with the stress-energy tensor
\begin{equation}
 T^\mn = \nabla_\eff^\mu\chi\nabla_\eff^\nu\chi - \left(\frac 1 2 \nabla_\alpha\chi\nabla^\alpha_\eff\chi + V(\chi)\right)g_\eff^\mn,
\end{equation}
where $\nabla_\eff^\mu \equiv g_\eff^\mn \nabla^\eff_\nu$ and $V(\chi)$ is the potential for the scalar field. The density and pressure associated to $\chi$ are
\begin{equation}
\rho_\chi = \frac{\dot\chi^2}{2N_\eff^2}+V(\chi), \qquad p_\chi = \frac{\dot\chi^2}{2N_\eff^2}-V(\chi). \label{eq:scalarrhop}
\end{equation}
The constraint (\ref{eq:bianchi-drgt}) now has a new ingredient: the lapse, $N_\eff$, which appears through the scalar field pressure.\footnote{The $\alpha_2$ theory studied in \rcite{deRham:2014naa} can be obtained by setting $\beta_0=-3$, $\beta_1=3/2$, $\beta_2=-1/2$, and $\beta_3=0$ \cite{Hassan:2011vm}. With this parameter choice, the Bianchi constraint (\ref{eq:bianchi-drgt}) reproduces eq. (5.8) of Ref.~\cite{deRham:2014naa}. \label{foot:alpha2}}

One can then use the Bianchi identity to solve for the lapse and substitute it into the Friedmann equation to obtain an equation for the cosmological dynamics that does not involve the lapse \cite{deRham:2014naa}. A simple way to substitute out the lapse is to use the relation, following straightforwardly from \cref{eq:bianchi-drgt},
\begin{equation}
\frac{\dot\chi^2}{2N_\eff^2} = V(\chi) + \frac{m^2M_\mathrm{Pl}^2a^2P(a)}{\alpha\beta a_\eff^2}, \label{eq:K}
\end{equation}
as the lapse only appears in the Einstein-frame Friedmann equation through $\dot\chi^2/2N_\eff^2$. This explains the result, first noticed in \rcite{deRham:2014naa}, that after solving for the lapse, the Friedmann equation loses its dependence on the kinetic term. Note however that we can also use \cref{eq:K} to solve for the potential, $V(\chi)$, and write the Einstein-frame Friedmann equation in a form that does not involve the potential. Of course, if we were to additionally use the continuity equation as discussed above, the Einstein-frame Friedmann equation would take the form of \cref{eq:FEgen} which contains neither the kinetic nor the potential term. 

Using \cref{eq:pa,eq:rhoa} we can find expressions for the kinetic and potential energies purely in terms of $a$,
\begin{align}
K(a) &= \frac{m^2M_\mathrm{Pl}^2a^3}{2\alpha a_\eff^3}\left(c_1a^{-1} + 2c_2a^{-2}+c_3a^{-3}\right), \label{eq:Ka}\\
V(a) &= -\frac{m^2M_\mathrm{Pl}^2a^3}{2\alpha a_\eff^3}\left(2d_0+d_1a^{-1}+ 2d_2a^{-2}+d_3a^{-3}\right), \label{eq:Va}
\end{align}
where $K \equiv \dot\chi^2/2N_\eff^2$, the $c_i$ are defined in \cref{eq:cidef}, and we have further defined
\begin{align}
d_0 &\equiv  \frac{\alpha}{\beta}  \beta_1, \nonumber \\
d_1 &\equiv \beta _1 + 5\frac{\alpha}{\beta}  \beta_2, \nonumber \\
d_2 &\equiv \beta _2 + 2\frac{\alpha}{\beta}  \beta_3, \nonumber \\
d_3 &\equiv \beta _3 - \frac{\alpha C}{m^2\Mp^2}.
\end{align}
Note that the terms proportional to $C$ include any possible pressureless matter component coupled to $g_\mn^\eff$. This integration constant will always appear when solving the continuity equation (\ref{eq:continuity}). The Friedmann equation is given by the generic \cref{eq:FEgen}. That is, we are left with the peculiar situation that the pressure, energy density, and Einstein-frame Friedmann equation are completely insensitive to the form of the scalar field potential. As discussed above, this lack of dependence on the details of the scalar field physics is illusory; the lapse does depend on $V(\chi)$ and $\dot\chi$, cf. \cref{eq:K}, and in turn the physical or Jordan-frame expansion history depends on the lapse, cf. \cref{eq:Heffdef}.

Let us briefly remark on a pair of important exceptions. The no-go theorem forbidding dynamical $a$ still applies when there is a scalar field present if either the potential does not depend on the lapse (such as a flat potential) or the field is not rolling. Let us rewrite \cref{eq:continuity} (which is equivalent to the Klein-Gordon equation) as
\begin{equation}
\frac{d}{dt}\left(\frac{\dot\chi^2}{2N_\eff^2} + V(\chi)\right) + 3\frac{\dot a_\eff}{a_\eff}\frac{\dot\chi^2}{N_\eff^2} = 0.
\end{equation}
If $V(\chi)$ is independent of $N_\eff$ then $\dot\chi^2/N_\eff^2$ cannot depend on $N_\eff$ and, by extension, neither can $p=\dot\chi^2/2N_\eff^2-V(\chi)$. In the specific case of $V(\chi)=\mathrm{const.}$ this is clearly true, and we find $\dot\chi^2/N_\eff^2 \propto a_\eff^{-6}$, so $p=p(a)$. Similarly, if the field is not rolling, $\dot\chi=0$, then it is clear from \cref{eq:scalarrhop} that $p$ loses its dependence on the lapse.

To conclude this section, when a scalar field is coupled to the effective metric, we avoid the no-go theorem and it is possible to have dynamical $a$, unless the potential does not depend on the lapse or the field is not rolling. This result agrees with and slightly generalises that presented in \rcite{deRham:2014naa,Gumrukcuoglu:2014xba}. In a realistic scenario, however, we will have not only a scalar field but also matter components present. We now turn to that scenario.

\section{Adding a Perfect Fluid}
\label{sec:scalandfluid}

We have seen that the no-go theorem on FLRW solutions in dRGT massive gravity continues to hold in the doubly-coupled theory if the only matter coupled to the effective metric is a perfect fluid whose energy density and pressure depend only on the scale factor. This complicates the question of computing dust-dominated or radiation-dominated solutions in massive gravity. One solution would be to treat the dust in terms of fundamental fields. Another would be to add an extra degree of freedom such as a scalar field. Its role is to introduce a lapse-dependent term into the Bianchi constraint (\ref{eq:bianchi-drgt}) and thereby avoid the no-go theorem.

It is this possibility which we study in this section. In \cref{sec:drgt-scalar} we examined the scalar-only case. Let us now include other matter components, such as dust or radiation, also coupled minimally to $g^\eff_\mn$. We assume that the density, $\rho_\mathrm{m}$, and pressure, $p_\mathrm{m}$, only depend on $a_\eff$.\footnote{As discussed above and in \rcite{deRham:2014naa}, in principle any dust or radiation is made of fundamental particles for which the stress-energy tensor does depend on the lapse. We introduce this effective-fluid description because it is the standard method of deriving cosmological solutions in nearly any gravitational theory and is thus an important tool for comparing to observations.} We can then write the total density and pressure as     
\begin{align}
\rho = K+V+\rho_\mathrm{m}, \nonumber\\
p = K-V+p_\mathrm{m},
\end{align}
so that
\begin{align}
K=\frac{\rho+p-(\rho_\mathrm{m}+p_\mathrm{m})}{2}, \nonumber \\
V=\frac{\rho-p-(\rho_\mathrm{m}-p_\mathrm{m})}{2}.
\label{eq:negK}
\end{align}
Note that \cref{eq:Ka,eq:Va} no longer hold, as they were derived without considering other matter, but \cref{eq:rhoa,eq:pa} are still valid and are crucial.

We would like to investigate the cosmological dynamics of this model. Rather than explicitly solving for the lapse and substituting it into the Friedmann equation for $H_\eff$, which leads to a very complicated result, we will take advantage of the known forms of $K(a_\eff)$ and $V(a_\eff)$, as well as the fact that $N_\eff$ only appears in $H_\eff$ and $K$ through the operator
\begin{equation}
\frac{d}{d\tau} = \frac{1}{N_\eff}\frac{d}{dt}.
\end{equation}
The physical Hubble rate is given by
\begin{equation}
H_\eff \equiv \frac{\dot a_\eff}{a_\eff N_\eff}=\frac{\alpha\dot a}{a_\eff N_\eff}.
\label{eq:heff2}
\end{equation}
Using the chain rule, we can write
\begin{equation}
\dot a=\frac{da}{dt}=\frac{da}{dV}\frac{dV}{d\chi}\frac{d\chi}{dt}=\frac{V'\dot\chi}{(dV/da)}, 
\end{equation}
where a prime denotes a derivative with respect to $\chi$. We also know that $\dot\chi=N_\eff\sqrt{2K}$, giving
\begin{equation}
\dot a=\frac{V'N_\eff\sqrt{2K}}{(dV/da)}, 
\end{equation}
which we can plug into \cref{eq:heff2} to obtain
\begin{equation}\label{eq:Heffscalar}
H^2_\eff =\frac{(V')^2 2K}{a^2_\eff (dV/da_\eff)^2}.
\end{equation}

This is the Friedmann equation for any universe with a scalar field rolling along a nonconstant potential. Every term in \cref{eq:Heffscalar} can be written purely in terms of $a_\eff$, allowing the full cosmological dynamics to be solved in principle. $K$ and $dV/da_\eff$ are given in terms of $a_\eff$ by \cref{eq:negK} [using \cref{eq:pa,eq:rhoa}]. $V'$ as a function of $a_\eff$ can be determined from the same equations once the form of $V(\chi)$ is specified. Note that while the lapse is not physically observable, its evolution in terms of $a$ can then be fixed by using \cref{eq:Heffdef} to find
\begin{equation}
\frac{N^2}{N_\eff^2}=2K\left(\frac{V'}{\alpha aH(dV/da_\eff)}\right)^2,
\end{equation}
where $H(a)$ is given by \cref{eq:FEgen}.

Assuming that the matter has a constant equation of state, we can use the known forms of $K(a)$ and $V(a)$ to find a relatively simple expression for the Friedmann equation up to $V'$,
\begin{equation}\label{eq:Hefffinal}
\left(\frac{H_\eff}{V'}\right)^2 = \frac{4\alpha^3\beta a_\eff^3\left(\mc_0 + \mc_1a_\eff + \mc_2 a_\eff^2 + \mc_\rho a_\eff^3\right)}{\left[3\mc_0 + 4\mc_1a_\eff + 5\mc_2 a_\eff^2 + 3(1-w)\mc_\rho a_\eff^3\right]^2},
\end{equation}
where for brevity we have defined
\begin{align}
\mc_0 &\equiv  \beta  \left[\alpha ^3 C+\beta ^2\beta _1+m^2\Mp^2 \left(3 \alpha  \left(\alpha  \beta _3-\beta  \beta_2\right) \right)\right], \nonumber \\
\mc_1 &\equiv -2 m^2\Mp^2 \left[\alpha  \left(\alpha  \beta _3-2 \beta  \beta _2\right)+ \beta^2\beta _1\right],  \nonumber \\
\mc_2 &\equiv m^2\Mp^2 \left(\beta  \beta _1-\alpha  \beta _2\right), \nonumber \\
\mc_\rho &\equiv -\alpha ^3 \beta  (1+w) \rho_\mathrm{m}.\label{eq:Hefffinalcoeff}
\end{align}
Notice that the right-hand side is a function of $a$ only.

Let us examine the past and future asymptotics of these cosmologies, taking into account radiation ($w=1/3$) in the former and dust ($w=0$) in the latter. At late times, taking $a_\eff\to\infty$ in \cref{eq:negK}, we find
\begin{align}
\frac{\dot\chi^2}{2N_\eff^2} &\xrightarrow{a_\eff\to\infty} \frac{m^2 M_p^2\left(\beta  \beta _1 - \alpha  \beta _2\right) }{2 \alpha ^3 \beta  a_\eff}, \\
V(\chi) &\xrightarrow{a_\eff\to\infty} -\frac{\beta _1 m^2 \Mp^2}{\alpha ^3 \beta }.
\end{align}
We see that the scalar field slows to a halt: $V(\chi)$ approaches a constant, while $d\chi/d\tau$, where $d\tau = N_\eff dt$ is the proper time, approaches zero. Taking the late-time limit of the Friedmann equation (\ref{eq:Hefffinal}), we obtain
\begin{equation}\label{eq:HefffinalLateTimes}
\left(\frac{H_\eff}{V'}\right)^2 \xrightarrow{a_\eff\to\infty} \frac{4\alpha^3\beta}{25c_2}a_\eff.
\end{equation}
Because $\chi$ approaches a constant $\chi_c$ at late times, $V'=(dV/d\chi)|_{\chi=\chi_c}$ contributes a constant to the Friedmann equation. Therefore we find that $H_\eff$ generically blows up, which is potentially disastrous behaviour. This implies a violation of the null energy condition. Notice also that there is no guarantee that $V=-\beta _1 m^2 \Mp^2/\alpha ^3 \beta$ is within the range of $V(\chi)$, assuming the scalar field potential is not set by gravitational physics. This may lead to further pathologies, as the form of $V(a)$ would be inconsistent with large values of $a_\eff$. As we discuss below, if $V'$ goes to 0 then, depending on the speed at which it does so, $H_\eff$ may be better behaved.

At early times, demanding the existence of a sensible radiation era leads to further problems. Assuming radiation couples to $g^\eff_\mn$, then $\rho_\mathrm{m} \sim a_\eff^{-4}$ with $p_\mathrm{m}=\rho_\mathrm{m}/3$. We have, cf. \cref{eq:negK}, that $2K = \rho + p - (\rho_\mathrm{m} + p_\mathrm{m})$, but, cf. \cref{eq:rhoa}, $\rho$ and $p$ do not have any terms scaling as steeply as $a_\eff^{-4}$. Therefore, in the presence of radiation, $\rho_\chi$ and $p_\chi$ pick up a \emph{negative} term going as $a_\eff^{-4}$ to exactly cancel out $\rho_\mathrm{m}$ and $p_\mathrm{m}$, leading to $K<0$ at sufficiently early times. From \cref{eq:Heffscalar} we see that this would lead to a negative $H_\eff^2$, and hence to an imaginary Hubble rate. Equivalently, we can take the early-time limit of \cref{eq:Hefffinal} to show
\begin{equation}\label{eq:HefffinalEarlyTimes}
\left(\frac{H_\eff}{V'}\right)^2 \xrightarrow{a_\eff\to0} -\frac{3}{4\rho_0}a_\eff^4,
\end{equation}
so that again we see (for a real potential) $H_\eff$ becoming imaginary.

How could these conclusions be avoided? We can reproduce sensible behaviour, but only if the potential is extremely contrived. At early times, we would need to arrange the scalar's dynamics so that $V'\to\infty$ ``before'' (i.e., at a later $a_\eff$ then) $K$ crosses zero.\footnote{The other obvious possibility, having $dV/da_\eff$ reach 0 before $K$ does, is impossible given the forms of $K(a)$ and $V(a)$.} We would then reach the initial singularity, $H_\eff\rightarrow\infty$, before the kinetic term turns negative.\footnote{This proposal has an interesting unexpected advantage: the Universe would begin at finite $a_\eff$, so a UV completion of gravity might not be needed to describe the Big Bang in the matter sector.} Moreover, we would need to tune the parameters of the theory so that $K=0$ happens at extremely early times, specifically before radiation domination. At intermediate times, $V'$ would need to scale in a particular way to [through \cref{eq:Hefffinal}] reproduce $H_\eff^2\sim a_\eff^{-4}$ and $H_\eff^2\sim a_\eff^{-3}$ during the radiation- and matter-dominated eras, respectively. Finally, in order to have $H_\eff\to\mathrm{const.}$ at late times, we see from \cref{eq:HefffinalLateTimes} that we would require $V'$ to decay as $a_\eff^{-1/2}$. We can construct such a potential going backwards by setting $H_\eff=H_{\Lambda\mathrm{CDM}}$ in \cref{eq:Hefffinal}, but there is no reason to expect such an artificial structure to arise from any fundamental theory. Even then we may still get pathological behaviour: $N_\eff$ diverges if at some point $H_\eff a_\eff = Ha$, cf. \cref{eq:Heffdef}.

\section{Mixed Matter Couplings}
\label{sec:mixedcouplings}

Before concluding, we briefly discuss a slightly different formulation which avoids some, but not all, of our conclusions. If we consider a scalar field and a perfect fluid, the avoidance of the no-go theorem on FLRW solutions only requires that the scalar field couple to $g^\eff_\mn$. In principle, all other matter could still couple to $g_\mn$. In fact, this is the theory that was studied in \rcite{deRham:2014naa}. This theory violates the equivalence principle in the scalar sector, but is not \emph{a priori} excluded, and will turn out to have slightly better cosmological behaviour. Moreover, there is a compelling theoretical reason to consider such couplings: matter loops would only generate a $g$-metric cosmological constant and would not destabilise the rest of the potential. However, the scalar field's energy would still contribute to the cosmological constant for $g^\eff_\mn$ and hence to all of the interaction terms unless, for example, this was forbidden by some symmetry. A massless scalar would be better behaved in this sense, but as we have shown above, such a scalar field will not avoid the no-go theorem because after integrating the Klein-Gordon equation, the pressure loses its dependence on the lapse.

Because the perfect fluid couples to $g_\mn$ and we derived the Bianchi constraint (\ref{eq:bianchi-drgt}) by taking the $g$-metric divergence of the Einstein equation, the constraint will now only contain $p_\chi$ rather than the total pressure, i.e.,
\begin{equation}
m^2M_\mathrm{Pl}^2a^2P(a) \dot a = \alpha\beta a_\eff^2p_\chi \dot a.
\end{equation}
This is the same constraint as in the scalar-only case discussed in \cref{sec:drgt-scalar}, so the scalar's kinetic and potential energies have the same forms, $K(a)$ and $V(a)$, as in \cref{eq:Ka,eq:Va}. The physical Hubble rate is now $H$, which after solving for the lapse is determined by the equation\footnote{Using the transformations to the $\alpha_2$ theory in \cref{foot:alpha2}, we recover eq. (5.9) of Ref.~\cite{deRham:2014naa}.}
\begin{equation}
3H^2=\frac{\rho_\mathrm{m}}{\Mp^2}+m^2 \left(c_0+3 c_1a^{-1}+ 3c_2a^{-2}+c_3a^{-3}\right),
\label{eq:drgt-friedconstrained}
\end{equation}
where the $c_i$ coefficients are defined in \cref{eq:cidef}. Because the scalar field does not have to respond to matter to maintain a particular form of $\rho(a)$ and $p(a)$, we no longer have pathological behaviour in the early Universe, where there will be a standard $a^{-4}$ evolution. Moreover, as was pointed out in \rcite{Gumrukcuoglu:2014xba}, there is late-time acceleration: as $\rho_\mathrm{m}\to0$, $3H^2\to m^2(\beta_0-(\alpha/\beta)\beta_1)$, which, if positive, leads to an accelerating expansion.

However, these are not always \emph{self}-accelerating solutions. We will demand two conditions for self-acceleration: that the late-time acceleration not be driven by a cosmological constant, and that it not be driven by $V(\chi)$, both of which can easily be accomplished without modifying gravity. In other words, we would like the effective cosmological constant at late times to arise predominantly from the massive graviton.

Let us start with the first criterion, the absence of a cosmological constant. Recall from \cref{sec:drgt} that we can write the dRGT interaction potential in terms of elementary symmetric polynomials of the eigenvalues of either $\mathbb X \equiv \sqrt{g^{-1}f}$ or $\mathbb K \equiv \mathbb I - \mathbb X$, with the strengths of the interaction terms denoted by the by-now familiar $\beta_n$ in the first case and by $\alpha_n$ in the latter. What is notable is that $\alpha_0\neq\beta_0$: the cosmological constant is not the same between these two parametrisations. Terms proportional to $\sdg$ arise from the other interaction terms when transforming from one basis to the other. In bigravity there is a genuine ambiguity as to how one defines the cosmological constant, and throughout this thesis, because we are concerned with cosmological solutions, we have chosen to identify the cosmological constant with the constant term appearing in the Friedmann equation for the physical metric. In massive gravity with a Minkowski reference metric, however, the presence of a Poincar\'e-invariant preferred metric allows for a more concrete definition of the cosmological constant.\footnote{We thank Claudia de Rham for helpful discussions on this point.} Consider expanding the metric as
\begin{equation}
g_\mn = \eta_\mn + 2h_\mn + h_{\mu\alpha}h_{\nu\beta}\eta^\mn.
\end{equation}
This expansion is useful because the metric is quadratic in $h_\mn$ but is fully nonlinear, i.e., we have not assumed that $h_\mn$ is small \cite{deRham:2014zqa}. In this language, the cosmological constant term, proportional to $\sdg$, can be eliminated by setting $\alpha_0 = \alpha_1 = 0$. Making this choice of parameter, and recalling, cf. \cref{eq:alphabeta}, that $\alpha_n$ and $\beta_n$ are related by \cite{Hassan:2011vm}
\begin{equation}
 \beta_n = (4-n)!\displaystyle\sum_{i=n}^{4}\frac{(-1)^{i+n}}{(4-i)!(i-n)!}\alpha_i,
\end{equation}
we find the effective cosmological constant can be expressed in terms of $\alpha_{2,3,4}$ by
\begin{align}
\Lambda_\eff &= \frac{m^2}{3}\left(\beta_0-\frac\alpha\beta\beta_1\right) \nonumber\\
&= \frac{m^2}{3}\left[3 \alpha _2 \left(2+\frac\alpha\beta\right)-\alpha _3 \left(4+ 3\frac\alpha\beta\right)+\alpha _4 \left(1+\frac\alpha\beta\right)\right]. \label{eq:drgt-effcc}
\end{align}

Part of this constant comes from the fixed behaviour of the scalar field potential.\footnote{Notice from \cref{eq:Ka} that, as in the case with a perfect fluid, the scalar field slows down to a halt at late times, so there is no contribution from the kinetic energy.} This piece is not difficult to single out: it consists exactly of the terms in \cref{eq:drgt-effcc} proportional to $\alpha/\beta$. Taking the late-time limit of \cref{eq:Va}, we can see that $V(\chi)$ asymptotes to
\begin{equation}
V(\chi) \xrightarrow{a_\eff\to\infty} -\frac{m^2\Mp^2\beta_1}{\alpha^3\beta}.
\end{equation}
Now consider the Friedmann equation in the form (\ref{eq:drgt-fried}) with, at late times, $\rho \to V$. We can define a cosmological-constant-like piece solely due to the late-time behaviour of $V$ given by
\begin{equation}
\Lambda_\chi \equiv \frac{\alpha V}{3\Mp^2}\left(\frac{a_\eff}{a}\right)^3 \xrightarrow{a_\eff\to\infty} \frac{m^2}{3}\frac\alpha\beta\left(3\alpha_2 - 3\alpha_3 + \alpha_4\right).
\end{equation}
Then \cref{eq:drgt-effcc} can simply be written in the form
\begin{equation}
\Lambda_\eff = \frac{m^2}{3}\left(6\alpha_2 - 4\alpha_3 + \alpha_4\right) + \Lambda_\chi = \frac{m^2}{3}\beta_0 + \Lambda_\chi,
\end{equation}
where in the last equality we mention that the residual term is nothing other than $m^2\beta_0/3$, which is simply a consistency check.

The modifications to gravity induced by the graviton mass therefore lead to a constant contribution to the Friedmann equations at late times, encapsulated in $m^2\beta_0/3$ (with $\alpha_0=\alpha_1=0$). In a truly self-accelerating universe, this term should dominate $\Lambda_\chi$. If it did not, the acceleration would be partly caused by the scalar field, and one could get the same end result in a much simpler way with, e.g., quintessence. For generic values of $\alpha_n$ and for $\beta\sim\mathcal{O}(1)$, both of these contributions are of a similar size and will usually have the same sign. To ensure self-accelerating solutions, one could, for example, tune the coefficients so that $3\alpha_2 - 3\alpha_3 + \alpha_4 = 0$ (the scalar field contributes nothing to $\Lambda_\eff$) or $3\alpha_2 - 3\alpha_3 + \alpha_4 < 0$ (the scalar field contributes negatively to $\Lambda_\eff$), or take $\beta\ll1$ (the scalar field contributes negligibly to $\Lambda_\eff$).

\section{Summary of Results}
\label{sec:dc-drgt-summary}

One can extend dRGT massive gravity by allowing matter to couple to an effective metric constructed out of both the dynamical and the reference metrics. The no-go theorem ruling out flat homogeneous and isotropic cosmologies in massive gravity \cite{D'Amico:2011jj} can be overcome when a scalar field is ``doubly coupled'' in such a way \cite{deRham:2014naa,Gumrukcuoglu:2014xba}. We have shown that this result is, unusually, dependent on the use of a fundamental field, such as a scalar field in the aforementioned references, as the no-go theorem is only avoided when the pressure of the matter coupled to $g^\eff_\mn$ depends on the cosmic lapse function. This lapse dependence is not present for the types of matter usually considered in late-time cosmological setups, such as radiation ($p\sim a_\eff^{-4}$) and dust ($p=0$), and therefore a universe containing \emph{only} such matter will still run afoul of the no-go theorem. While this may not be a strong physical criterion---cosmological matter is still built out of fundamental fields---it presents a sharp practical problem in relating the theory to cosmological observations. Furthermore, if one uses a scalar field to avoid the no-go theorem, it cannot live on a flat potential and must be rolling. The latter consideration would seem to rule out the use of the Higgs field to unlock massive cosmologies, as we expect it to reside in its minimum cosmologically.

Overall, in principle one can obtain observationally-sensible cosmologies in doubly-coupled massive gravity, but either a new degree of freedom must be included, such as a scalar field or some other matter source with a nontrivial pressure, or we must treat cosmological matter in terms of their constituent fields. Thus we cannot apply the standard techniques of late-time cosmology to this theory.

We have further shown that if dust and radiation are doubly coupled as well---which is necessary if we demand the scalar obey the equivalence principle---then the cosmologies generically are unable to reproduce a viable radiation-dominated era, and in the far future the Hubble rate diverges, rather than settling to a constant and producing a late-time accelerated expansion. These pathologies can only be avoided if the scalar field potential is highly contrived with tuned theory parameters, or dust and radiation do not doubly couple. In the latter case, there is generically late-time acceleration, but for much of the parameter space, this is in large part driven by the potential of the scalar field. In those cases the modification to general relativity may not be especially well motivated by cosmological concerns. Otherwise, the parameters of the theory need to be tuned to ensure that the theory truly self-accelerates.

It seems that dRGT massive gravity only has viable cosmological solutions---i.e., that evade the no-go theorems on existence \cite{D'Amico:2011jj} and stability \cite{DeFelice:2013awa}---if one either includes a scalar field or some other ``exotic'' matter with a lapse-dependent pressure (or possibly a pressure depending on $\dot a$) and couples it to the effective metric proposed in \rcite{deRham:2014naa} or goes beyond the perfect-fluid description of matter. Even if one includes a new scalar degree of freedom, significant pathologies arise if normal matter couples to the same effective metric. In all setups, the need for descriptions beyond a simple perfect fluid makes this theory unappealing from an observational standpoint.

We end with three small caveats. Notice that we have assumed that in unitary gauge for the St\"uckelberg fields, i.e., choosing coordinates such that $\eta_\mn=\operatorname{diag}(-1,1,1,1)$, the metric has the usual FLRW form (\ref{eq:FRW}). However, that form is arrived at by taking coordinate transformations of a more general homogeneous and isotropic metric, so that assumption may be overly restrictive.\footnote{We thank Fawad Hassan for pointing this out to us.} Equivalently, one could consider a more general, inhomogeneous and/or anisotropic, gauge for the St\"uckelberg fields.

We also note that if this theory does possess a ghost, even with a mass above the strong coupling scale, solutions to the nonlinear equations of motion could contain the ghost mode and therefore not be physical.\footnote{We thank Angnis Schmidt-May for discussions on this point.} However, a Hamiltonian analysis showed that the ghost does not appear around FLRW backgrounds \cite{deRham:2014naa}, suggesting that we have studied the correct cosmological solutions to any underlying ghost-free theory.

Finally, as discussed in \cref{chap:dc-background}, if one simply gives dynamics to the reference metric, we end up with a theory of doubly-coupled bigravity which treats the two metrics on completely equal footing and has been shown to produce observationally viable cosmologies.

\cleardoublepage

\thispagestyle{empty}

		\vspace*{\fill}
		
		\begin{flushright}
		{\Huge{ \bf Part II}\\
		Lorentz Violation}
		\vspace{2cm}
		
\begin{quote}
\textit{Einstein was a giant. His head was in the clouds, but his feet were on the ground.\\But those of us who are not that tall have to choose!}
\qauthor{Richard Feynman}
\end{quote}


		\end{flushright}
\vspace*{\fill}

\begin{savequote}[30pc]
The only thing that really worried me was the ether. There is nothing in the world more helpless and irresponsible and depraved than a man in the depths of an ether binge. And I knew we'd get into that rotten stuff pretty soon. Probably at the next gas station.
\qauthor{Hunter S. Thompson, \textit{Fear and Loathing in Las Vegas}}
\end{savequote}

\chapter{Lorentz Violation During Inflation}
\label{chap:aether}
\hrule
\vspace*{1cm}

For this final chapter, we move to the early Universe to ask what constraints we can put on the violation of Lorentz invariance during inflation. As discussed in \cref{sec:mg-aether}, we can use \emph{Einstein-aether theory} (\ae-theory) \cite{Jacobson:2000xp,Jacobson:2008aj} to model Lorentz violation in the boost sector, i.e., while maintaining rotational invariance on spatial hypersurfaces, at low energies. In \ae-theory, Lorentz invariance is spontaneously broken by the presence of a vector field nonminimally coupled to gravity. A Lagrange multiplier enforces the constraint that this vector, sometimes called the \emph{aether} and denoted here by $u^\mu$, be timelike and have fixed norm,
\begin{equation}
 u^\mu u_\mu = -m^2,
\end{equation}
where $m$ is a free parameter with mass dimension 1. This forces the aether to acquire a nonzero vacuum expectation value (VEV) at every point in spacetime, so at every point the aether picks out a timelike direction and hence defines a preferred reference frame. Note that \ae-theory is a vector-tensor theory of gravity and is, at the level of the theory, completely Lorentz invariant. It is the nontrivial constraint which ensures that Lorentz invariance, and specifically boost invariance, is always broken at the level of the solutions. \Ae-theory corresponds, for example, to the low-energy limit of Ho\v{r}ava-Lifschitz gravity, a well-studied candidate UV completion of general relativity which breaks the symmetry between time and space coordinates directly at the level of the action \cite{Horava:2009uw}.

In \cref{sec:aescalcoupling} we considered a generalisation of \ae-theory in which a scalar field, $\phi$, couples to the aether by allowing its potential, $V(\phi)$, to depend on the divergence or expansion of the aether, $\theta\equiv\nabla_{a}u^{a}$ \cite{Donnelly:2010cr,Barrow:2012qy,Sandin:2012gq}. This is of particular interest for cosmology because $\theta$ is related to the local Hubble expansion rate: the aether is forced by symmetry to align with the cosmic rest frame in a spatially homogeneous and isotropic background \cite{Carroll:2004ai,Jacobson:2008aj,Lim:2004js,Carruthers:2010ii}, and purely on geometric grounds we find $\theta=3mH$, with $H=\dot a/a$ is the cosmic-time Hubble parameter. The ability to use the expansion rate so freely in the field equations is a departure from general relativity and other purely metric theories: in such theories, $H$ is not a covariant scalar as it can only be defined in a coordinate-dependent way. Thus, this extension of pure \ae-theory opens up the interesting possibility of cosmological dynamics depending directly on the expansion rate in a way that is not allowed by general relativity or many modified gravity theories.

This coupling also allows the aether to directly affect cosmological dynamics at the level of the background. Recall from the discussion in \cref{sec:aecosmo} that this is not possible in ``pure'' \ae-theory, as the aether tracks the dominant matter source and hence can only rescale Newton's constant in the Friedmann equation, slowing down the expansion. By coupling the scalar field to $\theta$ in the way discussed above, one can obtain qualitative changes in the cosmological dynamics, cf. \cref{eq:friedmann,eq:friedmann2}. If we identify this scalar field with the inflaton, the coupling to the aether therefore modifies inflationary dynamics. In a simple case, it adds a driving force which can slow down or speed up inflation \cite{Donnelly:2010cr}. This theory with another simple form of the coupling is also closely related (up to the presence of transverse spin-1 perturbations) to $\Theta\mathrm{CDM}$, a dark energy theory in which the small cosmological constant is technically natural \cite{Blas:2011en,Audren:2013dwa}.

The type of coupling we have chosen---i.e., promoting $V(\phi)$ to $V(\theta,\phi)$---may seem restrictive, but it is in fact a reasonably general approach to coupling the aether to a scalar field. Any terms one can write down which do not fit in this framework would have mass dimension 5 or higher. Such terms would only be relevant at short distance scales and would not be power-counting renormalisable. Therefore, from an effective field theory approach, all the operators we wish to include are captured in $V(\theta,\phi)$, up to integration by parts. We refer the reader to \cref{sec:aescalcoupling} for a more in-depth discussion of the generality of this type of coupling. We will perform our analysis with the important assumptions that $\phi$ drives a period of slow-roll inflation and that its kinetic term is canonical, but otherwise leave its properties unrestricted. Hence we consider this theory to be a fairly general model of Lorentz violation in the inflaton sector.

Our aim is to explore the effects of such a coupling at the level of linear perturbations to a cosmological background, and in particular to find theoretical and observational constraints. For reasonable values of the coupling between the aether and the inflaton, these perturbations are unstable and can destroy the inflationary background. This places a constraint on the coupling which is several orders of magnitude stronger than the existing constraints. If the parameters of the theory are chosen to remove the instability, while satisfying existing constraints on the aether VEV, then the effects of the coupling on observables in the cosmic microwave background will be far below the sensitivity of modern experiments.

The remainder of this chapter is organised as follows. In \cref{sec:flatspace} we discuss the behaviour of linearised perturbations of the aether and the scalar around a (nondynamical) flat background, deriving a stability constraint (previously found by another method in \rcite{Donnelly:2010cr}) which provides a useful upper bound on the aether-scalar coupling. In \cref{sec:cosmperts}, we set up cosmological perturbations in this theory; the real-space perturbation equations can be found in \cref{app:realspace}. In \cref{sec:spin1perts} we examine the spin-1 cosmological perturbations of the aether and metric during a phase of quasi-de Sitter inflation. This demonstrates clearly the existence of a tachyonic instability, which we explore in some depth. In \cref{sec:spin0perts} we look at the spin-0 perturbations, finding the same instability and calculating the scalar power spectrum. Unusually, isocurvature modes do not appear to first order in a perturbative expansion around the aether norm. We give a worked example in \cref{sec:dj} which elucidates the arguments made for a general potential in the preceding sections, and conclude with a summary and discussion of our results in \cref{sec:summary}.

\section{Stability Constraint in Flat Space}
\label{sec:flatspace}

Before moving on to the main focus of this chapter, perturbations around a cosmological background, we briefly examine perturbation theory in flat space. Our goal is to derive a constraint on the coupling $V_{\theta\phi}$ by requiring that the aether and scalar perturbations be stable around a Minkowski background. This will set an upper limit relating the coupling to the effective mass of the scalar,
\begin{equation}
V_{\theta\phi}^{2}(0,0)\leq2c_{123}V_{\phi\phi}(0,0),
\end{equation}
where, we remind the reader, $c_{123}\equiv c_1+c_2+c_3$, and analogously for similar expressions. We will find this bound on $V_{\theta\phi}(0,0)^2$ useful when we examine the cosmological perturbations. This result complements and generalises a derivation in \rcite{Donnelly:2010cr}, which used different methods and selected a specific form of $V(\theta,\phi)$, and will take as a starting point the method utilised in \rcite{Lim:2004js} for pure \ae-theory.

We assume that the potential is analytic around $(\theta,\phi)=(0,0)$, because
if it diverges there the aether-scalar stress-energy tensor
(\ref{eq:fullsetensor}) will be nonzero and we
cannot have a Minkowski solution. We will also assume that $V(0,0)$ is
either vanishing or negligibly small; if not, then this contributes a
cosmological constant term to the stress-energy tensor, and our background is (anti-)de Sitter rather than
flat. Observations constrain such a term, barring a nonlinear screening
mechanism, to be very small.\footnote{The scalar field is canonical, coupled minimally to gravity, and not coupled at all to the matter sector, so we would not expect any screening mechanisms to be present in this theory.}

In flat space the field equations are solved by a constant-field configuration,
\begin{align}
\bar{u}^{\mu} &  =(m,0,0,0),\\
\bar{\lambda} &  =0,\\
\bar{\phi} &  =0.
\end{align}
We introduce small perturbations, $\{v^{\mu},\delta\lambda,\delta\phi\}$,
defined by
\begin{align}
u^{\mu} &  =\bar{u}^{\mu}+v^{\mu},\\
\lambda &  =\bar{\lambda}+\delta\lambda,\\
\phi &  =\bar{\phi}+\delta\phi.
\end{align}
Writing the action (\ref{eq:fullaction}) as
\begin{equation}
S=\int d^{4}x\mathcal{L},
\end{equation}
we expand the Lagrangian to quadratic order,
\begin{equation}
\mathcal{L}=\bar{\mathcal{L}}+\delta_{1}\mathcal{L}+\delta_{2}\mathcal{L},
\end{equation}
where $\delta_{1}\mathcal{L}$ and $\delta_{2}\mathcal{L}$ are of linear and
quadratic order, respectively. The background and linear Lagrangians recover
the background equations of motion, leaving us with the quadratic Lagrangian,
\begin{align}
\delta_{2}\mathcal{L}={} &  -c_{1}\partial_{\mu}v^{\nu}\partial^{\mu}v_{\nu
}-c_{2}(\partial_{\mu}v^{\mu})^{2}-c_{3}\partial_{\mu}v^{\nu}\partial_{\nu
}v_{\mu}+2\delta\lambda\left(  \bar{u}^{\mu}v_{\mu}\right) \nonumber\\
&  -\frac{1}{2}\partial_{\mu}\delta\phi\partial^{\mu}\delta\phi-\frac{1}
{2}\left[  V_{\theta\theta}(0,0)(\partial_{\mu}v^{\mu})^{2}+V_{\phi\phi
}(0,0)\delta\phi^{2}+2V_{\theta\phi}(0,0)\delta\phi(\partial_{\mu}v^{\mu
})\right]  ,\label{eq:quadlagfull}
\end{align}
whose variation yields the equations of motion of the perturbed variables.
From here we drop the $(0,0)$ evaluation on the derivatives of the potential
(although they remain implicit). The $\delta\lambda$ equation of motion is
\begin{equation}
\bar{u}^{\mu}v_{\mu}=0.
\end{equation}
It constrains the timelike component of the aether perturbation to vanish,
\begin{equation}
v^{0}=0.
\end{equation}
Inserting this result into \cref{eq:quadlagfull} and splitting $v^{i}$ into
spin-0 and spin-1 fields\footnote{The aether perturbation is in a reducible
subgroup of SO(3), so by decomposing $v^{i}$ like this we single out the real
dynamical degrees of freedom. Note also that, here and throughout this chapter, we will refer to scalar and vector modes of the aether as spin-0 and spin-1, respectively, so as not to confuse them with the scalar field $\phi$ and vector field $u^\mu$.} as
\begin{equation}
v^{i}=S^{i}+N^{i},
\end{equation}
where the spin-0 piece $S^{i}$ is the divergence of a scalar potential
($S^{i}=\partial^{i}\mathcal{V}$) and the spin-1 piece $N^{i}$ is transverse
to $S^{i}$ ($\partial_{i}N^{i}=0$), we find that the quadratic potential
decouples for these two pieces,
\begin{equation}
\delta_{2}\mathcal{L}=\mathcal{L}^{(0)}+\mathcal{L}^{(1)},
\end{equation}
where the spin-0 Lagrangian is
\begin{align}
\mathcal{L}^{(0)}={} &  c_{1}\dot{S}^{2}-c_{1}\partial_{i}S^{j}\partial
^{i}S_{j}-c_{2}(\partial_{i}S^{i})^{2}-c_{3}\partial_{i}S^{j}\partial_{j}
S^{i}\nonumber\\
&  +\frac{1}{2}\left(  \dot{\delta\phi}^{2}-\delta^{ij}\partial_{i}\delta
\phi\partial_{j}\delta\phi\right)  -\frac{1}{2}\left[  V_{\theta\theta
}(\partial_{i}S^{i})^{2}+V_{\phi\phi}\delta\phi^{2}+2V_{\theta\phi}\delta
\phi(\partial_{i}S^{i})\right] \label{eq:spin0lag}
\end{align}
and the spin-1 Lagrangian is
\begin{equation}
\mathcal{L}^{(1)}=c_{1}\dot{N}^{2}-c_{1}\partial_{i}N^{j}\partial^{i}
N_{j}.\label{eq:spin1lag}
\end{equation}
We have eliminated the cross-terms between the spin-0 and spin-1 pieces, and
the $c_{3}$ term in the spin-1 piece, using integration by parts.

Notice that a consequence of the spin-1 perturbation $N^{i}$ being
divergence-free is that the scalar-field coupling does not affect the spin-1
Lagrangian, because $\phi$ only couples to the aether through $\theta
=\nabla_{\mu}u^{\mu}$. In particular, this allows us to use the constraint
\begin{equation}
c_{1}>0
\end{equation}
from the start. This was derived in pure \ae-theory from requiring positivity
of the quantum Hamiltonian for both the spin-0 and spin-1 fields
\cite{Lim:2004js}, and is suggested by the fact that for $c_{1}\leq0$ the
kinetic terms for $S^{i}$ and $N^{i}$ in \cref{eq:spin0lag,eq:spin1lag} have
the wrong sign. Since this was proven to be true for the spin-1 perturbations
in \ae-theory and they remain unchanged in this extension of it, this
condition on $c_{1}$ must continue to hold.

Finally, we can vary the action with respect to our three perturbation
variables---$S^{i}$, $N^{i}$, and $\delta\phi$---to obtain the equations
of motion,
\begin{align}
\ddot{S}^{i}-\frac{c_{123}+\frac{1}{2}V_{\theta\theta}}{c_{1}}\partial
^{2}S^{i}-\frac{1}{2c_{1}}V_{\theta\phi}\delta^{ij}\partial_{j}\delta\phi &
=0,\\
\ddot{N}^{i}-\partial^{2}N^{i} &  =0,\\
\Box\delta\phi-V_{\phi\phi}\delta\phi-V_{\theta\phi}\partial_{i}S^{i} &  =0.
\end{align}
In \ae-theory, $\phi=0=V(\theta,\phi)$ and both aether equations are simply
wave equations with plane wave solutions \cite{Lim:2004js},
\begin{align}
S^{i}(\vec{k}) &  \propto e^{-ic_{s}^{(0)}kt+i\vec{k}\cdot\vec{x}
},\label{eq:spin0flat}\\
N^{i}(\vec{k}) &  \propto e^{-ic_{s}^{(1)}kt+i\vec{k}\cdot\vec{x}
},\label{eq:spin1flat}
\end{align}
with the propagation speeds for the spin-0 and spin-1 perturbations given by
\begin{align}
c_{s}^{(0)2} &  =\frac{c_{123}}{c_{1}},\\
c_{s}^{(1)2} &  =1.
\end{align}

The scalar coupling modifies the \ae-theory situation in two ways. First, $c_{123}$ is
shifted by $\frac{1}{2}V_{\theta\theta}$ evaluated at $(\theta=0,\phi=0)$
(remember that implicitly we are evaluating all the derivatives of $V$ there,
so they are just constants). This is to be expected: the expansion of the
potential around $(0,0)$ includes, at second order, the term $\frac{1}
{2}V_{\theta\theta}\delta\theta^{2}=\frac{1}{2}V_{\theta\theta}(\partial
_{i}S^{i})^{2}$, which can be absorbed into the $c_{2}$ term in the
(quadratic) Lagrangian by redefining $c_{2}\rightarrow c_{2}+\frac{1}
{2}V_{\theta\theta}$. We will find this same redefinition of $c_{2}$ appears
in the cosmological perturbation theory.

The second change from \ae-theory is more significant for the dynamics. When
$V_{\theta\phi}$ is nonzero---i.e., when the coupling between $u^{\mu}$ and
$\phi$ is turned on---it adds a source term to the wave equation for $S^{i}$
(the $N^{i}$ equation is unmodified because neither $\theta$ nor $\phi$
contain spin-1 pieces, as discussed above). Similarly, a $u^{\mu}$-dependent
source is added to the quadratic order Klein-Gordon equation for $\delta\phi$.

The equations of motion for $S^i$ and $\delta\phi$ are those of two coupled harmonic oscillators. To simplify these, we move to Fourier space, where the spin-0
degrees of freedom $S_{k}^{i}(t)=\partial^{i}\mathcal{V}_{k}(t)$ and
$\delta\phi_{k}(t)$ obey the coupled wave equations (dropping the $k$
subscripts and absorbing $\frac{1}{2}V_{\theta\theta}$ into $c_{2}$):
\begin{align}
\ddot{\mathcal{V}}+c_{s}^{(0)2}k^{2}\mathcal{V}-\frac{1}{2c_{1}}V_{\theta\phi
}\delta\phi &  =0,\\
\ddot{\delta\phi}+(k^{2}+V_{\phi\phi})\delta\phi-V_{\theta\phi}k^{2}
\mathcal{V} &  =0.
\end{align}
This system can be diagonalised\footnote{We thank the referee of \rcite{Solomon:2013iza} for this suggestion, which simplifies an earlier version of the calculation while obtaining the same result.} by defining
\begin{align}
\tilde{\mathcal{V}} &\equiv \mathcal{V} + \frac{V_{\theta\phi}}{2c_1\left(k^2+V_{\phi\phi}^2-\omega^2_-\right)}\delta\phi, \\
\tilde{\delta\phi} &\equiv \delta\phi + \frac{V_{\theta\phi} k^2}{c_s^{(0)2}k^2 - \omega^2_+}\mathcal{V},
\end{align}
where the $\omega_\pm$ are defined by
\begin{equation}
2\omega^2_\pm \equiv k^2\left(1+c_s^{(0)2}\right) + V_{\phi\phi}^2 \pm \sqrt{\left[k^2 \left(1-c_s^{(0)2}\right) + V_{\phi\phi}^2\right]^2 + \frac{2V_{\theta\phi}^2k^2}{c_1}}.
\end{equation}
Under this transformation, the equations of motion are simply
\begin{align}
\ddot{\tilde{\mathcal{V}}} + \omega^2_-\tilde{\mathcal{V}} &=0, \label{eq:tildev} \\
\ddot{\tilde{\delta\phi}} + \omega^2_+\tilde{\delta\phi} &=0. \label{eq:tildedphi}
\end{align}
Note that in the limit $V_{\theta\phi} \to 0$ where the two fields decouple, $\omega^2_+$ goes to $k^2+V_{\phi\phi}^2$, the squared frequency of a $\delta\phi$ mode, and $\omega^2_-$ goes to $c_s^2k^2$, the equivalent for $\mathcal{V}$ modes. We see that $\tilde{\mathcal{V}}$ and $\tilde{\delta\phi}$ are noninteracting, mixed modes which reduce to $\mathcal{V}$ and $\delta\phi$, respectively, in the absence of a scalar-aether coupling.

For stability, we require $\omega_\pm$ to be real, so that the solutions to \cref{eq:tildev,eq:tildedphi} are plane waves
rather than growing and decaying exponentials. Note that $\omega_+$ is manifestly real, so the $\tilde{\delta\phi}$ modes are always stable. It is the aether modes, $\tilde{\mathcal{V}}$,\footnote{Technically, the mixed aether-scalar modes which become arbitrarily close to the aether perturbations in the limit $V_{\theta\phi}\to0$.} which can be destabilised by the coupling to the scalar, while the reverse is not true. Stability imposes a constraint on $V_{\theta\phi}$,
\begin{equation}
V_{\theta\phi}^{2}\leq2c_{1}c_{s}^{(0)2}(k^{2}+V_{\phi\phi}),
\end{equation}
which, since we would like it to hold for arbitrarily large-wavelength modes
(i.e., arbitrarily small $k$), can be written, substituting back in the definition $c_{s}
^{(0)2}\equiv c_{123}/c_{1}$,
\begin{equation}
V_{\theta\phi}^{2}(0,0)\leq2c_{123}V_{\phi\phi}
(0,0),\label{eq:spin0constraint}
\end{equation}
where for clarity we have put back in the $(\theta,\phi)=(0,0)$ evaluation
which has been implicit. \Cref{eq:spin0constraint} constrains the coupling
between the aether and the scalar field in terms of the aether kinetic-term free parameters (or, equivalently, its no-coupling
propagation speed) and the effective mass of the scalar in flat space. It agrees with the
spin-0 stability constraint in \rcite{Donnelly:2010cr} which was derived in a
slightly different fashion for a specific form of $V(\theta,\phi
).$\footnote{Our notation is different than that used in \rcite{Donnelly:2010cr}
and as a result their constraint looks slightly different. They define the aether to be dimensionless and unit norm while we give it a norm $m$ with mass dimensions. To compensate for this, their $c_i$ are $16\pi Gm^2c_i$ in our notation. We have checked, translating between the two notations, that our constraint matches theirs.\label{foot:djnotation}} The $c_i$ are dimensionless, so we might expect them to naturally be $\mathcal{O}(1).$ Assuming this, \cref{eq:spin0constraint} roughly constrains the coupling $V_{\theta\phi}(0,0)$ to be less than the scalar field mass around flat space. Note that this constraint also implies
$c_{123}\geq0$,\footnote{Assuming that the scalar field is nontachyonic.}
which is the combined constraint from subluminal propagation and positivity of
the Hamiltonian of the spin-0 field in pure \ae-theory \cite{Lim:2004js}.

\section{Cosmological Perturbation Theory}
\label{sec:cosmperts}

The goal of this chapter is to explore the impact of the coupling between $\phi$ and
$u^{\mu}$ on small perturbations to a homogeneous and isotropic
cosmology. We will be particularly interested in a period of slow-roll
inflation driven by $\phi$. As has been explored in great depth over the past
three decades, a scalar field rolling slowly down its potential can lead to
cosmic inflation and all of the interesting cosmological consequences for
explaining the structure of the observed universe that follow from it
\cite{PeterUzan}. In this section we set up the linearised perturbation theory for the metric and the scalar around a homogeneous and isotropic universe. Recall that the background equations of motion, i.e., the Friedmann and Klein-Gordon equations, are presented in \cref{sec:aecosmo}.

\subsection{Perturbation Variables}

Let us consider linear perturbations about the FRW background (\ref{eq:ae-FRW}), defined by
\begin{equation}
ds^{2}=a^{2}(\tau)\left\{  -(1+2\Phi)d\tau^{2}-2B_{i}d\tau dx^{i}
+[(1+2\Psi)\delta_{ij}+2H_{Tij}]dx^{i}dx^{j}\right\},
\end{equation}
so the components of the perturbed metric are
\begin{align}
g_{00} &  =-a^{2}(1+2\Phi),\nonumber\\
g_{0i} &  =-a^{2}B_{i},\nonumber\\
g_{ij} &  =a^{2}[(1+2\Psi)\delta_{ij}+2H_{Tij}].
\end{align}
Inverting, and keeping terms to first-order we have
\begin{align}
g^{00} &  =-a^{-2}(1-2\Phi),\nonumber\\
g^{0i} &  =-a^{-2}B^{i},\nonumber\\
g^{ij} &  =a^{-2}[(1-2\Psi)\delta^{ij}-2H_{T}^{ij}].
\end{align}
Indices on spatial quantities like $B_{i}$ are raised and lowered with
$\delta_{ij}$. The Christoffel symbols are (with background parts in bold)
\begin{align}
\Gamma_{00}^{0} &  =\boldsymbol{\mathcal{H}}+\Phi^{\prime},\nonumber\\
\Gamma_{0i}^{0} &  =\Phi_{,i}-\mathcal{H}B_{i},\nonumber\\
\Gamma_{ij}^{0} &  =\left(  \boldsymbol{\mathcal{H}}(1+2\Psi)+\Psi^{\prime
}-2\mathcal{H}\Phi\right)  \delta_{ij}+B_{(i,j)}+(2\mathcal{H}H_{Tij}
+H_{Tij}^{\prime}),\nonumber\\
\Gamma_{00}^{i} &  =\Phi^{,i}-\mathcal{H}B^{i}-B^{\prime i},\nonumber\\
\Gamma_{0j}^{i} &  =\boldsymbol{\mathcal{H}\delta_{j}^{i}}+\delta
^{ik}B_{[j,k]}+\Psi^{\prime}\delta_{j}^{i}+H_{T}^{\prime}{}_{j}^{i}
,\nonumber\\
\Gamma_{jk}^{i} &  =\mathcal{H}B^{i}\delta_{jk}+\Psi_{,k}\delta_{j}^{i}
+\Psi_{,j}\delta_{k}^{i}-\Psi^{,i}\delta_{jk}+H_{T}{}_{j,k}^{i}+H_{T}{}
_{k,j}^{i}-H_{Tjk}{}^{,i},
\end{align}
where primes denote deriatives with respect to $\tau$. We do not reproduce the components of the Einstein tensor here; they can be found in the
literature \cite{Kodama:1985bj}.

The aether in the background has only $u^{0}=\frac{m}{a}$. Imposing the
constant-norm constraint, $u_{\mu}u^{\mu}=-m^{2}$, to first order the aether is given by
\begin{equation}
u^{\mu}=\frac{m}{a}\left(  (1-\Phi),V^{i}\right)  ,\label{eq:perturbedaether}
\end{equation}
where $V^i$ is the spatial perturbation to the aether. With lowered indices we have
\begin{equation}
u_{\mu}=ma\left(  -(1+\Phi),V_{i}-B_{i}\right).
\end{equation}
Taking the divergence of \cref{eq:perturbedaether} we can find the linearised expansion,
\begin{equation}
\theta=\frac{m}{a}\left[  3\mathcal{H}(1-\Phi)+3\Psi^{\prime}+V^{i}{}
_{,i}\right]. \label{eq:thetalin}
\end{equation}
Finally, the scalar field $\phi$ is split into a background piece and a small
perturbation,
\begin{equation}
\phi= \bar\phi+ \delta\phi,
\end{equation}
where $\bar\phi$ satisfies the Klein-Gordon equation in the background metric. Throughout this chapter we use overbars to denote background values.

\subsection{Linearised Equations of Motion}

In deriving the perturbation equations we will need to expand $V(\theta,\phi)$ around its background value. To second order, assuming $\delta\phi$ is similar in size to the metric perturbations, we have
\begin{equation}
V(\theta,\phi)=\bar{V}+\bar{V}_{\theta}\delta\theta+\bar{V}_{\phi}\delta
\phi+\frac{1}{2}\left[  \bar{V}_{\theta\theta}\delta\theta^{2}+\bar{V}
_{\phi\phi}\delta\phi^{2}+2\bar{V}_{\theta\phi}\delta\theta\delta\phi\right]
+O(\delta\theta^{3}),
\end{equation}
where, per \cref{eq:thetalin}, the linear piece of the expansion is given by
\begin{equation}\delta\theta  =\frac{m}{a}\left(  3\Psi^{\prime}-3\mathcal{H}\Phi+V^{i}
{}_{,i}\right).
\end{equation}
In deriving the linearised field equations we will need $V(\theta,\phi)$ and all of its derivatives up to first order,
\begin{align}
V(\theta,\phi) &  =\bar{V}+\bar{V}_{\theta}\delta\theta+\bar{V}_{\phi}
\delta\phi+O(\delta\theta^{2}),\nonumber\\
V_{\theta}(\theta,\phi) &  =\bar{V}_{\theta}+\bar{V}_{\theta\theta}
\delta\theta+\bar{V}_{\theta\phi}\delta\phi+O(\delta\theta^{2}),\nonumber\\
V_{\phi}(\theta,\phi) &  =\bar{V}_{\phi}+\bar{V}_{\phi\phi}\delta\phi+\bar
{V}_{\theta\phi}\delta\theta+O(\delta\theta^{2}),\nonumber\\
V_{\theta\theta}(\theta,\phi) &  =\bar{V}_{\theta\theta}+\bar{V}_{\theta
\theta\theta}\delta\theta+\bar{V}_{\theta\theta\phi}\delta\phi+O(\delta
\theta^{2}),\nonumber\\
V_{\theta\phi}(\theta,\phi) &  =\bar{V}_{\theta\phi}+\bar{V}_{\theta\theta
\phi}\delta\theta+\bar{V}_{\theta\phi\phi}\delta\phi+O(\delta\theta)^{2}.
\end{align}

The linearised equations of motion in real space are given in
\cref{app:realspace}. However, the symmetries of the FRW background allow us
to decompose the perturbations into spin-0, spin-1, and spin-2 components
\cite{Bardeen:1980kt}. In particular, because the background variables (including
the aether, which points only in the time direction) do not break the SO(3)
symmetry on spatial slices, these components conveniently decouple from each other. Hence we perform this
decomposition both to isolate the fundamental degrees of freedom and to make close contact with the rest of the literature on
cosmological perturbation theory.

We decompose the variables as
\begin{align}
\delta\phi &  =\sum_{k}\delta\phi_{k}Y^{(0)},\\
\Phi &  =\sum_{k}\Phi_{k}Y^{(0)},\\
\Psi &  =\sum_{k}\Psi_{k}Y^{(0)},\\
V^{i} &  =\sum_{k}\sum_{m=0,1}V_{k}^{(\pm m)}Y^{i}{}^{(\pm m)},\\
B^{i} &  =\sum_{k}\sum_{m=0,1}B_{k}^{(\pm m)}Y^{i}{}^{(\pm m)},\\
H_{T}^{ij} &  =\sum_{k}\sum_{m=0,1,2}H_{Tk}^{(\pm m)}{}Y^{ij}{}^{(\pm m)},
\end{align}
where $Y^{(0)}$, etc., are eigenmodes of the Laplace-Beltrami operator,
$\partial^{2}+k^{2}$ (see \rcite{Kodama:1985bj,Lim:2004js} for the forms of
these mode functions and some of their useful properties). From here on, we will drop
the $k$ subscripts. The spin-0, spin-1, and spin-2 perturbation equations can
then be found by plugging these expansions into the real space equations listed in \cref{app:realspace}.

\section{Spin-1 Cosmological Perturbations}

\label{sec:spin1perts}

We begin our analysis by focusing on the spin-1 perturbations. The spin-2 perturbations
are unmodified by the aether-scalar coupling because $V(\theta,\phi)$ only
contains spin-0 and spin-1 terms. The only physical spin-2 perturbations are the transverse and traceless parts of the metric
perturbation $H_{Tij}$, or gravitational waves, and they behave as they do in pure
\ae-theory \cite{Lim:2004js}. The spin-0 perturbations, discussed in \cref{sec:spin0perts}, are more complicated than the spin-1 perturbations due to the presence of $\delta\phi$ modes. The important physical result---the existence of unstable perturbations for large, otherwise experimentally-allowed regions of parameter space---will therefore be easier to see and understand in the context of the simpler spin-1 modes.

The only nontrivial spin-1 component of the aether field equation (\ref{eq:aethereom}) is $\nu=i$,
\begin{align}
&  \left\{  \left[  -2\frac{\alpha}{m^{2}}{\mathcal{H}}^{2}+\frac{\alpha
}{m^{2}}\frac{a^{\prime\prime}}{a}-c_{1}\frac{a^{\prime\prime}}{a}\right]
(B^{(\pm1)}-V^{(\pm1)})\right. \nonumber\\
&  +2c_{1}{\mathcal{H}}(V^{\prime(\pm1)}-B^{\prime(\pm1)})+\frac{1}{2}
(c_{3}-c_{1})k^{2}B^{(\pm1)}+c_{1}k^{2}V^{(\pm1)}\nonumber\\
&  \left.  -c_{13}\frac{k}{2}H_{T}^{\prime}{}^{(\pm1)}-c_{1}(B^{\prime
\prime(\pm1)}-V^{\prime\prime(\pm1)})\right. \nonumber\\
&  +\left.  \frac{1}{2}\left[  3\bar{V}_{\theta\theta}\left(  \frac
{a^{\prime\prime}}{a}-2\mathcal{H}^{2}\right)  +\frac{a}{m}\bar{V}_{\theta
\phi}\bar{\phi}^{\prime}\right]  \left(  B^{(\pm1)}-V^{(\pm1)}\right)
\right\}  Y_{i}^{(\pm1)}=0,\label{eq:spin1vi}
\end{align}
while the spin-1 perturbations of the stress-energy tensor are
\begin{align}
\delta T^{0}{}_{0}^{(\pm1)}={} &  0,\\
\delta T^{0}{}_{i}^{(\pm1)}={} &  \left\{  2\left(  \frac{m}{a}\right)
^{2}\left[  \left(  -2\frac{\alpha}{m^{2}}{\mathcal{H}}^{2}+\frac{\alpha
}{m^{2}}\frac{a^{\prime\prime}}{a}-c_{1}\frac{a^{\prime\prime}}{a}\right)
(V^{(\pm1)}-B^{(\pm1)})\right.  \right. \nonumber\\
&  \left.  -\left.  c_{1}a^{-2}\left[  a^{2}(V^{\prime(\pm1)}-B^{\prime(\pm
1)})\right]  ^{\prime}+\frac{1}{2}(c_{1}-c_{3})k^{2}(B^{(\pm1)}-V^{(\pm
1)})\right]  \right. \nonumber\\
&  +\left.  \frac{m}{a}\left[  \frac{3m}{a}\bar{V}_{\theta\theta}\left(
\frac{a^{\prime\prime}}{a}-2\mathcal{H}^{2}\right)  +\bar{V}_{\theta\phi}
\bar{\phi}^{\prime}\right]  \left(  V^{(\pm1)}-B^{(\pm1)}\right)  \right\}
Y_{i}^{(\pm1)},\label{eq:spin10i}\\
\delta T^{i}{}_{j}^{(\pm1)}={} &  2\left(  \frac{m}{a}\right)  ^{2}
c_{13}\left\{  a^{-2}\left[  a^{2}(-kV^{(\pm1)}+H^{\prime}{}_{T}^{(\pm
1)})\right]  ^{\prime}\right\}  Y^{i}{}_{j}{}^{(\pm1)},
\end{align}
where $\alpha=(c_{13}+3c_2)m^2$ is defined in \cref{eq:alphadef}. As a consistency check, these expressions reduce to those found in the
literature for a scalar field uncoupled to the aether \cite{Mukhanov:1990me}
(setting $V(\theta,\phi)=V(\phi)$) and for \ae-theory \cite{Lim:2004js} (setting $V(\theta
,\phi)=\alpha\theta^{2}$, with $c_{2}\rightarrow c_{2}+\alpha$). For
convenience, from here on we will absorb $\frac{1}{2}\bar{V}_{\theta\theta}$ into $c_{2}$ and indicate
the change with a tilde, e.g.,
\begin{equation}
\tilde{\alpha}\equiv\left(  c_{1}+3c_{2}+c_{3}+\frac{3}{2}\bar{V}
_{\theta\theta}\right)  m^{2}\text{,}
\end{equation}
and similarly for quantities like $\tilde G_{c}$. While this is convenient
notation we should remember that $\bar{V}_{\theta\theta}$ and hence all tilded
quantities are not necessarily constant, although they are nearly so during a
slow-roll phase.\footnote{A nonconstant $\bar{V}_{\theta\theta}$ requires
cubic or higher order terms in the potential. For the quadratic Donnelly-Jacobson potential discussed in \cref{sec:dj}, $\bar{V}
_{\theta\theta}$ is constant and can be freely set to zero by absorbing it
into $c_{2}$.}

We should first note that due to its direct coupling to the aether, the scalar
field \emph{does} source spin-1 perturbations, which is impossible in the
uncoupled case as the scalar field itself contains no spin-1 piece. In pure
\ae-theory the spin-1 perturbations decay away as $a^{-1}$ \cite{Lim:2004js}.
We may wonder if the scalar-vector coupling can counteract this and generate a
nondecaying spin-1 spectrum.

Using the gauge freedom in the spin-1 Einstein equations, we choose to work in
a gauge where $H_{T}^{{(\pm1)}}=0$; that is, we foliate spacetime with
shear-free hypersurfaces. The $i$--$j$ Einstein equation in the spin-1 case is
unmodified from the \ae-theory case \cite{Lim:2004js} and gives a constraint
relating the shift $B^{{(\pm1)}}$ and the spin-1 aether perturbation
$V^{{(\pm1)}}$,
\begin{equation}
B^{{(\pm1)}}=\gamma V^{{(\pm1)}}, \label{eq:bveq}
\end{equation}
where
\begin{equation}\gamma\equiv16\pi Gm^{2}c_{13}.
\end{equation}
It is tempting to notice the similarities between the $v=i$ aether field equation (\ref{eq:spin1vi}) and
the $0$--$i$ Einstein equation (\ref{eq:spin10i}), but this is just hinting at
the underlying redundancy between the two equations. Indeed, using
\cref{eq:spin1vi} to eliminate the scalar field term in \cref{eq:spin10i} just
leaves us with an identity. This is because, due to the constraint
equation (\ref{eq:bveq}), the two perturbations $B$ and $V$ are related, and hence (by the
Bianchi identities) these two equations have to contain the same content.
We choose to use the $0$--$i$ Einstein equation to derive our equation of
motion for the spin-1 perturbations. In this equation, the scalar field
couples to the vector perturbations of the aether and the metric via $\frac
{m}{a}\bar{V}_{\theta\phi}\bar{\phi}^{\prime}$. In the quadratically-coupled
potential of Donnelly and Jacobson, which we discuss in detail in \cref{sec:dj}, the coupling $\bar V_{\theta\phi}$ is
exactly constant. In general, we will take $\bar{V}_{\theta\phi}$ to be
constant to first order in slow roll.

Inserting the constraint into the $0$--$i$ Einstein equation we find
\begin{align}
&  \left(  2\frac{\alpha}{m^{2}}{\mathcal{H}}^{2}-\frac{\alpha}{m^{2}}
\frac{a^{\prime\prime}}{a}+c_{1}\frac{a^{\prime\prime}}{a}\right)  V^{(\pm
1)}\nonumber\\
&  +\frac{1}{2}\left[  (c_{1}-c_{3})+\frac{c_{13}}{1-\gamma}\right]
k^{2}V^{(\pm1)}+c_{1}(2{\mathcal{H}}V^{(\pm1)\prime}+V^{(\pm1)\prime\prime})\nonumber\\
&  -\frac{1}{2}\left[  3\bar{V}_{\theta\theta}\left(  \frac{a^{\prime\prime}
}{a}-2\mathcal{H}^{2}\right)  +\frac{a}{m}\bar{V}_{\theta\phi}\bar{\phi
}^{\prime}\right]  V=0.\label{eq:spin10iv}
\end{align}
Following \rcite{Lim:2004js}, we define $\xi=aV^{{(\pm1)}}$ to eliminate the
first-derivative terms, so \cref{eq:spin10iv} becomes
\begin{equation}
\xi^{\prime\prime}+c_{s}^{{(\pm1)}2}k^{2}\xi+\left(  A\frac{\tilde{\alpha}
}{m^{2}c_{1}}-\frac{1}{2}\frac{a}{mc_{1}}\bar{V}_{\theta\phi}\bar{\phi
}^{\prime}\right)  \xi=0,\label{eq:xieom}
\end{equation}
where the no-coupling sound speed $c_{s}^{{(\pm1)}}$ is the de Sitter propagation speed
of the spin-1 aether and metric perturbations when the coupling to the scalar
field is absent \cite{Lim:2004js},
\begin{equation}
c_{s}^{{(\pm1)}2}=\frac{1}{2}\left[  (1-c_{3}/c_{1})+\frac{1+c_{3}/c_{1}
}{1-\gamma}\right].
\end{equation}
The background quantity $A$ is defined by
\begin{align}
A &\equiv 2\mathcal{H}^{2}-\frac{a^{\prime\prime}}{a} \nonumber \\
&={\mathcal{H}}^{2}-{\mathcal{H}}^{\prime} \nonumber \\
&=-a\left(  \frac{{\mathcal{H}}}{a}\right)'
\end{align}
and vanishes in exact de Sitter space.

\subsection{Slow-Roll Limit}

\label{sec:vecslowroll}

The equation of motion (\ref{eq:xieom}) for the spin-1 aether and metric perturbations is difficult to solve in full generality. It was solved in
pure de Sitter space ($A=0$) in \ae-theory (i.e., in the absence of the scalar field) in \rcite{Lim:2004js}. In that limit,
\cref{eq:xieom} is a wave equation with real frequency, so $\xi$ was found to be
oscillatory. Therefore, in \ae-theory the spin-1 shift perturbation, $B^{{(\pm1)}}=\gamma\xi/a$,
decays exponentially,\footnote{Here and in the rest of this chapter, ``exponential'' growth or decay should be taken to mean exponential in cosmic time, or as a power law in conformal time.} leaving the post-inflationary universe devoid of spin-1 perturbations. To investigate whether the inflaton coupling term will change this conclusion, let us solve \cref{eq:xieom} in the slow-roll limit.

We define the slow-roll parameters, $\varepsilon$ and $\eta$, in the usual way,
\begin{alignat}{2}
\varepsilon &=-\frac{\dot{H}}{H^{2}} &&=1-\frac{{\mathcal{H}}^{\prime}}{{\mathcal{H}}^{2}},\\
\eta &=\phantom{-}\frac{\dot{\varepsilon}}{H\varepsilon} &&=\frac{\varepsilon
^{\prime}}{{\mathcal{H}}\varepsilon},
\end{alignat}
where for completeness we have included both the cosmic-time and conformal-time definitions. Slow-roll inflation occurs whenever $\varepsilon,\eta\ll1$. In this limit, both parameters are constant at first order and we can find
\begin{align}
a &  \approx-\frac{1}{H\tau}(1+\varepsilon),\\
{\mathcal{H}} &  \approx-\frac{1}{\tau}(1+\varepsilon
).\label{eq:generalshslowroll}
\end{align}
Taking conformal-time derivatives we can calculate the background quantity defined above,
\begin{equation}
A={\mathcal{H}}^{2}-{\mathcal{H}}^{\prime} \approx \frac{\varepsilon}{\tau^{2}}.
\end{equation}
Note that during slow roll, $H$ is approximately constant but $\sh$ is not; therefore, even though we are working in conformal time, we will often choose to write the equations of motion and their solutions in terms of $H$, treating it as a free parameter which measures the energy scale of inflation.

Using these relations, as well as the Klein-Gordon equation in the slow-roll
limit and the fact that ${\mathcal{H}}=aH$, we can write the $\xi$ equation of
motion to first order in slow roll as
\begin{equation}
\xi^{\prime\prime}+c_{s}^{{(\pm1)}2}k^{2}\xi+\frac{1}{\tau^{2}}\left(
\frac{\tilde{\alpha}}{m^{2}c_{1}}\varepsilon+\frac{1}{6mc_{1}}\frac{1+2\varepsilon
}{H^{3}}\bar{V}_{\theta\phi}\bar{V}_{\phi}\right)  \xi
=\mathcal{O}(\varepsilon^2).\label{eq:spin1xislowrollo1}
\end{equation}
$\bar{V}(\theta,\phi)$ and its derivatives will be constant at leading order in
the slow-roll parameters, so if we ignore $\mathcal{O}(\varepsilon)$ terms then $\bar{V}_{\phi}$ and $\bar{V}_{\theta\phi}$ in \cref{eq:spin1xislowrollo1} are constants. Our equation of motion for the spin-1 perturbations can then be written simply as
\begin{equation}
\xi'' + c_s^{\pmvec2}k^2\xi - \frac{\Lambda}{\tau^2}\xi = 0\label{eq:spin1xislowroll}
\end{equation}
where we have defined the constant
\begin{equation}
\Lambda \equiv -\frac{\bar{V}_{\phi}\bar{V}_{\theta\phi}}{6mc_{1}H^{3}}+\mathcal{O}(\varepsilon).\label{eq:lambdadef}
\end{equation}
We have also assumed that $\Lambda$ dominates $\tilde{\alpha}\varepsilon
/(m^{2}c_{1})\sim\varepsilon$ (which is $\sim\mathcal{O}(10^{-2})$ \cite{Ade:2013zuv,Ade:2013uln}), the term from pure \ae-theory. In principle this need not be
true, if the coupling term $\bar{V}_{\theta\phi}$ were extraordinarily small. If the aether-scalar
coupling is to do anything interesting, then $\bar{V}_{\theta\phi}$ must be
larger than that, so we will continue to assume that it is.

Finally, let us identify the effective mass of the spin-1 aether perturbation, $V=\xi/a$. Writing the equation of motion (\ref{eq:spin1xislowroll}) in terms of $V$, switching to cosmic time, and working to leading order in slow roll, we find
\begin{equation}
 \ddot V + 3H\dot V + (2-\Lambda)H^2V + \frac{c_s^2k^2}{a^2}V \approx 0.
\end{equation}
Outside the sound horizon, $c_sk\ll a$, we find that $V$ has an effective mass in the slow-roll r\'egime given by
\begin{equation}
M_\mathrm{eff}^2 = (2-\Lambda) H^2 \approx 2H^2-\frac{\bar{V}_{\phi}\bar{V}_{\theta\phi}}{6mc_{1}H}. \label{eq:meffsq}
\end{equation}
We expect a tachyonic instability to develop for negative $M_\mathrm{eff}^2$, i.e., for $\Lambda>2$. We proceed to demonstrate that just an instability arises.

\subsection{Full Solution for the Vector Modes}

Noticing the similarity between \cref{eq:spin1xislowroll} and the usual
Mukhanov-Sasaki equation \cite{Mukhanov:1990me}, which has solutions in terms of Bessel functions, we
change variables to $g \equiv x^{-1/2}\xi$ with $x\equiv-c_{s}^{{(\pm1)}}k\tau$ to recast
\cref{eq:spin1xislowroll} as Bessel's equation for $g(x)$,
\begin{equation}
x^{2}\frac{d^{2}g}{dx^{2}}+x\frac{dg}{dx}+\left(  x^{2}-\nu^{2}\right)  g=0,
\end{equation}
with the order $\nu$ given by
\begin{equation}
\nu^{2}\equiv\frac{1}{4}+\Lambda.
\end{equation}
Depending on the sign and magnitude of $\Lambda$, the order $\nu$ can be real
or imaginary. We will find it convenient to write the general solution in
terms of the Hankel functions as
\begin{equation}
\xi=\frac{\sqrt{\pi}}{2}\sqrt{-\tau}\left[  \alpha_{k}H_{\nu}^{(1)}
(-c_{s}^{{(\pm1)}}k\tau)+\beta_{k}H_{\nu}^{(2)}(-c_{s}^{{(\pm1)}}
k\tau)\right]  .
\end{equation}
To determine the values of the Bogoliubov coefficients, $\alpha_{k}$ and
$\beta_{k}$, we need to match this solution in the subhorizon limit, $-c_{s}^{{(\pm1)}}k\tau\rightarrow\infty$, to the quantum vacuum state of the
aether perturbations in flat space. This is desirable because we can
assume that, at such short wavelengths, these modes do not ``see'' the cosmic expansion. In section~IV.B of \rcite{Lim:2004js} the quantum
mode functions $N_{k}$ for the aether perturbation $v^{i}$ were demonstrated
to satisfy
\begin{equation}
N_{k}=\frac{1}{4\sqrt{|c_{1}|k}}e^{-ikt}.\label{eq:flatspacemode}
\end{equation}
This function is related to $\xi$ by $N_{k}=\frac{m}{a}V=\frac{m}{a^{2}}\xi$.
The mode $N_{k}$ is defined in Minkowski space, where $a\equiv1$ with
$t\equiv\tau,$ so we only need to modify it by a factor of $m$ to obtain $\xi$. Using the
asymptotic formula
\begin{equation}
\lim_{-c_{s}^{{(\pm1)}}k\tau\rightarrow\infty}H_{\nu}^{(1,2)}(-c_{s}^{{(\pm
1)}}k\tau)=\sqrt{\frac{2}{\pi}}\frac{1}{\sqrt{-c_{s}^{{(\pm1)}}k\tau}}e^{\mp
i(c_{s}^{{(\pm1)}}k\tau+\delta)},
\end{equation}
with $\delta=\frac{\pi}{2}(\nu+1/2)$, we find that in the subhorizon limit,
\begin{equation}
\xi\rightarrow\frac{1}{\sqrt{2c_{s}^{{(\pm1)}}k}}\left[  \alpha_{k}
e^{-i(c_{s}^{{(\pm1)}}k\tau+\delta)}+\beta_{k}e^{+i(c_{s}^{{(\pm1)}}
k\tau+\delta)}\right]  .
\end{equation}
Matching to \cref{eq:flatspacemode}, and ignoring the unimportant phase
factors $e^{\pm i\delta}$, we see that we need
\begin{align}
\alpha_{k} &  =\frac{1}{4m}\sqrt{\frac{2}{|c_{1}|}},\\
\beta_{k} &  =0,
\end{align}
where we have (consistently) put in some factors of $c_{s}^{{(\pm1)}}$ which
do not appear in the flat-space calculation because it ignores gravity,
but would have appeared if we had included gravity.\footnote{To see this,
consider \cref{eq:xieom} in the case $a=1$, which is the spin-1 perturbation
equation in flat space with gravitational perturbations turned on. Since
this requires $\phi=0$, the equation of motion (\ref{eq:xieom}) just becomes
$\xi^{\prime\prime}+c_{s}^{{(\pm1)}2}k^{2}\xi=0 $. This has the same solution
as we found in the case with gravity turned off in \cref{sec:flatspace}, but
with the sound speed modified, as expected.} Substituting in this value of
$\alpha_{k}$, we find the full solution for the spin-1 perturbation
\begin{equation}
V^\pmvec = \frac{1}{a} \sqrt{\frac{\pi}{2}}\frac{1}{4m\sqrt{|c_1|}} \sqrt{-\tau} H_\nu^{(1)}(-c_s^\pmvec k\tau) .\label{eq:vksol}
\end{equation}
As a consistency check, if we turn off the scalar-aether coupling, we have
$\nu=1/2$, and (up to an irrelevant phase of $-\pi/2$) we recover equation
(91) in \rcite{Lim:2004js}.

\subsection{Tachyonic Instability}

\label{sec:instability}

On superhorizon scales, the Hankel functions behave as
\begin{equation}
\lim_{c_{s}^{{(\pm1)}}k\tau\rightarrow0}H_{v}^{(1)}(-c_{s}^{{(\pm1)}}
k\tau)=\frac{i}{\pi}\Gamma(\nu)\left(  \frac{-c_{s}^{{(\pm1)}}k\tau}
{2}\right)  ^{-\nu}.
\end{equation}
Plugging this into \cref{eq:vksol}, we see that the large-scale vector perturbations to
the aether and metric depend on time as
\begin{equation}
V^{{(\pm1)}}\sim a^{-1}\sqrt{-\tau}(-\tau)^{-\nu}\sim(-\tau)^{\frac{3}{2}-\nu}.
\end{equation}

When the aether-scalar coupling (proportional to $\Lambda$) is small or absent, such that $-1/4<\Lambda<2$, the vector perturbations decay and are unobservable, as in pure \ae-theory \cite{Lim:2004js}. If $\Lambda$ is outside that range, then the coupling is large enough to change the nature of the vector perturbations. The coupling has two possible effects, depending on its sign. If $\nu$ is imaginary ($\Lambda < -1/4$), then the vector modes are both oscillatory and decaying.\footnote{Recall that, during inflation, $\tau$ runs from $-\infty$ to 0.} This corresponds to a large coupling which significantly damps the perturbations. On the other hand, the vector modes will experience runaway growth if $3/2-\nu$ is real and negative, or $\Lambda>2$. In this case the coupling is large, but with the opposite sign to the previous case, and this large coupling drives runaway production of aether modes. This is precisely the tachyonic instability we anticipated in \cref{sec:vecslowroll}, as it results from the aether perturbations acquiring an imaginary effective mass.

Since this growth is exponential (in cosmic time, or in number of $e$-folds), it seems quite probable that this growing vector mode will overwhelm the slow-roll background solution and therefore lead to an instability. In this subsection we will calculate the growth of a single vector mode and compare it to the background evolution.

In order to maintain a homogeneous and isotropic background spacetime, the
time-space term in the stress-energy tensor must be zero at the level of the
background ($\bar T^{0}{}_i = 0$). The spin-1 perturbations do contribute to these terms in the
stress-energy tensor (\ref{eq:spin10i}) through terms proportional to
$V_{k}^{{(\pm1)}}Y_{i,k}^{{(\pm1)}}$. In particular, we will focus on the
scalar-aether coupling term
\begin{equation}
T^{0}{}_{i,k} = \ldots + m\bar{V}_{\theta\phi}\frac{\bar{\phi}^{\prime}}{a}
V_{k}^{{(\pm1)}}Y_{i,k}^{{(\pm1)}} + \ldots,
\end{equation}
which we will write as
\begin{equation}
T^{0}{}_{i,k}\supset m\bar{V}_{\theta\phi}\frac{\bar{\phi}^{\prime}}{a}
V_{k}^{{(\pm1)}}Y_{i,k}^{{(\pm1)}}. \label{eq:T0isingleterm}
\end{equation}
While we focus on this term for simplicity, we note that there are many other terms in $T^0{}_i$ which receive contributions from the vector modes, and some may even be larger than the term in \cref{eq:T0isingleterm}.

Our strategy will be to focus on a single mode, picking one of the larger modes available to us. Because $V_{k}$ grows with decreasing $k$, we choose a mode which crosses the sound horizon at some early conformal time $\tau_{i}$. Such a mode has wave number
\begin{equation}
k=\frac{1}{-c_{s}^{{(\pm1)}}\tau_{i}}.
\end{equation}
The perturbation $V_{k}^{{(\pm1)}}$ is given by \cref{eq:vksol}, which for a
superhorizon perturbation becomes
\begin{equation}
V_{k}^{{(\pm1)}}(N)=-\frac{i}{2\pi}\frac{H}{4m\sqrt{|c_{1}|}}\Gamma(\nu
)2^{\nu}\left(  -\tau_{i}\right)  ^{\frac{3}{2}}e^{\left(  \nu-\frac{3}
{2}\right)  N},
\end{equation}
where $N$ is the number of $e$-folds after the mode crossed the sound horizon.

The mode function $Y^{(\pm1)}_{i,k}$ is given by \cite{Lim:2004js}
\begin{equation}
Y^{(\pm1)}_{i,k}(\vec{x}) = \frac{1}{\sqrt{2}k}\left[ \left( \vec{k}\times
\vec{n}\right) _{i} \mp i \left( \vec{k}\times\vec{n}\right) _{i}\right]
e^{i\vec{k}\cdot\vec{x}},
\end{equation}
where $\vec{n}$ is a unit vector orthogonal to $\vec{k}$. We can always choose
three orthogonal coordinates such that $k^{i} = k\delta^{i}{}_{1}$ and
$n^{i}=\delta^{i}{}_{2}$, so the mode function is
\begin{equation}
Y^{(\pm1)}_{i,k}(\vec{x}) = \frac{1\mp i}{\sqrt{2}}e^{i\vec{k}\cdot\vec{x}
}\delta^{3}{}_{i}.
\end{equation}
This oscillates throughout space; we will choose $\vec{x}$ such that
$\operatorname{Re}\left[ (i\pm1)e^{i\vec{k}\cdot\vec{x}}\right] $ has its
maximum value of 1. (The other terms in $V_{k}^{(\pm1)} Y^{(\pm1)}_{i,k}$ are
all manifestly real.)

Therefore, this particular mode has a contribution to the $0$--$i$ component of
the stress-energy tensor which includes a term
\begin{equation}
T^{0}{}_{i,k}\supset-m\bar{V}_{\theta\phi}\frac{\bar{\phi}^{\prime}}{a}
\frac{1}{2\pi}\frac{H}{4m\sqrt{|c_{1}|}}\Gamma(\nu)2^{\nu}\left(  -\tau
_{i}\right)  ^{\frac{3}{2}}e^{\left(  \nu-\frac{3}{2}\right)  N}\frac{1}
{\sqrt{2}}\delta^{3}{}_{i}.
\end{equation}
Using the slow-roll equation for $\bar{\phi}$, $3\sh\bar\phi' \approx -a^2 \bar V_\phi$, we can write this as
\begin{equation}
T^{0}{}_{i,k}\supset\frac{\bar{V}_{\phi}\bar{V}_{\theta\phi}}{24\pi
\sqrt{|c_{1}|}}\Gamma(\nu)2^{\nu-\frac{1}{2}}\left(  -\tau_{i}\right)
^{\frac{3}{2}}e^{\left(  \nu-\frac{3}{2}\right)  N}\delta^{3}{}_{i}.
\end{equation}
Comparing this to the background $0$--$0$ component of $T^{\mu}{}_{\nu}$,
$\bar{T}^{0}{}_{0}=\bar{\rho}=3H^{2}/8\pi\tilde G_{c}$, we find
\begin{equation}
\frac{T^{0}{}_{i,k}}{\bar{T}^{0}{}_{0}}\supset\frac{\tilde G_{c}\bar{V}_{\phi}\bar
{V}_{\theta\phi}}{9H^{2}\sqrt{|c_{1}|}}\Gamma(\nu)2^{\nu-\frac{1}{2}}\left(
-\tau_{i}\right)  ^{\frac{3}{2}}e^{\left(  \nu-\frac{3}{2}\right)  N}
\delta^{3}{}_{i}.\label{eq:t0icomp}
\end{equation}
Using the slow-roll Friedmann equation we could rewrite this purely in terms
of the potential as
\begin{equation}
\frac{T^{0}{}_{i,k}}{\bar{T}^{0}{}_{0}}\supset\frac{1}{24\pi\sqrt{|c_{1}|}}\frac
{\bar{V}_{\phi}\bar{V}_{\theta\phi}}{\bar{V}-\bar{\theta}\bar{V}_{\theta}
}\Gamma(\nu)2^{\nu-\frac{1}{2}}\left(  -\tau_{i}\right)  ^{\frac{3}{2}
}e^{\left(  \nu-\frac{3}{2}\right)  N}\delta^{3}{}_{i}.\label{eq:t0icomppot}
\end{equation}

The key feature here is the exponential dependence on $N$ for $\nu>3/2$ (the
condition we found above for $V_{k}^{{(\pm1)}}$ to grow exponentially in cosmic time). While
the derivatives of the potential in the numerator of \cref{eq:t0icomppot}
should be a few orders of magnitude smaller than the potential in the
denominator due to slow roll, this is likely to be dwarfed by
the exponential dependence on the number of $e$-folds, which even for the bare
minimum length of inflation, $N\sim50$--$60$, will be very large. Moreover, as
we will see in \cref{sec:dj}, $\nu$ can in principle be larger than $3/2$ even
by several orders of magnitude, hence the other terms with exponential
dependence on $\nu$, as well as the gamma function, can be quite large as well.

Therefore, when $\nu>3/2$ the vector modes will generically drive the
off-diagonal term in the stress-energy tensor far above the background
density. This does not necessarily mean that isotropy is violated. As discussed in \cref{sec:spin0perts}, the same
physical process that drives $V_{k}^{{(\pm1)}}$ will similarly pump
energy into the spin-0 piece, $V_{k}^{(0)}$, which affects the
perturbations to $\bar{T}^{0}{}_{0}$ as well as $\bar{T}^{0}{}_{i}$. Consequently, background
homogeneity and isotropy could still hold, but the slow-roll solution to the
background Friedmann equations which we perturbed would be invalid. Either way
our inflationary background becomes dominated by the perturbations.

Note that this calculation was done for a single mode, albeit one of the
largest ones available because $V_{k}$ grows for smaller $k$. Integrating over
all modes produced during inflation would of course exacerbate the
instability.

We will explore this instability in greater quantitative detail in
\cref{sec:dj}, where we examine a specific potential for which we can
elucidate the constraints on $\bar{V}_{\phi}$ and $\bar{V}_{\theta\phi}$.

\subsection{What Values Do We Expect for \texorpdfstring{$\Lambda$}{Lambda}?}

\label{sec:lambdavalues}

$V^{{(\pm1)}}$ has an effective mass-squared (\ref{eq:meffsq}) which depends on both the theory's free parameters and derivatives of the scalar potential, and can be of either sign. When it is negative, the aether modes are tachyonic and $V^\pmvec$ contains an exponentially growing mode. This occurs when the parameter $\Lambda$, defined in \cref{eq:lambdadef}, satisfies $\Lambda>2$. To leading order in the slow-roll parameters, $\Lambda$ is written in terms of several free parameters:
$c_{1},m,H,$ and the potential derivatives $\bar{V}_{\theta\phi}$ and $\bar
{V}_{\phi}$,
\begin{equation}
\Lambda \equiv -\frac{\bar{V}_{\phi}\bar{V}_{\theta\phi}}{6mc_{1}H^{3}
}+\mathcal{O}(\varepsilon).
\end{equation}
Hence $\Lambda$ can span a fairly large range of orders of magnitude. However, there are several existing constraints on these parameters, most of
which constrain several of them in terms of each other.

There are two things we can do to clarify our expression for $\Lambda$. We generally expect that for a slow-roll phase, $\ddot{\phi}\ll3H\dot{\phi}$
and $\frac{1}{2}\dot{\phi}^{2}\ll M_{\mathrm{Pl}}^{2}H^{2}$, where the Planck
mass is given as usual by
\begin{equation}
M_{\mathrm{Pl}}^{-2}=8\pi G=8\pi G_{c}\cdot\mathcal{O}(1).
\end{equation}
We will rewrite the second inequality in terms of a slow-roll
parameter, $\zeta$, as
\begin{equation}
\frac{1}{2}\dot{\phi}^{2}  =M_{\mathrm{Pl}}^{2}H^{2}\zeta, \qquad\zeta \ll1.
\end{equation}
Using the slow-roll Friedmann and Klein-Gordon equations, we
can then rewrite $\Lambda$ as
\begin{equation}
\Lambda = \operatorname{sgn}(\dot{\phi})\frac{\zeta^{1/2}}{\sqrt{2}c_{1}
}\left(  \frac{m}{M_{\mathrm{Pl}}}\right)  ^{-1}\frac{\bar{V}_{\theta\phi}}
{H}+\mathcal{O}(\varepsilon).
\end{equation}
Next, we can redefine the coupling $\bar V_{\theta\phi}$ using the flat-space stability constraint
(\ref{eq:spin0constraint}) that we derived in \cref{sec:flatspace}. Let us define a dimensionless coupling $\sigma$ by
\begin{equation}
V_{\theta\phi}^{2}(0,0)\equiv2c_{123}M_{0}^{2}\sigma,
\end{equation}
where $M_{0}^{2}=V_{\phi\phi}(0,0)$ is the effective mass-squared of the
scalar around a Minkowski background, so that the stability constraint is simply
\begin{equation}
\sigma\leq1.
\end{equation}
Therefore, we have
\begin{equation}
\Lambda = \operatorname{sgn}(\dot{\phi})\zeta^{1/2}\sigma^{1/2}\frac
{c_{s}^{(0)}}{\sqrt{c_{1}}}\frac{\bar{V}_{\theta\phi}}{V_{\theta\phi}
(0,0)}\frac{M_{0}}{H}\left(  \frac{m}{M_{\mathrm{Pl}}}\right)  ^{-1}
+\mathcal{O}(\varepsilon).\label{eq:lambda}
\end{equation}

The instability occurs when $\Lambda>2$. Let us first examine
\cref{eq:lambda} to see the conditions under which it is positive. Most of the terms are
manifestly positive. Positivity of the Hamiltonian for spin-1 perturbations in
flat space requires $c_{1}\geq0$ \cite{Lim:2004js}.\footnote{This was derived in pure \ae-theory. However, recall from
\cref{sec:flatspace} that the spin-1 modes in flat space are unaffected by
the scalar-aether coupling: spin-1 perturbations are by construction divergence-free, so $\theta$ only contains spin-0 aether perturbations. Hence $c_{1}$ must still be positive.} Tachyonic stability of the scalar requires $M_{0}$ to be
real and positive. The timelike constraint on the aether requires that $m$ be
positive as well. Putting this all together, we find
\begin{equation}
\Lambda = \operatorname{sgn}(\dot{\phi})\frac{\bar{V}_{\theta\phi}}
{V_{\theta\phi}(0,0)}\underbrace{\zeta^{1/2}
\vphantom{\frac{c_s^{(0)}}{\sqrt{c_1}}}}_{>0}\underbrace{\sigma^{1/2}
\vphantom{\frac{c_s^{(0)}}{\sqrt{c_1}}}}_{>0}\underbrace{\frac{c_{s}^{(0)}
}{\sqrt{c_{1}}}}_{>0}\underbrace{\frac{M_{0}}{H}
\vphantom{\frac{c_s^{(0)}}{\sqrt{c_1}}}}_{>0}\underbrace{\left(  \frac
{m}{M_{\mathrm{Pl}}}\right)  ^{-1}\vphantom{\frac{c_s^{(0)}}{\sqrt{c_1}}}}
_{>0}+\mathcal{O}(\varepsilon),
\end{equation}
implying that in order for $\Lambda$ to be positive, $\dot{\phi}$ and the
coupling $\bar{V}_{\theta\phi}$ need to have the same sign. This is not
difficult to achieve in practice; in the quadratic potential of
\rcite{Donnelly:2010cr} (see \cref{sec:dj} for more discussion), it amounts to
requiring that $\phi$ and $\bar{V}_{\theta\phi}$ have opposite signs, which is
true for a large space of initial conditions leading to inflating trajectories.

Next we need to see under which conditions $|\Lambda|$ can be $\mathcal{O}(1)$
or greater. We have assumed that the scalar slow-roll parameter $\zeta$ is
small. In particular, in the absence of the aether, $\zeta$ is equal to
$\varepsilon$, which observations constrain to be $\sim\mathcal{O}(10^{-2})$ \cite{Ade:2013zuv}. It
therefore seems sensible that $\zeta^{1/2}$ should be small but not terribly
small, perhaps $\sim\mathcal{O}(10^{-1})$ or so.

Similarly, the scalar-aether coupling, $\sigma$, is constrained by the flat-space stability of the spin-0 modes to be strictly less than 1. However,
we do not want to consider couplings so small as to be uninteresting, so we
may choose the coupling to be as close to $\sigma=1$ as is allowed. Therefore,
$\sigma^{1/2}$ ought to be smaller, but need not be too much smaller, than 1.

Written in the form of \cref{eq:lambda}, the value of $\Lambda$ is sensitive
to how $V_{\theta\phi}$ and $V_{\phi\phi}$ differ between a quasi-de Sitter
inflationary background and a Minkowski background. In the quadratic potential
$V(\theta,\phi)=\frac{1}{2}M^{2}\phi^{2}+\mu\theta\phi$ which we discuss in
\cref{sec:dj}, both of these are constant, although one could construct inflationary potentials for which this is not true. The effective
mass of the scalar during inflation, $M=\bar{V}_{\phi\phi}^{1/2}$, should be
less than the Hubble rate in order to produce perturbations. Putting all this
together, we are left with
\begin{equation}
\Lambda = \operatorname{sgn}(\dot{\phi})\underbrace{\zeta^{1/2}
\vphantom{\frac{c_s^{(0)}}{\sqrt{c_1}}}}_{<1}\underbrace{\sigma^{1/2}
\vphantom{\frac{c_s^{(0)}}{\sqrt{c_1}}}}_{\leq1}\underbrace{\frac{c_{s}^{(0)}
}{\sqrt{c_{1}}}}_{\mathcal{O}(1)}\underbrace{\frac{\bar{V}_{\theta\phi}
}{V_{\theta\phi}(0,0)}\frac{M_{0}}{M}}_{\mathcal{O}(1)?}\underbrace{\frac
{M}{H}\vphantom{\frac{c_s^{(0)}}{\sqrt{c_1}}}}_{\ll1}\underbrace{\left(
\frac{m}{M_{\mathrm{Pl}}}\right)  ^{-1}
\vphantom{\frac{c_s^{(0)}}{\sqrt{c_1}}}}_{\gg1?}+\mathcal{O}(\varepsilon). \label{eq:lambdaoom}
\end{equation}
We can see that in order for $\Lambda$ to be larger than 2, the aether VEV,
$m$, needs to be at least a few orders of magnitude smaller than the Planck
scale. $m$ is effectively the Lorentz symmetry-breaking mass scale. It can
therefore be quite a bit smaller than the Planck mass, although if it were below the scale of collider experiments, any couplings to matter could displace the aether from its VEV and Lorentz-violating effects could be visible.

There are several experimental and observational results suggesting that $m/M_\mathrm{Pl}$ should be quite small. Here we briefly discuss three strong constraints, arising from big bang nucleosynthesis, solar-system tests, and the absence of gravitational \v{C}erenkov radiation, as well as a possible caveat.

As mentioned in \cref{sec:mg-aether}, the gravitational constant appearing in the Friedmann equations, $\gc$, and the gravitational constant appearing in the Newtonian limit, $G_{N}$, are both displaced from the ``bare'' gravitational constant, $G$, by a factor that is, schematically, $1+c_i(m/M_\mathrm{Pl})^2$. The primordial abundances of light elements such as helium and deuterium probe the cosmic expansion rate during big bang nucleosynthesis, which depends on $\gc$ through the Friedmann equations. Therefore, by comparing this to $G_N$ measured on Earth and in the solar system, $c_im^2$ can be constrained. Assuming the $c_i$ are $\mathcal{O}(1)$,\footnote{As the $c_i$ are dimensionless parameters, this is perfectly reasonable. Note that even if $m$ were order $M_\mathrm{Pl}$ or larger and the constraints discussed in here are actually constraints on the smallness of the $c_i$, $\Lambda$ still depends on these parameters as $c_1^{-1/2}$.} the BBN constraint implies $m/M_\mathrm{Pl}\lesssim10^{-1}$ \cite{Carroll:2004ai}.

Slightly better constraints on $\gc/G_N$ come from the cosmic microwave background (CMB) \cite{Robbers:2007ca,Bean:2010zq}. The tightest bound, $|G_N/\gc-1|<0.018$ at $95\%$ confidence level, was computed using CMB data (WMAP7 and SPT) and the galaxy power spectrum (WiggleZ) in a theory closely related to the one described in this chapter, and should hold generally for \ae-theory at the order-of-magnitude level \cite{Audren:2013dwa}. These constrain $m/M_\mathrm{Pl}$ to be no greater than a few percent.

There are yet stronger bounds on $m/\Mp$ through constraints on the preferred-frame parameters, $\alpha_{1,2}$, in the parametrised post-Newtonian (PPN) formalism. These coefficients scale, to leading order, as $c_i(m/M_\mathrm{Pl})^2$ \cite{Foster:2005dk,Jacobson:2008aj}. The observational bounds $\alpha_1\lesssim10^{-4}$ and $\alpha_2\lesssim4\times10^{-7}$ therefore imply $m/M_\mathrm{Pl} \lesssim 6\times10^{-4}$. Recent pulsar constraints on $\alpha_{1,2}$ are even stronger than this \cite{Shao:2013wga}, although they are derived in the strong-field r\'egime and thus might not be directly applicable to the weak-field \ae-theory results. Similarly, recent binary pulsar constraints on Lorentz violation \cite{Yagi:2013qpa} constrain $m/\Mp\lesssim10^{-1}$, assuming $c_i \sim\mathcal{O}(1)$.

The strongest constraints come from the absence of ``gravitational \v{C}erenkov radiation.'' Because the aether changes the permeability of the vacuum, coupled aether-graviton modes may travel subluminally, despite being nominally massless. Consequently, high-energy particles moving at greater speeds can emit these massless particles, in
analogy to the usual \v{C}erenkov radiation. This emission causes high-energy
particles to lose energy, and at an increasing rate for higher-energy particles.
Among the highest-energy particles known are cosmic rays, which travel
astronomical distances and hence could degrade drastically due to
such gravitational \v{C}erenkov effects. Such a degradation has, however, not been observed; this generically constrains
$m/M_{\mathrm{Pl}}<3\times10^{-8}$ \cite{Elliott:2005va}.

We should note that these constraints can be side-stepped if certain convenient exact relationships hold among the $c_{i}$, although crucially they cannot all be avoided in this way simultaneously without allowing for superluminal propagation of the aether modes \cite{Jacobson:2008aj}. The PPN parameters $\alpha_{1,2}$ are identically zero when $c_3=0$ and $2c_1=-3c_2$. The BBN constraint is automatically satisfied by requiring $2c_1 + 3c_2 + c_3$ to vanish, as this sets $\gc=G_N$ \cite{Carroll:2004ai}. Note that the PPN cancellations imply the BBN cancellation, though the reverse is not necessarily true.\footnote{The conditions for PPN and BBN to cancel can be relaxed by including a $c_4$ term which describes a quartic aether self-interaction. We have ignored such a term in order to simplify the theory, although like the other three terms, it is permitted when that the aether equations of motion are demanded to be second order in derivatives. When $c_4 \neq0$, the vanishing of $\alpha_{1,2}$ continues to imply that the BBN constraints are satisfied.} The \v{C}erenkov constraints vanish if all five dynamical gravitational (metric and aether) degrees of freedom propagate exactly luminally. This happens when $c_3=-c_1$ and $c_2=c_1/(1-2c_1)$ \cite{Elliott:2005va}. Note that while $\alpha_2=0$ in this parameter subspace, $\alpha_1 = -8c_1(m/M_\mathrm{Pl})^2$, which would place a constraint on $m/M_\mathrm{Pl}$ of order $10^{-2}$. It is worth mentioning that the \v{C}erenkov constraints on $m$ will also be avoided if the mode speeds for some of the aether-metric modes are superluminal. This includes a two-dimensional parameter subspace in which the PPN and BBN constraints are automatically satisfied \cite{Jacobson:2008aj}. Whether superluminal propagation is acceptable in \ae-theory is somewhat controversial. It is a metric theory of gravity, so superluminality should imply violations of causality, including propagation of energy around closed timelike curves \cite{Elliott:2005va,Lim:2004js}. However, this may be seen as an \textit{a posteriori} demand, and some authors (see, e.g., \rcite{Jacobson:2008aj}) do not require it.

It is unclear what fundamental physical principle, if any, would cause the $c_i$ to cancel in any of the aforementioned ways. Hence it seems to be a fairly general result that $m$ must be several orders of magnitude below the Planck scale. If $m/M_\mathrm{Pl}$ is small enough compared to $M/H$ and the other small parameters appearing in \cref{eq:lambdaoom}, $\Lambda$ can easily be above 2 and the aether-inflaton coupling runs a serious danger of causing an instability. For a given $m/\Mp$, this places a constraint on the size of the coupling $\bar V_{\theta\phi}$. We will discuss this constraint more quantitatively in \cref{sec:dj} for a specific choice of the potential.

\section{Spin-0 Cosmological Perturbations: Instability and Observability}
\label{sec:spin0perts}

Let us now consider the spin-0 perturbations. Before getting bogged down in calculational details, we first summarise this section. The spin-0 equations
are complicated by the addition of $\delta \phi $ modes which add a new degree of freedom. In order to tackle
these equations, we use the smallness of $m/M_{\mathrm{Pl}}$, discussed in 
\cref{sec:lambdavalues}, to solve the perturbations order-by-order, along
the lines of the approach in \rcite{Blas:2011en}. At lowest order in $m/M_{
\mathrm{Pl}}$, the perturbations $\Phi$ and $\delta \phi$ have the same
solutions as in the standard slow-roll inflation in general relativity.
These can be substituted into the $\xi$ equation of motion to solve for $
\xi $ at lowest order, which we then substitute back into the $\Phi $ and $
\delta \phi $ equations at $\mathcal{O}(m/M_{\mathrm{Pl}})$.

The instability found in the spin-1 perturbations reappears, and occurs in
essentially the same region of parameter space. We then assume that the
parameters are such that this instability is absent, in which case $\xi$ is
roughly constant. We solve for the metric perturbation $\Phi$ and find that
neither its amplitude nor its scale-dependence are significantly changed from
the standard slow-roll case. In particular, we calculate two key
inflationary observables: the scale-dependence of the $\Phi$ power spectrum, 
$n_s$, and the tensor-to-scalar ratio, $r$.

Surprisingly, the first corrections due to the aether-scalar coupling enter
at $\mathcal{O}(m^2/M_\mathrm{Pl}^2)$. Up to first order in $m/M_\mathrm{Pl}$,
the aether-scalar coupling has no effect on cosmic perturbations on
superhorizon scales, assuming that $m/M_\mathrm{Pl}$ is small compared to
unity and that the perturbations are produced during a slow-roll quasi-de
Sitter phase. A corollary of this is that superhorizon isocurvature modes, a
generic feature of coupled theories, are not produced by the aether-scalar
coupling up to $\mathcal{O}(m^2/M_\mathrm{Pl}^2)$. Because of the smallness of 
$m/M_\mathrm{Pl}$, any deviations to $n_s$ and $r$ caused by the
aether-scalar coupling are unobservable to the present and near-future
generations of CMB experiments.

Since the pure \ae-theory terms in the perturbed Einstein equations carry
two powers of $u^{\mu }$ (which is proportional to $m$) and so only begin to
contribute at $\mathcal{O}(m^2/M^2_{\mathrm{Pl}})$, we will not recover the
cosmological perturbation results of pure \ae-theory by taking any limits,
as we only work in this section to $\mathcal{O}(m/M_{\mathrm{Pl}})$. The
effects of \ae-theory on the spin-0 perturbations are mild, amounting
essentially to a rescaling of the power spectrum amplitude that is $\mathcal{O}(m^2/M^2_{\mathrm{Pl}})$ and is degenerate with $\varepsilon $ \cite
{Lim:2004js}.

\subsection{The Spin-0 Equations of Motion}

In order to eliminate nonphysical degrees of freedom, we need to specify a
choice of coordinate system with no remaining gauge freedom. We choose to
work in Newtonian gauge, where $B^{(0)}=H_{T}^{(0)}=0$. The equations of
motion are relatively simple in this gauge, and the perturbation $\Phi $ has
a simple interpretation as the relativistic generalisation of the Newtonian
gravitational potential \cite{Mukhanov:1990me}. Hereafter we will drop the
spin-0 superscripts.

The $0$--$0$, $0$--$i$, and $i$--$i$ Einstein equations, respectively, are 
\begin{align}
4\pi\tilde G_c\left(-\bar\phi'\delta\phi'-a^2\bar V_\phi\delta\phi\right) &= \left(3\sh^2 - A\right)\Phi - 3\mathcal{H}\Psi' - \frac{\tilde G_c}{G}k^2\Psi - 8\pi\tilde G_cc_1m^2k^2\Phi \nonumber \label{eq:spin000} \\
&\hphantom{{}=} + 8\pi\tilde G_cc_1m^2k(V'+\mathcal{H}V) - 8\pi\tilde G_c\tilde\alpha\mathcal{H}kV \nonumber \\
&\hphantom{{}=} + 4\pi\tilde G_cma\bar V_{\theta\phi}(\bar\phi'\Phi-3\mathcal{H}\delta\phi) \\
\frac{1}{8\pi G}(k\mathcal{H}\Phi-k\Psi') &= \frac{k}{2}\bar\phi'\delta\phi - \tilde\alpha AV +c_1m^2a^{-1}(ak\Phi)' \nonumber \label{eq:spin00i} \\
&\hphantom{{}=} - c_1m^2\frac{\xi''}{a} + \frac{1}{2}ma\bar V_{\theta\phi}\bar\phi'V \\
4\pi\tilde G_c\left(\bar\phi'\delta\phi'-a^2\bar V_\phi\delta\phi\right) &= \left(3\sh^2 - A\right)\Phi + \mathcal{H}\Phi' - 2\mathcal{H}\Psi' - \Psi'' - \frac{8\pi\tilde G_cm^2}{\gamma}\tilde c_{123}k^2(\Phi+\Psi) \label{eq:spin0ij} \nonumber \\
&\hphantom{{}=} + 4\pi\tilde G_c\frac{3m^3}{a}A\bar V_{\theta\theta\theta}\left(3\Psi^\prime-3\mathcal{H}\Phi+kV\right). \nonumber \\
&\hphantom{{}=} - 4\pi\tilde G_cma\left[\bar V_{\theta\phi}(3\mathcal{H}\delta\phi+\delta\phi') + \bar V_{\theta\phi\phi}\bar\phi'\delta\phi\right] \nonumber \\
&\hphantom{{}=} + 4\pi\tilde G_cm^2\bar V_{\theta\theta\phi}\left[3A\delta\phi-\bar\phi'(3\Psi'-3\mathcal{H}\Phi+kV)\right].
\end{align}
The off-diagonal $i$--$j$ Einstein equation, unmodified by the coupling between
the aether and scalar, gives a constraint, 
\begin{equation}
k^{2}(\Phi +\Psi )=\gamma a^{-2}(a^{2}kV)^{\prime },  \label{eq:offdiagspin0}
\end{equation}
where $\gamma \equiv 16\pi Gm^{2}c_{13}$ was defined in \cref{sec:spin1perts}. We may eliminate $\Psi $ and its derivatives by the constraint (\ref{eq:offdiagspin0}) and its conformal-time derivatives, 
\begin{align}
\Psi' &= \gamma(ak)^{-1}\left(\xi''-A\xi\right)-\Phi', \label{eq:offdiagspin0der1} \\
\Psi'' &= \gamma(ak)^{-1}\left(\xi'''-\mathcal{H}\xi''-A\xi'+A\xi\left(\mathcal{H}-\frac{A'}{A}\right)\right)-\Phi'', \label{eq:offdiagspin0der2}
\end{align}
where, as for the spin-1 perturbations, we have defined $\xi \equiv aV$ and $
A\equiv{\mathcal{H}}^{2}-{\mathcal{H}}^{\prime }$. Note the presence of third
derivatives of $\xi $ in the expression for $\Psi ^{\prime \prime }$, which
could severely complicate the Einstein equations at $\mathcal{O}(m^2/M_{\mathrm{Pl}}^2)$.

Finally, the $\nu =i$ aether equation of motion is, using 
\cref{eq:offdiagspin0,eq:offdiagspin0der1,eq:offdiagspin0der2}, 
\begin{equation}
\xi ^{\prime \prime }+\frac{\tilde{c}_{123}m^{2}}{c_{1}m^{2}+\tilde{\alpha}
\gamma }k^{2}\xi +\left( \frac{\tilde{\alpha}(1-\gamma )A-\frac{1}{2}ma\bar{V
}_{\theta \phi }\bar{\phi}^{\prime }}{c_{1}m^{2}+\tilde{\alpha}\gamma }
\right) \xi =\frac{c_{1}m^{2}+\tilde{\alpha}}{c_{1}m^{2}+\tilde{\alpha}
\gamma }k(a\Phi )^{\prime }-\frac{1}{2}\frac{ma^{2}\bar{V}_{\theta \phi }}{
c_{1}m^{2}+\tilde{\alpha}\gamma }k\delta \phi  \label{eq:xispin0eom}
\end{equation}
where, as before, tildes indicate the usual \ae-theory constants modified
by appropriate factors of $\frac{1}{2}\bar{V}_{\theta \theta }$.

We can perform a consistency check by observing that these reduce to $\delta
T^{\mu }{}_{\nu }$ for a single scalar field in general relativity \cite
{Mukhanov:1990me} when the aether is turned off (in the limit $m\rightarrow
0 $), as well as $\delta T^{\mu }{}_{\nu }$ and the $\xi $-equation of
motion in \ae-theory \cite{Lim:2004js} in the limit $V(\theta ,\phi
)\rightarrow V(\phi )$.

\subsection{The Instability Returns}

\label{sec:spin0instability}

To lowest order in $m/M_{\mathrm{Pl}}$, the constraint equation (\ref
{eq:offdiagspin0}) tells us simply that the anisotropic stress vanishes: $
\Psi =-\Phi $. Taking this into account, the $0$--$i$ Einstein equation at
lowest order in $m/M_{\mathrm{Pl}}$ is 
\begin{equation}
\left( a\Phi \right) ^{\prime }=4\pi Ga\bar{\phi}^{\prime }\delta \phi .
\label{eq:0iozero}
\end{equation}
The $\nu =i$ aether equation of motion (\ref{eq:xispin0eom}) is, dropping
terms of $\mathcal{O}(m^{2}/M_{\mathrm{Pl}}^{2})$, 
\begin{equation}
\xi ^{\prime \prime }+c_{s}^{(0)2}k^{2}\xi -\frac{a\bar{V}_{\theta \phi }
\bar{\phi}^{\prime }}{2mc_{1}}\xi =\left( 1+\frac{\alpha }{c_{1}m^{2}}
\right) k(a\Phi )^{\prime }-\frac{a^{2}\bar{V}_{\theta \phi }k}{2mc_{1}}
\delta \phi
\end{equation}
where 
\begin{equation}
c_{s}^{(0)2}=\frac{{c}_{123}m^{2}}{c_{1}m^{2}+\alpha \gamma }=\frac{c_{123}}{
c_{1}}\left(1+\mathcal{O}\left( \frac{m^2}{M^2_{\mathrm{Pl}}}\right)\right)
\end{equation}
is the same spin-0 sound speed as in flat space (cf. \cref{sec:flatspace})
to first order in $m/M_{\mathrm{Pl}}$. In de Sitter space this becomes,
using \cref{eq:0iozero} to replace $(a\Phi )^{\prime }$ with $\delta \phi $, 
\begin{equation}
\xi ^{\prime \prime }+c_{s}^{(0)2}k^{2}\xi -\frac{\bar{V}_{\theta \phi }\dot{
\bar{\phi}}}{2mc_{1}H^{2}}\frac{\xi }{\tau ^{2}}=\frac{k}{H^{2}}\left[
\left( 1+\frac{\alpha }{c_{1}m^{2}}\right) 4\pi G\dot{\bar{\phi}}-\frac{1}{2}
\frac{\bar{V}_{\theta \phi }}{mc_{1}}\right] \frac{\delta \phi }{\tau ^{2}},
\label{eq:spin0xieom}
\end{equation}
to lowest order in the slow-roll parameters and $m/M_{\mathrm{Pl}}$.

Combined with the perturbed Klein-Gordon equation, $\xi $ and $\delta \phi $
obey coupled oscillator equations. However, to zeroth order in $m/M_{\mathrm{
Pl}},$ the scalar field is unaffected by the aether perturbations,\footnote{The aether coupling will still enter the perturbed Klein-Gordon equation at
this order through the potential terms.} so on superhorizon scales $\delta
\phi $ is constant up to slow-roll corrections, resulting in the standard
nearly scale-invariant power spectrum. This is consistent with the flat-space case discussed in \cref{sec:flatspace}, where it was found that the coupling to the aether does not destabilise the scalar modes. Taking $\delta\phi$ to be constant and restricting to superhorizon scales, $
c_{s}^{(0)}k\tau \ll 1$, \cref{eq:spin0xieom} is solved by
\begin{equation}
\xi =C_{+}\tau ^{n_{+}}+C_{-}\tau ^{n_{-}}+k\delta \phi \left[ \dot{\bar{\phi
}}^{-1}-\left( 1+\frac{\alpha }{c_{1}m^{2}}\right) \frac{m}{M_{\mathrm{Pl}}}
\frac{c_{1}}{\bar{V}_{\theta \phi }M_{\mathrm{Pl}}}\right] , \label{eq:spin0xisol}
\end{equation}
where $C_{\pm }$ are arbitrary constants, and 
\begin{equation}
n_{\pm }=\frac{1}{2}\pm \sqrt{\frac{1}{4}+\frac{\bar{V}_{\theta \phi }\dot{
\bar{\phi}}}{2mc_{1}H^{2}}}.
\end{equation}
As with the spin-1 perturbations, the spin-0 piece of $V=\xi /a$ can either
grow or decay exponentially (in cosmic time). In this case it will grow if 
\begin{equation}
\frac{\bar{V}_{\theta \phi }\dot{\bar{\phi}}}{2mc_{1}H^{2}}>2.
\end{equation}
This is exactly the same as the condition, $\Lambda >2$, for the spin-1
modes to be unstable. The real condition for instability may be slightly
different, as $\Lambda >2$ could violate our assumption that $m/M_{\mathrm{Pl
}}$ is small; however, the additional $\mathcal{O}(m^2/M^2_{\mathrm{Pl}})$
terms would only change some multiplicative factors, and not by orders
of magnitude.

As in the spin-1 case, we can most easily see the effect of unstable aether
modes on the metric perturbations through the off-diagonal $i$--$j$ Einstein
equation (\ref{eq:offdiagspin0}). If $V$ blows up exponentially then so will 
$\Phi +\Psi $, and the metric perturbations will overwhelm the
FLRW background.

\subsection{The Small-Coupling Limit}

From here on we will assume that the aether perturbations are stable, so that
\begin{equation}
\Lambda \equiv \frac{\bar{V}_{\theta \phi }\dot{\bar{\phi}}}{2mc_{1}H^{2}}<2.
\end{equation}
This can be further split into two dominant cases, $|\Lambda|\ll1$ and $\Lambda<-1/4$. There are regions in parameter space which are not covered by these cases, such as $\Lambda \sim 1$, but these are likely to be highly fine-tuned as many of the parameters which enter $\Lambda$ have no relationship to each other \textit{a priori}. Consequently we should consider various values of $\Lambda$ on an order-of-magnitude basis.

$|\Lambda| \ll 1$ corresponds to the limit where the coupling $|\bar V_{\theta\phi}|$ is small compared to the mass scale $c_1mH^2/\dot{\bar\phi}$. Assuming that the background relations for the slow-roll parameters hold as in GR (which we will explore more rigorously in \cref{sec:dj} for a particular potential), then we have $\varepsilon = 4\pi G\dot{\bar{\phi}}^{2}/H^{2}$ up to $\mathcal{O}(m/\Mp)$, and this limit can be written as
\begin{equation}
\frac{|\bar V_{\theta\phi}|}{H} \ll \frac{c_1}{\sqrt\varepsilon}\frac{m}{\Mp}.
\end{equation}
In this limit, the term $C_- \tau^{n_-}$ in \cref{eq:spin0xisol} is constant up to slow-roll corrections, as is the term proportional to $\delta\phi$, while $C_+ \tau^{n_+}$ decays as $\tau$. 

This case should be qualitatively similar to \ae-theory as it makes the aether-scalar coupling very small. However, we might be worried by the appearance of a $\bar V_{\theta\phi}^{-1}$ in \cref{eq:spin0xisol}. The limit $\bar V_{\theta\phi} \to 0$ does smoothly go to \ae-theory. The aether perturbation $\xi$ only appears, to $\mathcal{O}(m/\Mp)$, in the $0$--$i$ Einstein equation,
\begin{equation}
\Phi' + \sh\Phi = 4\pi G \bar\phi' \left(\delta\phi + \frac{m\bar V_{\theta\phi}}{k}\xi\right).
\end{equation}
The $\bar V_{\theta\phi}$ in the $\mathcal{O}(m/\Mp)$ term will cancel out the problematic $\bar V_{\theta\phi}^{-1}$ in the solution for $\xi$. Taking $\Lambda \to 0$ and substituting in the solution (\ref{eq:spin0xisol}), this becomes
\begin{equation}
\Phi' + \sh\Phi \approx 4\pi G\bar\phi'\delta\phi\left[1 - \frac{m^2}{\Mp^2}c_1\left(1+\frac{\alpha}{c_1m^2}\right)\right].
\end{equation}
The corrections enter at $\mathcal{O}(m^2/\Mp^2)$ and are negligible for the purposes of this analysis. Therefore the limit $|\Lambda| \ll 1$ should only differ from \ae-theory at $\mathcal{O}(m^2/\Mp^2) \lesssim 10^{-15}$.

It is worth mentioning that for small but finite $\Lambda$ there will be new effects on extremely large scales, $k \lesssim \bar V_{\theta\phi}$. These may or may not be observable, depending on the scales covered during inflation.

\subsection{The Large-Coupling Limit: The \texorpdfstring{$\Phi$}{Phi} Evolution Equation}

One interesting case is left: a large coupling with opposite sign to $\dot{\bar\phi}$, or $\Lambda < 1/4$. We will consider this for the rest of this section. However, we should mention that the sign of $\dot{\bar\phi}$ depends on initial conditions, and if this sign condition were not satisfied, then (as discussed in \cref{sec:spin0instability}) the aether-scalar coupling would drive a severe tachyonic instability. Hence such a large coupling may not be an ideal part of a healthy inflationary theory.

In this large-coupling case, both of the $\tau ^{\pm }$ terms in \cref{eq:spin0xisol} are decaying and we are left with
\begin{align}
\xi &= \frac{k\delta \phi }{\dot{\bar{\phi}}} \left(1+\mathcal{O}\left( \frac{m}{M_{
\mathrm{Pl}}}\right)\right) \label{eq:xisolspin0}\\
&\approx \frac{\sqrt{4\pi G}}{H\varepsilon ^{1/2}}k\delta \phi,
\end{align}
where we have dropped the $\mathcal{O}(m/\Mp)$ terms in the coefficient of the $\delta\phi$ term. Recall that \cref{eq:spin0xisol,eq:xisolspin0} have been derived for superhorizon perturbations in the slow-roll limit. Hence we will focus our analysis on superhorizon scales, and while we will leave the behaviour of the scale factor unspecified in this subsection, it is worth keeping in mind that our results may not be valid in spacetimes that are not quasi-de Sitter. Using this solution for $\xi$, we can write the $0$--$i$ Einstein equation to $\mathcal{O}(m/M_{\mathrm{Pl}})$ as 
\begin{equation}
\Phi ^{\prime }+{\mathcal{H}}\Phi =4\pi G\left( \bar{\phi}^{\prime }+ma\bar{V
}_{\theta \phi }\right) \delta \phi .  \label{eq:0iom1}
\end{equation}
It is an interesting result that we can write the $0$--$i$ Einstein equation in geometrical terms as 
\begin{equation}
\Phi ^{\prime }+{\mathcal{H}}\Phi =A\delta \phi /\bar{\phi}^{\prime }
\label{eq:0iom1Aform}
\end{equation}
to both zeroth and first order in $\mathcal{O}(m/M_{\mathrm{Pl}})$. This
does not hold, however, to higher orders, and might not hold away from
quasi-de Sitter space or on subhorizon scales.

Next we solve the metric perturbation $\Phi $ to $\mathcal{O}(m/M_{\mathrm{Pl
}})$. Our master equation is the sum of the $0$--$0$ and $i$--$i$ Einstein
equations, dropping a $k^{2}\Phi $ term which is negligible on superhorizon
scales, 
\begin{align}
-8\pi Ga^{2}\bar{V}_{\phi }\delta \phi ={}& \Phi ^{\prime \prime }+6{
\mathcal{H}}\Phi ^{\prime }+2\left( 3{\mathcal{H}}^{2}-A\right) \Phi  \notag
\\
& +4\pi Gma\bar{V}_{\theta \phi }\left( \bar{\phi}^{\prime }\Phi -6{\mathcal{
H}}\delta \phi -\delta \phi ^{\prime }\right)  \notag \\
& -4\pi Gma\bar{V}_{\theta \phi \phi }\bar{\phi}^{\prime }\delta \phi + \ldots  \label{eq:00+ii}
\end{align}
where we have dropped terms of $\mathcal{O}(m^2/\Mp^2)$ and higher.

\Cref{eq:00+ii} becomes an evolution equation for $\Phi$ after using \cref{eq:0iom1} to rewrite the pieces proportional to $\delta\phi$ and $\delta\phi'$ in terms of $\Phi$ and its derivatives. We can also use the background equations of motion to remove $\bar V_\phi$ in favour of other coefficients appearing in \cref{eq:00+ii}. For this latter step, we start with the
background relation (using \cref{eq:friedmann,eq:friedmann2} and assuming $\phi$ is gravitationally dominant) 
\begin{equation}
A={\mathcal{H}}^{2}-{\mathcal{H}}^{\prime }=4\pi G\left( \bar{\phi}^{\prime
2}+ma\bar{V}_{\theta \phi }\bar{\phi}^{\prime }\right) .
\end{equation}
Taking the conformal-time derivative, we find (dropping $\mathcal{O}(m^2/M_{\mathrm{Pl}}^2)$ terms, as we do throughout) that 
\begin{equation}
\frac{A^{\prime }}{A}=\left( 2-\frac{ma\bar{V}_{\theta \phi }}{\bar{\phi}
^{\prime }}\right) \frac{\bar{\phi}^{\prime \prime }}{\bar{\phi}^{\prime }}+
\frac{ma\bar{V}_{\theta \phi }{\mathcal{H}}}{\bar{\phi}^{\prime }}+ma\bar{V}
_{\theta \phi \phi }.
\end{equation}
Using the background Klein-Gordon equation, we obtain an expression for $\bar V_\phi$ which includes contributions from the aether-scalar coupling,
\begin{equation}
-2a^{2}\bar{V}_{\phi }=\left( \frac{A^{\prime }}{A}+4{\mathcal{H}}\right) 
\bar{\phi}^{\prime }+ma\bar{V}_{\theta \phi }\left( \frac{1}{2}\frac{
A^{\prime }}{A}-{\mathcal{H}}\right) -ma\bar{V}_{\theta \phi \phi }\bar{\phi}
^{\prime }.  \label{eq:AVeq}
\end{equation}
In deriving the previous two expressions we have made use of the assumption that
\begin{equation}
 ma\frac{\bar V_{\theta\phi}}{\bar\phi'} \sim \varepsilon^{-1/2}\frac m\Mp \frac{\bar V_{\theta\phi}}{H} \ll 1.
\end{equation}

We can immediately use \cref{eq:AVeq,eq:0iom1} to remove $\bar{V}_{\phi }$ and the $\delta \phi$ terms from \cref{eq:00+ii}. We can also take the conformal-time derivative of \cref{{eq:0iom1}} to find, using \cref{eq:AVeq,eq:0iom1}, an expression for $\delta \phi'$,
\begin{equation}
4\pi G\bar{\phi}^{\prime }\delta \phi ^{\prime }=\Phi ^{\prime \prime
}+\left( {\mathcal{H}}-\frac{1}{2}\frac{A^{\prime }}{A}\right) \Phi ^{\prime
}+\left( {\mathcal{H}}^{2}-\frac{1}{2}\frac{A^{\prime }}{A}{\mathcal{H}}
-A\right) \Phi ,
\end{equation}
where we have dropped the $\mathcal{O}(m/M_{\mathrm{Pl}})$ term as $\delta
\phi ^{\prime }$ only appears in \cref{eq:00+ii} at that order. Using these relations, as well as the definition of $A$, the sum of the $0$--$0$
and $i$--$i$ perturbed Einstein equations (\ref{eq:00+ii}) becomes 
\begin{align}
&\Phi^{\prime \prime }+ \left(2{\mathcal{H}} - \frac{A^{\prime }}{A}
\right)\Phi^{\prime }+ \left(2{\mathcal{H}}^2 -2A - \frac{A^{\prime }}{A}{
\mathcal{H}}\right)\Phi  \notag \\
={}&\frac{ma\bar V_{\theta\phi}}{\bar\phi^{\prime }}\left[\Phi^{\prime
\prime }+\left(2{\mathcal{H}}-\frac{A^{\prime }}{A}\right)\Phi^{\prime
}+\left(2{\mathcal{H}}^2-2A-\frac{A^{\prime }}{A}{\mathcal{H}}\right)\Phi
\right].
\end{align}
Simplifying, we find the evolution equation for $\Phi$ to $\mathcal{O}(m/M_
\mathrm{Pl})$, 
\begin{equation}
\Phi'' + \left(2\sh - \frac{A'}{A}\right)\Phi' + \left(2\sh^2 -2A -
\frac{A'}{A}\sh\right)\Phi = 0. \label{eq:phievol}
\end{equation}

This is a surprising result. This is exactly the equation obeyed by $\Phi $
in single-field slow-roll inflation in the absence of a coupling to any
other fields \cite{Mukhanov:1990me}. Coupling to new fields generically
introduces source terms to this equation, signalling the introduction of
isocurvature modes. We have therefore shown that, to first order in $m/M_{\mathrm{Pl}}$, the scalar-aether coupling does not produce any isocurvature modes on superhorizon scales during slow-roll inflation.

What would happen if we included higher-order terms? The pure \ae-theory
terms do not change \cref{eq:phievol} \cite{Lim:2004js}. This is
understandable because the aether tracks the background energy density,
precluding the production of isocurvature modes. However, we have introduced
new coupling terms in the Einstein equations at $\mathcal{O}(m^2/M_{\mathrm{Pl}}^2)$ and higher which could potentially produce isocurvature modes. It is currently unclear whether the unusual cancellations that led to the result (\ref{eq:phievol}) will hold at these orders.

The solution to \cref{eq:phievol} is well-known \cite{Mukhanov:1990me}, 
\begin{equation}
\Phi = C\left(1 - \frac{{\mathcal{H}}}{a^2}\int a^2d\tau\right),
\label{eq:phisoln}
\end{equation}
where $C$ is a constant. The remarkable fact that the $0$--$i$ Einstein
equation can be written in the form (\ref{eq:0iom1Aform}) to either zeroth
or first order in $m/M_\mathrm{Pl}$ means that to first order, the
relationship between $\Phi$ and $\delta\phi$ is the same as in the case
without the aether. Using \cref{eq:phisoln} to find $a^{-1}(a\Phi)^{\prime }$
and plugging that into \cref{eq:0iom1Aform}, we can determine the constant $
C $, 
\begin{equation}
C = \frac{aH}{\bar\phi'}\delta\phi,
\end{equation}
exactly as in general relativity.

The amplitude of $\delta \phi $ is determined by quantising it in a
(quasi-)de Sitter background on subhorizon scales, $k\gg aH$, and imposing a
Bunch-Davies vacuum state. The variable $\delta \phi $ is coupled to the spin-0 aether
perturbations, as discussed in \cref{sec:flatspace}, and its dispersion
relation is modified by $\mu \equiv V_{\theta \phi }(0,0)$. However, the flat-space stability condition constrains this to be less that the flat-space mass of the scalar, $M_0 \equiv V_{\phi \phi }^{1/2}(0,0)$, up to an $
\mathcal{O}(1)$ factor. Therefore, if the initial conditions are set at
scales $k\gg M_0$ (which follows from $k\gg aH$ since $M_0 \ll H$), then $k\gg
\mu $ as well, and the scalar at these scales behaves as it does in the case
with no aether. We therefore see that the scalar and metric perturbations, $\delta
\phi $ and $\Phi $, are exactly the same as in general relativity up to $\mathcal{
O}(m/M_{\mathrm{Pl}})$.

\subsection{The Large-Coupling Limit: CMB Observables}

Let us finally connect these calculations to observations. As mentioned
at the beginning of this section, the two key inflationary observables currently accessible to CMB
experiments are the spectral index of the primordial power spectrum, $n_s$,
and the tensor-to-scalar ratio, $r$.

We have seen that, surprisingly, neither of these will be affected by the
aether-scalar coupling at $\mathcal{O}(m/M_\mathrm{Pl})$. Any new effects
must therefore enter at earliest at $\mathcal{O}(m^2/M_\mathrm{Pl}^2)$. To
discuss these effects, we split $\Phi$ into zeroth-, first-, and
second-order pieces, 
\begin{equation}
\Phi = \Phi_\mathrm{GR} + \left(\frac{m}{M_\mathrm{Pl}}\right)^2\Phi_2 +
\ldots .  \label{eq:phiexpansion}
\end{equation}
Using this expansion, the power spectrum of $\Phi$ is 
\begin{equation}
P_\Phi = \langle\Phi^2\rangle = \langle\Phi_\mathrm{GR}^2\rangle + 2\left(
\frac{m}{M_\mathrm{Pl}}\right)^2\langle\Phi_\mathrm{GR}\Phi_2\rangle + \ldots.
\label{eq:phipowerspec}
\end{equation}
The deviation from scale-invariance, $n_s$, is defined by 
\begin{equation}
n_s - 1 = \frac{d \ln \Delta_\Phi^2}{d\ln k},
\end{equation}
where the dimensionless power spectrum is 
\begin{equation}
\Delta_\Phi^2 = \frac{k^3}{2\pi} P_\Phi.
\end{equation}
In GR, the deviation from scale-invariance is $-2\varepsilon-\eta$. Using
the results 
\begin{align}
\frac{d\ln\Phi_\mathrm{GR}^2}{d\ln k} &= -3-2\varepsilon-\eta, \\
\frac{d\ln\Phi_2^2}{d\ln k} &= -3 + \left(n_s-1\right)_2,
\end{align}
where $\left(n_s-1\right)_2$ is the spectral index of $\Phi_2$, and assuming
that $\Phi_2$ is not too much larger than $\Phi_\mathrm{GR}$, the spectral
index to second order in $m/M_\mathrm{Pl}$ is given by 
\begin{equation}
n_s - 1 = -2\varepsilon - \eta + \left(\frac{m}{M_\mathrm{Pl}}\right)^2\frac{
\Phi_2}{\Phi_0}\left[2\varepsilon + \eta + \left(n_s-1\right)_2\right] +
\ldots.
\end{equation}
Finally, we consider the tensor-to-scalar ratio, $r$, defined by 
\begin{equation}
r=\frac{\Delta _{t}^{2}}{\Delta _{\Phi }^{2}},
\end{equation}
where $\Delta _{t}^{2}$ is the dimensionless power spectrum of the spin-2
perturbations, $H_{Tk}^{(\pm 2)}$. Pure \ae-theory effects contribute a
constant rescaling to the tensor spectrum which only becomes important at $\mathcal{O}(m^2/M_{\mathrm{Pl}}^2)$ \cite{Lim:2004js}. Recall that the
coupling between the aether and $\phi $, however, has no effect on the
tensor perturbations as none of the coupling terms contain spin-2 pieces, so
the tensor spectrum $\Delta _{t}^{2}$ is unchanged apart from the
aforementioned (small) rescaling. Therefore, $r$ is modified by a factor 
\begin{equation}
\frac{r}{r_{\mathrm{GR}}}=\frac{\Delta _{\Phi _{0}}^{2}}{\Delta _{\Phi }^{2}},
\end{equation}
where $r_{\mathrm{GR}}$ is the tensor-to-scalar ratio in the absence of the
aether-inflaton coupling. Using the expansion (\ref{eq:phiexpansion}), we
find that the corrections to $r$ are small, 
\begin{equation}
\frac{r}{r_{\mathrm{GR}}}=1-2\left( \frac{m}{M_{\mathrm{Pl}}}\right) ^{2}
\frac{\Phi _{2}}{\Phi _{0}}+\ldots .
\end{equation}

What size are the corrections to $n_{s}-1$ and $r$? As
discussed in \cref{sec:lambdavalues}, $m/M_{\mathrm{Pl}}$ is no larger than $
\sim \mathcal{O}(10^{-7})$, barring any special cancellations among the $
c_{i}$. We constructed the expansion of $\Phi $ so that $\Phi _{2}$ is comparable in size to $\Phi _{0}$. We assume that there are no
effects such as instabilities at $\mathcal{O}(m/M_{\mathrm{Pl}})$ which
would cause this construction to fail (the one instability that we have found
in the spin-0 modes, discussed in \cref{sec:spin0instability}, has been
assumed to vanish, by making the coupling either very small or of the opposite sign to $\dot{\bar\phi}$). The Planck sensitivity to $r$ is about $10^{-1}$, and
about $10^{-2}$ to $n_{s}-1$ \cite{Ade:2013zuv,Ade:2013uln}.

We see that the first corrections to $\Phi$ enter at $\mathcal{O}(m^2/M_\mathrm{Pl}^2)$. This is constrained by other experiments to be a tiny
number, placing any coupling between $\phi$ and $\theta$, which is not
already ruled out, far outside the current and near future windows of CMB
observability.

\section{Case Study: Quadratic Potential}

\label{sec:dj}

\subsection{Slow-Roll Inflation: An Example}

\label{sec:djsr}

The arguments so far have been made for a general potential $V(\theta,\phi)$ with
only minimal assumptions. In order to be more quantitative, we will now look
more closely at a particular form of the potential for which the
inflationary dynamics are known and relatively simple.

The Donnelly-Jacobson potential \cite{Donnelly:2010cr} contains all terms
relevant to the dynamics at quadratic order in the fields and is given by 
\begin{equation}
V(\theta ,\phi )=\frac{1}{2}M^{2}\phi ^{2}+\mu \theta \phi .  \label{eq:DJ}
\end{equation}
A term proportional to $\theta $ would contribute a total derivative to the
action and hence be nondynamical (note that the potential enters the
Friedmann equation through $V-\theta V_{\theta }$, not $V$ itself), while a
term proportional to $\theta ^{2}$ could be absorbed into $c_{2}$ and would
only renormalise $G_{c}$. We take $\mu >0$ as the theory is invariant under
the combined symmetry $\mu \rightarrow -\mu $ and $\phi \rightarrow -\phi $.
Any dynamics with $\mu <0$ can be obtained by flipping the sign of $\phi $.

This is simply $m^2\phi ^{2}$ chaotic inflation with an extra force that pushes 
$\phi $ towards negative values \cite{Donnelly:2010cr}. In the case where the scalar field has no mass term, $\phi $ possesses exact shift symmetry, $\phi \rightarrow \phi +
\mathrm{const.},$ and this theory essentially becomes $\Theta \mathrm{CDM}$, a
dark energy theory in which $\mu $ is related to the dark energy scale and,
importantly, is protected from radiative corrections by the existence of a
discrete symmetry \cite{Blas:2011en,Audren:2013dwa}. Interestingly, in the
special case where the aether is hypersurface-orthogonal, this theory also
admits a candidate UV completion in the consistent nonprojectable extension 
\cite{Blas:2009qj,Blas:2009ck,Blas:2010hb} of Ho\v{r}ava-Lifschitz gravity 
\cite{Horava:2009uw}. In that case, however, the spin-1 modes we have
discussed vanish. This is because the aether can be written as the
(normalised) gradient of a scalar field corresponding to a global time
coordinate, so it possesses no spin-1 modes. A similar coupling was also
considered in \rcite{Libanov:2007mq}.

The equations of motion, in conformal time, are 
\begin{align}
{\mathcal{H}}^{2}& =\frac{4\pi G_{c}}{3}a^{2}\left( M^{2}\phi ^{2}+\phi
^{\prime 2}a^{-2}\right) ,  \label{eq:djfriedmann} \\
{\mathcal{H}}^{\prime }& =\frac{4\pi G_{c}}{3}a^{2}\left( M^{2}\phi
^{2}-2\phi ^{\prime 2}a^{-2}-3\frac{m\mu }{a}\phi ^{\prime }\right) , \\
0& =\phi ^{\prime \prime }+2{\mathcal{H}}\phi ^{\prime 2}M^{2}\phi +3{
\mathcal{H}}m\mu a.  \label{eq:kg}
\end{align}
Normally, we can obtain a slow-roll inflationary solution to leading order
by neglecting $\dot{\phi}^{2}=a^{-2}\bar{\phi}^{\prime 2}$ in the Friedmann
equation (\ref{eq:djfriedmann}) and $\ddot{\bar{\phi}}$\footnote{We cannot just drop $\bar{\phi}^{\prime \prime }$ as it contains a term like 
$H\dot{\phi}$. It is easiest to drop the second derivative piece from the
cosmic time scalar evolution equation and \emph{then} move to conformal
time.} in the scalar evolution equation (\ref{eq:kg}). The same applies in
this theory; we now briefly justify this.

A slow-roll inflationary phase requires $H$ to be changing slowly, and for
inflation to be successful it needs to last at least 50--60 $e$-folds.
This is guaranteed by making sure the slow-roll parameters 
\begin{alignat}{2}
\varepsilon & =-\frac{\dot{H}}{H^{2}} & & =1-\frac{{\mathcal{H}}^{\prime }}{{
\mathcal{H}}^{2}}, \\
\eta & =\phantom{-}\frac{\dot{\varepsilon}}{H\varepsilon } & & =\frac{
\varepsilon ^{\prime }}{{\mathcal{H}}\varepsilon },
\end{alignat}
are both very small compared to unity. For convenience we will work in
cosmic time ($t=\int ad\tau $) here. The slow-roll parameters are 
\begin{align}
\varepsilon & =\frac{4\pi G_{c}}{H^{2}}\left( \dot{\phi}^{2}+m\mu \dot{\phi}
\right) , \\
\eta & =2\left[ \varepsilon +\frac{\ddot{\phi}}{H\dot{\phi}}\left( \frac{2
\dot{\phi}+m\mu }{2\dot{\phi}+2m\mu }\right) \right] .
\end{align}
Defining 
\begin{align}
\delta & \equiv \frac{4\pi G_{c}\dot{\phi}^{2}}{3H^{2}}, \\
\lambda & \equiv \frac{\ddot{\phi}}{3H\dot{\phi}}, \\
\gamma & \equiv \frac{M}{H}\frac{\mu }{\mu _{c}},
\end{align}
where 
\begin{equation}
\mu _{c}\equiv \frac{1}{\sqrt{12\pi G_{c}}}\frac{M}{m},  \label{eq:mucdef}
\end{equation}
we can calculate the slow-roll parameters, 
\begin{align}
\varepsilon & =3\delta +\gamma \delta ^{1/2}, \\
\eta & =-3\lambda \left( \frac{6\delta ^{\frac{1}{2}}+\gamma }{6\delta ^{
\frac{1}{2}}+2\gamma }\right) .
\end{align}

The usual slow-roll conditions, $\dot{\phi}^{2}\ll H^{2}$ and $\ddot{\phi}\ll 3H\dot{\phi}$, are equivalent to $\delta \ll 1$ and $\lambda \ll 1$,
respectively. We generally expect $M<H$ in order for the inflaton to produce
perturbations. As we will see below, both the requirement that inflation end and the stability considerations discussed
in \cref{sec:flatspace} impose $\mu <\mu _{c}$. When combined, these
conditions imply $\gamma <1$. So, under these reasonable assumptions on $M$
and $\mu $, in order to ensure $\varepsilon \ll 1$ and $\eta \ll 1$ we
simply need $\dot{\phi}^{2}\ll H^{2}$ and $\ddot{\phi}\ll 3H\dot{\phi}$ as
usual. Note, however, that the usual identifications of $\varepsilon $ and $
\eta $ in terms of the potential will be changed if the scalar-aether
coupling is large enough for $\gamma $ to be comparable to $\delta ^{1/2}$.

In the slow-roll limit, the Friedmann and Klein-Gordon equations are,
respectively, 
\begin{align}
\mathcal{H} &  \simeq\sqrt{\frac{4\pi G_{c}}{3}}Ma|\bar{\phi}
|,\label{eq:shslowroll}\\
\phi^{\prime} &  \simeq-\sqrt{\frac{1}{12\pi G_{c}}}Ma\left(
\operatorname{sgn}(\phi)+\frac{\mu}{\mu_{c}}\right)
.\label{eq:phiprimeslowroll}
\end{align}
Notice the appearance of $\mu_{c}$ defined above. During slow-roll, it is
related to the inflationary dynamics by 
\begin{equation}
\mu_{c}=\frac{M^{2}|\phi|}{\theta}.
\end{equation}
The value of $\mu /\mu _{c}$ is physically significant because it determines
the stability of the slow-roll solution. The number of $e$-folds that
inflation lasts tends to infinity as $\mu \rightarrow \mu _{c}$, which
corresponds to exact de Sitter expansion; for $\mu >\mu _{c}$ the slow-roll
solution is unstable and grows without bound \cite{Donnelly:2010cr}. We will therefore always consider inflationary solutions with $\mu <\mu _{c}$.

There is an additional constraint on $\mu/\mu_{c}$ from the spin-0 stability
constraint (\ref{eq:spin0constraint}). Substituting the definition of $\mu
_{c}$ into this gives the constraint 
\begin{equation}
\frac{\mu^{2}}{\mu_{c}^{2}}\leq24\pi G_{c}m^{2}c_{123}=\frac{24\pi
Gm^{2}c_{123}}{1+8\pi G\alpha}.  \label{eq:muconstraint}
\end{equation}
The same constraint was derived along similar lines in \rcite{Donnelly:2010cr}.\footnote{As discussed in \cref{foot:djnotation}, our action and potential differ from
those in \rcite{Donnelly:2010cr} because we give the aether units of mass
while their aether is dimensionless. Taking the different definitions of $
c_{i}$, $m$, and $\mu$ into account, our constraint agrees with theirs.}
Since $c_{123}\leq1$ and $\alpha\geq0$ (see \cref{sec:flatspace}, as well as 
\rcite{Lim:2004js,Carroll:2004ai}), this is more restrictive than simply $
\mu<\mu_{c}$, unless $m$ is comparable to, or greater than, the Planck scale---a possibility that seems to be ruled out by experiments, as discussed in 
\cref{sec:lambdavalues}. Since experiments suggest $m/\Mp \lesssim 10^{-7}$, $\mu/\mu_c$ must be so small that inflationary dynamics would be effectively unchanged by the coupling, unless cancellations among the $c_i$ conspire to weaken the bounds on $m$.

\subsection{The Instability Explored}

Specialising to the Donnelly-Jacobson potential and using the slow-roll
equations (\ref{eq:shslowroll}) and (\ref{eq:phiprimeslowroll}), we can write the spin-1
equation of motion (\ref{eq:spin1xislowroll}) to first order in the
slow-roll parameters as 
\begin{equation}
\xi^{\prime\prime}+c_{s}^{{(\pm1)}2}k^{2}\xi-\frac{\Lambda}{\tau^{2}}\xi=0,
\label{eq:djxieq}
\end{equation}
with $\Lambda$ given by 
\begin{align}
\Lambda & \equiv -\frac{1}{2}\frac{\mu\mu_{c}}{c_{1}H^{2}}\left(
\operatorname{sgn}(\phi)+\frac{\mu}{\mu_{c}}\right)  +\mathcal{O}
(\varepsilon),\nonumber\\
&  =-\frac{M^{2}}{H^{2}}\left(  c_{s}^{(0)2}\sigma+\operatorname{sgn}(\mu
\phi)\frac{c_{s}^{(0)}\sqrt{\sigma}}{\sqrt{3c_{1}}}\sqrt{\frac{M_{\mathrm{Pl}
}^{2}}{m^{2}}+c_{13}+3c_{2}}\right)  +\mathcal{O}(\varepsilon
).\label{eq:lambdadefdj}
\end{align}
Here, as in \cref{sec:lambdavalues}, we have defined the dimensionless coupling 
$\sigma$ by 
\begin{equation}
\mu^{2}=2c_{123}M^{2}\sigma=24\pi G_{c}m^{2}c_{123}\sigma\mu_{c}^{2},
\label{eq:djsigdef}
\end{equation}
so that flat-space stability of the spin-0 modes implies $\sigma\leq1$.

As with the general case, the solution (\ref{eq:vksol}) to \cref{eq:djxieq}
is written in terms of the first Hankel function of order $\nu$, where 
\begin{equation}
\nu^{2} \equiv\frac{1}{4} + \Lambda
\end{equation}
Repeating the analysis of \cref{sec:instability}, we pick a single mode
which leaves the sound horizon at some conformal time $\tau _{i}$, which we
could take to be the start of inflation. We pick a mode which crosses the
horizon early because $V_{k}(\tau )$ is largest at small $k$ (with $\tau $
held fixed), so this is one of the larger superhorizon modes available. We
want to calculate the contribution of this mode to the $0$--$i$ component of the stress-energy tensor. If it exceeds the background energy density,
then this would indicate a violation of isotropy and signal an instability in the background solution which we found in \cref{sec:djsr}.

Using the slow-roll scalar equation and our expression (\ref
{eq:phiprimeslowroll}) for $\bar{\phi}^{\prime }$, we find 
\begin{align}
\bar{V}_{\phi }\bar{V}_{\theta \phi }& =M\mu H\sqrt{\frac{3}{4\pi G_{c}}}
\left( \operatorname{sgn}(\phi )+\frac{\mu }{\mu _{c}}\right)  \notag \\
& =M^{2}H\left( 6mc_{123}\sigma +\operatorname{sgn}(\mu \phi )\sqrt{12c_{123}\sigma 
\left[ M_{\mathrm{Pl}}^{2}+(c_{13}+3c_{2})m^{2}\right] }\right) .
\end{align}
We can substitute this directly into \cref{eq:t0icomp} to find one of the
terms in the contribution that this mode makes to $T^{0}{}_{i}$, 
\begin{equation}
T^{0}{}_{i,k}/\bar{T}^{0}{}_{0}\supset \frac{c_{s}^{(0)}}{12\sqrt{3}\pi }
\frac{M}{H}\frac{M}{\sqrt{M_{\mathrm{Pl}}^{2}+\alpha }}\left( \operatorname{sgn}(\mu
\phi )\sqrt{\sigma }+\sqrt{3c_{123}}\sigma \frac{m}{\sqrt{M_{\mathrm{Pl}
}^{2}+\alpha }}\right) \Gamma (\nu )2^{\nu -\frac{1}{2}}\left( -\tau
_{i}\right) ^{\frac{3}{2}}e^{\left( \nu -\frac{3}{2}\right) N}\delta
^{3}{}_{i}.  \label{eq:djt0icomp}
\end{equation}
We can now get a more quantitative handle on the argument made in 
\cref{sec:instability}. Assuming $\nu >3/2$, the exponential in $(\nu -3/2)N$
is likely to overwhelm the other terms within the 50--60 or more $e$-folds
that will occur after $\tau _{i},$ which we take to be near the start of
inflation. While several terms in \cref{eq:djt0icomp} are likely to be
several orders of magnitude smaller than unity, including $M/H$, $m/M_{
\mathrm{Pl}}$,\footnote{A requirement for $\nu $ to be greater than $3/2$ in the first place.} and
possibly $M/M_{\mathrm{Pl}}$, it is unlikely that these could be so small as
to overwhelm the exponential terms and the gamma function. Hence, for $\nu
>3/2,$ we expect that the slow-roll background solution we found in 
\cref{sec:djsr} is unstable, rapidly dominated by perturbations in the
aether field generated by its coupling to the inflaton.

In \cref{sec:lambdavalues} we found that $\nu$ can surpass $3/2$, even by
several orders of magnitude, if the aether VEV, $m,$ is suitably small
compared to the Planck scale. Armed with a specific form for the potential,
we now briefly clarify that argument and use it to place constraints on the aether-scalar coupling parameter, $\mu$.

If $\nu >3/2$ then $\Lambda >2$, where $\Lambda $ is defined in 
\cref{eq:lambdadefdj}. It is not difficult to check that this is the same as
the $\Lambda $ we discussed for a general potential, \cref{eq:lambdadef},
which we wrote in various forms in \cref{sec:lambdavalues}. There, we found
that for $\Lambda $ to be positive we needed $\mu \dot{\phi}$ to be
positive. With the Donnelly-Jacobson potential, we have an expression for $
\dot{\phi}$, \cref{eq:phiprimeslowroll}. From there we see that $\mu \dot{\phi
}$ is only positive (assuming $\mu <\mu _{c}$) when $\mu \phi $ is negative.
We will take $\mu $ to be positive and then ask if $\phi $ can be negative
(the opposite case is trivial, as the theory has combined $\mu \rightarrow
-\mu $, $\phi \rightarrow -\phi $ symmetry). This is not at all uncommon,
and depends only on initial conditions. The dynamics for this inflationary
model are encapsulated in $(\phi ,\dot{\phi})$ phase portraits for a range
of $\mu /\mu _{c}$ in \rcite{Donnelly:2010cr}. Per \cref{eq:djsigdef}, $\mu
/\mu _{c}$ is of order $(m/M_{\mathrm{Pl}})\sigma ^{1/2}$. Because
observations suggest $m\ll M_{\mathrm{Pl}}$ (see \cref{sec:lambdavalues}), $
\mu $ should be very small compared to $\mu _{c}$ even when $\sigma $
approaches unity. Hence, the phase portrait for $\mu =0$ in \rcite
{Donnelly:2010cr} will be very close to the dynamics we are interested in.
In the exact $\mu =0$ case, there are as many inflating paths with $\phi <0$
as $\phi >0$, because when $\mu =0$, the equations for $\phi $ and $\dot{\phi
}$ have combined $\phi \rightarrow -\phi $ and $\dot{\phi}\rightarrow -\dot{
\phi}$ symmetry. The next phase portraits show a tendency, increasing with $
\mu $, for inflating paths to live in the $\phi >0$ half of the phase plane.
Since $\mu \ll \mu _{c}$, nearly half of all initial conditions leading to
viable inflation have $\mu \phi <0$.

Considering each piece in $\Lambda$ on an order-of-magnitude basis, and
taking $\operatorname{sgn}(\mu\phi)=-1$, we have 
\begin{equation}
\Lambda = -\underbrace{\frac{M^{2}}{H^{2}}}_{\ll1}\left( \underbrace {
c_{s}^{(0)2}\sigma\vphantom{\frac{c_s^{(0)}\sqrt{\sigma}}{\sqrt{3c_1}}}}
_{\lesssim1}-\underbrace{\frac{c_{s}^{(0)}\sqrt{\sigma}}{\sqrt{3c_{1}}}}_{
\mathcal{O}(1)}\sqrt{\frac{M_{\mathrm{Pl}}^{2}}{m^{2}}+c_{13}+3c_{2}}\right)
+\mathcal{O}(\varepsilon).
\end{equation}
Evidently, $\Lambda$ will be greater than $2$ if the smallness of $m$
compared to the Planck scale exceeds the (square of the) smallness of the
scalar mass, $M$, compared to the Hubble scale, 
\begin{equation}
\frac{M_{\mathrm{Pl}}}{m}\gtrsim\frac{2\sqrt{3c_{1}}}{c_{s}^{(0)}\sqrt{
\sigma }}\left( \frac{M}{H}\right) ^{-2},
\end{equation}
where we have assumed that $m/\Mp\ll1$. While $M/H$ should be small, there are no limits on how small $m/M_{\mathrm{
Pl}}$ should be before the collider scale, and moreover, as discussed in 
\cref{sec:lambdavalues}, there are already likely to be stringent
experimental constraints on $m/M_{\mathrm{Pl}}$ (although these tend to
depend on the $c_{i}$ not cancelling out in particular ways).

The tachyonic instability discussed here and in \cref{sec:spin1perts} is absent when $\mu$ and $\phi$ have the same sign. In this case, the coupling only serves to dampen aether perturbations. For the Donnelly-Jacobson potential, what remains is effectively just $m^2\phi^2$ inflation. If the signs of $\mu$ and $\phi$ are different, or if we were to demand that inflation be viable for all initial conditions, then the absence of this instability puts a very strong constraint on the magnitude of $\mu$,
\begin{equation}
\frac{|\mu|}{H} \lesssim 2\sqrt{6}c_1\frac{m}{\Mp}\left(\frac{M}{H}\right)^{-2}.
\end{equation}
From the background dynamics, we expect $(M/H)^2 \approx 3\varepsilon + \mathcal{O}(m/\Mp) \sim \mathcal{O}(10^{-2})$, while the absence of gravitational \v{C}erenkov radiation constrains $m/\Mp \lesssim \mathcal{O}(10^{-7})$, in the absence of certain cancellations among the $c_i$. Thus the constraint on $\mu$ is of the order
\begin{equation}
\frac{|\mu|}{H} \lesssim \mathcal{O}(10^{-5}).
\end{equation}
This should be compared to the previous strongest constraint on $\mu$, the flat-space stability constraint discussed in \rcite{Donnelly:2010cr} and \cref{sec:flatspace},
\begin{equation}
\frac{|\mu|}{H} < \sqrt{2c_{123}} \sim \mathcal{O}(1).
\end{equation}

\section{Summary of Results}

\label{sec:summary}

We have examined cosmological perturbations in a theory of single-field,
slow-roll inflation coupled to a vector field that spontaneously breaks
Lorentz invariance, looking both to explore the effects of such a coupling on inflationary cosmology and to place constraints on it. The particular model is Einstein-aether theory, a
theory of a fixed-norm timelike vector called the ``aether,''
coupled to a canonical scalar field by allowing its potential to depend on the
divergence of the aether, $\theta =\nabla _{\mu }u^{\mu }$. In a homogeneous
and isotropic cosmology, $\theta$ is related directly to the Hubble rate by $H=\theta/3m$. This construction allows $H$ to play a role in cosmological dynamics
that it cannot in general relativity, where it is not a spacetime scalar. Moreover, it is a fairly general model of coupling between a fixed-norm vector and a scalar field. In particular, while many couplings can be written down which are not captured by a potential $V(\theta,\phi)$, all such terms have mass dimension 5 or higher and therefore fall outside the scope of the low-energy effective theory. Similarly, Einstein-aether theory is the most general low-energy theory which violates boost Lorentz invariance in the gravitational sector \cite{ArmendarizPicon:2010mz}.

Around a slow-roll inflationary background, this theory can possess a tachyonic instability. The instability is present if the norm of the aether, effectively the Lorentz symmetry breaking scale, is sufficiently small compared to the Planck mass, and the aether-scalar coupling is suitably large. In this region of parameter space, assuming a technical requirement on the initial conditions, scalar and vector perturbations both grow exponentially, destroying the inflationary background. Demanding the absence of this instability for generic initial conditions places a constraint on the coupling which is significantly stronger than the existing constraints, which are based on the stability of the perturbations around flat space and the viability of a slow-roll solution. Hence this constraint is by far the strongest on an aether-scalar coupling to date, with the assumption that the scalar drives a slow-roll inflationary period.

The root of the instability is the smallness of the aether VEV, $m$, compared to the Planck mass. The noncoupled terms in the aether Lagrangian each have two factors of $u^\mu$, so these aether terms will come with a factor of $(m/\Mp)^2$ in the Einstein equations. Terms involving two or more $\theta$ derivatives of the scalar field potential will also enter the Einstein equations with these factors or higher. However, terms associated with the coupling $V_{\theta\phi}$, which only has one aether derivative, will only have one power of $m/\Mp$ and so will generically be larger (depending on the size of $V_{\theta\phi}$) than the other aether-related terms. In the aether equation of motion, this coupling term will be a power of $\Mp/m$ larger than the other terms for the same reason. When the coupling is sufficiently large, it is exactly this term that drives the instability.

If the instability is absent, then observables in the CMB are unaffected by the coupling at the level of observability of current and near-future experiments: the corrections are smaller than $\mathcal{O}(10^{-15})$. This is due partly to the smallness of the aether norm relative to the Planck scale, but is exacerbated by the presence of unusual cancellations. Solving for the spin-0 perturbations order by order in the aether VEV, $m/\Mp$, no isocurvature modes are produced at first order. This is unexpected, as isocurvature modes are a generic feature of multi-field theories. Stronger yet, the perturbations are completely unchanged at first order in $m/\Mp$ from the case without any aether at all. This is largely a result of unexpected cancellations which hint at a deeper physical mechanism.

It is unclear whether these unexpected conclusions hold to higher orders in $m/\Mp$. At $\mathcal{O}(m^2/\Mp^2)$, several new coupling terms enter the perturbed Einstein equations, \cref{eq:spin000,eq:spin00i,eq:spin0ij}, with a qualitatively different structure to the terms which appear at $\mathcal{O}(m/\Mp)$. The possibility therefore remains that the isocurvature modes that one would expect from the multiple interacting scalar degrees of freedom might re-emerge at this level. If they do, they would be severely suppressed relative to the adiabatic modes.

Beyond perhaps an extreme fine-tuning, there does not seem to be a subset of the parameter space in which observable vector perturbations are produced without destroying inflation. Even if such modes could be produced, they do not freeze out on superhorizon scales and are sensitive to the uncertain physics, such as reheating, between the end of inflation and the beginning of radiation domination. Therefore any observational predictions for vector modes would be strongly model-dependent. Nonetheless, it should be stressed that the line between copious vector production (that quickly overcomes the background) and exponentially decaying vector production is so thin, as it depends on unrelated free parameters, that there is no reason to expect this theory would realise it.

While we made these arguments for a general potential, we also looked at a
specific, simple worked example, the potential of Donnelly and Jacobson \cite{Donnelly:2010cr}. This potential includes all dynamical terms at quadratic order, and amounts to $m^2\phi^2$ chaotic inflation with a coupling to the aether that
provides a driving force. It contains many of the terms allowed for the aether and scalar up to dimension 4.\footnote{One could also add a tadpole term proportional to $\phi$ and a term proportional to $\phi^2\theta$. The latter would effectively promote the coupling $\mu$ to $\mu + \mathrm{const.}\times\phi$, so during slow-roll inflation the effective $\mu$ would still be roughly constant.} The constraint this places on the coupling $\mu \equiv V_{\theta\phi}$,
\begin{equation}
\frac{|\mu|}{H} \lesssim 2\sqrt{6}c_1\frac{m}{\Mp}\left(\frac{M}{H}\right)^{-2} \lesssim \mathcal{O}(10^{-5}),
\end{equation}
is stronger by several orders of magnitude than the the next best constraint \cite{Donnelly:2010cr},
\begin{equation}
\frac{|\mu|}{H} < \sqrt{2c_{123}} \sim \mathcal{O}(1).
\end{equation}
It is worth emphasising again the two conditions for our constraint to hold. First, the scalar must drive a period of slow-roll inflation. Second, the instability can be avoided if $\mu$ and $\phi$ have the same sign. Consequently, the new constraint applies only if we demand that inflation be stable for all initial conditions. If such a coupling were to exist, this constraint could be seen as a lower bound on $m$, to be contrasted to the many upper bounds on $m$ in the literature.

\begin{savequote}[30pc]
If it be true that good wine needs no bush, 'tis true that a good play needs no epilogue; yet to good wine they do use good bushes, and good plays prove the better by the help of good epilogues.
\qauthor{Rosalind, \textit{As You Like It}, Epilogue}
\end{savequote}

\chapter{Discussion and Conclusions}
\label{chap:conclusion}
\hrule
\vspace*{1cm}

Our concern throughout this thesis has been the use of cosmology as a laboratory for testing gravity. In the first part, we focused on the theoretical and cosmological implications of endowing the graviton with a finite mass, leading to theories either of \emph{massive gravity}, in which there is a single, massive graviton, or \emph{massive bigravity}, in which two gravitons, one massless and the other massive, interact with each other. We have studied both of these variants, while focusing on bigravity as it is a better setting for studying Friedmann-Lema\^{i}tre-Robertson-Walker cosmologies. In the second part, we turned our attention to theories of gravity which violate Lorentz invariance, using the early-Universe inflationary era to place constraints on Lorentz violation. Below we will summarise in more detail the problems we have sought to address and the results obtained in this thesis, before ending by examining the implications of this work for future studies of modified gravity.

\section{Problems Addressed in This Thesis}

The expansion of the Universe is accelerating. If we assume that the Standard Model of particle physics, perhaps augmented by one or more massive dark matter particles, describes the matter content of the Universe, then general relativity and, indeed, our intuitive notions of gravity suggest that the expansion should slow down, as galaxies and dark matter particles exert their gravitational pulls on one another. To accommodate the surprising acceleration, one must either give up on the assumption that we are using the right description of matter, and introduce an ``exotic'' dark energy, or explore the possibility that gravity is modified from general relativity at extraordinarily large distances.

The simplest example of such a modification, which is to simply allow for a nonzero cosmological constant, fails in one crucial respect: the cosmological constant receives quantum corrections which are many orders of magnitude larger than the value required to explain the accelerating Universe. This motivates theories of modified gravity which self-accelerate and are \emph{technically natural}, i.e., in which whatever parameter value and theory structure lead to the accelerated expansion are not destabilised by radiative corrections. Ghost-free massive gravity, particularly in its bimetric form, appears to be such a theory: the special potential structure, necessary to eliminate the dangerous Boulware-Deser ghost, and the small graviton mass both seem to be stable against quantum corrections \cite{deRham:2013qqa}, while bigravity in particular can easily accommodate self-accelerating solutions which are in as good agreement with background observations as $\Lambda$CDM is \cite{vonStrauss:2011mq,Akrami:2012vf,Konnig:2013gxa}.

The expansion histories predicted by bigravity can be very close to $\Lambda$CDM and to each other. To break this degeneracy, it is necessary to go beyond the FLRW assumption and consider the evolution of perturbations. In so doing, we can derive predictions for the formation and growth of cosmic structure, and thereby open up new avenues of testing the theory. This is especially important now, as the advent of next-generation galaxy surveys like {\sc Euclid} will allow observations of structure formation to place constraints on modified gravity and dark energy comparable to or exceeding those from purely geometric observations. Moreover, by expanding our study to linear perturbations we can test the stability of the FLRW solutions. Indeed, results in the literature prior to the work undertaken in this thesis had found evidence for cosmological instabilities in massive bigravity \cite{Comelli:2012db,DeFelice:2014nja,Comelli:2014bqa}. It is therefore crucial to undertake a detailed analysis of precisely when cosmological solutions in bigravity are and are not linearly stable.

The presence of two metric tensors in a theory brings with it significant conceptual challenges. In general relativity, the metric tensor is not just a dynamical field which obeys a particular equation of motion. It has a deep physical role as the geometry of spacetime. In a bimetric theory, where there is a second dynamical metric and in which the action treats both metrics on equal footing (up to parameter interchanges), there is a danger of demoting the metric from its geometric office. In most studies of bigravity, this issue is swept under the rug by only coupling matter to one metric. This can be identified as the physical metric, as matter follows its geodesics, while the other is simply a rank-2 tensor coupled to it in order to modify gravity. The geometric appearance of this second metric, such as the fact that its kinetic term is given by the Ricci scalar, is then something of a coincidence, not a sign that we should assign any particular geometric interpretation to this field. These considerations immediately raise the question of whether matter can be consistently coupled to both metrics, and, if so, whether we must give up on a geometric understanding of gravity in the process.

Turning towards early times and high energies, Lorentz invariance is one of the most fundamental ingredients in the best-tested physical theories, but may have to be given up in order to resolve the seemingly-intractable problem of quantum gravity. Lorentz violation is very well constrained in the Standard Model \cite{Mattingly:2005re}, but much less so in other sectors of physics, such as gravity, inflation, and dark matter. In this thesis we have sought to improve our understanding of the first two, by allowing a Lorentz-violating vector field to couple both to the metric and to a slowly-rolling scalar field. Such interactions would be expected to arise if physics is Lorentz-violating, barring some symmetry forbidding them, and so an exploration of inflationary dynamics and perturbations can provide a strong test of the effects and sizes of such couplings.

\section{Summary of Original Results}

\subsection{Massive Gravity and Bigravity}

\Cref{chap:bigravity-stability,chap:bigravity-subhorizon} dealt with cosmological perturbation theory in singly-coupled massive bigravity. In \cref{chap:bigravity-stability} we set up the formalism for perturbations and investigated the linear stability of FLRW solutions. By employing a carefully-selected gauge intended to ease the process of integrating out auxiliary degrees of freedom, we reduced the ten Einstein and conservation equations to a system of two equations for the two independent, dynamical perturbation variables. At small scales, the coefficients of this system become effectively constant, allowing us to determine the eigenfrequencies of the system and thereby identify stable and unstable models. Only one corner of bigravity, the infinite-branch $\beta_1\beta_4$ model, turns out to be linearly stable at all times.

In \cref{chap:bigravity-subhorizon} we applied our perturbation formalism to observations, studying the evolution of structure in the subhorizon and quasistatic approximation. This limit is applicable for most modes observable on linear scales and is of great utility for deriving testable predictions, as it reduces many differential equations to algebraic ones. In bigravity, this allows us to directly relate every metric perturbation to the density contrast, $\delta$, which in turn obeys an evolution equation that can be solved numerically. Doing this, we obtained predictions for structure growth for each one- and two-parameter bimetric model with viable background evolution, written in terms of three common modified-gravity parameters: the growth rate and index, $f$ and $\gamma$; the modification to Newton's constant, $Q$; and the anisotropic stress, $\eta$. Note that the quasistatic approximation, necessary to ensure that time derivatives of the fields drop out, under the assumption that they are of order Hubble, is not always applicable for unstable models. In these cases, the results can be applied (up to possible changes in initial conditions for the growth rate) at later times when modes are stable. The predictions hold straightforwardly for the stable infinite-branch $\beta_1\beta_4$ model, where we find large deviations from general relativity, which should be easily measurable by {\sc Euclid}, at all times and for all parameters. This provides a rare example of a modified-gravity model which can be unambiguously ruled out against $\Lambda$CDM.

In \cref{chap:dc-finsler,chap:dc-background,chap:dc-drgt} we deal with doubly-coupled bigravity from both observational and theoretical standpoints. In \cref{chap:dc-finsler} we studied a simple example of a theory in which the coupling of matter to both metrics removes any notion of an effective physical metric for all but the simplest fields. As an example of the new structures which might need to be used to describe spacetime in such a theory, we showed that the dynamics of point-particles can be described in terms of a Finsler manifold, a non-Riemannian structure in which there is an effective metric which depends, in addition to the two gravitational metrics, on the particle's own velocity.

The particular double coupling studied in \cref{chap:dc-finsler} suffers from the Boulware-Deser ghost and therefore should be seen as a toy model, rather than a potentially-realistic description of the Universe \cite{Yamashita:2014fga,deRham:2014naa}. An improvement can be made by coupling matter to a specially-selected effective metric for which the ghost seems to reappear only at energies outside the domain of validity of the theory \cite{deRham:2014naa}. This new double coupling formed the basis of \cref{chap:dc-background,chap:dc-drgt}. In the former, we studied the effects of this coupling in bigravity, deriving the Friedmann equations, comparing the simplest models to data, and investigating certain special parameter r\'egimes. These models can agree well with observations; if they have improved stability properties, they will increase the space of cosmologically-viable bimetric models.

In the latter chapter, we investigated this double coupling in the context of single-metric massive gravity. This theory famously suffers from a no-go theorem ruling out dynamical, flat FLRW solutions around a Minkowski reference metric \cite{D'Amico:2011jj}. Other choices of spatial curvature or reference metric lead to instabilities \cite{Gumrukcuoglu:2011ew,Gumrukcuoglu:2011zh,Vakili:2012tm,DeFelice:2012mx,Fasiello:2012rw,DeFelice:2013awa}. The no-go theorem is circumvented by the double coupling \cite{deRham:2014naa}, but we found that there is still a serious problem in comparing these theories to observations. The avoidance of the no-go theorem relies crucially on the use of fundamental fields as the sources for Einstein's equations. The usual description of matter on large scales by an effective fluid, such as pressureless dust, is insufficient and, in the absence of any other fields, leads back to nondynamical cosmologies just as in singly-coupled massive gravity. This can only be avoided either by adding new degrees of freedom, such as a scalar field, or by treating dust as a field. We examined in further depth the case in which a scalar field is added to the theory, coupling it to the effective metric in order to unlock dynamical solutions. If the matter fluid also couples to the effective metric, then significant pathologies arise due to the highly-constrained nature of the theory. If it couples instead to the gravitational metric, one can indeed obtain sensible late-time acceleration, but for many parameter choices it is driven as much by the scalar field's potential as it is by the massive graviton. This is because the avoidance of the no-go theorem requires the scalar field to have a potential, and the theory's constraints force it to evolve along the potential in a fixed way throughout cosmic history.

\subsection{Lorentz-Violating Gravity}

In \cref{chap:aether} we studied Einstein-aether theory, a general low-energy model of boost violation in the gravitational sector, during inflation. Einstein-aether is a vector-tensor theory in which a vector field, or aether, couples to gravity nonminimally while a Lagrange multiplier imposes the constraint that it be fixed-norm and timelike. This spontaneously picks out a preferred frame at each point in spacetime.

If such a field exists, it should also be coupled to a scalar inflaton, unless forbidden by symmetries of either field. In the case of single-field, slow-roll inflation, this does not appear to be the case. We showed that all operators up to mass dimension 4 coupling the scalar to the aether can be included simply by allowing the scalar's potential to depend on the divergence of the aether. This model had previously been introduced as a way of allowing the cosmic dynamics to depend directly on the Hubble rate, which is not a spacetime scalar in general relativity.

We derived the cosmological perturbations for Einstein-aether theory coupled in this way to a scalar field, and calculated the evolution of spin-0 and spin-1 perturbations during inflation. In both cases, sufficiently large couplings lead to a tachyonic instability for nearly half of all inflationary trajectories. Demanding the absence of this instability for all initial conditions leads to new constraints on the size of the aether-scalar coupling which is at least five orders of magnitude stronger than the previous best constraint.

\section{Outlook}

What is the next step for modified gravity? Progress on both the observational and theoretical fronts is due to be made over the next decade. Upcoming data from {\sc Euclid}, the Dark Energy Survey, and the Square Kilometre Array, among others, will measure both the geometry and the structure of our Universe with unprecedented precision. This will allow for more precise tests of modified gravity, especially in the r\'egime of perturbations. Euclid, for example, will be able to measure the anisotropic stress, $\eta$, within about 10\% if it is not strongly scale-dependent \cite{Amendola:2013qna}. Recall from \cref{chap:bigravity-subhorizon} that the one linearly stable bimetric model, the infinite-branch $\beta_1\beta_4$ model, has an anisotropic stress that deviates from GR by at least 30\%, and as much as 50\%, at all times. Clearly we are on the verge of entering a new experimental era in modified gravity.

On the theoretical side, much more remains to be done. The dark energy problem has not yet reached the status of mass generation in the Standard Model, in which the Higgs mechanism was for decades a clear frontrunner solution, or inflation, where a single, slowly-rolling scalar field is the consensus best available framework. There is no ``theory to beat'' in modified gravity and dark energy. Observationally, a cosmological constant is simple and fits the available data, but its theoretical issues may be too problematic to overcome. The simplest extension of general relativity is, in a sense, scalar-tensor theories, but none of these theories has emerged as being clearly better than the rest; instead, we have the Horndeski theory, an extremely general framework that covers all scalar-tensor theories with second-order (i.e., healthy) equations of motion \cite{Horndeski:1974wa}. While massive gravity and bigravity are closely related to Horndeski theory---the helicity-0 mode of the massive graviton in the decoupling limit is of the Horndeski form---they are ultimately a qualitatively different approach to modifying general relativity. So is Einstein-aether, and any number of other theories with, e.g., higher dimensions or nonlocality. None of these has emerged as a front-runner; ideally, we would demand self-acceleration, technical naturalness, agreement with observations, and some degree of aesthetic virtue. While massive gravity, and its bimetric extension in particular, checks off many of these boxes, the difficulty in obtaining stable cosmological solutions in all but a handful of nonminimal models continues to provide impetus to find something better.

It is impossible to guess what developments may emerge out of the aether.\footnote{Pun intended.} Science is notorious for pulling out its most significant discoveries when least expected and when the research of the day seemed not to be leading up to them at all. However, we can make some educated guesses as to which directions may soon be extended and, perhaps, provide fertile ground for new opportunity. Recently, healthy scalar-tensor theories beyond Horndeski have been discovered \cite{Zumalacarregui:2013pma,Gleyzes:2014dya,Gleyzes:2014qga}. The presence of nontrivial constraints means that, while the equations of motion contain higher derivatives, the propagating degrees of freedom obey second-order equations. As discussed in \cref{sec:pm}, a partially-massless theory of gravity would immediately satisfy practically all of the requirements we would impose on a theory of modified gravity, as its gauge symmetry would both determine and protect a small value of the cosmological constant. Whether a partially-massless theory in fact exists has been a topic of intense debate, with no resolution reached to date \cite{Hassan:2012gz,deRham:2012kf,deRham:2013wv,Joung:2014aba,Hassan:2014vja,Garcia-Saenz:2014cwa}. If a theory is found which possesses the partially-massless symmetry nonlinearly around all backgrounds, it would unquestionably be a boon for modified gravity.

Let us examine some of the possible directions opened up by the work described in this thesis. In \cref{chap:bigravity-stability}, we found that most of the bimetric models with viable cosmological backgrounds suffer from an instability at early times, leaving behind only a rather small and nonminimal corner of the parameter space. This means that, for any given mode, linear perturbation theory breaks down. It means nothing more and nothing less. At the moment this is a technical rather than a fundamental difficulty. While the instability may signal a breakdown of the background, it may also be the case that the instability is cured at higher orders. It is already known that nonlinear effects make spatial gradients of the helicity-0 mode of the massive graviton shallower; this is the Vainshtein mechanism which eliminates the fifth force and reduces physics to general relativity in regions like the solar system \cite{Vainshtein:1972sx}. It is conceivable that similar effects will work to cure the large time derivatives of cosmological perturbations at higher orders. However, this will require either a study of higher-order perturbation theory in bigravity or the development of a more sophisticated treatment of nonlinear effects. Going to higher orders is not simple, as the complicated theory even at linear order shows. Even then, there is no guarantee that a Vainshtein-like mechanism, or any other effect which would cure the instability, would arise at second order. Note that N-body simulations, which are commonly used to study modified-gravity effects in highly nonlinear r\'egimes, usually rely on the quasistatic approximation which fails precisely in these unstable models. Indeed, by taking the quasistatic approximation, one could never have discovered the instability: the quasistatic equations, presented in \cref{chap:bigravity-subhorizon}, show no sign of the instability as the relevant terms are taken to vanish. Instead we needed, in \cref{chap:bigravity-stability}, to use the full cosmological perturbation equations from the start. To date, only tentative steps have been taken towards removing the dependence on the quasistatic limit, and only for a very simple class of scalar-tensor theories with few of the complications of massive bigravity \cite{Llinares:2013qbh,Llinares:2013jua}. It seems, then, that understanding whether the cosmological instability dooms most bimetric models and, if not, how to calculate observables will require the development of new methods to better understand nonlinear perturbations.

Throughout \cref{chap:dc-finsler,chap:dc-background,chap:dc-drgt} we circled the question of how to construct a healthy, doubly-coupled bimetric theory, i.e., a sensible physical theory that treats both metrics on entirely equal footing. Two options immediately jump to mind. One is to simply include minimally-coupled matter Lagrangians for each metric with the same matter, as we studied in \cref{chap:dc-finsler} and was introduced in \rcite{Akrami:2013ffa}. This was later shown to reintroduce the Boulware-Deser ghost at arbitrarily-low energies \cite{Yamashita:2014fga,deRham:2014naa}. Another possibility is to couple matter to the nonlinear generalisation of the massless mode,
\begin{equation}
\mathcal{G}_\mn = \frac{1}{M_g^2+M_f^2}\left(M_g^2g_\mn + M_f^2 f_\mn\right).
\end{equation}
Such a coupling was both introduced and shown to possess the Boulware-Deser ghost in \rcite{Hassan:2012wr}.

In \cref{chap:dc-background} we discussed the only known double coupling in which the ghost does not re-emerge at low energies, introduced in \rcite{deRham:2014naa} and derived separately and extended to multimetric setups in \rcite{Noller:2014sta}. While, as we have shown in \cref{chap:dc-background,chap:dc-drgt}, this coupling leads to a rich phenomenology, the status of the ghost has been somewhat contentious and it is not yet clear exactly at which scale it appears \cite{deRham:2014naa,Hassan:2014gta,deRham:2014fha}. The consensus seems to be that the mass of the ghost is at least at the strong-coupling scale of the theory, and possibly parametrically larger. If the latter, then one could argue that this is ghost-free taken as an effective field theory, since the ghostly operators are outside the theory's r\'egime of validity \cite{deRham:2014naa,deRham:2014fha}. Indeed, it is not hard to conceive of the ghost being cured by higher-energy operators which, by definition, are ignored below the strong-coupling scale. However, even if the effective theory is healthy, the nonlinear massive gravity we are using might not correctly describe it.\footnote{We thank Angnis Schmidt-May for helpful discussions on the following points.} Consider, as a very simple example, the fourth-order equation of motion
\begin{equation}
\epsilon \ddddot{x} + \ddot{x} + \omega^2 x = 0. \label{eq:4thorderx}
\end{equation}
This has a ghost, by virtue of Ostrogradsky's theorem, and if the limit $\epsilon\to0$ is taken in the equation of motion, the ghost disappears as the equation becomes second-order. However, if we take the same limit in the \emph{solutions} to \cref{eq:4thorderx}, we do not obtain the correct solutions to the second-order equation of motion: the extra two modes do not decouple from the theory.\footnote{More precisely, for nonzero $k$ the frequency of the additional modes goes to infinity in the limit $\epsilon\to0$, and the actual value of the limit is not well-defined. In a theory where all modes have positive energy, infinite-frequency modes are impossible to excite with finite energy. An infinite-frequency negative-energy mode, which is what we have here, would however become even easier to produce \cite{Woodard:2006nt}.} For more details on this example, we refer the reader to section~3 of \rcite{Woodard:2006nt}.

It is therefore not clear whether the presence of a ghost in doubly-coupled massive gravity and bigravity, even if we assume it is not excited at the energy scales for which we are solving, would lead to the same solutions as the ghost-free low-energy theory would. This should not be a problem with the FLRW solutions, for which the absence of the ghost at all scales has been proven \cite{deRham:2014naa}, but is a sign that we need to continue to search for a doubly-coupled theory which is truly free of the Boulware-Deser ghost. At present it is unclear whether such a theory exists and, if so, what form it will take. If the coupling is not defined by minimally coupling matter to an effective metric, then the problems and potential solutions discussed in \cref{chap:dc-finsler} may turn out to be quite relevant.

However, a bird in the hand is worth two in the bush, and if we momentarily set aside our pining for a fully ghost-free doubly-coupled theory, we will notice that we have a double coupling, discussed in \cref{chap:dc-background}, which at the very least is good for cosmological solutions. Circumstantial evidence suggests that the new matter coupling should allow for cosmologically-stable models in a much larger region of the parameter space than the singly-coupled theory does. Recall from \cref{chap:bigravity-stability} that the instability which plagues most singly-coupled models appears specifically for small $y$. This is a problem in any finite-branch model (and most models only have viable solutions on the finite branch) because the quartic equation (\ref{eq:quartic-mg}) requires $y=0$ at early times. We have seen that in the doubly-coupled theory, $y$ starts at $\beta/\alpha$ and hence is always nonzero if $y_c\neq0$. While the modified Einstein equations will change the perturbation behaviour, if the rule of thumb that instabilities occur for small $y$ holds, then double coupling should open up many more stable models. Moreover, recall, cf. \cref{eq:eigenfreq_b2}, that the $\beta_2$ model is always stable in singly-coupled bigravity. The problem with this model is that it is ruled out by observations and theoretical conditions at the background level. In the doubly-coupled theory, the $\beta_2$ model has an acceptable background as long as $\beta/\alpha$ is above a threshold value, cf. \cref{fig:comb}. Finally, in one simple example, when $y=y_c$ at all times and as a result the metrics are proportional, we know that the perturbations must be well-behaved. We found that the effective Friedmann equation in this case reduces to that of $\Lambda$CDM. It can, moreover, be shown that for \emph{any} solutions to the doubly-coupled theory in which $g_\mn$ and $f_\mn$ are related by a conformal factor, the theory reduces to general relativity,\footnote{In particular, matter couples only to the massless spin-2 field.} and that this equivalence to general relativity extends to linear perturbations \cite{Schmidt-May:2014xla}. This implies that the perturbations around the conformal cosmological solutions we have found must be the same as in general relativity, and hence are stable. A full investigation of the cosmological perturbations in this theory is therefore very well motivated in the search for cosmologically-viable models of bigravity.

\cleardoublepage
\thispagestyle{empty}

 \vspace*{\fill}

\begin{quote}
\textit{We are all in the gutter, but some of us are looking at the stars.}
\qauthor{Oscar Wilde, \textit{Lady Windermere's Fan}}
\end{quote}
\vspace*{\fill}

\cleardoublepage

\thispagestyle{empty}

	\vspace*{\fill}
		\begin{flushright}
		{\Huge{ \bf Appendices}}
		
		\vspace{2cm}

\begin{quote}
\textit{God does not care about our mathematical difficulties. He integrates empirically.}
\qauthor{Albert Einstein}
\end{quote}

				\end{flushright}

			\vspace*{\fill}

\appendix


\chapter{Deriving the Bimetric Perturbation Equations}
\label{app:perteqs}

In \cref{sec:linearisedeqs} we presented the full linearised Einstein and fluid conservation equations for massive bigravity in a general gauge and without making a choice of the time coordinate (i.e., with a general lapse). Since these equations are arrived at by a fairly lengthy calculation, in this appendix we detail their derivation.

The perturbations of the Einstein tensor are standard and can be found in, e.g., \rcite{Mukhanov:1990me}. In order to calculate the fluid conservation equations we only need to know the linearised Christoffel symbols. For the $g$ metric, these are
\begin{align}
\Gamma^0_{00} &= \frac{\dot N}{N} + \frac{1}{2}\dot{E}_g \nonumber\\
\Gamma^0_{0 i} &= \partial_i\left(\frac{1}{2}E_g + \frac{\dot a}{N} F_g \right) \nonumber\\
\Gamma^0_{ij} &= \frac{a^2}{N^2}\left[\left(H(1+A_g) + \frac{1}{2}\dot{A}_g -HE_g\right)\delta_{ij} + \frac{1}{2}\partial_i\partial_j\dot{B}_g + H\partial_i\partial_jB_g \right] - \frac{a}{N}\partial_i\partial_j F_g \nonumber\\
\Gamma^i_{00} &= \frac{N}{a}\partial^i\left(\frac{1}{2}\frac{N}{a}E_g + \dot{F}_g + HF_g\right) \nonumber\\
\Gamma^i_{0j} &= \left(H + \frac{1}{2}\dot{A}_g\right)\delta^i{}_j + \frac{1}{2}\partial^i\partial_j\dot{B}_g \nonumber\\
\Gamma^i_{jk} &= \frac{1}{2}\left(\delta^i{}_j \partial_kA_g + \delta^i{}_k \partial_j A_g - \delta_{jk}\partial^iA_g + \partial^i\partial_j\partial_kB_g\right) - \frac{\dot a}{N}\delta_{jk}\partial^iF_g.
\end{align}
Note that in background, only $\Gamma^0_{00}$, $\Gamma^0_{ij}$, and $\Gamma^i_{0j}$ are nonzero. Similarly we can find the $f$-metric Christoffel symbols, $\tilde \Gamma^\mu_{\nu\rho}$,
\begin{align}
\tilde \Gamma^0_{00} &= \frac{\dot X}{X} + \frac{1}{2}\dot{E}_f \nonumber\\
\tilde \Gamma^0_{0 i} &= \partial_i\left(\frac{1}{2}E_f + \frac{\dot Y}{X} F_f \right) \nonumber\\
\tilde \Gamma^0_{ij} &= \frac{Y^2}{X^2}\left[\left(K(1+A_f) + \frac{1}{2}\dot{A}_f -KE_f\right)\delta_{ij} + \frac{1}{2}\partial_i\partial_j\dot{B}_f + K\partial_i\partial_jB_f \right] - \frac{Y}{X}\partial_i\partial_j F_f \nonumber\\
\tilde \Gamma^i_{00} &= \frac{X}{Y}\partial^i\left(\frac{1}{2}\frac{X}{Y}E_f + \dot{F}_f + KF_f\right) \nonumber\\
\tilde \Gamma^i_{0j} &= \left(K + \frac{1}{2}\dot{A}_f\right)\delta^i{}_j + \frac{1}{2}\partial^i\partial_j\dot{B}_f \nonumber\\
\tilde \Gamma^i_{jk} &= \frac{1}{2}\left(\delta^i{}_j \partial_kA_f + \delta^i{}_k \partial_j A_f - \delta_{jk}\partial^iA_f + \partial^i\partial_j\partial_kB_f\right) - \frac{\dot Y}{X}\delta_{jk}\partial^iF_f.
\end{align}

The bulk of the work lies in calculating the perturbations of the mass term. We will focus on deriving the linearised field equations, i.e., calculating the matrices $Y^\mu_{(n)\nu}$, rather than the second-order action. The metric determinants to linear order are
\begin{align}
\operatorname{det} g &=  -N^2a^6\left(1 + E_g + 3A_g + \nabla^2B_g\right), \\
\operatorname{det} f &=  -X^2Y^6\left(1 + E_f + 3A_f + \nabla^2B_f\right).
\end{align}
The matrix $\mathbbm{X} = \sqrt{g^{-1}f}$ is defined in terms of the two metrics as
\begin{equation}
\mathbbm{X}^\mu{}_\rho \mathbbm{X}^\rho{}_\nu \equiv g^{\mu\rho}f_{\rho\nu}. \label{eq:xdef}
\end{equation}
Its background value is simply
\begin{align}
\mathbbm{X}^0{}_0 &= x, \nonumber \\
\mathbbm{X}^i{}_j &= y \delta^i{}_j.
\end{align}
Using this we can solve \cref{eq:xdef} to first order in perturbations to find
\begin{align}
\mathbbm{X}^0{}_0 &= x\left(1 + \frac{1}{2}\Delta E\right), \nonumber \\
\mathbbm{X}^0{}_i &= \frac{1}{x+y}\frac{Y}{N}\left(y\partial_i F_g - x\partial_i F_f\right), \nonumber \\
\mathbbm{X}^i{}_0 &= \frac{1}{x+y}\frac{X}{a}\left(y\partial^i F_f - x \partial^i F_g\right), \nonumber \\
\mathbbm{X}^i{}_j &= y\left[\left(1+\frac{1}{2}\Delta A\right)\delta^i{}_j + \frac{1}{2}\partial^i\partial_j\Delta B \right].
\end{align}
The trace of this is
\begin{equation}
[\mathbbm{X}] = x\left(1 + \frac{1}{2}\Delta E\right) + y\left[3\left(1+\frac{1}{2}\Delta A\right) + \frac{1}{2}\nabla^2\Delta B\right].
\end{equation}
Similarly we can solve for the matrix $\mathbbm{Y} = \sqrt{f^{-1}g}$, although we do not write its components here as they can be found by simply substituting $(N, a, {}_g)$ with $(X, Y, {}_f)$ and vice versa.\footnote{It may also be calculated explicitly or by using the fact that $\mathbbm{Y}$ is simply the matrix inverse of $\mathbbm{X}$, which can be easily inverted to first order.}

We now need the matrices $\mathbbm{X}^2$ and $\mathbbm{X}^3$ and their traces in order to compute the matrices appearing in the mass terms of the Einstein equations. For $\mathbbm{X}^2$ we find
\begin{align}
(\mathbbm{X}^2)^0{}_0 &= x^2(1 + \Delta E), \nonumber \\
(\mathbbm{X}^2)^0{}_i &= \frac{Y}{N}\left(y\partial_i F_g - x\partial_i F_f\right), \nonumber \\
(\mathbbm{X}^2)^i{}_0 &= \frac{X}{a}\left(y\partial^i F_f - x\partial^iF_g\right), \nonumber \\
(\mathbbm{X}^2)^i{}_j &= y^2 \left[ (1+\Delta A)\delta^i{}_j + \partial^i\partial_j\Delta B\right],
\end{align}
with trace
\begin{equation}
\left[\mathbbm{X}^2\right] = x^2 (1 + E_f - E_g) + y^2 \left[ 3(1 + \Delta A) + \nabla^2\Delta B\right].
\end{equation}
$\mathbbm{X}^3$ is given by
\begin{align}
(\mathbbm{X}^3)^0{}_0 &= x^3\left(1 + \frac{3}{2}\Delta E\right), \nonumber \\
(\mathbbm{X}^3)^0{}_i &= \frac{Y}{N}\left(x + y - \frac{xy}{x+y}\right) \left(y\partial_i F_g - x \partial_i F_f\right), \nonumber \\
(\mathbbm{X}^3)^i{}_0 &= \frac{X}{a}\left(x + y - \frac{xy}{x+y}\right) \left(y\partial^i F_f - x \partial^i F_g\right), \nonumber \\
(\mathbbm{X}^3)^i{}_j &= y^3 \left[\left(1+\frac{3}{2}\Delta A\right)\delta^i{}_j + \frac{3}{2}\partial^i\partial_j\Delta B \right],
\end{align}
with trace
\begin{equation}
\left[\mathbbm{X}^3\right] = x^3\left(1 + \frac{3}{2}\Delta E\right) + y^3 \left[3\left(1+\frac{3}{2}\Delta A\right)+ \frac{3}{2}\nabla^2\Delta B \right].
\end{equation}
$\mathbbm{Y}^2$ and $\mathbbm{Y}^3$ can be determined trivially from these.

Having calculated these we can determine the matrices $Y^\mu_{(n)\nu}(\sqrt{g^{-1}f})$ and $Y^\mu_{(n)\nu}(\sqrt{f^{-1}g})$. Two helpful intermediate results are
\begin{align}
\frac{1}{2}\left([\mathbbm{X}]^2 - [\mathbbm{X}^2]\right) &= y^2 \left[3(1 + \Delta A) + \nabla^2 \Delta B\right] \nonumber \\
&\hphantom{{}=} + xy\left[3\left(1+\frac{1}{2}\left(\Delta A + \Delta E\right)\right) + \frac{1}{2}\nabla^2 \Delta B\right], \\
\frac{1}{6}\left([\mathbbm{X}]^3 - 3[\mathbbm{X}][\mathbbm{X}^2] + 2[\mathbbm{X}^3]\right) &= y^3\left(1 + \frac{3}{2}\Delta A + \frac{1}{2}\nabla^2\Delta B\right) \nonumber \\
&\hphantom{{}=} + xy^2\left(3\left(1 + \Delta A + \frac{1}{2}\Delta E\right) + \nabla^2\Delta B\right).
\end{align}
To obtain those intermediate results and the $0$--$0$ and $i$--$j$ components of the $Y$ matrices, it saves a lot of algebra to write the traces, $0$--$0$ components, and $i$--$j$ components of the various $\mathbbm{X}$ matrices in terms of
\begin{align}
c_1 &= x, & c_2 &= 3y, \nonumber \\
 \delta_1 &= \frac{1}{2}\Delta E, & \delta_2 &= \frac{1}{2}\Delta A + \frac{1}{6}\nabla^2\Delta B, && \delta_3{}^i{}_j = \left(\frac{1}{2}\partial^i\partial_j - \frac{1}{6}\delta^i{}_j\nabla^2\right)\Delta B.
\end{align}
Finally, the matrices $Y^\mu_{(n)\nu}(\sqrt{g^{-1}f})$ defined in \cref{eq:ymatdef} are given by
\begin{itemize}

\item $n=0$:
\begin{equation}
Y^\mu_{(0)\nu}\left(\sqrt{g^{-1}f}\right) = \delta^\mu{}_\nu,
\end{equation}

\item $n=1$:
\begin{align}
Y^0_{(1)0}\left(\sqrt{g^{-1}f}\right) &= -y\left[3\left(1+\frac{1}{2}\Delta A\right) + \frac{1}{2}\nabla^2\Delta B\right], \nonumber \\
Y^0_{(1)i}\left(\sqrt{g^{-1}f}\right) &= \frac{1}{x+y}\frac{Y}{N}\left(y\partial_i F_g - x \partial_i F_f\right), \nonumber \\
Y^i_{(1)0}\left(\sqrt{g^{-1}f}\right) &= \frac{1}{x+y}\frac{X}{a}\left(y\partial^i F_f - x \partial^i F_g\right), \nonumber \\
Y^i_{(1)j}\left(\sqrt{g^{-1}f}\right) &= -x\left(1 + \frac{1}{2}\Delta E\right)\delta^i{}_j \nonumber\\
&\hphantom{{}=} - 2y\left[\left(1+\frac{1}{2}\Delta A\right)\delta^i{}_j + \frac{1}{4}\left(\delta^i{}_j\nabla^2 - \partial^i\partial_j\right)\Delta B\right],
\end{align}

\item $n=2$:
\begin{align}
Y^0_{(2)0}\left(\sqrt{g^{-1}f}\right) &= y^2 \left[ 3(1 + \Delta A) + \nabla^2\Delta B\right], \nonumber \\
Y^0_{(2)i}\left(\sqrt{g^{-1}f}\right) &= -\frac{2y}{x+y} \frac{Y}{N}\left(y\partial_i F_g - x \partial_i F_f\right), \nonumber \\
Y^i_{(2)0}\left(\sqrt{g^{-1}f}\right) &= -\frac{2y}{x+y} \frac{X}{a}\left(y\partial^i F_f - x \partial^i F_g\right), \nonumber \\
Y^i_{(2)j}\left(\sqrt{g^{-1}f}\right) &= y^2\left[(1 + \Delta A)\delta^i{}_j + \frac{1}{2}\left(\delta^i{}_j\nabla^2 - \partial^i\partial_j\right)\Delta B\right] \nonumber \\
&\hphantom{{}=} + 2xy\left[ \left(1 + \frac{1}{2}(\Delta A + \Delta E)\right)\delta^i{}_j + \frac{1}{4} \left(\delta^i{}_j \nabla^2 - \partial^i\partial_j\right)\Delta B \right],
\end{align}

\item $n=3$:
\begin{align}
Y^0_{(3)0}\left(\sqrt{g^{-1}f}\right) &= -y^3\left[1 + \frac{3}{2}\Delta A + \frac{1}{2}\nabla^2\Delta B\right], \nonumber \\
Y^0_{(3)i}\left(\sqrt{g^{-1}f}\right) &= \frac{y^2}{x + y} \frac{Y}{N} \left(y\partial_i F_g - x \partial_i F_f\right), \nonumber \\
Y^i_{(3)0}\left(\sqrt{g^{-1}f}\right) &= \frac{y^2}{x+y} \frac{X}{a}\left(y\partial^i F_f - x \partial^i F_g\right), \nonumber \\
Y^i_{(3)j}\left(\sqrt{g^{-1}f}\right) &= -xy^2\left[\left(1 + \Delta A + \frac{1}{2}\Delta E\right)\delta^i{}_j + \frac{1}{2}\left(\delta^i{}_j\nabla^2 - \partial^i\partial_j\right)\Delta B\right].
\end{align}

\end{itemize}
Plugging these into the field equations, (\ref{eq:Einsteingeng}) and (\ref{eq:Einsteingenf}), we obtain the full perturbation equations presented in \cref{sec:linearisedeqs}.

\chapter{Explicit Solutions for the Modified Gravity Parameters}
\label{app:hcoeff}

As discussed in \cref{sec:subhorizon-results}, the modified gravity parameters $Q$ and $\eta$ in singly-coupled massive bigravity have the Horndeski form,
\begin{align}
\eta &= h_2\left(\frac{1 + k^2h_4}{1+k^2h_5}\right), \\
Q &= h_1\left(\frac{1 + k^2h_4}{1+k^2h_3}\right),
\end{align}
while the growth of structures can be written in these terms as
\begin{equation}
\ddot \delta + H\dot\delta - \frac{1}{2}\frac{Q}{\eta}\frac{a^2\bar\rho}{M_g^2}\delta = 0.
\end{equation}

Hence the five $h_i$ coefficients allow us to determine all three modified gravity parameters we consider in \cref{chap:bigravity-subhorizon}. They are given explicitly by
\begin{align}
h_1 &= \frac{1}{1+y^2}, \\
h_2 &= -\frac{\left(1+y^2\right) \left(\beta _1+3 \beta _3 y^4+\left(6 \beta _2-2 \beta _4\right) y^3+3 \left(\beta _1-\beta _3\right) y^2\right)}{-\beta   _1+\left(3 \beta _2-\beta _4\right) y^5+\left(6 \beta _1-9 \beta _3\right) y^4+\left(3 \beta _0-15 \beta _2+2 \beta _4\right) y^3+\left(3 \beta _3-7   \beta _1\right) y^2}, \\
h_3 &= -\frac{y^2}{h_6}\frac{3}{1+y^2}\bigg[\beta _3^2 y^7+\left(4 \beta _2 \beta _3-2 \beta _3 \beta _4\right) y^6+3 \left(2 \beta _1-3 \beta_3\right) \beta _3 y^5 +\left(4 \beta _0 \beta _3-19 \beta _2 \beta _3-\beta _4 \beta _3+2 \beta _1 \beta _4\right) y^4  \nonumber \\
   &\hphantom{{}=-y^2\bigg[} +\left(-3 \beta _1^2-18 \beta_2^2-6 \beta _3^2+4 \beta _0 \beta _2+2 \beta _2 \beta _4\right) y^3 +3 \left(\left(\beta _0-3 \beta _2\right) \beta _3+\beta _1 \left(\beta _4-5 \beta_2\right)\right) y^2  \nonumber \\
   &\hphantom{{}=-y^2\bigg[} +\left(-7 \beta _1^2+2 \beta _3 \beta _1+2 \left(\beta _0-3 \beta _2\right) \beta _2\right) y -\beta _1 \left(\beta _0+\beta _2\right)\bigg], \\
h_4 &= \frac{y^2}{h_6}\bigg[2 \beta _3 \beta _4 y^6+2 \left(3 \beta _2^2-\beta _4 \beta _2-3 \left(\beta _1-2 \beta _3\right)
   \beta _3\right) y^5 +\left(\beta _1 \left(6 \beta _2-4 \beta _4\right)+3 \beta _3 \left(-2 \beta _0+9 \beta _2+\beta _4\right)\right) y^4 \nonumber\\
   &\hphantom{{}=y^2\bigg[}+2 \left(3 \beta
   _1^2-2 \beta _3 \beta _1+18 \beta _2^2+9 \beta _3^2-3 \beta _0 \beta _2-3 \beta _2 \beta _4\right) y^3 +\left(37 \beta _1 \beta _2+27 \beta _3 \beta _2-9
   \beta _0 \beta _3-9 \beta _1 \beta _4\right) y^2 \nonumber\\
   &\hphantom{{}=y^2\bigg[}+2 \left(10 \beta _1^2-3 \beta _3 \beta _1-3 \left(\beta _0-3 \beta _2\right) \beta _2\right) y + 3 \beta _1 \left(\beta _0+\beta _2\right)\bigg], \\ 
h_5 &= -\frac{y^2}{h_6}\frac{1}{h_7}\bigg[4 \beta _3^2 \beta _4^2 y^{11}+\beta _3 \left(24 \beta _4 \beta _2^2+\left(9 \beta _3^2-8 \beta _4^2\right) \beta _2+3 \beta _3 \left(19 \beta _3-8 \beta
   _1\right) \beta _4\right) y^{10} \nonumber\\
   &\hphantom{{}=-\frac{y^2}{h_6}\frac{1}{h_7}\bigg[} +2 \left(18 \beta _2^4-12 \beta _4 \beta _2^3+\left(117 \beta _3^2-36 \beta _1 \beta _3+2 \beta _4^2\right) \beta _2^2+6
   \beta _3 \left(4 \beta _1+5 \beta _3\right) \beta _4 \beta _2 \right. \nonumber\\
   &\hphantom{{}=-\frac{y^2}{h_6}\frac{1}{h_7}\bigg[+2(} \left.+\beta _3 \left(99 \beta _3^3-81 \beta _1 \beta _3^2+18 \beta _1^2 \beta _3-3 \beta _4^2
   \beta _3-12 \beta _0 \beta _4 \beta _3-8 \beta _1 \beta _4^2\right)\right) y^9 \nonumber\\
   &\hphantom{{}=-\frac{y^2}{h_6}\frac{1}{h_7}\bigg[}-\left(72 \beta _3 \left(\beta _2-\beta _4\right) \beta _1^2+\left(-72
   \beta _2^3+72 \beta _4 \beta _2^2+\left(117 \beta _3^2-16 \beta _4^2\right) \beta _2+\beta _3^2 \left(85 \beta _4-72 \beta _0\right)\right) \beta _1 \right. \nonumber\\
   &\hphantom{{}=-\frac{y^2}{h_6}\frac{1}{h_7}\bigg[-(} \left.+9
   \beta _3 \left(-60 \beta _2^3+8 \beta _0 \beta _2^2+12 \beta _4 \beta _2^2-96 \beta _3^2 \beta _2+19 \beta _0 \beta _3^2+5 \beta _3^2 \beta
   _4\right)\right) y^8 \nonumber\\
   &\hphantom{{}=-\frac{y^2}{h_6}\frac{1}{h_7}\bigg[}-2 \left(36 \beta _3 \beta _1^3-\left(54 \beta _2^2-36 \beta _4 \beta _2+69 \beta _3^2+8 \beta _4^2\right) \beta _1^2 \right. \nonumber\\
   &\hphantom{{}=-\frac{y^2}{h_6}\frac{1}{h_7}\bigg[-2(} \left.-2 \beta _3
   \left(123 \beta _2^2-76 \beta _4 \beta _2+3 \left(13 \beta _3^2+\beta _4 \left(4 \beta _0+\beta _4\right)\right)\right) \beta _1 \right.\nonumber\\
   &\hphantom{{}=-\frac{y^2}{h_6}\frac{1}{h_7}\bigg[-2(} \left.-3 \left(72 \beta _2^4-36
   \beta _4 \beta _2^3+\left(255 \beta _3^2+4 \beta _4^2\right) \beta _2^2-21 \beta _3^2 \beta _4 \beta _2-9 \beta _3^4+6 \beta _0^2 \beta _3^2 \right.\right. \nonumber\\
   &\hphantom{{}=-\frac{y^2}{h_6}\frac{1}{h_7}\bigg[-2(-3(} \left.\left.+\beta _0
   \left(-12 \beta _2^3+4 \beta _4 \beta _2^2-93 \beta _3^2 \beta _2+3 \beta _3^2 \beta _4\right)\right)\right) y^7 \nonumber\\
   &\hphantom{{}=-\frac{y^2}{h_6}\frac{1}{h_7}\bigg[}+\left(24 \left(3 \beta _2-2 \beta
   _4\right) \beta _1^3+\beta _3 \left(-72 \beta _0+507 \beta _2-77 \beta _4\right) \beta _1^2 \right. \nonumber\\
   &\hphantom{{}=-\frac{y^2}{h_6}\frac{1}{h_7}\bigg[+(}\left.+\left(876 \beta _2^3-508 \beta _4 \beta _2^2+600 \beta _3^2
   \beta _2+48 \beta _4^2 \beta _2-3 \beta _3^2 \beta _4+\beta _0 \left(-72 \beta _2^2+48 \beta _4 \beta _2-69 \beta _3^2\right)\right) \beta _1 \right. \nonumber\\
      &\hphantom{{}=-\frac{y^2}{h_6}\frac{1}{h_7}\bigg[+(} \left. +3 \beta _3
   \left(24 \beta _2 \beta _0^2+\left(-228 \beta _2^2+16 \beta _4 \beta _2+9 \beta _3^2\right) \beta _0+9 \beta _2 \left(48 \beta _2^2-4 \beta _4 \beta _2-7
   \beta _3^2\right)\right)\right) y^6 \nonumber\\
      &\hphantom{{}=-\frac{y^2}{h_6}\frac{1}{h_7}\bigg[}+2 \left(18 \beta _1^4+45 \beta _3 \beta _1^3+\left(477 \beta _2^2-36 \beta _0 \beta _2-170 \beta _4 \beta _2+14 \beta
   _3^2+9 \beta _4^2\right) \beta _1^2 \right.\nonumber\\
      &\hphantom{{}=-\frac{y^2}{h_6}\frac{1}{h_7}\bigg[+2(} \left.+3 \beta _3 \left(126 \beta _2^2-42 \beta _0 \beta _2+6 \beta _4 \beta _2-3 \beta _3^2+2 \beta _0 \beta _4\right)
   \beta _1 \right. \nonumber\\
      &\hphantom{{}=-\frac{y^2}{h_6}\frac{1}{h_7}\bigg[+2(} \left.+6 \left(\beta _0-3 \beta _2\right) \beta _2 \left(-15 \beta _2^2+3 \beta _0 \beta _2+2 \beta _4 \beta _2+6 \beta _3^2\right)\right)
   y^5 \nonumber \\
      &\hphantom{{}=-\frac{y^2}{h_6}\frac{1}{h_7}\bigg[} +\left(\left(441 \beta _2-79 \beta _4\right) \beta _1^3+\beta _3 \left(-33 \beta _0-8 \beta _2+33 \beta _4\right) \beta _1^2 \right.\nonumber\\
         &\hphantom{{}=-\frac{y^2}{h_6}\frac{1}{h_7}\bigg[+(} \left.+\left(648 \beta _2^3-156
   \beta _0 \beta _2^2-60 \beta _4 \beta _2^2+9 \beta _3^2 \beta _2+9 \beta _0 \beta _3^2\right) \beta _1+36 \left(\beta _0-3 \beta _2\right) \beta _2^2
   \beta _3\right) y^4 \nonumber \\
      &\hphantom{{}=-\frac{y^2}{h_6}\frac{1}{h_7}\bigg[}+2 \beta _1 \left(39 \beta _1^3-26 \beta _3 \beta _1^2+\left(167 \beta _2^2-15 \beta _4 \beta _2+15 \beta _3^2-3 \beta _0 \left(\beta
   _2+\beta _4\right)\right) \beta _1-12 \beta _0 \beta _2 \beta _3\right) y^3 \nonumber\\
   &\hphantom{{}=-\frac{y^2}{h_6}\frac{1}{h_7}\bigg[}+\beta _1 \left(36 \beta _2^3+9 \beta _1 \beta _3 \beta _2+3 \beta _0 \left(3
   \beta _1^2-5 \beta _3 \beta _1-4 \beta _2^2\right)+\beta _1^2 \left(112 \beta _2-9 \beta _4\right)\right) y^2 \nonumber\\
   &\hphantom{{}=-\frac{y^2}{h_6}\frac{1}{h_7}\bigg[}+2 \beta _1^2 \left(11 \beta _1^2-3 \beta _3
   \beta _1+12 \beta _2^2\right) y+3 \beta _1^3 \left(\beta _0+\beta _2\right)\bigg],
\end{align}
where we have introduced two additional coefficients, $h_6$ and $h_7$, defined as
\begin{align}
h_6 &=  3 m^2a^2 \left(1+y^2\right) \left(\beta _1+\beta _3 y^2+2 \beta _2 y\right) \left(\beta _1^2+\beta _3 \beta _4 y^5+\left(3 \beta _2^2-\beta _4 \beta _2-3
   \left(\beta _1-2 \beta _3\right) \beta _3\right) y^4 \right.\nonumber\\
   &\hphantom{{}=\bigg[} \left. +\left(3 \beta _1 \beta _2+12 \beta _3 \beta _2-3 \beta _0 \beta _3-2 \beta _1 \beta _4\right)
   y^3+\left(3 \beta _1^2+\beta _3 \beta _1-3 \left(\beta _0-3 \beta _2\right) \beta _2\right) y^2+5 \beta _1 \beta _2 y\right), \\
h_7 &= \frac{\left(\beta _1+y \left(2 \beta _2+\beta _3 y\right)\right) \left(3 \beta _0 y^3+y^2 \left(3 \beta _2 y \left(y^2-5\right)+\beta _3 \left(3-9
   y^2\right)-\beta _4 y \left(y^2-2\right)\right)+\beta _1 \left(6 y^4-7 y^2-1\right)\right)}{1+y^2}.
\end{align}
While this notation is inspired by Refs. \cite{Amendola:2012ky,Amendola:2013qna}, we have defined $h_{1,6,7}$ differently.

\chapter{Transformation Properties of the Doubly-Coupled Bimetric Action}
\label{app:sym}

Here we describe the transformation properties of the action (\ref{eq:actionunscaled}) and how they determine the number of physically-relevant parameters for the doubly-coupled bigravity theory discussed in \cref{chap:dc-background}.

\section{Rescaling the Action}
Let us write the action as
\begin{align}
S & =  -\frac{M_{g}^{2}}{2}\int d^{4}x\sqrt{-\det g}R\left(g\right)-\frac{M_{f}^{2}}{2}\int d^{4}x\sqrt{-\det f}R\left(f\right) \nonumber \\
 &  \hphantom{{}=}+m^2M_g^2\int d^{4}x\sqrt{-\det g}V\left(\sqrt{g^{-1}f};\beta_n\right) +\int d^{4}x\sqrt{-\det g_\eff}\mathcal{L}_{m}\left(g_\eff,\Phi\right),
\end{align}
where
\begin{equation}
V\left(\sqrt{g^{-1}f};\beta_n \right) = \sum_{n=0}^{4}\beta_{n}e_{n}\left(\sqrt{g^{-1}f}\right)
\end{equation}
is the usual dRGT interaction potential and satisfies
\begin{equation}
\sqrt{-\det g}V\left(\sqrt{g^{-1}f};\beta_{n}\right)=\sqrt{-\det f}V\left(\sqrt{f^{-1}g};\beta_{4-n}\right).
\end{equation}
Due to this property, the action is invariant under
\begin{equation}
g_\mn\leftrightarrow f_\mn, \qquad M_g \leftrightarrow M_f, \qquad \alpha \leftrightarrow \beta, \qquad \beta_n \rightarrow \beta_{4-n}, 
\end{equation}
since the effective metric
\begin{equation}
g^\eff_\mn = \alpha^2 g_\mn +2\alpha\beta g_{\mu\alpha} \mathbb{X}^\alpha_\nu + \beta^2 f_\mn
\label{eq:appgeff}
\end{equation}
is also invariant under this transformation, as shown below. Because the overall scaling of the action is unimportant, there is a related transformation which keeps the action invariant, but only involves the ratio of $M_g$ and $M_f$,
\begin{equation}
\frac{M_{g}}{M_{f}}g_{\mn}\leftrightarrow\frac{M_{f}}{M_{g}}f_{\mn},\qquad\left(\frac{M_{f}}{M_{g}}\right)^{4-n}\beta_{n}\rightarrow\left(\frac{M_{f}}{M_{g}}\right)^{n}\beta_{4-n},\qquad\frac{M_{f}}{M_{g}}\alpha^{2}\leftrightarrow\frac{M_{g}}{M_{f}}\beta^{2}.
 \end{equation}
These transformations reflect a \emph{duality} of the action since they map one set of solutions, with a given set of parameters, to another set of solutions. 

Not all of the parameters $M_g$, $M_f$, $\alpha$, $\beta$, and $\beta_n$ are physically independent. In effect, we can rescale these parameters, together with $g_\mn$ and $f_\mn$, to get rid of either $M_g$ and $M_f$ \emph{or} $\alpha$ and $\beta$. In the end, only the ratio between $M_g$ and $M_f$, or $\alpha$ and $\beta$, together with $\beta_n$, are physically meaningful. The two parameter choices are physically equivalent and can be mapped to one another. We now describe the two scalings that give rise to the two parameter choices.

Under the scalings
\begin{align}
&& g_{\mu\nu}&\rightarrow\alpha^{-2}g_{\mu\nu}, & f_{\mu\nu}&\rightarrow\beta^{-2}f_{\mu\nu}, & M_{g}^{2}&\rightarrow\alpha^{2}M_{g}^{2}, && \nonumber \\
&& M_{f}^{2}&\rightarrow\beta^{2}M_{f}^{2}, &  m^2&\rightarrow\alpha^2m^2, & \beta_{n}&\rightarrow\left(\frac{\beta}{\alpha}\right)^{n}\beta_{n}, &&
\end{align}
the effective metric becomes
\begin{equation}
g^\eff_\mn = g_\mn + 2 g_{\mu\alpha} \mathbb{X}^\alpha_\nu + f_\mn
\end{equation}
while the action becomes
\begin{align}
S & =  -\frac{M_{g}^{2}}{2}\int d^{4}x\sqrt{-\det g}R\left(g\right)-\frac{M_{f}^{2}}{2}\int d^{4}x\sqrt{-\det f}R\left(f\right) \nonumber \\
 &  \hphantom{{}=}+m^2M_g^2\int d^{4}x\sqrt{-\det g}\sum_{n=0}^{4}\beta_{n}e_{n}\left(\sqrt{g^{-1}f}\right) \nonumber \\
 &  \hphantom{{}=} +\int d^{4}x\sqrt{-\det g_\eff}\mathcal{L}_{m}\left(g_\eff,\Phi\right).
\end{align}
The effective metric is thus uniquely defined in this parameter framework, while the ratio between $M_g$ and $M_f$ is the free parameter (in addition to the $\beta_n$). For this choice of scaling, the action is invariant under
\begin{equation}
g_{\mu\nu}\leftrightarrow f_{\mu\nu},\quad\beta_{n}\rightarrow\beta_{4-n},\qquad M_{g}\leftrightarrow M_{f},
\end{equation}
or, more generally,
\begin{equation}
\frac{M_{g}}{M_{f}}g_{\mu\nu}\leftrightarrow\frac{M_{f}}{M_{g}}f_{\mu\nu},\qquad\left(\frac{M_{f}}{M_{g}}\right)^{4-n}\beta_{n}\rightarrow\left(\frac{M_{f}}{M_{g}}\right)^{n}\beta_{4-n}.
\end{equation}

If, instead, we apply the scalings
\begin{align}
&& g_{\mu\nu}&\rightarrow\frac{M_\eff^{2}}{M_{g}^{2}}g_{\mu\nu},& f_{\mu\nu}&\rightarrow\frac{M_\eff^{2}}{M_{f}^{2}}f_{\mu\nu}, & \beta_{n}&\rightarrow\left(\frac{M_{f}}{M_{g}}\right)^{n}\beta_{n},&& \nonumber \\
&& m^2&\rightarrow m^{2}\frac{M_{g}^{4}}{M_\eff^{4}}, & \alpha^{2}&\rightarrow\frac{M_{g}^{2}}{M_\eff^{2}}\alpha^{2},&\beta^{2}&\rightarrow\frac{M_{f}^{2}}{M_\eff^{2}}\beta^{2},&&
\label{eq:Meffscaling}
\end{align}
then the effective metric is still of the form (\ref{eq:appgeff}), while the action becomes
\begin{align}
S & =  -\frac{M_\eff^{2}}{2}\int d^{4}x\sqrt{-\det g}R\left(g\right)-\frac{M_\eff^{2}}{2}\int d^{4}x\sqrt{-\det f}R\left(f\right) \nonumber \\
 &  \hphantom{{}=}+m^{2}M_\eff^2 \int d^{4}x\sqrt{-\det g}\sum_{n=0}^{4}\beta_{n}e_{n}\left(\sqrt{g^{-1}f}\right) \nonumber \\
 &  \hphantom{{}=} +\int d^{4}x\sqrt{-\det g_\eff}\mathcal{L}_{m}\left(g_\eff,\Phi\right).
\end{align}
For this choice of scaling, only the ratio between $\alpha$ and $\beta$, together with the $\beta_n$, is physically important (the effective coupling $M_\eff^2$ can be absorbed in the normalisation of the matter content). Under this form, the action is invariant under
\begin{equation}
g_{\mu\nu}\leftrightarrow f_{\mu\nu},\qquad\beta_{n}\rightarrow\beta_{4-n},\qquad\alpha\leftrightarrow\beta.
\end{equation}
To move from the framework with $M_g$ and $M_f$ to the one with $\alpha$ and $\beta$, one simply performs the rescaling
\begin{align}
&&M_g^2&\rightarrow \frac{M_\eff^2}{\alpha^2}, & M_f^2 &\rightarrow \frac{M_\eff^2}{\beta^2}, & g_\mn &\rightarrow \alpha^2 g_\mn, &&\nonumber \\
&&f_\mn&\rightarrow \beta^2 f_\mn, & \beta_n &\rightarrow \left(\frac{\alpha}{\beta}\right)^n \beta_n, & m^4&\rightarrow \frac{m^2M_\eff^2}{\alpha^4}.&&
\end{align}

Each of the parameter frameworks has its advantages. In the $M_g$ and $M_f$ framework, there is a {\it unique} effective metric, and it is the relative coupling strengths that determine the physics. In the $\alpha$ and $\beta$ framework, we have one single gravitational coupling, $M_\eff^2$, and the singly-coupled limits are more apparent in the effective metric. Note that the ratio between $\alpha$ and $\beta$ only appears in the matter sector, whereas in the $M_g$ and $M_f$ formulation their ratio appears in both the matter sector and interaction potential.  

\section{Symmetry of the Effective Metric}

In this section, we show that the effective metric is symmetric under the interchanges
\begin{equation}
g_\mn \leftrightarrow f_\mn, \qquad \alpha \leftrightarrow \beta.
\end{equation}
In order to do this, we take advantage of the fact that $g_{\mu\alpha} \mathbb{X}^\alpha_\nu$ is symmetric, i.e., $g \mathbb{X} = \mathbb{X}^T g$, as shown in \rcite{Hassan:2012wr}. We will find it useful to discuss the metrics in terms of their vielbeins, since we are dealing with square-root matrices and vielbeins are, in a sense, ``square roots'' of their respective metrics. We use Greek letters for spacetime indices and Latin letters for Lorentz indices. The $g$- and $f$-metric vielbeins are defined by
\begin{align}
 g_\mn &\equiv \eta_{ab} e^a_\mu e^b_\nu, \\
 f_\mn &\equiv \eta_{ab} L^a_\mu L^b_\nu,
\end{align}
while the inverse metrics are given by $g^\mn = \eta^{ab} e_a^\mu e_b^\nu$ and similarly for $f^\mn$. The vielbeins of $g^\mn$ are inverses of the vielbeins for $g_\mn$, $e^a_\mu e_b^\mu = \delta^a_b$ and $e^a_\mu e^\nu_a = \delta^\nu_\mu$, and again similarly for the $f_\mn$ vielbeins.

We will assume the symmetry condition (also called the Deser-Van Nieuwenhuizen gauge condition)
\begin{equation}
e_a^\mu L_{b \mu} = e^\mu_bL_{a \mu}, \label{eq:symmcond} 
\end{equation}
where Lorentz indices are raised and lowered with the Minkowski metric. It is likely, though it has not yet been proven, that this condition holds for all physically-relevant cases. In four dimensions, it holds when $g^{-1}f$ has a real square root (proven in \rcite{Deffayet:2012zc}, where it was conjectured that this result is valid also in higher dimensions). Assuming this condition, then it has been shown \cite{Gratia:2013gka} that the square-root matrix is given by
\begin{equation}
 \mathbb{X}^\mu_\nu = e^\mu_aL^a_\nu. \label{eq:sqrtviel}
\end{equation}
The inverse of this is clearly
\begin{equation}
(\mathbb{X}^{-1})_{\nu}^{\mu}=L_{a}^{\mu}e_{\nu}^{a},
\end{equation}
since then
\begin{equation}
\mathbb{X}_{\alpha}^{\mu}(\mathbb{X}^{-1})_{\nu}^{\alpha}=e_{a}^{\mu}L_{\alpha}^{a}L_{b}^{\alpha}e_{\nu}^{b}=e_{a}^{\mu}e_{\nu}^{a}=\delta_{\nu}^{\mu}.
\end{equation}
The form of the inverse then implies
\begin{equation}
\left(\sqrt{f^{-1}g}\right)^{-1}=\sqrt{g^{-1}f}, \label{eq:Xinv}
\end{equation}
which will be a useful property when showing the symmetry of the effective metric. We also have
\begin{equation}
g_{\mu\alpha}\mathbb{X}_{\nu}^{\alpha}=e_{a\mu}e_{\alpha}^{a}e_{b}^{\alpha}L_{\nu}^{b}=e_{a\mu}L_{\nu}^{a}.
\end{equation}
In order to show that $g \mathbb{X} = \mathbb{X}^T g$, we must thus have
\begin{equation}
e_{a\mu}L_{\nu}^{a}=e_{a\nu}L_{\mu}^{a}.
\end{equation}
Notice that this is not exactly the same as eq.~(\ref{eq:symmcond}),
since in the first case we contract over spacetime indices, whereas
here we contract over Lorentz indices. The two symmetry conditions
are, however, equivalent, as discussed in detail in \cite{Hoek:1982za}.

An alternative way of seeing that $g\mathbb{X}=\mathbb{X}^Tg$ is as follows. Since
\begin{equation}
f_{\mu\alpha}\mathbb{X}_{\nu}^{\alpha}=L_{\mu}^{a}L_{a\alpha}e_{b}^{\alpha}L_{\nu}^{b}=L_{\mu}^{a}L_{b\alpha}e_{a}^{\alpha}L_{\nu}^{b}=f_{\nu\alpha}\mathbb{X}_{\mu}^{\alpha},
\end{equation}
we have
\begin{equation}
f\mathbb{X}=\mathbb{X}^{T}f.
\label{eq:fXsym}
\end{equation}
But $f = g\mathbb{X}^2$, so \cref{eq:fXsym} can also be written $g\mathbb{X}^{3}=\mathbb{X}^{T}g\mathbb{X}^{2}$, which implies
\begin{equation}
g\mathbb{X}=\mathbb{X}^{T}g.
\label{eq:gXsym}
\end{equation}

Using this property it is straightforward to show that the effective metric is symmetric under the interchange of the two metrics. The effective metric we study was introduced in \rcite{deRham:2014naa} in the form
\begin{equation}
g_{\mu\nu}^\eff=\alpha^{2}g_{\mu\nu}+2\alpha\beta g_{\alpha(\mu}\mathbb{X}_{\nu)}^{\alpha}+\beta^{2}f_{\mu\nu}.
\end{equation}
Due to the symmetry property (\ref{eq:gXsym}), we can write this without the explicit symmetrisation,
\begin{equation}
g_{\mu\nu}^\eff=\alpha^{2}g_{\mu\nu}+2\alpha\beta g_{\alpha\mu}\mathbb{X}_{\nu}^{\alpha}+\beta^{2}f_{\mu\nu}.
\end{equation}
Suppose now that we do the transformation
\begin{equation}
g_\mn\leftrightarrow f_\mn,\qquad\alpha\leftrightarrow\beta. \label{eq:dualtrans}
\end{equation}
The effective metric becomes
\begin{equation}
\label{eq:gefftrans}
g_{\mu\nu}^\eff=\alpha^{2}g_{\mu\nu}+2\alpha\beta f_{\mu\alpha}(\sqrt{f^{-1}g})_{\nu}^{\alpha}+\beta^{2}f_{\mu\nu}.
\end{equation}
This can be brought into the original form for $g^\eff_\mn$ using the matrix property
\begin{equation}
f\left(\sqrt{g^{-1}f}\right)^{-1}=gg^{-1}f\left(\sqrt{g^{-1}f}\right)^{-1}=g\sqrt{g^{-1}f}.
\end{equation}
Combining this with \cref{eq:Xinv} we get
\begin{equation}
f\sqrt{f^{-1}g}=g\sqrt{g^{-1}f}.
\end{equation}
Applying this to \cref{eq:gefftrans} we see that the effective metric is invariant under the duality transformation (\ref{eq:dualtrans}). This ensures that the entire Hassan-Rosen action treats the two metrics on entirely equal footing when matter couples to $g_\mn^\eff$. Note that this duality does not hold for the single-metric (dRGT) massive gravity as it is broken by the kinetic sector.

\chapter{Einstein-Aether Cosmological Perturbation Equations in Real Space}

\label{app:realspace}

In this appendix, we present the real-space equations of motion for the linear cosmological perturbations in Einstein-aether theory coupled to a scalar field as described in \cref{chap:aether}.

We have for the $\nu=0$ component of the aether field equation (\ref {eq:aethereom})
\begin{align}
&-6(c_{13}+2c_2)\sh^2\Phi+6c_2\left(\frac{a''}{a}\right)\Phi \nonumber \\
&+\sh\left[(2c_1+c_2)V^i{}_{,i}+c_3(V^i{}_{,i}+ B^i{}_{,i})+3c_2\Phi'+3(2c_{13}+c_2)\Psi'\right] \nonumber \\
&-c_3(\Phi^{,i}{}_i-B{}^i{}_{,i}'+V{}^i{}_{,i}')-c_2(V^i{}_{,i}'+3\Psi'')+a^2\delta \lambda_\textrm{\ae} \nonumber \\ 
&+ \frac{1}{2}\bar V_{\theta\theta}\left[6\left(\frac{a''}{a}-2\mathcal{H}^2\right)\Phi + 3\mathcal{H}(\Phi^\prime+\Psi^\prime)-3\Psi''+\mathcal{H}V^i{}_{,i} - V^i{}_{,i}^\prime \right] \nonumber \\
& - \frac{3}{2}\frac{m}{a}\bar V_{\theta\theta\theta}\left(\frac{a''}{a}-2\mathcal{H}^2\right)\left(3\Psi^\prime-3\mathcal{H}\Phi+V^i{}_{,i}\right) \nonumber \\
& + \frac{1}{2}\frac{a}{m}\left[\bar V_{\theta\phi}(\bar\phi'\Phi -\delta\phi') - \bar V_{\theta\phi\phi}\bar\phi'\delta\phi\right] \nonumber \\
& - \frac{1}{2}\bar V_{\theta\theta\phi}\left[\bar\phi'(3\Psi' - 3\mathcal{H}\Phi + V^i{}_{,i}) + 3\left(\frac{a''}{a}-2\mathcal{H}^2\right)\delta\phi \right] = 0,
\end{align}
and the $\nu=i$ component is
\begin{align}
&-\left[2\frac{\alpha}{m^2}\sh^2-\frac{\alpha}{m^2}\left(\frac{a''}{a}\right)+c_1\left(\frac{a''}{a}\right)\right](B_i-V_i) \nonumber \\
&+\sh\left[(c_1+\frac{\alpha}{m^2})\Phi_{,i}+2c_1(V_i'-B_i')\right] \nonumber \\
&+\frac{1}{2}(-c_3+c_1)B_{[i,j]}{}^j-c_1V_{i,j}{}^j-c_{23}V^j{}_{,ij} \nonumber \\
&-c_{13} h'{}^j{}_{i,j}+c_1\Phi'_{,i}-\frac{\alpha}{m^2} \Psi'_{,i}-c_1(B_i''-V_i'') \nonumber \\
& +\frac{1}{2}\bar V_{\theta\theta}\left[3\left(\frac{a''}{a}-2\mathcal{H}^2\right)(B_i-V_i)-3\Psi^\prime_{,i}+3\mathcal{H}\Phi_{,i}-V^j{}_{,ij}\right] \nonumber \\
& + \frac{1}{2}\frac{a}{m}\bar V_{\theta\phi}\left(\bar\phi'(B_i-V_i)-\delta\phi_{,i}\right)=0.
\end{align}
The combined aether-scalar stress energy tensor (\ref{eq:fullsetensor}) has perturbations
\begin{align}
\delta T^0{}_0 ={}& 2\frac{m^2}{a^2}\Big\{-3(c_{13}+3c_2)\sh^2\Phi+c_1a^{-1}\left[a(B^i-V^i)_{,i}\right]' \nonumber \\
& +(c_{13}+3c_2)\sh(V^i{}_{,i}+3\Psi')+c_1\Phi^{,i}{}_i\Big\} \nonumber \\
& + \frac{1}{a^2}\left(\bar\phi'^2\Phi-\bar\phi'\delta\phi'-a^2\bar V_\phi\delta\phi\right) \nonumber \\
&+ \frac{3m^2}{a^2}\mathcal{H}\bar V_{\theta\theta}\left(3\Psi^\prime-3\mathcal{H}\Phi+V^i{}_{,i}\right) + \frac{3m}{a}\mathcal{H}\bar V_{\theta\phi}\delta\phi, \\
\delta T^0{}_i ={}& 2\frac{m^2}{a^2}\left\{\left[-2(c_{13}+3c_2)\sh^2+(3c_2+c_3)\left(\frac{a''}{a}\right)\right](V_i-B_i) \right. \nonumber \\
&-c_1a^{-2}\left[a^2(V_i'-B_i')\right]'-c_1a^{-1}(a\Phi_{,i})' \nonumber \\
& \left. +\frac{1}{2}(-c_1+c_3)\left[(B_i-V_i)^{,j}{}_j-(B^j-V^j)_{,ij}\right]\right\} \nonumber \\
& - \frac{1}{a^2}\bar\phi'\delta\phi_{,i} + \frac{3m^2}{a^2}\bar V_{\theta\theta}\left(\frac{a''}{a}-2\mathcal{H}^2\right)(V_i-B_i) + \frac{m}{a}\bar V_{\theta\phi}\bar\phi'(V_i-B_i), \\
\delta T^i{}_j ={}& 2\frac{m^2}{a^2}\left\{(c_{13}+3c_2)\left[\sh^2-2\left(\frac{a''}{a}\right)\right]\Phi\delta^i{}_j-(c_{13}+3c_2)\sh\Phi'\delta^i{}_j \right. \nonumber \\
& +\left.a^{-2}\bigg[a^2(c_2V^k{}_{,k}\delta^i{}_j+(c_{13}+3c_2)\Psi'\delta^i{}_j +\frac{1}{2}c_{13}(V^i{}_{,j}+V_j{}^{,i}+2h^i{}_j')\bigg]'\right\} \nonumber \\
& + \left\{- \frac{1}{a^2}\left(\bar\phi'^2\Phi-\bar\phi'\delta\phi'+a^2\bar V_\phi\delta\phi\right)\right. \nonumber \\
& + \left.\frac{m^2}{a^2}\bar V_{\theta\theta}\left[3\left(\mathcal{H}^2-2\frac{a''}{a}\right)\Phi-3\mathcal{H}\Phi^\prime+a^{-2}\left(a^2\left(3\Psi^\prime+V^k{}_{,k}\right)\right)^\prime\right]\right. \nonumber\\
& + \left.\frac{3m^3}{a^3}\bar V_{\theta\theta\theta}\left(\frac{a''}{a}-2\mathcal{H}^2\right)\left(3\Psi^\prime-3\mathcal{H}\Phi+V^{k}{}_{,k}\right)\right. \nonumber \\
& + \left.\frac{m}{a}\left[\bar V_{\theta\phi}(3\mathcal{H}\delta\phi+\delta\phi'-\bar\phi'\Phi) + \bar V_{\theta\phi\phi}\bar\phi'\delta\phi\right]\right. \nonumber \\
& + \left. \frac{m^2}{a^2}\bar V_{\theta\theta\phi}\left(3\left(\frac{a''}{a}-2\mathcal{H}^2\right)\delta\phi+\bar\phi'(3\Psi'-3\mathcal{H}\Phi+V^k{}_{,k})\right)\right\}\delta^i{}_j.
\end{align}

We can do a consistency check by choosing $V(\theta,\phi) = \frac{1}{2}
\beta\theta^2 + V(\phi)$. This corresponds to pure \ae -theory, with $c_2$
rescaled to $c_2 + \beta$, and a scalar field coupled only to gravity. The
cosmological perturbations in that model are presented in \cite{Lim:2004js}.
Our equations agree with the literature in this limit, as we would expect.

	\renewcommand\chapterheadstartvskip{\vspace*{2\baselineskip}}

{\renewcommand{\baselinestretch}{.01}
\begin{singlespace}

	\bibliographystyle{JHEP}
	\bibliography{bibliography}

\end{singlespace}

\end{document}